\documentclass[11pt]{article}
\usepackage{ifpdf}
\usepackage{geometry} 
\usepackage[parfill]{parskip}    
\ifpdf
\usepackage[pdftex]{graphicx}
\else
\usepackage[dvips]{graphicx}
\fi
\usepackage{amssymb}
\usepackage{epstopdf}
\usepackage{subfigure}
\usepackage{fancyhdr}
\usepackage{algorithms/algorithmic}
\usepackage{algorithms/algorithm}
\usepackage{array}
\usepackage{url}
\usepackage{comment}

\ifpdf
\usepackage[pdftex]{hyperref}
\else
\usepackage[dvips]{hyperref}
\fi

\title{SMART: A statistical framework for optimal design matrix generation with application to fMRI}
\author{Gautam Pendse\thanks{To whom correspondence should be addressed. e-mail: gpendse@mclean.harvard.edu}$^{\,\,\, 1,5}$ \and Adam Schwarz$^{2,5}$ \and Richard Baumgartner$^{3,5}$ \and Alexandre Coimbra$^{4,5}$ \and David Borsook$^{1,5}$ \and Lino Becerra$^{1,5}$ \\
\mbox{}\\ \\ \\ \\ 
$^1$ Imaging and Analysis Group (IMAG), Harvard Medical School \\ \\
$^2$ Translational Imaging Group, Lilly Research Laboratories\\ \\
$^3$ Biometrics Research, Merck Research Laboratories \\ \\
$^4$ Imaging Department, Merck Research Laboratories \\ \\
$^5$ Imaging Consortium for Drug Development (ICD)}
\date{Mar 10, 2009}
\begin{document}

\maketitle

\newpage 

\section{Abstract}
The general linear model (GLM) is a well established tool for analyzing functional magnetic resonance imaging (fMRI) data. Most fMRI analyses via GLM proceed in a massively univariate fashion where the same design matrix is used for analyzing data from each voxel.  A major limitation of this approach is the locally varying nature of  signals of interest  as well as associated confounds. This local variability results in a potentially large bias and uncontrolled increase in variance for the contrast of interest. The main contributions of this paper are two fold (1) We develop a statistical framework called SMART that enables estimation of an optimal design matrix while explicitly controlling the bias variance decomposition over a set of potential design matrices and  (2) We develop and validate a numerical algorithm for computing optimal design matrices for general fMRI data sets. The implications of this framework include the ability to match optimally the magnitude of underlying signals to their true magnitudes while also matching the "null" signals to zero size thereby optimizing both the sensitivity and specificity of signal detection. By enabling the capture of multiple profiles of interest using a single contrast (as opposed to an F-test) in a way that optimizes for both bias and variance enables the passing of first level parameter estimates and their variances to the higher level for group analysis which is not possible using F-tests. We demonstrate the application of this approach to \textit{in vivo} pharmacological fMRI data capturing the acute response to a drug infusion, to task-evoked, block design fMRI and to the estimation of a haemodynamic response function (HRF) response in event-related fMRI. Although developed with motivation from fMRI, our framework is quite general and has potentially wide applicability to a variety of disciplines.

\section{Introduction}
General linear models (GLMs) with Gaussian noise are very popular tools for fMRI model-based analyses \cite{GLM:1995}. The design matrix (DM) for GLM analysis is usually  based on the stimulus paradigm used during the experiment. With each column or explanatory variable (EV) of the DM is associated a parameter estimate (PE) measuring the strength of that EV in the overall model fit. The investigator defines linear contrasts of interest to extract meaningful values reflecting aspects of the brain's response to the applied paradigm. Since fMRI data is composed of thousands of measured timeseries across different points or voxels in the brain ($\sim$30000 voxels is typical), GLM based analysis for fMRI proceeds in a massively univariate way, meaning that the same DM is used to analyze all voxels. 

One very attractive property of the PE's estimated using GLM is that they are unbiased and of minimum variance if the DM is correctly specified (Gauss-Markov theorem) \cite{Hocking:book}. However, the exact mechanism underlying fMRI signal generation is extremely complex and in fMRI data the 'true' signal of interest is often superimposed with various artifactual signals due to physiology, motion and possible scanner effects. Moreover, the true temporal profile of the signal of interest may not be constant across brain regions or subjects; that is, there might be a range of temporal response profiles induced by the same paradigm. Thus, the assumption of a correctly specified model using a single DM for all voxels often does not hold in real fMRI data. In view of this fact, when using a GLM framework to analyze such data, one must have a good handle on the bias and variance of imperfect PE's calculated using the mis-specified DM. This is an extremely important point that cannot be ignored in fMRI analysis especially because the implications of Gauss-Markov theorem do not hold for mis-specified DMs. If the bias and variance introduced in the PE's at the first (individual subject) level are uncontrolled then misleading results can be obtained when generalizing to a group of subjects.

Small modeling misspecifications can be corrected to a certain extent using simple approaches, for example, by adding the derivative of the main EV to the DM to capture small temporal shifts \cite{Friston:1998}, \cite{Schwarz:2007}. However, these additional EV's are mostly based on heuristics and can still result in uncontrolled bias and variance of the resulting PE's. Basis function approaches result in more flexibility by allowing arbitrary response shapes to be matched via appropriately specified regressors \cite{FLOBS:2004}. It is possible to achieve a low variance fit to the data using these basis functions but it is difficult to define a meaningful contrast of interest that captures the underlying signal amplitude. In addition, the PE's estimated are again not controlled for bias or variance. Moreover, group analysis is non-trivial in this context as the first-level F-test results cannot simply be propagated to the group level.

In this article, we wish to derive a general theoretical basis that enables computation of optimal DM's  for GLM analyses as well as provide an algorithm that is practical to implement for practitioners. First, we develop an algorithmic framework called SMART (SiMultaneous biAs vaRiance and Residual opTimization) to derive automatically both a meaningful contrast and a DM simultaneously or, given a specified contrast derive a DM suitable for use at all voxels that models the set of all potential DM's in an optimal way, capturing a wide range of potential signals of interest while controlling for both the bias and variance of the signal amplitude measure. This explicit optimization will automatically optimize both the sensitivity and specificity of detecting signals of interest in the data. Second, we apply the framework to specific case studies arising from both pharmacological challenge and task-evoked fMRI experiments. We apply the optimal design matrices to the real fMRI data and demonstrate the more robust detection of fMRI responses \textit{in vivo}.

\section{Development of SMART}

\subsection[Performance measure]{Performance measure for a single design matrix}
To start let us assume that the true model generating the data is
\begin{equation}\label{1}
y = X \beta + \varepsilon
\end{equation}
where $X \in \mathbf{R}^{n \times q}$, $\beta \in \mathbf{R}^q$, $y \in \mathbf{R}^n$ and $\varepsilon \sim N(0, \sigma^2 I_n)$. 
[If the noise is non-white then the same discussion in this section applies after an initial pre-whitening step. A complication is that there might be interaction between pre-whitening and model misspecification.]

Unfortunately we do not know what $X$ is and so we use a design matrix $Z \in \mathbf{R}^{n \times p}$ for analyzing the data generated by the above model. 
\begin{equation}\label{2}
y = Z \gamma + \varepsilon_1
\end{equation}
where $\varepsilon_1 \sim N(0,\sigma_1^2 I_n)$. 
The usual GLM estimates are:
\begin{equation}\label{glm1}
\hat{\gamma} = (Z^T Z)^{-1} Z^T y
\end{equation}
and
\begin{equation}\label{glm2}
\hat{Cov(\hat{\gamma})} = \hat{\sigma}_1^2 \, (Z^T Z)^{-1}
\end{equation}
where
\begin{equation}
\hat{\sigma}_1^2 = \frac{(y - Z\hat{\gamma})^T(y - Z \hat{\gamma})}{n - p}
\end{equation}

For a contrast of interest $c_X \in \mathbf{R}^q$ for the true model \ref{1}, let $c_Z \in \mathbf{R}^p$  be the corresponding contrast of interest in the proposed model \ref{2}. 

It can be shown that the following holds under the true model:

\begin{equation} \label{3}
E(\hat{\gamma}) = (Z^T Z)^{-1} Z^T X \beta
\end{equation}

\begin{equation} \label{5}
\frac{\hat{\sigma}_1^2}{\sigma^2} \sim \frac{\chi^2(n-p, \Delta)}{n - p}
\end{equation}

where 
\begin{equation} \label{6}
\Delta = \frac{ \beta^T X^T P_Z X \beta } {\sigma^2}
\end{equation}
and
\begin{equation} \label{7}
P_Z = I_n - Z (Z^T Z)^{-1} Z^T
\end{equation}

When $Z = X$ in the above then we recover the usual GLM quantities.

To answer the question, how does $c_Z^T \hat{\gamma}$ compare with $c_X^T \hat{\beta}$ we define the following

Define the normalized contrast bias $C_b$ (assuming $\frac{\beta}{\sigma} \neq 0$) as follows:
\begin{equation}\label{8}
C_b = \frac{ c_Z^T E(\hat{\gamma}) - c_X^T \beta }{c_X^T \beta} = \frac{ c_Z^T (Z^T Z)^{-1} Z^T X (\beta/\sigma) } { c_X^T (\beta/\sigma)} - 1
\end{equation}
This measures bias as a fractional change in the PE of interest from the true value. When $\frac{\beta}{\sigma} = 0$, $C_b$ becomes undefined. In this case we define it as the numerator in the above equation, i.e., $C_b = c_Z^T (Z^T Z)^{-1} Z^T X (\beta/\sigma) $

Define the normalized model variance bias $V_b$ as follows:
\begin{equation}\label{9}
V_b = E \left(\frac{\hat{\sigma}_1^2}{\sigma^2}\right) - 1 =  \frac{ \beta^T X^T P_Z X \beta } {(n-p) \sigma^2}
\end{equation}

Define the normalized contrast variance change with respect to the Gauss-Markov estimate as follows:
\begin{equation}\label{10}
CV_{\Delta} = \frac{ E(\hat{\sigma}_1^2/\sigma^2) c_Z^T (Z^T Z)^{-1} c_Z } {   c_X^T (X^T X)^{-1} c_X  } - 1
\end{equation}

The test statistic of interest is:
\begin{equation}\label{tstatdef}
T(\hat{\gamma}, \hat{\sigma_1}; Z; c_Z) =  \frac{ c_Z^T (\hat{\gamma}/\hat{\sigma_1}) } {  \sqrt{c_Z^T \, (Z^T Z)^{-1} \,  c_Z }} \sim \frac{\mathbf{N}\left( \frac{ c_Z^T (Z^T Z)^{-1} Z^T X (\beta/\sigma) } { \sqrt{ c_Z^T (Z^T Z)^{-1} c_Z} } , 1 \right)}{ \sqrt{ \frac{\chi^2(n - p, \Delta)}{n  - p} } }
\end{equation}

Since $\hat{\gamma}$ and $\hat{\sigma_1}$ are independent:

\begin{equation}\label{12}
E(T) = E \left( \frac{ c_Z^T (\hat{\gamma}/\hat{\sigma_1}) } {  \sqrt{c_Z^T \, (Z^T Z)^{-1} \,  c_Z }} \right)
= \frac{ c_Z^T (Z^T Z)^{-1} Z^T X (\beta/\sigma) } {  \sqrt{c_Z^T \, (Z^T Z)^{-1} \,  c_Z } \sqrt{ 1 +   \frac{ \beta^T X^T P_Z X \beta } {(n-p) \sigma^2}  }  }
\end{equation}

Ideally we would like the misspecified model to perform well, i.e, $c_Z^T \hat{\gamma}$ be as close to $c_X^T \beta$ as possible and at the same time have as small \textit{estimated} variance as possible. This bias-variance tradeoff is captured in the function
\begin{equation}\label{13}
F = \frac{\hat{\sigma}_1^2}{\sigma^2} c_Z^T (Z^T Z)^{-1} c_Z + \frac{1}{\sigma^2}(E(c_Z^T \hat{\gamma} - c_X^T \beta))^2
\end{equation}
Note that this function captures simultaneously not only the bias and variance of the contrast of interest but also the full model residual (via $\hat{\sigma}_1$). This fact will be exploited in the next section.
The expected value of the above under the true model can be written as:
\begin{equation}\label{biasvariance}
E(F) = (V_b + 1)  c_Z^T (Z^T Z)^{-1} c_Z + (c_X^T \beta/\sigma)^2 C_b^2
\end{equation}

All the above definitions are functions of the signal to noise ratio ($\beta/\sigma$) in the data.

\subsection{Calculating optimal design matrices}
In this section we will set up the optimization problem that will enable us to compute optimal design matrices for arbitrary data sets. Ideally we would like the PE's from our optimal design matrix to have nice properties such as a low bias, low variance as well as a "low residual" overall model fit. It can be seen that \ref{13} will be small when a candidate
DM satisfies these ideals as compared to another that does not. Hence \ref{13} is a joint performance measure that captures all attributes of interest in one function for a given design matrix $X$. How do we generalize this concept to enable good performance of the optimal DM over a range of candidate DMs?
 Suppose our data is expected to contain the $m$ design matrices $X_1,X_2,\ldots,X_m$. Matrix $X_i$ is of size $n \times p_i$, where $p_i$ is the number of regressors in $X_i$. Suppose noisy data is generated from $X_i$ at SNR $\beta_i/\sigma_i$ and suppose that the contrast of interest for $X_i$ is $c_{X_i}$.
Expanding \ref{biasvariance} we get:

\begin{eqnarray}\label{eqnf}
f(Z, c_Z; X_i ; \frac{\beta_i}{\sigma_i} ; c_{X_i}) = c_Z^T (Z^T Z)^{-1} c_Z \left[ 1 + tr \left( P_Z \frac{ X_i \beta_i \beta_i^T X_i^T }{ (n - p_i) \sigma_i^2} \right) \right] \\ 
+ \frac{1}{\sigma_i^2} c_Z^T (Z^T Z)^{-1} Z^T X_i \beta_i \beta_i^T X_i^T Z (Z^T Z)^{-1} c_Z - \frac{2}{\sigma_i^2} c_Z^T (Z^T Z)^{-1} Z^T X_i \beta_i \beta_i^T c_{X_i} +  (c_{X_i}^T \beta_i/\sigma_i)^2 \nonumber
\end{eqnarray}
 
 Suppose weights $w_1, w_2, \ldots, w_m$ measure the frequency of occurance of each DM $X_i$ in the data such that higher values of $w_i$ indicate a higher frequency and $\sum_{i = 1}^m w_i = 1$.
 The objective function of interest is the mean performance measure over all design matrices. Hence, we define the following composite objective function (leaving off the multiplier $\frac{1}{\sum_{i = 1}^m w_i }$):
 
 \begin{equation}\label{composite_objective}
G(Z, c_Z) = \sum_{i = 1}^m w_i f(Z, c_Z; X_i ; \frac{\beta_i}{\sigma_i} ; c_{X_i})
 \end{equation}
  
 Define the quantities:
 \begin{equation}
 \Sigma = 
 \left(
 \begin{array}{ccc}
\frac{1}{n - p_1} & 0 & \ldots \\
\vdots & \ddots & \vdots \\
0 & \ldots & \frac{1}{n - p_m} 
\end{array}
\right)
 \end{equation}
 
 \begin{equation}
 H = \left(\frac{  \sqrt{w_1}  X_1 \beta_1}{\sigma_1}, \ldots,  \frac{ \sqrt{w_i} X_i \beta_i}{\sigma_i},\ldots,   \frac{\sqrt{w_m} X_m \beta_m}{\sigma_m} \right)
 \end{equation} 
 and
\begin{equation}
 \ell = \left(  \frac{\sqrt{w_1} c_{X_1}^T \beta_1}{\sigma_1}, \ldots,  \frac{ \sqrt{w_i} c_{X_i}^T \beta_i}{\sigma_i}, \ldots,  \frac{ \sqrt{w_m} c_{X_m}^T \beta_m}{\sigma_m} \right)
 \end{equation}
  
  With these definitions the composite objective \ref{composite_objective} can be written as:
  
  \begin{eqnarray}\label{objfn}
 G(Z, c_Z) = c_Z^T (Z^T Z)^{-1} c_Z \left[  \sum_{i = 1}^m w_i + tr \left( P_Z H \Sigma H^T \right) \right] \\ + \, c_Z^T (Z^T Z)^{-1} Z^T H H^T Z (Z^T Z)^{-1} c_Z - 2 c_Z^T (Z^T Z)^{-1} Z^T H \ell +  \sum_{i = 1}^m w_i (c_{X_i}^T \beta_i/\sigma_i)^2
  \end{eqnarray}
  
  In general, one can put constraints on the columns of $Z$ (e.g., fixing certain columns) such as:
  \begin{equation}
  Z A = B,
  \end{equation}
  
  where $A \in R^{p \times q}$ and $B \in R^{n \times q}$ are fixed matrices. Similar constraints can be imposed on the contrast vector:
  
  \begin{equation}
  C c_Z = d
  \end{equation}
  
  where $C \in R^{r \times p}$ is a fixed matrix and $d \in R^{r}$ is a fixed vector.
  
  Our goal is to minimize the composite objective function $G(Z, c_Z)$ that measure the weighted bias/variance decomposition over all potential DMs in the data.
 Hence, the complete optimization problem is written out as:
  
  \begin{eqnarray}
  {\hat{Z}, \hat{c_Z}} = \mbox{ arg min }_{Z, c_Z} G(Z, c_Z) \\
  \mbox{ s.t. } ZA = B \\
  \mbox{ s.t. } C c_Z = d
  \mbox{ s.t. } \mbox{ rank}(Z) = p
  \end{eqnarray}
  
  The last constraint above simply fixes the rank of $Z$ or the number of independent columns in $Z$.

\subsection{Local control of bias and variance}
It is straightforward to extend the concepts developed above to attain a local control of bias-variance decomposition i.e., to weigh the contribution of bias and variance terms to the overall performance measure for each DM $X_i$. The first step is modifying \ref{eqnf} to accomodate user defined bias/variance weighting by introducing a parameter
$\phi_i$ for each $X_i$ and rewriting the performance measure for $X_i$ as follows:
\begin{eqnarray}\label{eqnfbiasvarwtd}
f(Z, c_Z; X_i ; \frac{\beta_i}{\sigma_i} ; c_{X_i}; \phi_i) = 2 \phi_i \left( c_Z^T (Z^T Z)^{-1} c_Z \left[ 1 + tr \left( P_Z \frac{ X_i \beta_i \beta_i^T X_i^T }{ (n - p_i) \sigma_i^2} \right) \right] \right) \\ 
+ (2 - 2\phi_i) \left( \frac{1}{\sigma_i^2} c_Z^T (Z^T Z)^{-1} Z^T X_i \beta_i \beta_i^T X_i^T Z (Z^T Z)^{-1} c_Z - \frac{2}{\sigma_i^2} c_Z^T (Z^T Z)^{-1} Z^T X_i \beta_i \beta_i^T c_{X_i}  + (c_{X_i}^T \beta_i/\sigma_i)^2 \right) \nonumber
\end{eqnarray}

The parameter $\phi_i \in (0, 1)$ controls the relative importance of the bias and variance terms. When $\phi_i = 0.5$, both terms are equally weighted as in \ref{eqnf}. Higher values of $\phi_i$ give higher weight to the variance term and lower values of $\phi_i$ give higher weight to the bias term. The composite objective function for local bias-variance weighting is defined as before:

 \begin{equation}\label{composite_objectivewtd}
G_\phi(Z, c_Z) = \sum_{i = 1}^m w_i f(Z, c_Z; X_i ; \frac{\beta_i}{\sigma_i} ; c_{X_i}; \phi_i)
 \end{equation}

Define diagonal matrices $\Phi_V$ and $\Phi_B$ as follows:

\begin{equation}
 \Phi_V = 
 \left(
 \begin{array}{ccc}
2 \phi_1 & 0 & \ldots \\
\vdots & \ddots & \vdots \\
0 & \ldots & 2 \phi_m 
\end{array}
\right)
 \end{equation}
and
\begin{equation}
 \Phi_B = 
 \left(
 \begin{array}{ccc}
2 - 2 \phi_1 & 0 & \ldots \\
\vdots & \ddots & \vdots \\
0 & \ldots & 2 - 2 \phi_m 
\end{array}
\right)
 \end{equation}

 With these definitions $G_\phi$ can be written as:
  
 \begin{eqnarray}\label{objfn}
G_\phi(Z, c_Z) = c_Z^T (Z^T Z)^{-1} c_Z \left[  \sum_{i = 1}^m 2 \phi_i w_i + tr \left( P_Z H \Phi_V \Sigma H^T \right) \right] \\ + \, c_Z^T (Z^T Z)^{-1} Z^T H \Phi_B H^T Z (Z^T Z)^{-1} c_Z - 2 c_Z^T (Z^T Z)^{-1} Z^T H \Phi_B \ell + \sum_{i = 1}^m w_i (2 - 2 \phi_i) (c_{X_i}^T \beta_i/\sigma_i)^2
 \end{eqnarray}
 
 This approach can be easily extended to the simultaneous optimization of multiple contrasts using the function:
 
 \begin{equation}\label{objfn_multicon}
 G_{\phi}(Z,c_{Z_1},\ldots,c_{Z_q}) = \sum_{s = 1}^q G_{\phi}(Z,c_{Z_s})
 \end{equation}

\section{Algorithm}

\subsection{Implementation}
In this section we describe simplified optimization strategy that seems to work for the nature of the problem under consideration. Basically it involves simple gradient descent steps with adaptive step sizes. This practical algorithm is summarized in \hyperlink{alg1}{Algorithm 1}. 

\begin{figure}[htbp]
\begin{center}
\includegraphics[width = 160mm]{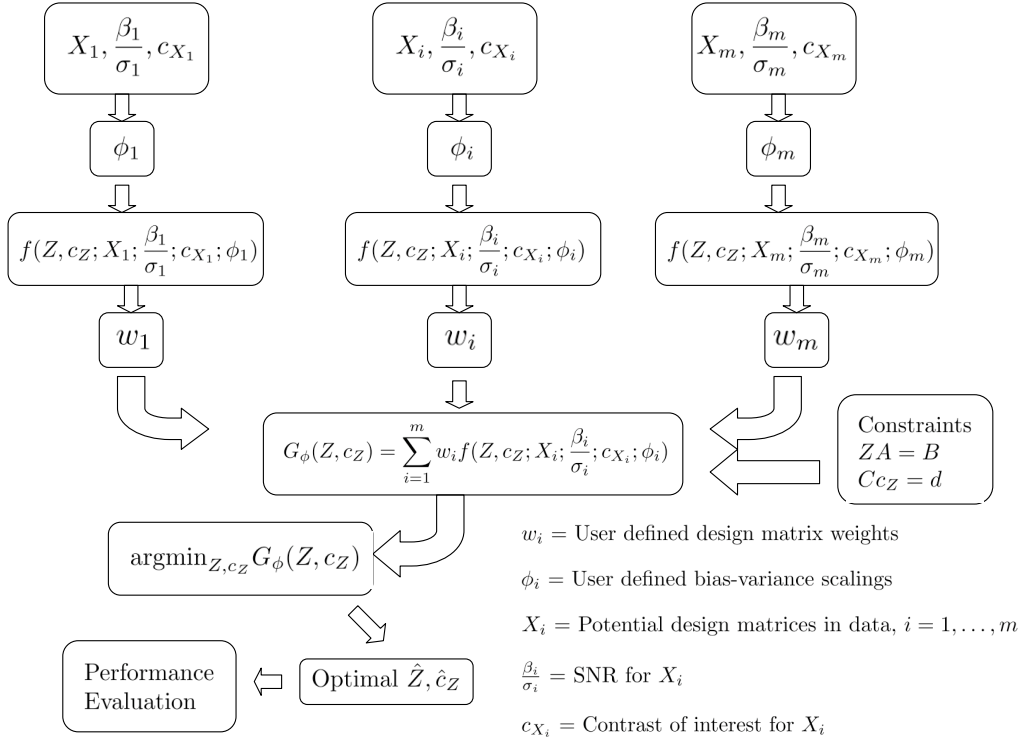}
\caption{Flowchart for computing optimal design matrices. Design matrices $X_i$, contrasts of interest $c_{X_i}$ and signal to noise ratios $\frac{\beta_i}{\sigma_i}$ are locally weighted using bias-variance weighting $\phi_i$ to compute performance measures $f(Z, c_Z; X_i; \frac{\beta_i}{\sigma_i}; c_{X_i}, \phi_i)$. 
These performance measures are combined using weights $w_i$ to form the objective function $G_\phi(Z, c_Z)$. Function $G_\phi(Z,c_Z)$ is then optimized with respect to $Z$ and $c_Z$ subject to user defined constraints to yield the optimal design matrices $\hat{Z}$ and $\hat{c_Z}$. A post-processing step then generates performance curves for user inspection.}
\label{flowchart}
\end{center}
\end{figure}

\begin{algorithm}\hypertarget{alg1}{ }
\begin{algorithmic}[1]
\REQUIRE Problem variables $H$, $\Sigma$, $\ell$, $A$, $C$
\REQUIRE Algorithmic variables $\alpha_0 \in (0, 10^{-3})$, $\theta \in (1, 5]$ and $\eta_1, \eta_2 \in (0, 10^{-6})$
\REQUIRE The size of optimal DM, $p$ and Initial point $Z_0$, $c_{Z_0}$ satisfying $Z_0 A = B$ and  $C c_{Z_0} = d$
\ENSURE Outputs are the optimal DM, $\hat{Z}$, the optimal contrast, $\hat{c_Z}$ and the optimal objective function, $\hat{F}$
\STATE Compute orthogonal projectors $P_A = I_p - A (A^T A)^{-1} A^T$ and $P_{C^T} = I_p - C^T (C C^T)^{-1} C$
\STATE $found = 0$, $j = 0$
\WHILE{ $found = 0 $ }
\STATE Let $S_j = \frac{\partial G}{\partial Z}(Z_j, c_{Z_j})$ and $T_j= \frac{\partial G}{\partial c_Z}(Z_j, c_{Z_j})$

\STATE $success = 0$

	\WHILE { $success = 0$ }

		\STATE $Z_{j + 1} = Z_j - \alpha_j S_j P_A$
		\STATE $c_{Z_{j + 1}} = c_{Z_j} - \alpha_j P_{C^T} T_j$
		\STATE $F_j = G(Z_j, c_{Z_j})$ and $F_{j + 1} = G(Z_{j + 1}, c_{Z_{j + 1}})$

		\IF{ $F_{j + 1} < F_{j}$ }
			\STATE $\alpha_{j+1} = \theta \alpha_j$
			\STATE $success = 1$
		\ELSE
			\STATE $\alpha_{j} = \alpha_j / \theta$
		\ENDIF
	
	\ENDWHILE
	
	\IF { $||F_{j + 1} - F_j|| \le \eta_1$ or $\alpha_{j + 1} \le \eta_2$ }
	      \STATE $found = 1$
	\ELSE
	\STATE $j = j + 1$
	\ENDIF
	
\ENDWHILE
\RETURN $\hat{Z} = Z_{j + 1}$, $\hat{c_Z} = c_{Z_{j + 1}}$ and $\hat{F} = F_{j + 1}$
\end{algorithmic}
\caption{Algorithm for optimizing DM}
\end{algorithm}

The gradients of $G_{\phi}$ are given by: (see the appendix \ref{gradient_derivation} for detailed derivation)
\begin{eqnarray}\label{gradobjfn}
\frac{\partial G}{\partial Z} = -Z (Z^T Z)^{-1} (2 c_Z c_Z^T) (Z^T Z)^{-1} \left[\sum_{i = 1}^m 2 \phi_i w_i + \mbox{tr}\left(P_Z H \Phi_V \Sigma H^T\right) \right] \\ \nonumber
- 2 (c_Z^T (Z^T Z)^{-1} c_Z) P_Z H \Phi_V \Sigma H^T Z (Z^T Z)^{-1} \\ \nonumber
-2 Z (Z^T Z)^{-1} c_Z c_Z^T (Z^T Z)^{-1} Z^T H \Phi_B H^T Z (Z^T Z)^{-1} \\ \nonumber
-2 Z (Z^T Z)^{-1} (Z^T H \Phi_B H^T Z) (Z^T Z)^{-1} c_Z c_Z^T (Z^T Z)^{-1} \\ \nonumber
+ 2 H\Phi_B H^T Z (Z^T Z)^{-1} c_Z c_Z^T (Z^T Z)^{-1} \\ \nonumber
+ 2 Z (Z^T Z)^{-1} Z^T H \Phi_B \ell c_Z^T (Z^T Z)^{-1} \\ \nonumber
+ 2 Z (Z^T Z)^{-1} c_Z \ell^T \Phi_B^T H^T Z (Z^T Z)^{-1} \\ \nonumber
-2 H \Phi_B \ell c_Z^T (Z^T Z)^{-1} \nonumber
\end{eqnarray}

\begin{eqnarray}\label{gradobjfn2}
\frac{\partial G}{\partial c_Z} = 2 (Z^T Z)^{-1} c_Z \left[\sum_{i = 1}^m 2 \phi_i w_i + \mbox{tr}\left(P_Z H \Phi_V \Sigma H^T\right) \right] \\ \nonumber
+ 2 (Z^T Z)^{-1} Z^T H \Phi_B H^T Z (Z^T Z)^{-1} c_Z \\
-2 (Z^T Z)^{-1} Z^T H \Phi_B \ell \nonumber
\end{eqnarray}

 When optimizing over multiple contrasts as per \ref{objfn_multicon}, the gradients are given by:
 \begin{equation}
\frac{ \partial{G_\phi} }{\partial Z} = \sum_{s = 1}^q \frac{ \partial G_{\phi}(Z,c_{Z_s}) }{\partial Z} \mbox{ and } \frac{ \partial{G_\phi} }{\partial c_{Z_r}} = \frac{ \partial G_{\phi}(Z,c_{Z_r}) } { \partial c_{Z_r}}
 \end{equation}

\subsection{Validation}

We validate the practical approach by comparing optimal solutions from the more sophisticated solver with the ones produced using the practical solver. 
Our state-of-the-art optimization solver (see appendix) was used to solve the validation test problems.The optimization core uses an augmented lagrangian algorithm (inspired by the implementation in LANCELOT package \cite{Conn:1991}, \cite{LANCELOT:1992}) to solve equality constrained problems. Inequality constraints are handled by first transforming them to equality constraints via slack variables and solving the resulting bound constrained optimization problem. Some features of interest are as follows:
\begin{enumerate}
\item A Trust region based approach \cite{More:1983} is used to generate search directions at each step (for both equality constrained and inequality constrained problems).
\item For equality constraints only, the subproblems above are solved using a conjugate gradient approach (Newton-CG -Steihaug) \cite{Steihaug:1983} that is fast and accurate even for large problems and can handle both positive definite and indefinite Hessian approximations. If both equality and inequality constraints are present then we solve the trust region problem with a non-linear gradient projection technique \cite{byrd95limited} followed by subspace optimization using Newton-CG-Steihaug.
\item A symmetric rank 1 (SR1) quasi-Newton approximation to the Hessian \cite{Conn:SR1} is used which is known to generate good Hessian approximations for both convex and non-convex problems. As suggested in \cite{Nocedal:book} we do the update also on the rejected steps to gather curvature information about the function. We provide options for BFGS \cite{Broyden:1970} especially for convex problems and an option for preconditioning the CG iterations. We also implement limited memory variants of SR1 and BFGS for large problems.
\item Our algorithm accepts vectorized constraints so that multiple constraints can be programmed simultaneously. Only gradient information is required. Hessian information is optional but not required.
\end{enumerate}

We tested the performance of our algorithm using standard optimization benchmarks from the GAMS performance benchmark problems (\url{http://www.gamsworld.org/performance}, \cite{COPS3}). The appendix \ref{optimization_solver} provides more technical details of the algorithm.

\subsubsection{Validation Test A}
Motivated by a practical data-set that we later describe we consider for illustration purposes the case study when the profiles of interest are shifted relative to a base profile by variable units and our goal is to simultaneously capture all responses with a single design matrix. To test and validate the optimization framework, we used  the basic design matrix $X_0$ from figure \ref{fig1}. $m = 50$ expected design matrices were proposed with

\begin{eqnarray} \label{sampleDM}
X_i(:,1) = X_0(:,1) \mbox{ shifted right by } i \mbox{ timepoints } \\
X_i(:,2) = X_0(:,2)
\end{eqnarray}
 
 We chose $\frac{\beta_i}{\sigma_i} = [1,0.5]^T $ and $c_{X_i} = [1,0]^T, \forall i$. The weights were chosen as $w_i = 1, \forall i$ to reflect the equal likelihood of observing any $X_i$. We chose $\phi_i = 0.5, \forall i$ in this validation test.
 
 The rank  of $Z$ was chosen to be $4$ and the matrix $A$ was chosen as $A = [e_1, e_2]$ where $e_1 \in R^4$ is a unit vector with $1$ at position 1 and zeros elsewhere. Similarly for $e_2$. The matrix $B$ was chosen as $X_0$ to fix the first two columns of $Z$ to those of $X_0$. 
 
 $C$ was chosen as the identity matrix $I_4$ and $d$ was set to $[1,0,0,0]^T$ to fix the contrast $c_Z$.  
  
\begin{figure}[htbp]
\begin{center}
\includegraphics[width = 120mm, angle = -90]{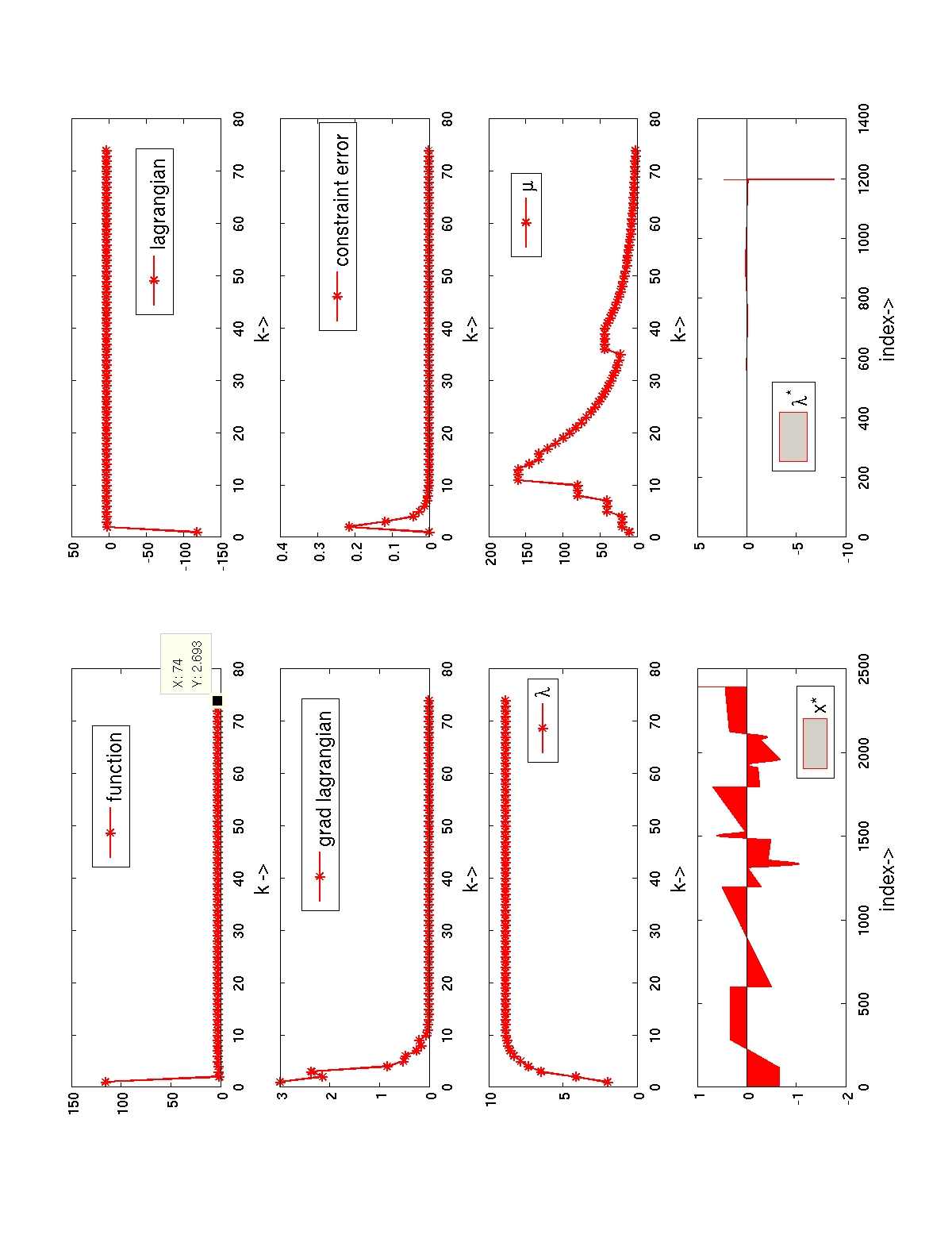}
\caption{Validation Test A: Convergence diagnostics for the advanced solver. Figure shows the evolution of objective function, the Lagrangian, norm of the gradient of the Lagrangian, norm of the constraint satisfaction error, norm of the Lagrange multipliers and progress monitoring parameter over algorithm iterations. The last row shows the optimal solution (i.e., the DM and contrast) displayed as a vector and the optimal Lagrange multipliers for the chosen constraints on the columns of $Z$ and the contrast $c_Z$. The optimal objective of $G(\hat{Z}, \hat{c_Z}) = 2.693$ was attained in 74 iterations.}
\label{convdiag}
\end{center}
\end{figure}
    
\begin{figure}[htbp]
\centering
\begin{tabular}{cc}
\subfigure[Initial design matrix $Z_0$]
{
\hspace{-2cm}
\label{fig14a}
\includegraphics[width = 70mm, angle = -90]{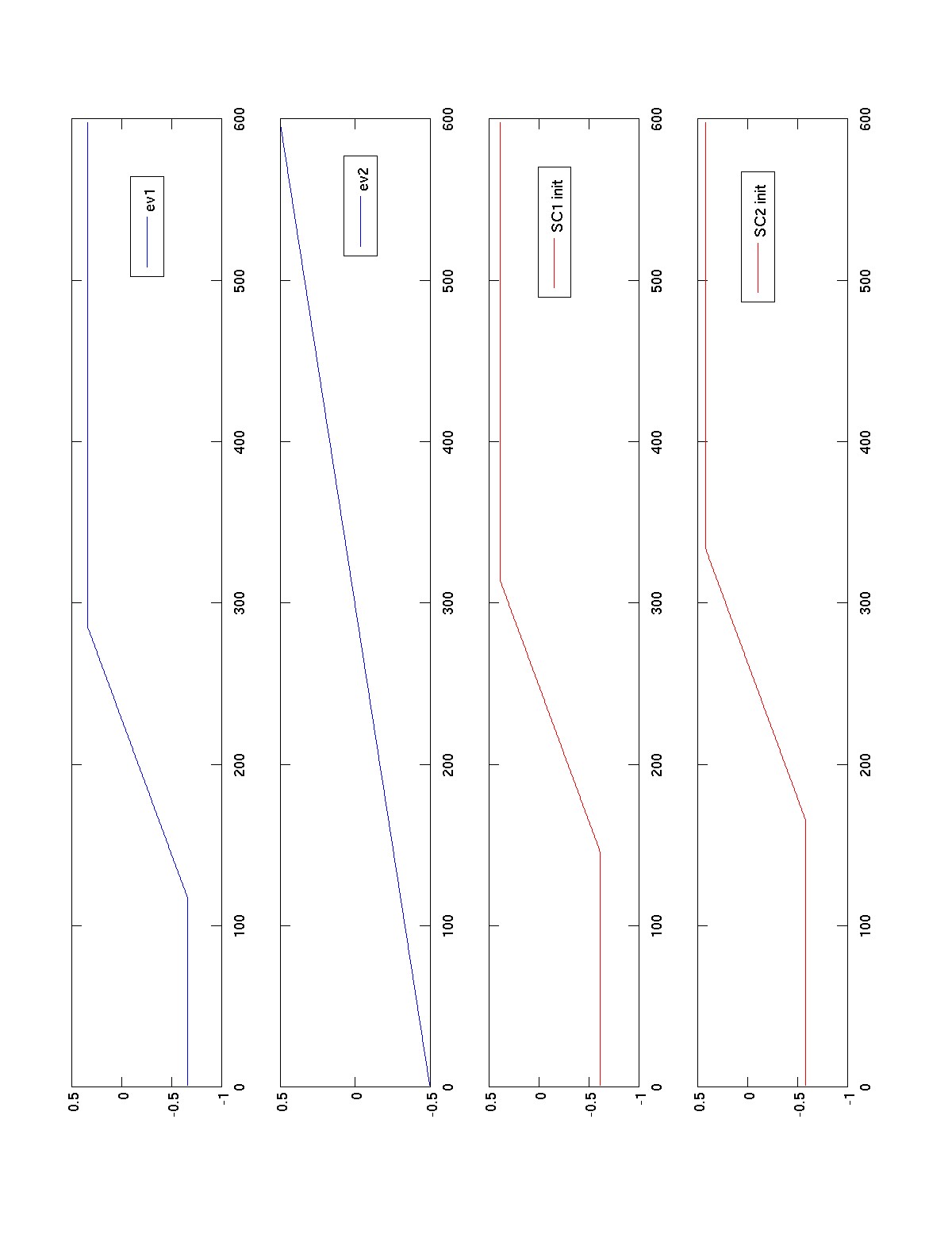}
}

&

\subfigure[Optimal design matrix $\hat{Z}$]
{
\label{fig14b}
\includegraphics[width = 70mm, angle =  -90]{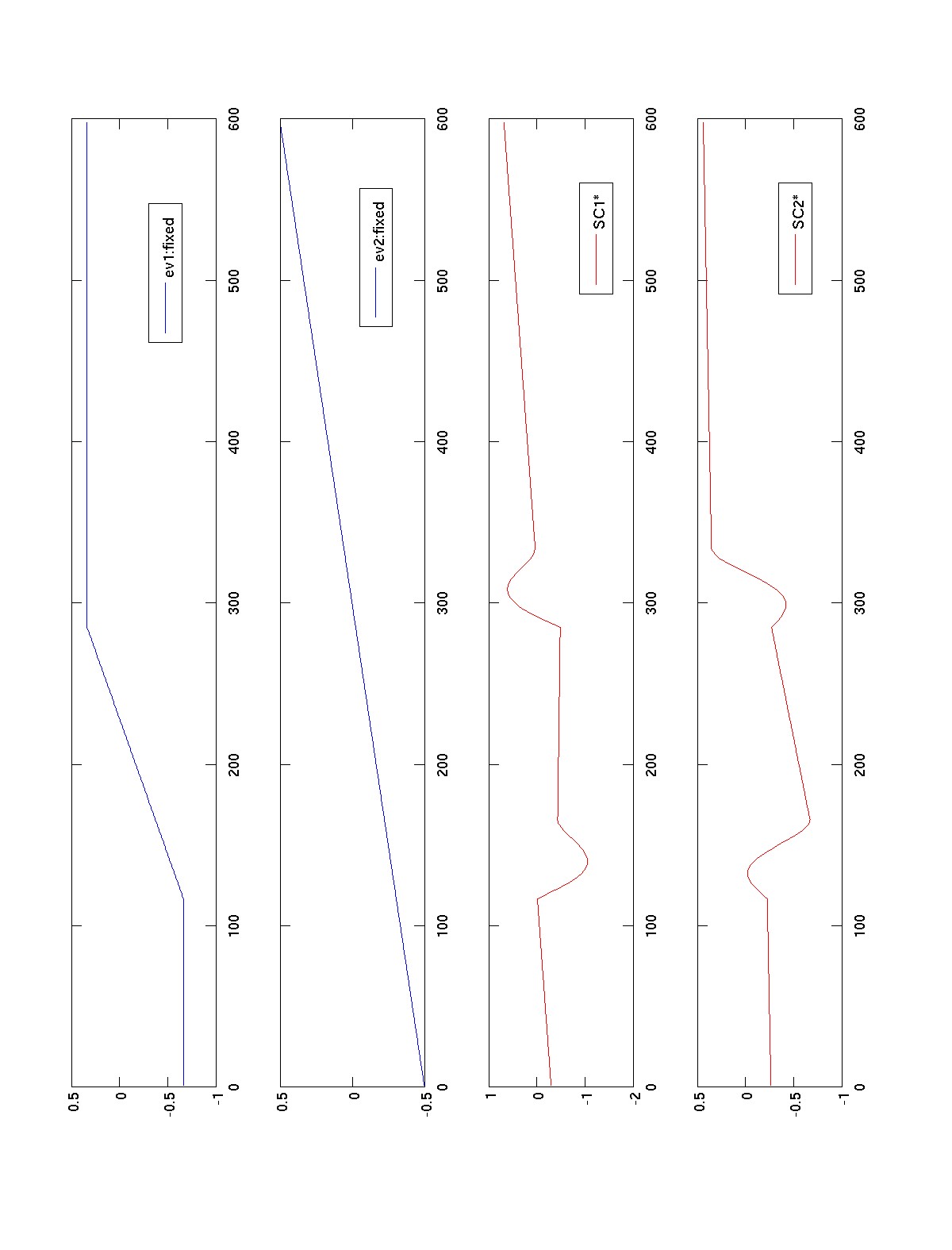}
}

\end{tabular}
\caption{  Validation Test A: (a) Initial design matrix $Z_0$  and (b) Optimal design matrix $\hat{Z}$ ($m = 50$). The first two EVs were constrained to their shift 0 values. The contrast $\hat{c_Z}$ was constrained to be $[1,0,0,0]^T$.}
\label{fig14}
\end{figure}

\subsubsection{Validation Test B}

\begin{figure}[htbp]
\begin{center}
\includegraphics[width = 120mm, angle = -90]{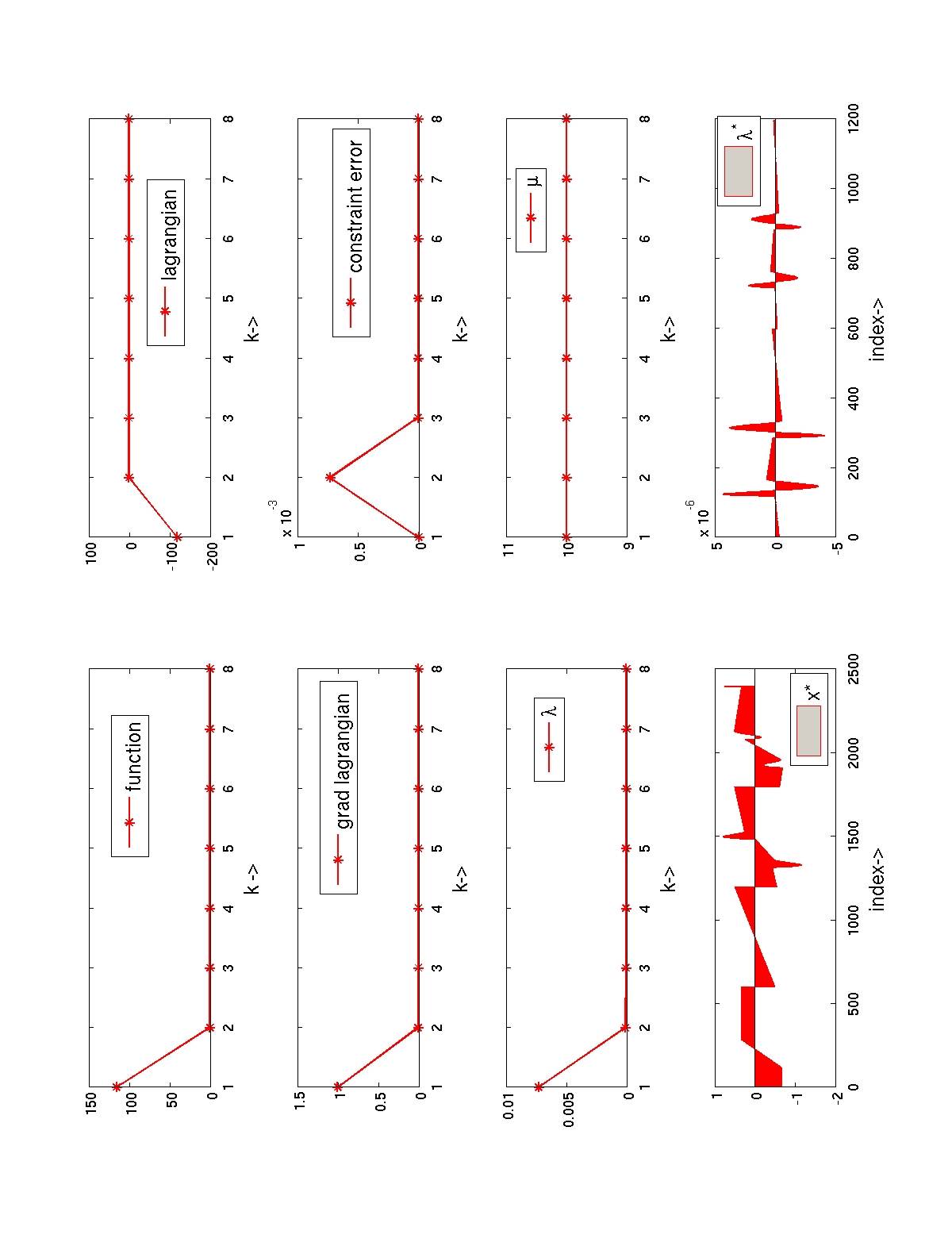}
\caption{Validation Test B: Convergence diagnostics for the advanced solver. Figure shows the evolution of objective function, the Lagrangian, norm of the gradient of the Lagrangian, norm of the constraint satisfaction error, norm of the Lagrange multipliers and progress monitoring parameter over algorithm iterations. The last row shows the optimal solution (i.e., the DM and contrast) displayed as a vector and the optimal Lagrange multipliers for the chosen constraints on the columns of $Z$ and the contrast $c_Z$. The optimal objective of $G(\hat{Z}, \hat{c_Z}) = 0.2655$ was attained in 8 iterations.}
\label{convdiag2}
\end{center}
\end{figure}

\begin{figure}[htbp]
\centering
\begin{tabular}{cc}
\subfigure[Initial design matrix $Z_0$]
{
\hspace{-2cm}
\label{fig15a}
\includegraphics[width = 70mm, angle = -90]{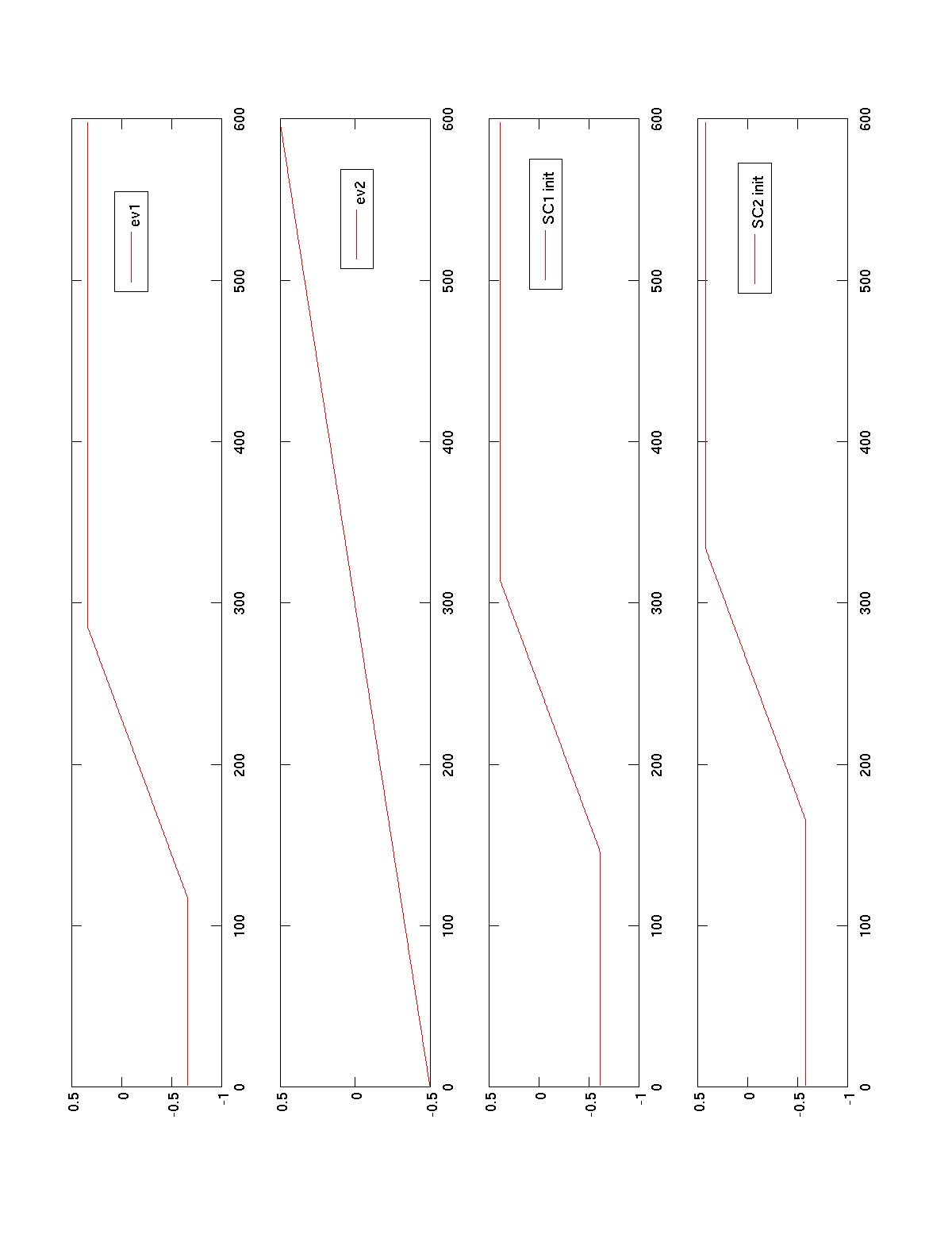}
}

&

\subfigure[Optimal design matrix $\hat{Z}$]
{
\label{fig15b}
\includegraphics[width = 70mm, angle = -90]{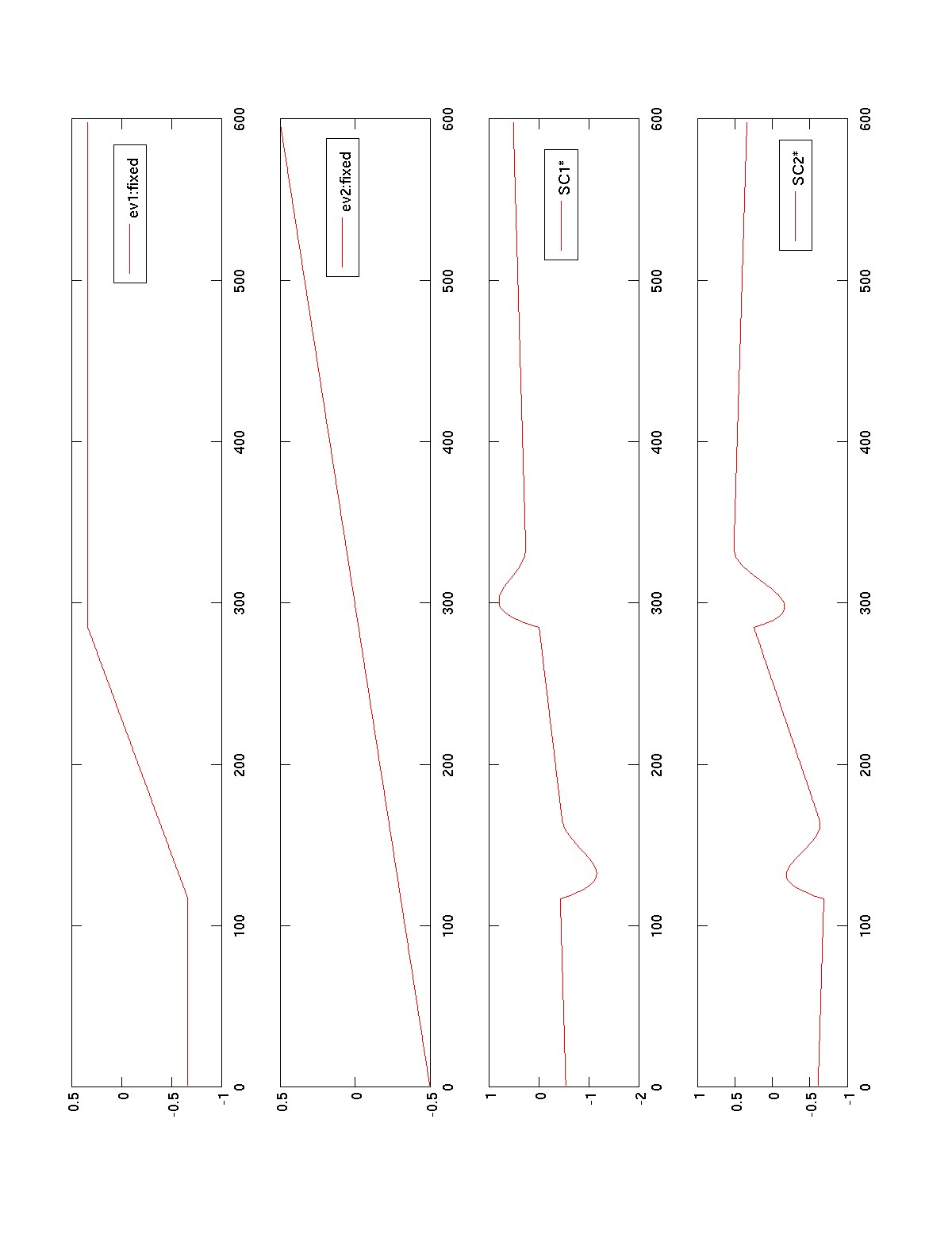}
}

\end{tabular}
\caption{  Validation Test B: (a) Initial design matrix $Z_0$  and (b) Optimal design matrix $\hat{Z}$ ($m = 50$). The first two EVs were constrained to their shift 0 values. The contrast $\hat{c_Z}$ was left unconstrained.}
\label{fig15}
\end{figure}

In this case, the contrast vector $c_Z$ was left unconstrained. Everything else is the same as in Validation Test A. Convergence diagnostics and optimal $\hat{Z}$ for this case are shown in Figure \ref{convdiag2} and Figure \ref{fig15} respectively. The optimal contrast was determined to be $\hat{c_Z} = [0.73519; 0.47890; 0.76016; 0.75789]$.

\begin{figure}[htbp]
\centering
\begin{tabular}{cc}
\subfigure[Initial design matrix $Z_0$]
{
\hspace{-2cm}
\label{fig16a}
\includegraphics[width = 65mm, angle = -90]{arxiv_figs/test_run_uncon_init_designmat.jpg}
}

&

\subfigure[Optimal design matrix $\hat{Z}$]
{
\label{fig16b}
\includegraphics[width = 65mm, angle = -90]{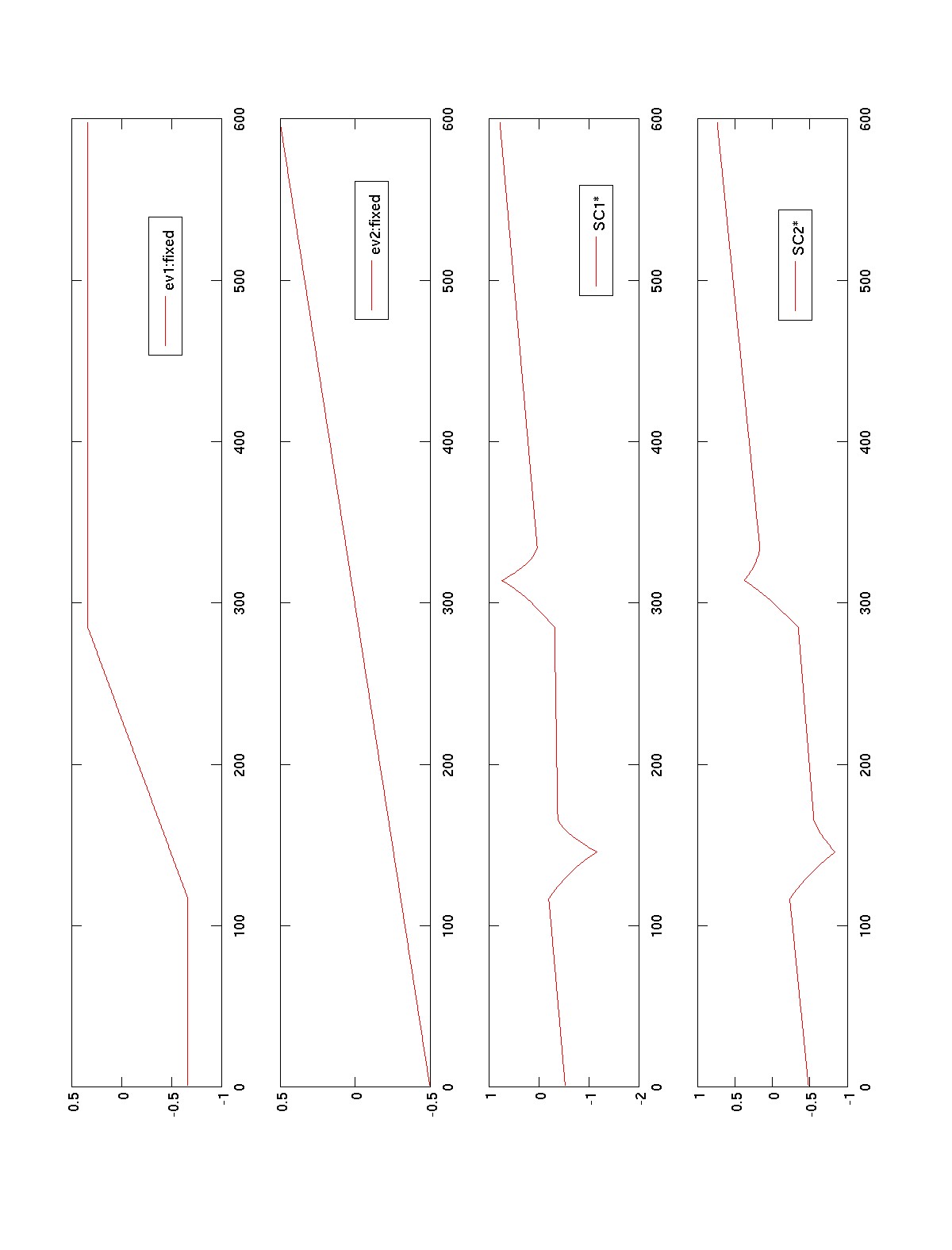}
}

\end{tabular}
\caption{  Validation Test A: (a) Initial design matrix $Z_0$  and (b) Optimal design matrix $\hat{Z}$  ($m = 50$). The first two EVs were constrained to their shift 0 values. The contrast $\hat{c_Z}$ was constrained to be $[1;0;0;0]$. The problem was solved using \textbf{Algorithm 1}. The optimal objective was $G(\hat{Z}, \hat{c_Z}) = 2.693490$.}
\label{fig16}
\end{figure}

\begin{figure}[htbp]
\centering
\begin{tabular}{cc}
\subfigure[Initial design matrix $Z_0$]
{
\hspace{-2cm}
\label{fig17a}
\includegraphics[width = 65mm, angle = -90]{arxiv_figs/test_run_uncon_init_designmat.jpg}
}

&

\subfigure[Optimal design matrix $\hat{Z}$]
{
\label{fig17b}
\includegraphics[width = 65mm, angle = -90]{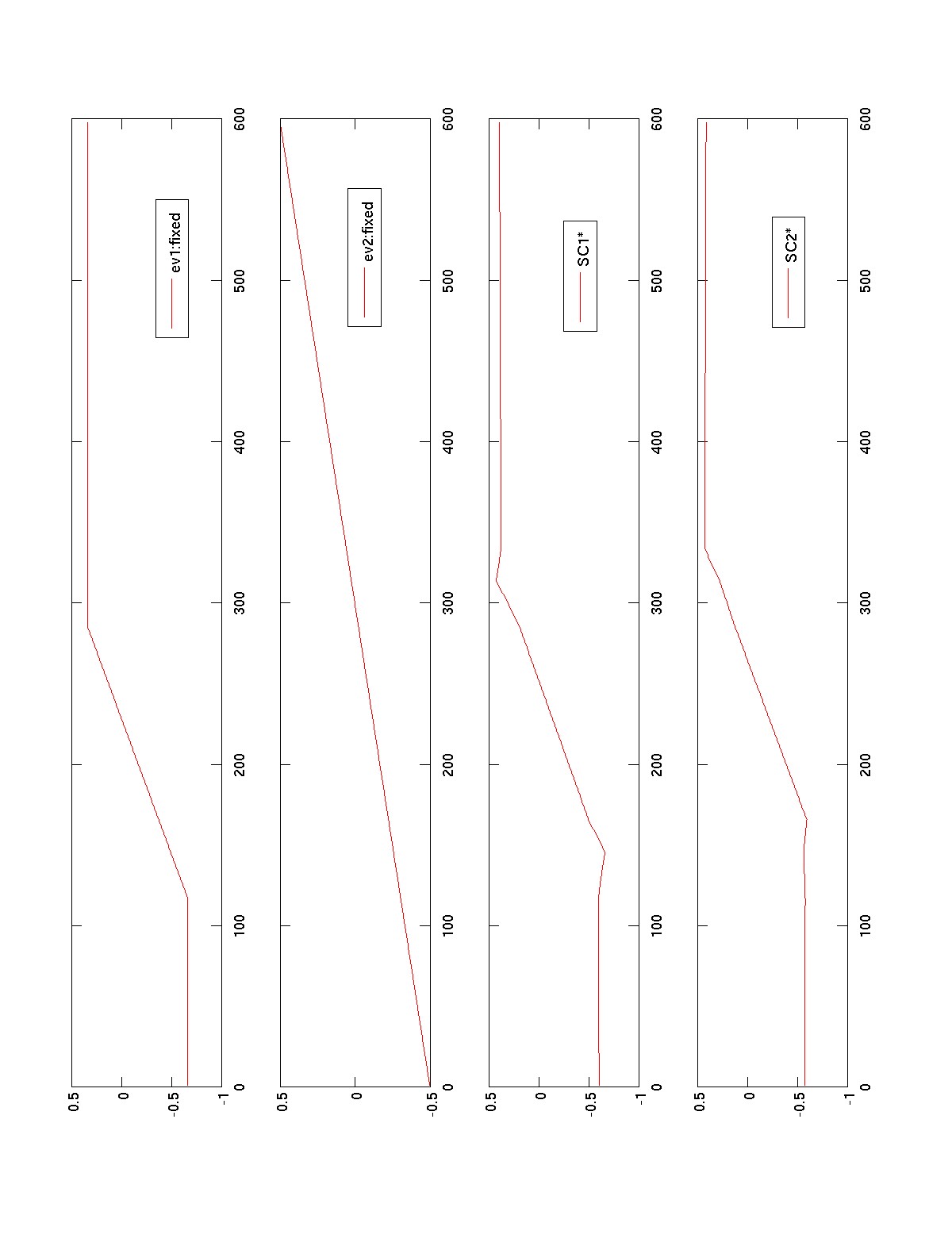}
}

\end{tabular}
\caption{  Validation Test B: (a) Initial design matrix $Z_0$  and (b) Optimal design matrix $\hat{Z}$ ($m = 50$). The first two EVs were constrained to their shift 0 values. The contrast $\hat{c_Z}$ was left unconstrained. The problem was solved using \textbf{Algorithm 1}. The optimal objective was $G(\hat{Z}, \hat{c_Z}) = 0.265502$}
\label{fig17}
\end{figure}

\begin{table}[htdp]
\begin{center}
\begin{tabular}{|c|c|c|}
\hline
 & Exact & Algorithm 1 \\
 \hline
Case a & 2.693467 & 2.693490 \\
\hline
Case b & 0.265484 & 0.265502 \\
\hline
\end{tabular}
\end{center}
\caption{Optimal objective values for the Exact algorithm and Algorithm 1 for Case a and Case b.}
\label{table1}
\end{table}%

The optimal contrast using Algorithm 1 was found to be $\hat{c_Z} = [0.73523;0.47890;0.75704;0.75844]$.

\section{Algorithmic issues}
\subsection{Note on initialization}\label{initializationstrategy}
It is well known that when finding a local solution to an optimization problem as we do here, the choice of an initial point could have an impact on the estimated local solution. We acknowledge that there could be many interesting initialization strategies.
Here we propose one such strategy for initialization of $Z$ and $c_{Z}$. We recommend a heuristic strategy for initialization of the primary column in $Z$. First we try to find a vector $v$ that is closest to primary columns in $X_i$ in the following sense:

\begin{equation}\label{findv}
\mbox{ argmin }_v \sum_{i = 1}^m  ||\left( c^T_{X_i} \frac{\beta_i}{\sigma_i} \right) \left(  X_i  c_{X_i} \right) - \left( c^T_{X_i} \frac{\beta_i}{\sigma_i} \right)  v)||_2^2
\end{equation}

The solution to \ref{findv} is given by:
\begin{equation}
\hat{v} = \frac{ \sum_{i = 1}^m \left( c^T_{X_i} \frac{\beta_i}{\sigma_i} \right) \left(  X_i c_{X_i} \right)}{ \sum_{i = 1}^m  \left( c^T_{X_i} \frac{\beta_i}{\sigma_i} \right)}
\end{equation}

Next we form the $n \times m$ matrix $M = [M_1,\ldots,M_i,\ldots M_m]$ of residuals with $i$th column:
\begin{equation}
M_i = \left [\left( c^T_{X_i} \frac{\beta_i}{\sigma_i} \right) \left(  X_i c_{X_i} \right) - \left( c^T_{X_i} \frac{\beta_i}{\sigma_i} \right)  \hat{v} \right]
\end{equation}

Next we take the singular value decomposition of $M$ to get:
\begin{equation}
M = U_M \Sigma_M V_M^T
\end{equation}
where $U_M \in \mathbf{R}^{n \times m}$ is a matrix of left singular vectors, $V_M \in \mathbf{R}^{m \times m}$ is a matrix of right singular vectors and $\Sigma_M$ holds the singular values of $M$.

For optimizing a $p$ column matrix $Z$ we choose (in matlab notation) the following initialization:
\begin{eqnarray}
Z_0(:,1) = \hat{v} \\
Z_0(:,2:p) = U_M(:,1:(p-1)) \\
c_{Z_0} = [1;0;\ldots;0] \mbox{ ($p$ rows) }
\end{eqnarray}
Thus the vector $\hat{v}$ is used to initialize the primary column of $Z_0$ and the columns of $U_M$ are used to initialize the 
non-primary columns of $Z_0$. The next step is to modify $Z_0$ and $c_{Z_0}$ so that they satisfy the constraints $C c_{Z_0} = d$ and $Z_0 A = B$.

\begin{eqnarray}
Z_0 = Z_0 + (B - Z_0 A) (A^T A)^{-1} A^T \\
c_{Z_0} = c_{Z_0} + C^T (C C^T)^{-1} (d - C c_{Z_0})
\end{eqnarray}

This is not the only way to initialize $Z$, there can be many other strategies. In fact we have not used this initialization strategy in many of the examples presented here precisely to illustrate this point.

\subsection{Estimating the size of $Z$}

By accounting for expected variations in the shape and size of the response and then optimizing for a design matrix $Z$ via the solution of an inverse problem that explicitly controls for bias and variance we 
automatically avoid overfitting in this framework. The framework also allows for inclusion of "null" data that is not simply Gaussian noise but is some structured signal such as a drift (see Example 3) to explicitly instruct the optimization process to "equate" it to "no signal" during optimization. 
Why then is it important to choose the size of $Z$? One reason is to maximize the degrees of freedom available for subsequent first level or higher level statistical tests. We propose the following strategy for choosing the "optimal" number of columns in $Z$.

\begin{algorithm}\hypertarget{alg2}{ }
\begin{algorithmic}[1]
\REQUIRE Problem variables $H$, $\Sigma$, $\ell$, $A$, $C$
\REQUIRE Algorithmic variables $\alpha_0 \in (0, 10^{-3})$, $\theta \in (1, 5]$ and $\eta_1, \eta_2 \in (0, 10^{-6})$
\REQUIRE Initial choice $p = p_0$ and $p_{max}$ the maximum value of $p$
\REQUIRE User chosen accuracy cutoff $R_c \in (0.5, 1)$ with a default value of $R_c = 0.95$ ($95\%$ cutoff)
\ENSURE Outputs are the optimal number of columns in the DM $p_{opt}$
\STATE Set $j = 1$
\FOR{ $p = p_0$  to $p_{max}$ }
\REQUIRE $Z_0$, $c_{Z_0}$ satisfying $Z_0 A = B$ and  $C c_{Z_0} = d$ such that $Z_0$ and $c_Z$ have $p$ columns
\STATE Run \hyperlink{alg1}{Algorithm 1} to estimate $\hat{Z}$, $\hat{c_Z}$ and $\hat{F}$ for the current value of $p$
\STATE Set $p_{est}(j) = p$
\STATE Set $F(j) = \hat{F}$, $Z(j) = \hat{Z}$ and $c_Z(j) = \hat{c_Z}$
\STATE $j = j + 1$
\ENDFOR

\STATE Compute $F_{max} = max_{j} F(j)$ and $F_{min} = min_{j} F(j)$

\STATE Set $j = 1$
\FOR{ $p = p_0$ to $p_{max}$} 
\STATE $R(j) = ( F_{max} - F(j) ) / ( F_{max} - F_{min} )$
\STATE $j = j + 1$
\ENDFOR
\STATE Calculate the minimum $j$ meeting the cutoff, $\hat{j} = min \{ j : R(j) \ge R_c \}$ 
\RETURN $p_{opt} = p_{est}(\hat{j})$
\end{algorithmic}
\caption{Choosing optimal number of columns in $Z$}
\end{algorithm}

The basic idea in \hyperlink{alg2}{Algorithm 2} is to run \hyperlink{alg1}{Algorithm 1} for a range of values of $p$ and choose a value of $p$ that achieves a user chosen reduction in the objective function value relative to the maximum possible reduction over all values of $p$. Please note that the strategy for choosing the number of columns proposed here is by no means the only one. For example, it could also involve reduction in the model variance $V_b$ below a specified user value. If the $R$ vs $p$ curve has a local maximum as opposed to a monotonic increase then the location of this maximum also is a reasonable choice for the optimal size of $Z$. In principle one can correct for variability due to initialization using \hyperlink{alg3}{Algorithm 3} that essentially runs \hyperlink{alg2}{Algorithm 2} for a number of initializations.

\begin{algorithm}\hypertarget{alg3}{ }
\begin{algorithmic}[1]
\REQUIRE Problem variables $H$, $\Sigma$, $\ell$, $A$, $C$
\REQUIRE Algorithmic variables $\alpha_0 \in (0, 10^{-3})$, $\theta \in (1, 5]$ and $\eta_1, \eta_2 \in (0, 10^{-6})$
\REQUIRE Initial choice $p = p_0$ and $p_{max}$ the maximum value of $p$
\REQUIRE User chosen accuracy cutoff $R_c \in (0.5, 1)$ with a default value of $R_c = 0.95$ ($95\%$ cutoff)
\REQUIRE User chosen number of trials $n_{iter}$
\ENSURE Outputs are the optimal number of columns in the DM $p_{opt} (n_{iter})$ over $n_{iter}$ runs
\FOR { $j = 1$ to $n_{iter}$ }
\STATE Run \hyperlink{alg2}{Algorithm 2} to get $p_{opt}$ for this run
\STATE Set $p_{est}(j) = p_{opt}$
\ENDFOR
\STATE $p_{opt}(n_{iter}) = \mbox{Median}_j p_{est}(j)$
\RETURN $p_{opt}(n_{iter})$
\end{algorithmic}
\caption{Choosing optimal number of columns in $Z$ - Sensitivity to Initialization}
\end{algorithm}

\begin{figure}[htbp]
\centering
\begin{tabular}{cc}
\subfigure[$p$ versus $\hat{F}$]
{
\label{fig18a}
\includegraphics[width = 70mm, angle = -90]{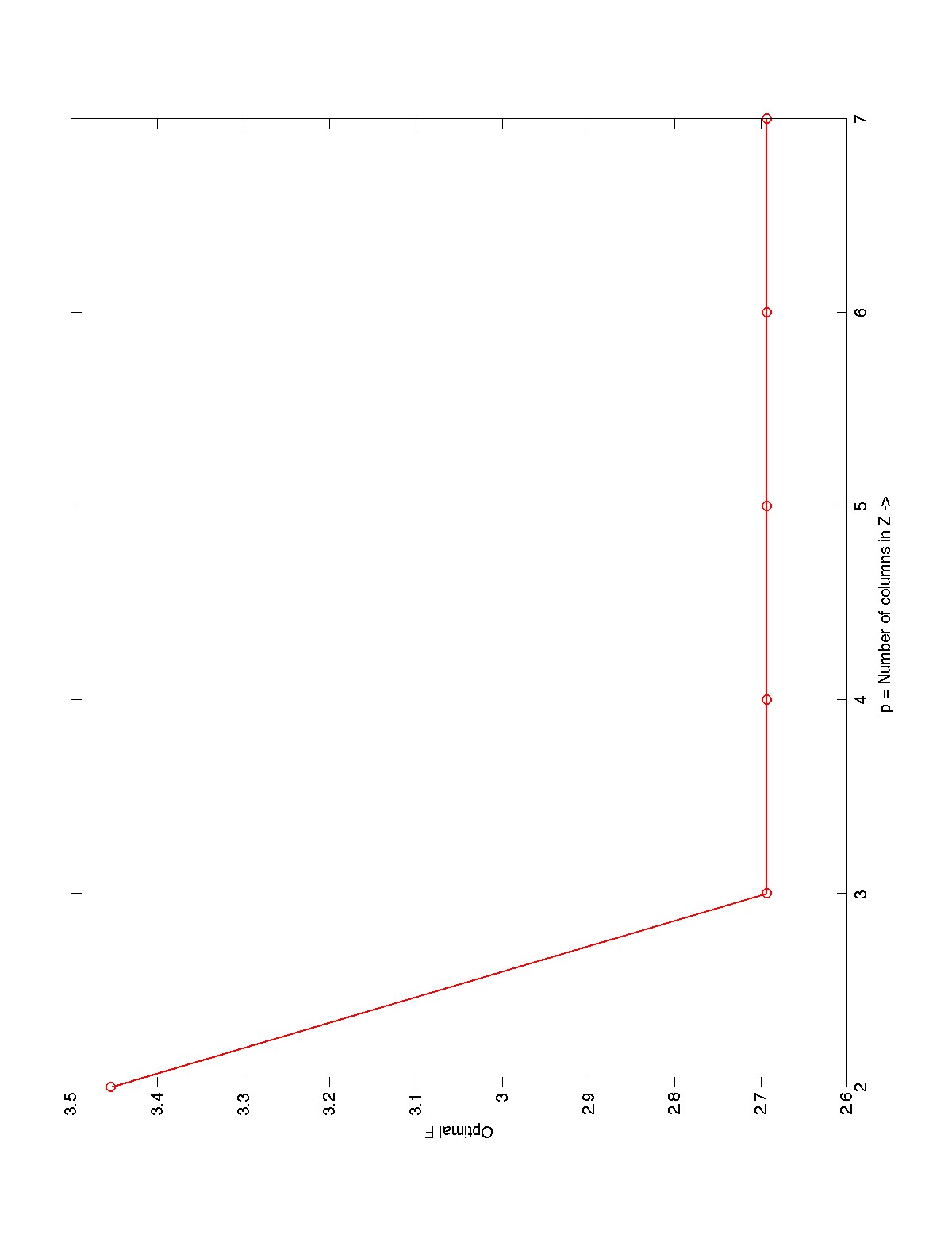}
}

&

\subfigure[$p$ versus $\hat{F}$ zoomed in]
{
\label{fig18b}
\includegraphics[width = 70mm, angle = -90]{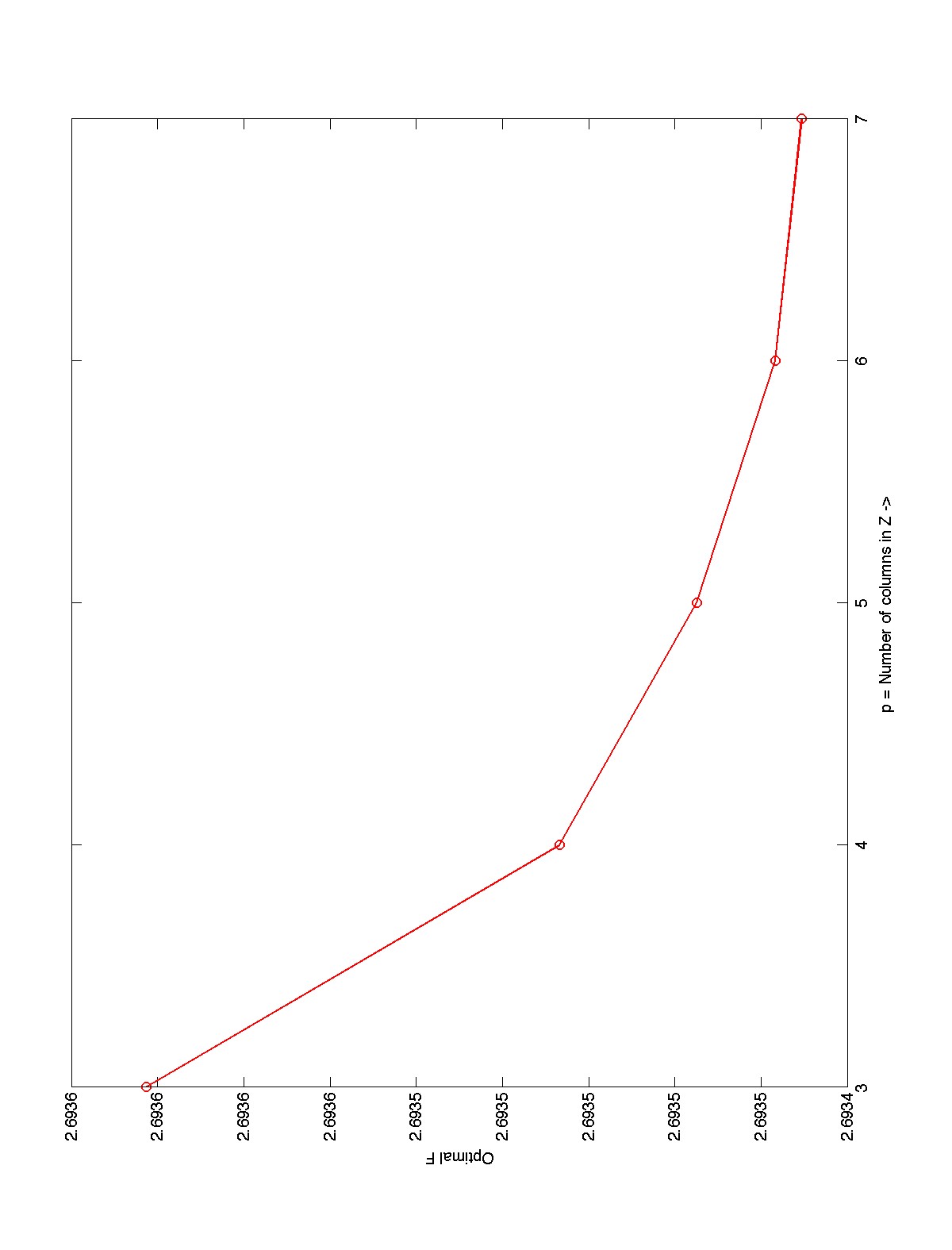}
}

\end{tabular}
\caption{  Illustration of the procedure to estimate the size of $Z$ using data from Validation Test A (a) Number of columns $p$ versus the optimal objective $\hat{F}$ (b) Same as (a) but showing the slow decline of the tail in (a)}
\label{fig18}
\end{figure}

\begin{figure}[htbp]
\centering
\includegraphics[width = 75mm, angle = -90]{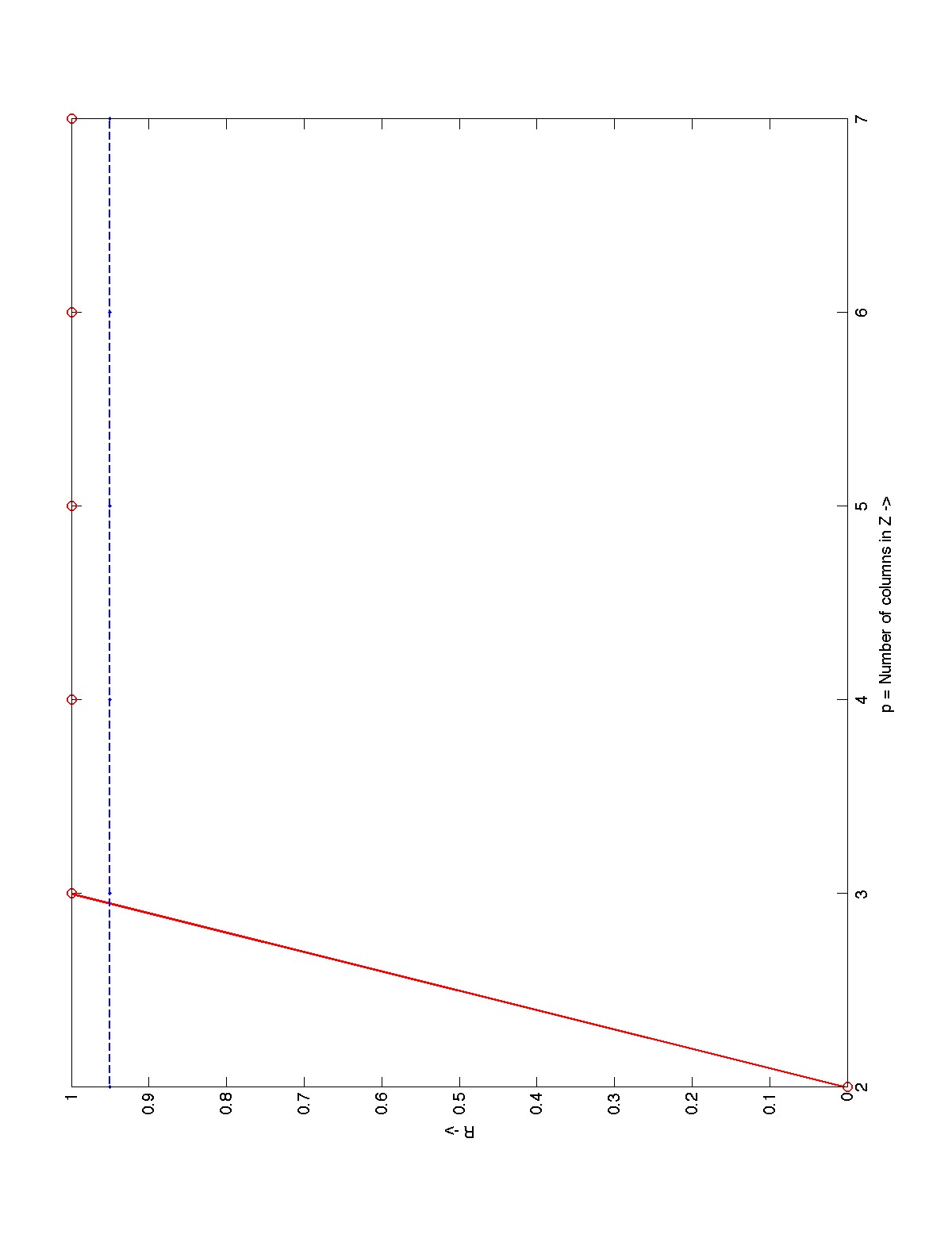}
\caption{ Data from Validation Test A showing $R$ versus $p$ curve and the cutoff at 0.95 indicating that $p_{opt} = 3$}
\label{fig19}
\end{figure}

We ran \hyperlink{alg2}{Algorithm 2} for Validation Test A and the results are shown in Figures \ref{fig18}-\ref{fig19}. It was found that $p_{opt} = 3$ using a cutoff of $R_c = 0.95$ for this case study.

\subsection{Intelligent choice of $\phi_i$}\label{intelligentphi}
 So far we have not discussed sensible choices of $\phi_i$ for design matrix $X_i$. Is a constant $\phi_i$ for all $i$ the best choice? In this section we wish to motivate some heuristic rules for selecting $\phi_i$ based on user defined objectives.
 Equation \ref{eqnfbiasvarwtd} can be re-written using definitions of $C_b$ and $CV_{\Delta}$ for $X_i$ from \ref{8} and \ref{10} as follows:
 \begin{equation}\label{eqnfbiasvarwtd_rewritten}
 f(Z, c_Z; X_i ; \frac{\beta_i}{\sigma_i} ; c_{X_i}; \phi_i) =  \left\{ (2 \phi_i) c_{X_i}^T (X_i^T X_i)^{-1} c_{X_i}  \right\} (CV_{\Delta i}   + 1)  + \left\{ (2 - 2 \phi_i) \left(\frac{c_{X_i}^T \beta_i}{\sigma_i}\right)^2 \right\} C^2_{bi} 
 \end{equation}
 
 The contribution of fractional contrast bias $C_{bi}^2$ to the objective function is controlled by its multiplier $\left\{ (2 - 2 \phi_i) \left(\frac{c_{X_i}^T \beta_i}{\sigma_i}\right)^2 \right\}$. Similarly, the contribution of fractional variance change w.r.t Gauss-Markov estimate term $( CV_{\Delta i} + 1) $ is controlled by its multiplier
 $  \left\{ (2 \phi_i) c_{X_i}^T (X_i^T X_i)^{-1} c_{X_i}  \right\} $. 
 \subsubsection{Choice A}
 Suppose the user wants to put $k$ times more emphasis on the $C_{bi}^2$ term compared to $( CV_{\Delta i} + 1) $ term. Then a sensible choice would be:
 \begin{equation}
 \left\{ (2 - 2 \phi_i) \left(\frac{c_{X_i}^T \beta_i}{\sigma_i}\right)^2 \right\} = k \left\{ (2 \phi_i) c_{X_i}^T (X_i^T X_i)^{-1} c_{X_i}  \right\}
 \end{equation}
 
 Solving for $\phi_i$ and denoting the calculated value by $\phi_i(k)$ to indicate $k$ times more emphasis on $C_{bi}^2$ term compared to $ (CV_{\Delta i} + 1)$ term, we get:
 \begin{equation}
 \phi_i(k) = \frac{ \left(\frac{c_{X_i}^T \beta_i}{\sigma_i}\right)^2 }{ \left(\frac{c_{X_i}^T \beta_i}{\sigma_i}\right)^2 + k \, c_{X_i}^T (X_i^T X_i)^{-1} c_{X_i} }
 \end{equation}
 
 It is clear that as $k \to \infty$, $\phi_i(k) \to 0$ and as $k \to 0$, $\phi_i(k) \to 1$. This behavior is as expected but interestingly it is non-linear. In particular, choosing $k = 1$ i.e., equal emphasis on $C_{bi}^2$ and $(CV_{\Delta i} + 1)$ terms does not necessarily imply $\phi_i(1) = 0.5$.
 Choosing sensible values of $k$ boils down to sifting through the optimal solutions and picking ones that make practical sense. In general one wants $C^2_{bi} \sim 10^{-3}$ and $CV_{\Delta} \sim 0$. Thus it makes sense to choose $k = 10^3$ to make both terms in the optimization of a similar magnitude while
 satisfying reasonable practical objectives.
 
 \subsubsection{Choice B}
Suppose the user wants to put $k$ times more emphasis on the $|C_{bi}|$ term compared to $( CV_{\Delta i} + 1) $ term. In this case a sensible choice would be:
\begin{equation}
 \sqrt{(2 - 2 \phi_i)} \left| \left(\frac{c_{X_i}^T \beta_i}{\sigma_i}\right) \right| = k \left\{ (2 \phi_i) c_{X_i}^T (X_i^T X_i)^{-1} c_{X_i}  \right\}
 \end{equation}
Squaring both sides and solving the resulting quadratic equation and discarding the negative root results in:
\begin{equation}
\phi_i(k) = \frac{ -1 + \sqrt{1 + 4 \, a(k)} }{2 \, a(k)}
\end{equation}
where
\begin{equation}
a(k) = \frac{ 2 \left\{ c_{X_i}^T (X_i^T X_i)^{-1} c_{X_i} \right\}^2 k^2 }{ \left( \frac{c_{X_i}^T \beta_i}{\sigma_i} \right)^2 }
\end{equation}
It can be shown that as $k \to \infty, a \to \infty, \phi_i(k) \to 0$ and as $k \to 0, a \to 0, \phi_i(k) \to 1$. Again, this behavior is to be expected. As before, choosing $k = 1$ i.e., equal emphasis on $|C_{bi}|$ and $(CV_{\Delta i} + 1)$ terms does not imply $\phi_i(1) = 0.5$. Practically speaking one wants $|C_{bi}| \sim 10^{-2}$ and $CV_{\Delta} \sim 0$ and hence a reasonable value of $k$ for choice B would be $k = 10^2$ to make the two terms comparable in magnitude during the optimization process.

\begin{figure}[htbp]
\begin{center}
\includegraphics[width = 75mm, angle = -90]{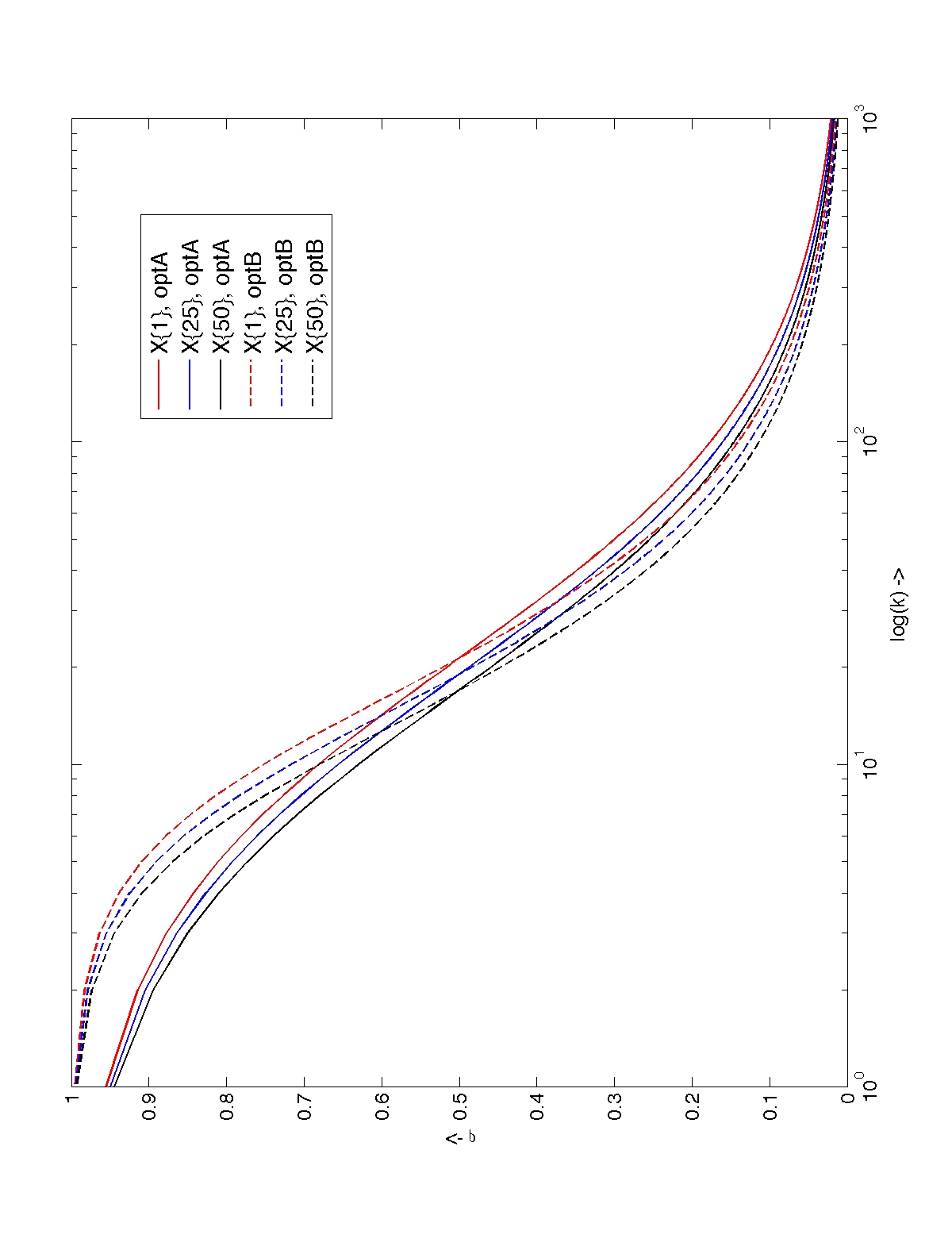}
\caption{Plot of $\log k$  vs $\phi$ for design matrices $X_1$, $X_{25}$ and $X_{50}$ from Example 1. This figure shows that as $k \to \infty$, $\phi \to 0$ and as $k \to 0$, $\phi \to 1$ but the relationship is non-linear, i.e., choosing $\phi = 0.5$ does not give equal emphasis to bias and variance terms ($k \neq 1$). }
\label{phioptionAB}
\end{center}
\end{figure}

An example of the non-linear relationship between $\log k$ and $\phi$ is shown in Figure \ref{phioptionAB}.

\section{Case Studies}
In this section we present performance curves for the estimated optimal DM's for six example cases. In particular, we look at the contrast bias $C_b$, the model variance bias $V_b$ and contrast variance change w.r.t  the Gauss-Markov estimate $CV_{\Delta}$ as key performance measures.

In the first 3 examples, $X_i, \frac{\beta_i}{\sigma_i}, c_{X_i}$ were chosen as in \ref{sampleDM}. In the Example 4 additional $X_i$ were added to those in Examples 1-3 for illustration purposes. Example 5 deals with a new set of $X_i$ derived from the standard block design used in fMRI experiments. Example 6 illustrates the application of proposed technique to the problem of capturing variable shapes of the Haemodynamic response function (HRF) in fMRI.

\subsection{Example 1}
\begin{figure}[htbp]
\centering
\begin{tabular}{cc}

\subfigure[]
{
\hspace{-2cm}
\label{case1a}
\includegraphics[width = 75mm, angle = -90]{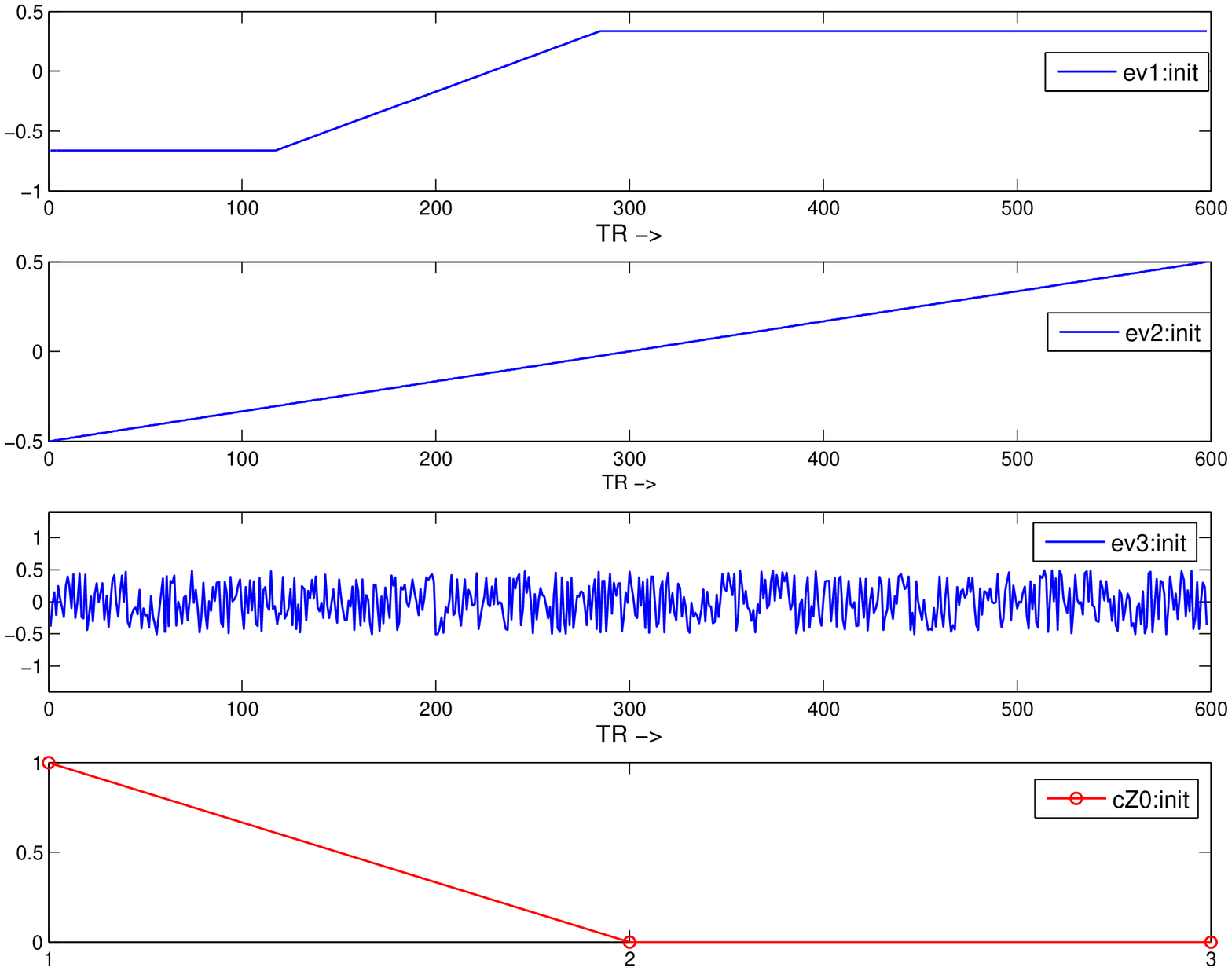}
}

&

\subfigure[]
{
\hspace{-1cm}
\label{case1b}
\includegraphics[width = 75mm, angle = -90]{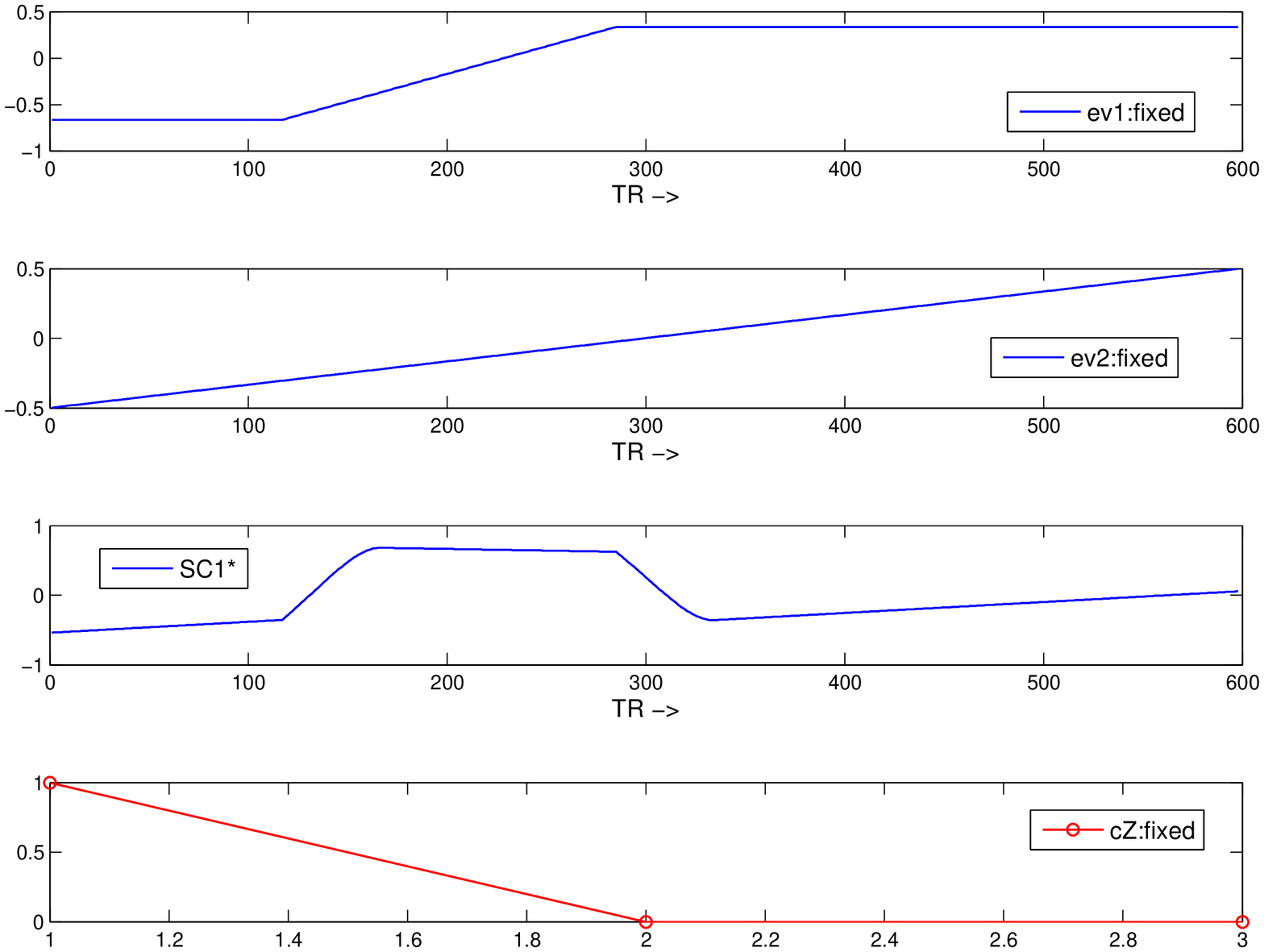}
}

\\

\subfigure[]
{
\hspace{-2cm}
\label{case1c}
\includegraphics[width = 75mm, angle = -90]{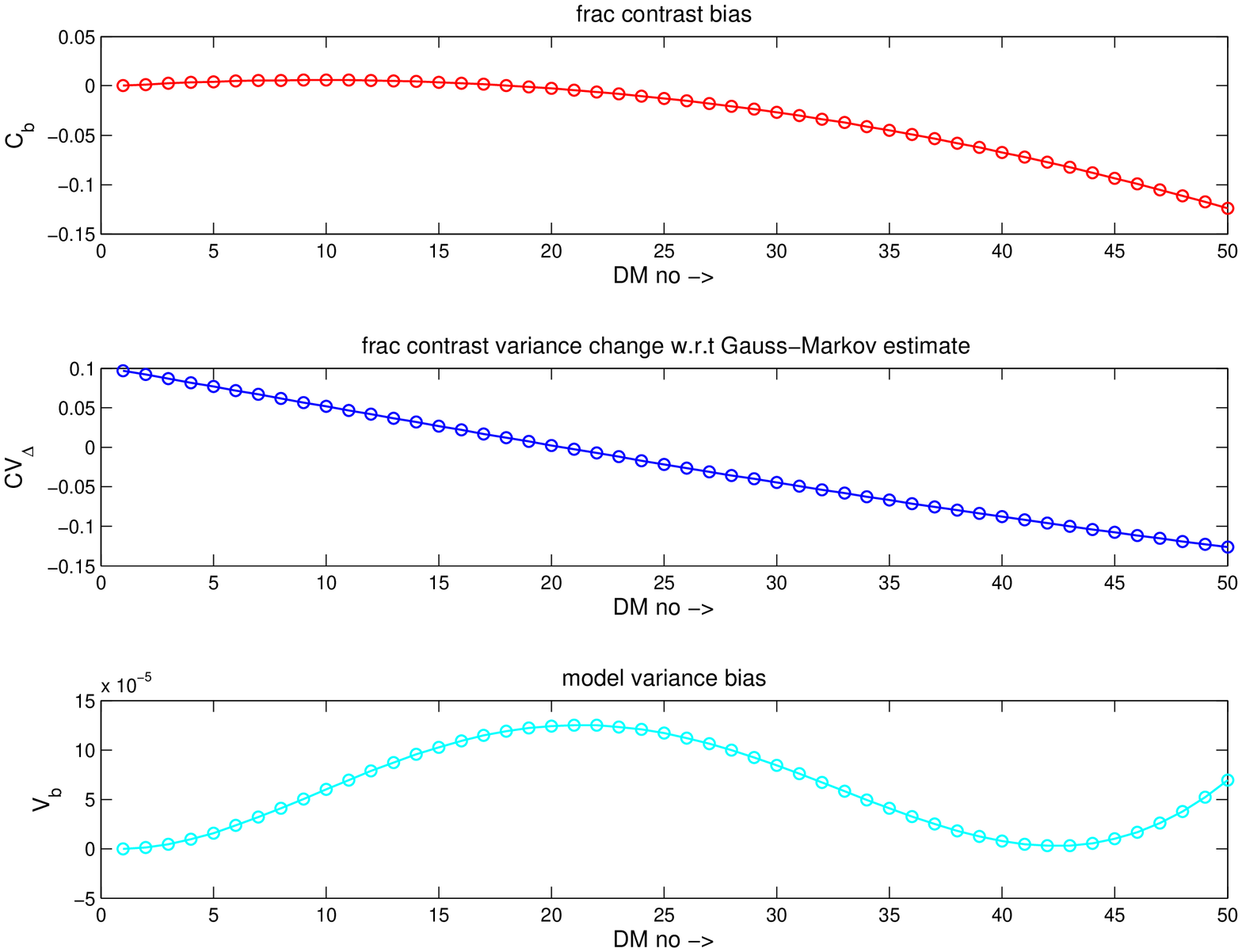}
}

&

\subfigure[]
{
\hspace{-1cm}
\label{case1d}
\includegraphics[width = 75mm, angle = -90]{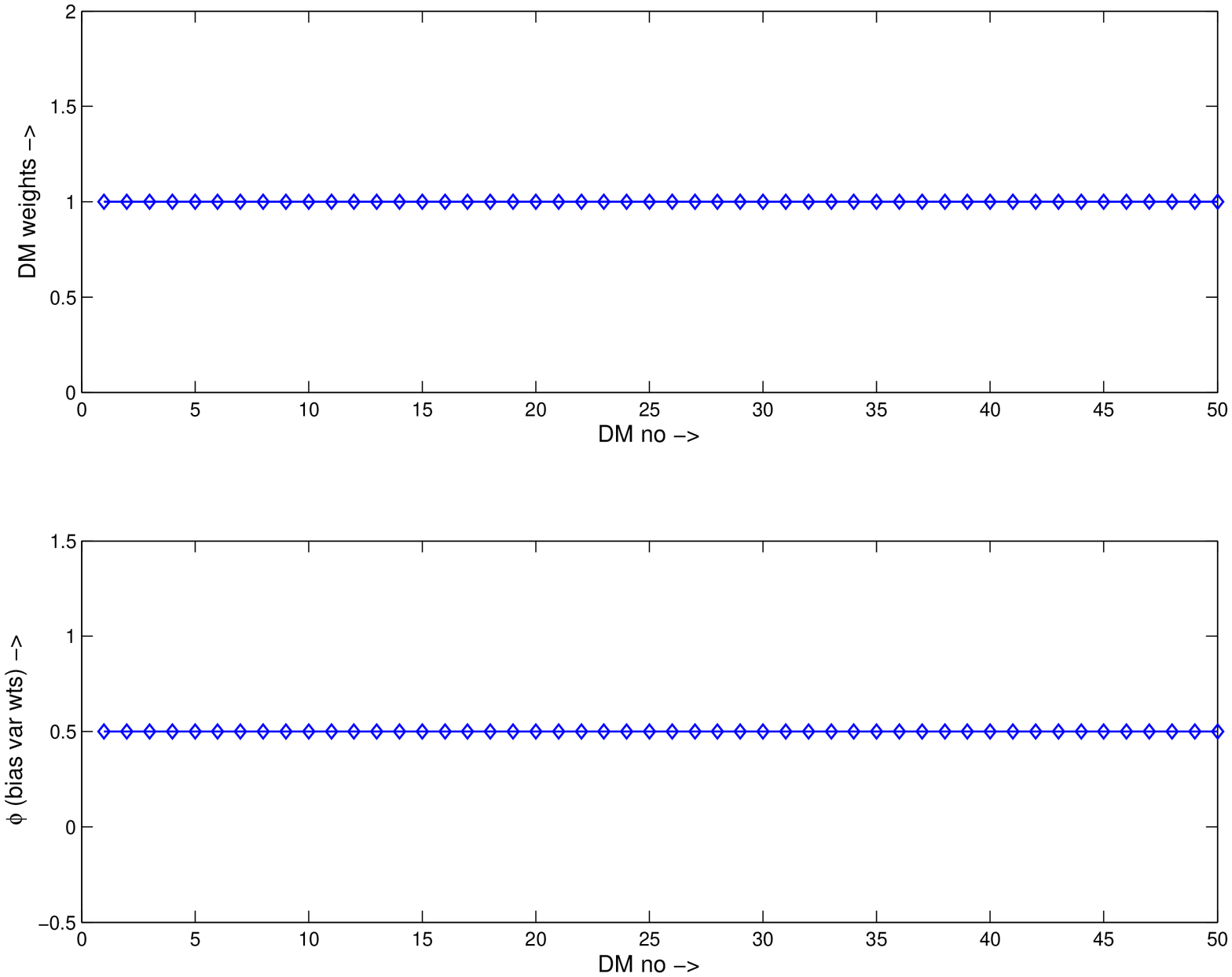}
}

\end{tabular}
\caption{ Example 1: (a) Initial design matrix (DM) along with random initialization of the 3rd column. The first two columns were fixed at their initial values and the contrast was fixed at [1;0;0] (b) Estimated optimal DM. Notice how the 3rd column converges to a non-random profile (c) Performance curves showing the fractional contrast bias $C_b$, contrast variance change w.r.t Gauss-Markov estimate $CV_{\Delta}$ and model variance bias $V_b$ (d) For all $i$, the DM weights $w_i = 1$ implying equal likelihood of observing any of the specified DM's and bias-variance scalings $\phi_i = 0.5$ for all $i$ implying the default weighting of bias and variance. }
\label{case1A}
\end{figure}

\begin{figure}[htbp]
\centering
\begin{tabular}{cc}

\subfigure[]
{
\hspace{-2cm}
\label{case1e}
\includegraphics[width = 75mm, angle = -90]{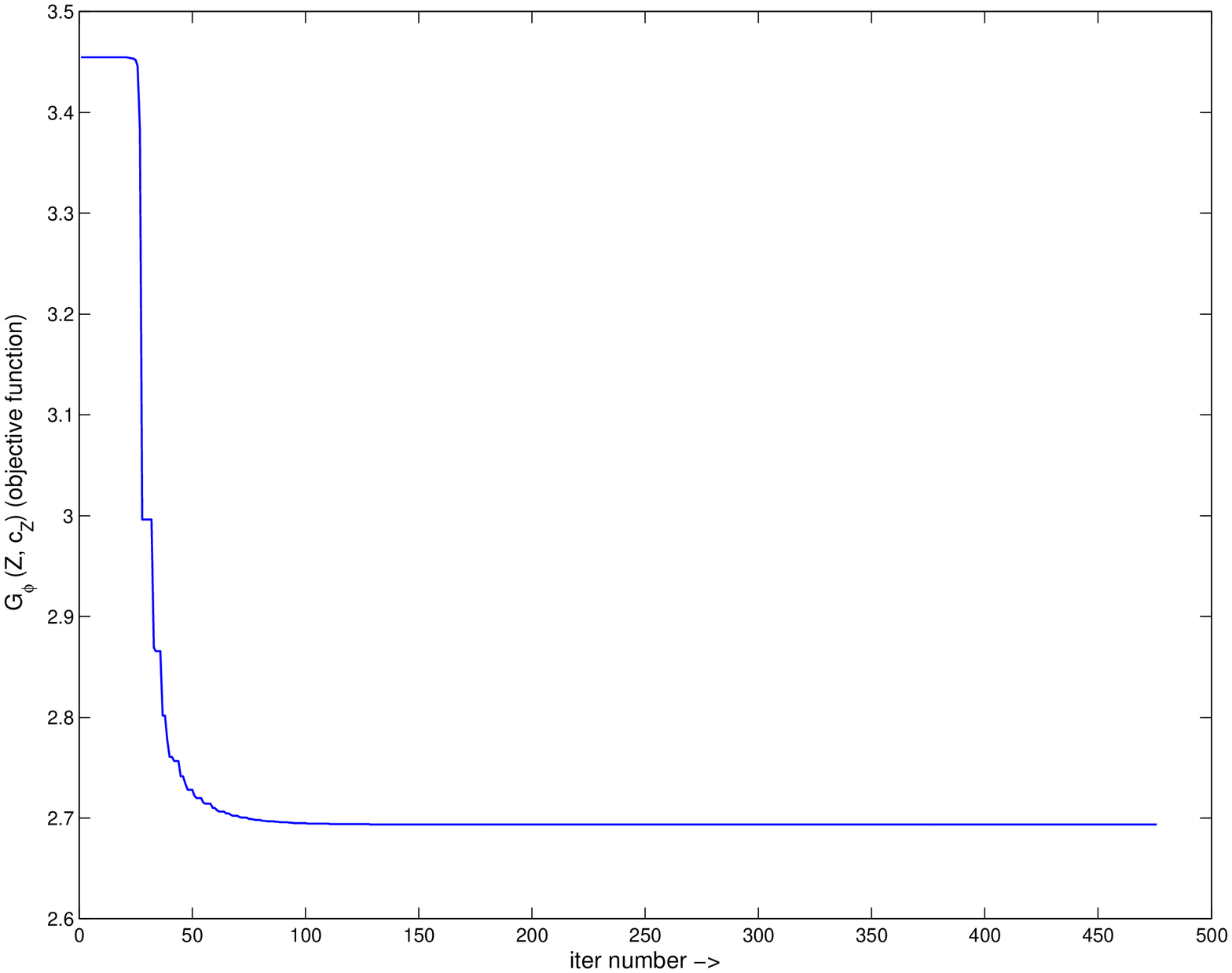}
}

&

\subfigure[]
{
\hspace{-1cm}
\label{case1f}
\includegraphics[width = 75mm, angle = -90]{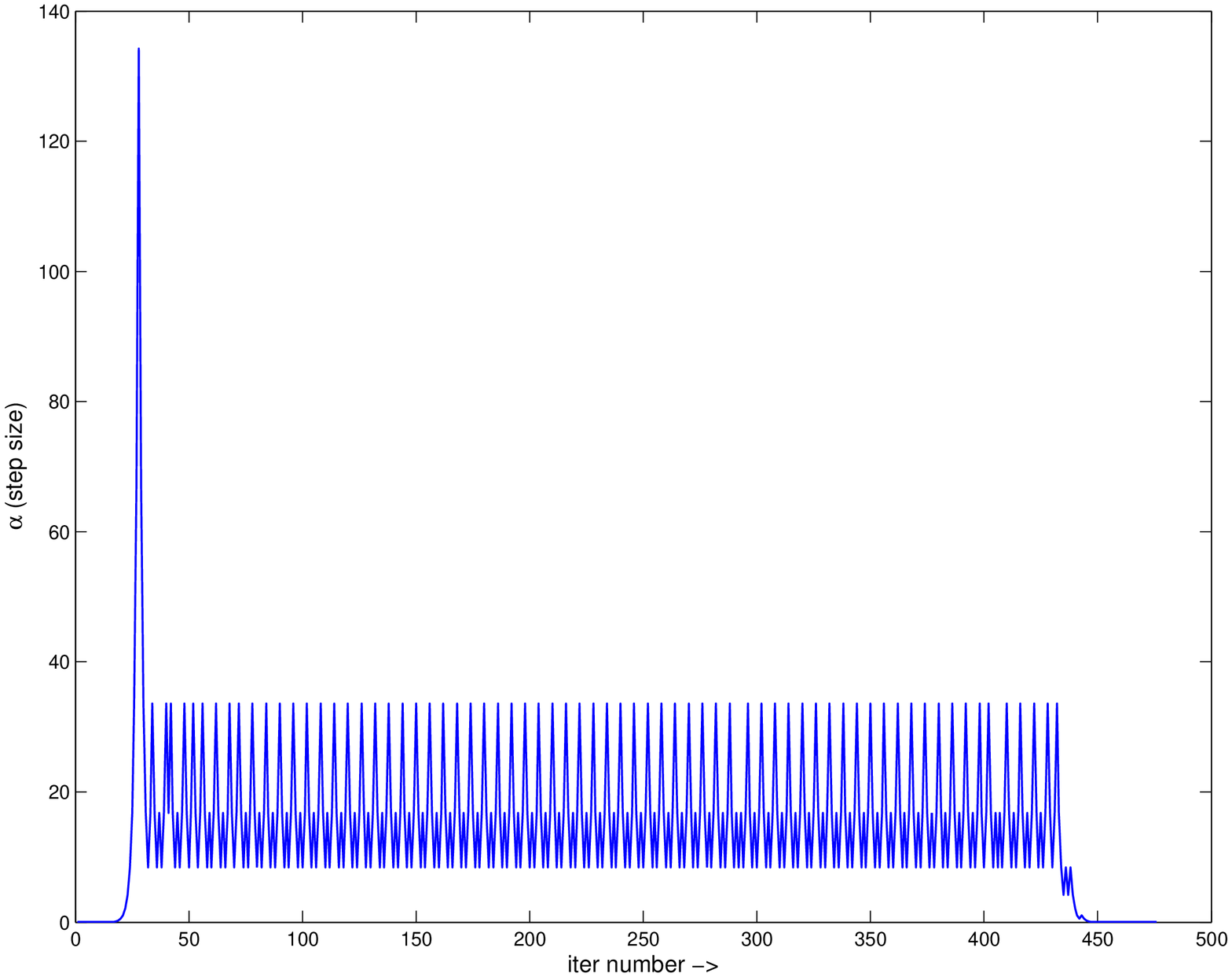}
}

\\

\subfigure[]
{
\hspace{-2cm}
\label{case1g}
\includegraphics[width = 75mm, angle = -90]{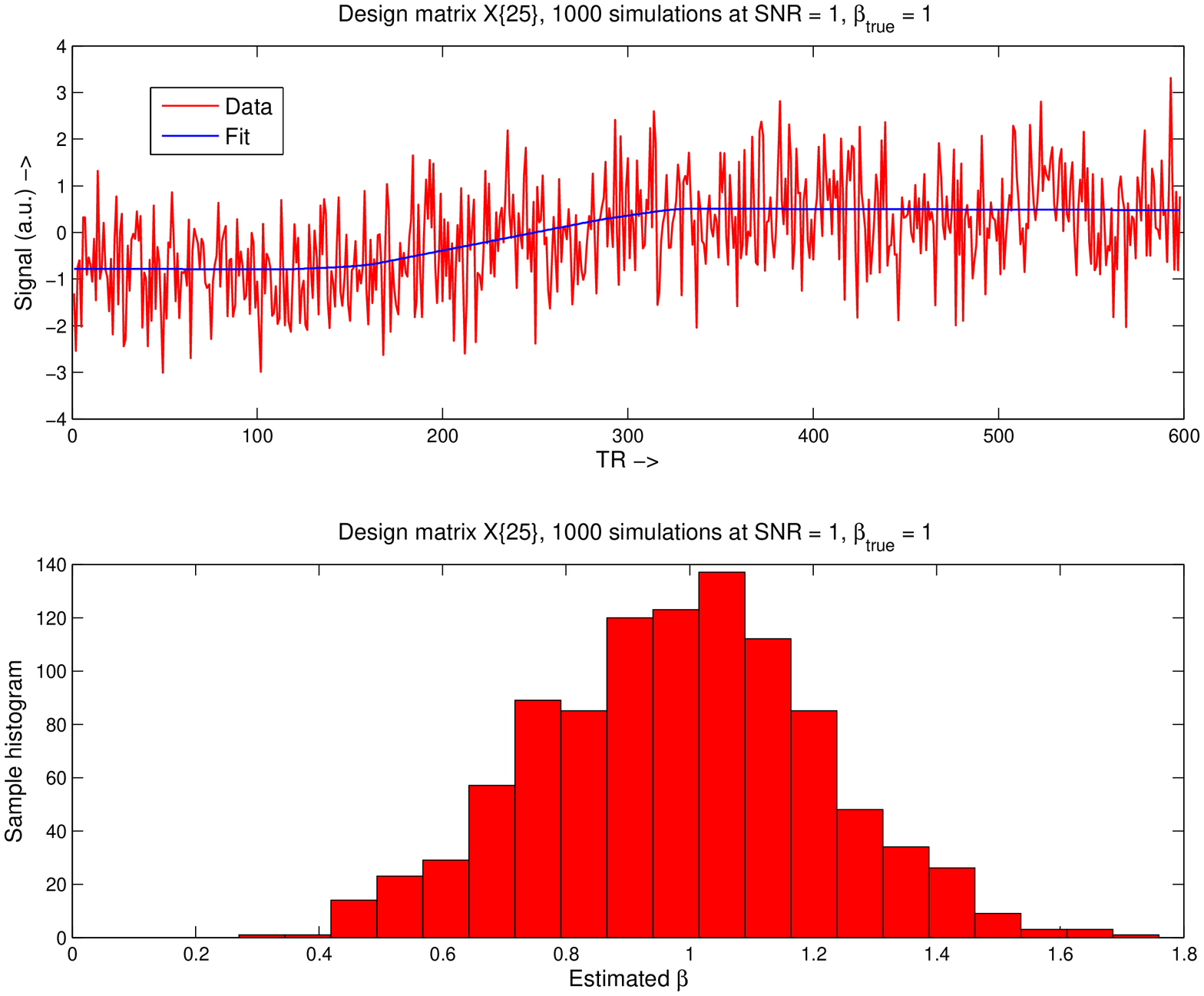}
}

&

\subfigure[]
{
\hspace{-1cm}
\label{case1h}
\includegraphics[width = 75mm, angle = -90]{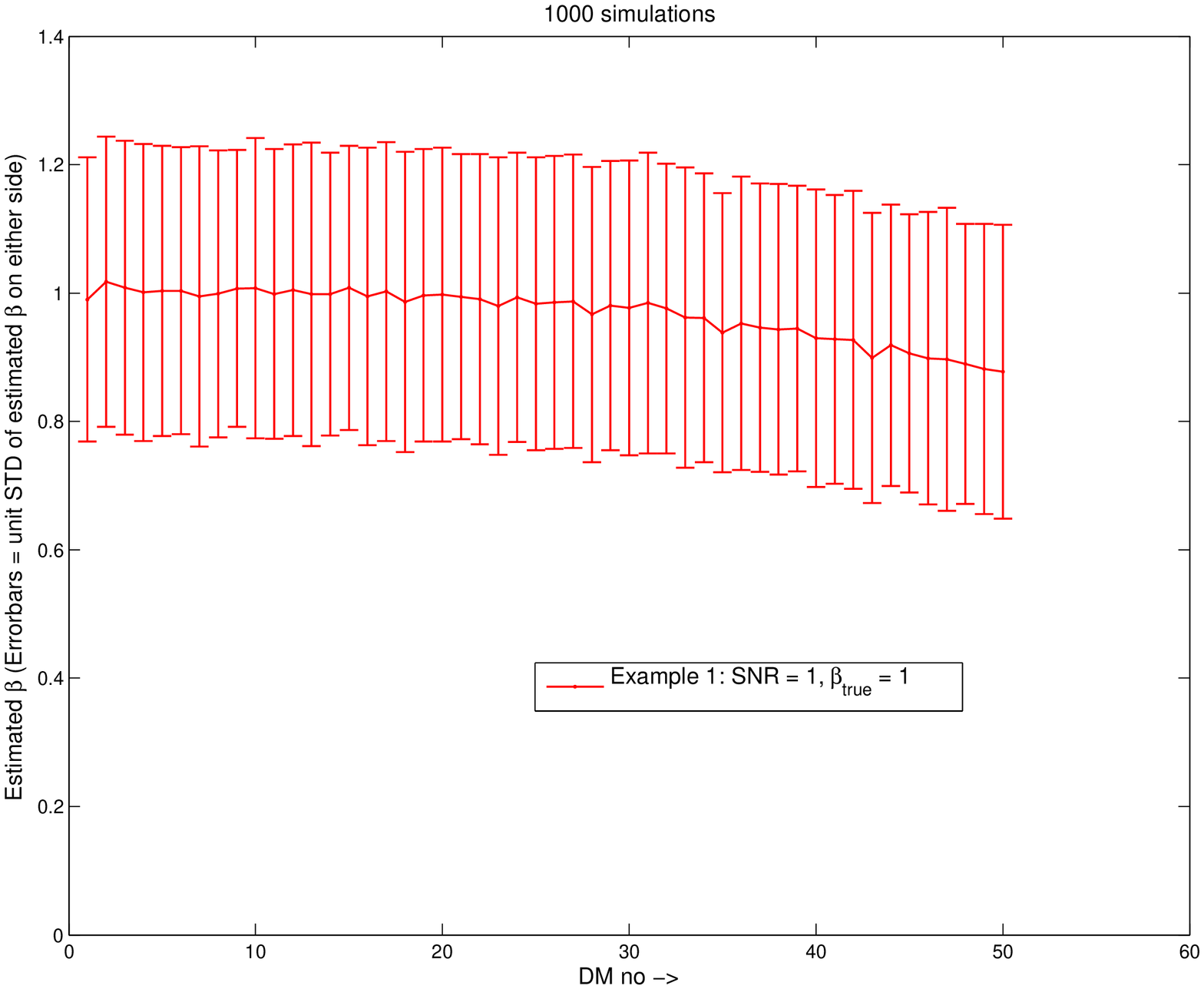}
}

\end{tabular}
\caption{  Example 1: (a) Figure showing the evolution of objective function values $G_\phi(Z,c_Z)$ over algorithm iterations. Notice how the function value stabilizes as convergence is reached (b) Figure showing the variation in the step size $\alpha$ over algorithm iterations. Step size controlling parameter $\theta$ in Algorithm 1 was set to $\theta = 2$. For each design matrix (DM) $X_i$ entered into optimization, 1000 simulated data-sets were generated at SNR $\frac{\beta_i}{\sigma_i}$. A GLM analysis was run on each of these data-sets using the optimized DM. Figure (c) shows an example of simulated data for DM $X_{25}$ at SNR $\frac{\beta_{25}}{\sigma_{25}}$ and the GLM fit using the optimal DM. It also shows the distribution of $c_Z^T \hat{\gamma}$ over 1000 simulations. Figure (d) is a summary errorbar plot showing $\hat{E}(c_Z^T \hat{\gamma})$ over 1000 simulations for data generated from each DM. The error bars represent unit standard deviation of $c_Z^T \hat{\gamma}$ (\textbf{not} standard deviation of $\hat{E}(c_Z^T \hat{\gamma})$ )  to quantify the variance in estimation via simulation.}
\label{case1B}
\end{figure}

In this example, we used the default weighting of bias and variance by choosing $\phi = 0.5$. The first two columns of $Z$ were fixed as before and the contrast $c_Z$ was fixed at [1;0;0]. We also chose $w_i = 1$ to give equal weights to all design matrices. The unconstrained column in $Z$ was initialized randomly with elements drawn from a uniform distribution $U(0,1)$. The results are shown in Figures \ref{case1A} and \ref{case1B}.

\subsection{Example 2}
\begin{figure}[htbp]
\centering
\begin{tabular}{cc}

\subfigure[]
{
\hspace{-2cm}
\label{case1Aa}
\includegraphics[width = 75mm, angle = -90]{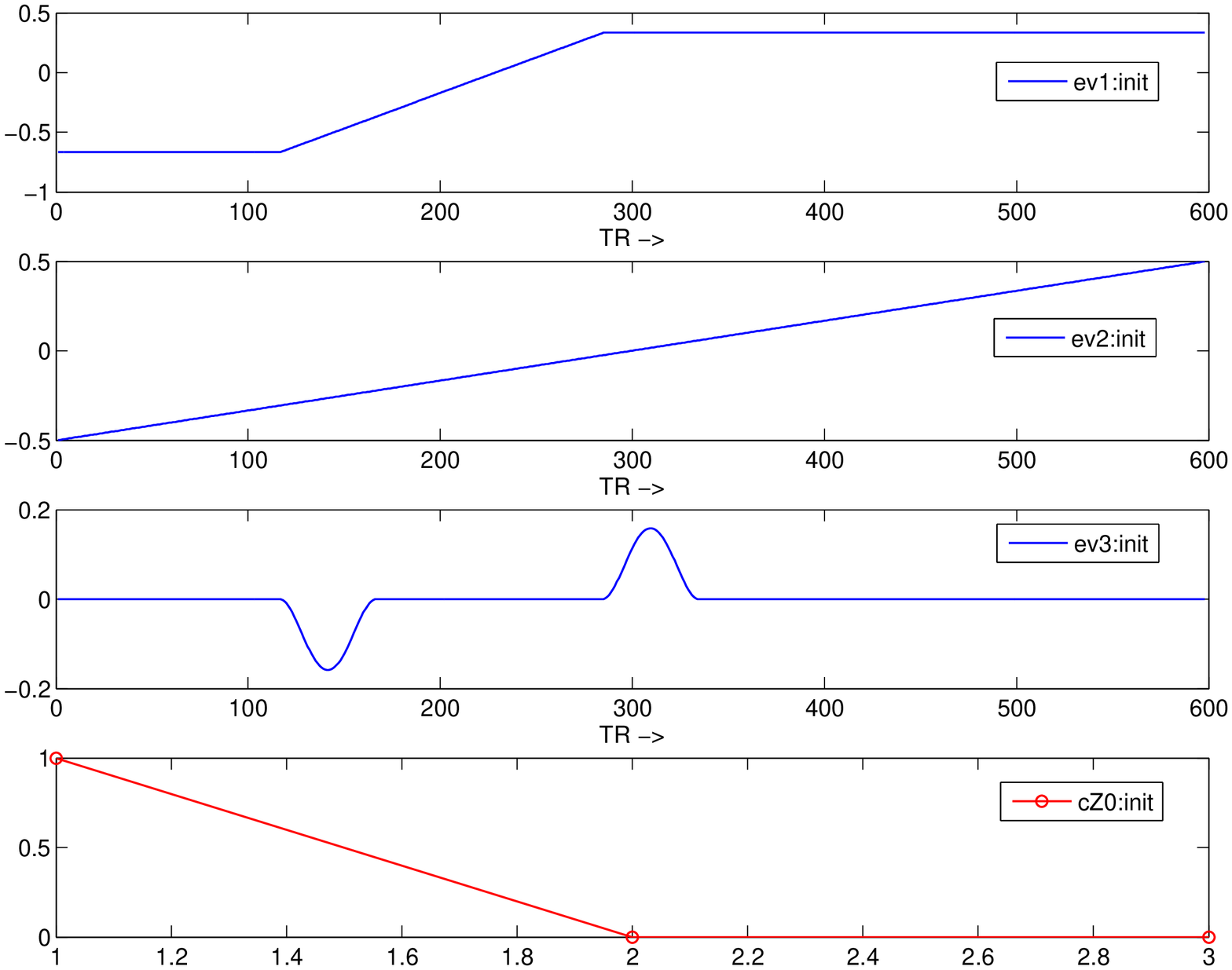}
}

&

\subfigure[]
{
\hspace{-1cm}
\label{case1Ab}
\includegraphics[width = 75mm, angle = -90]{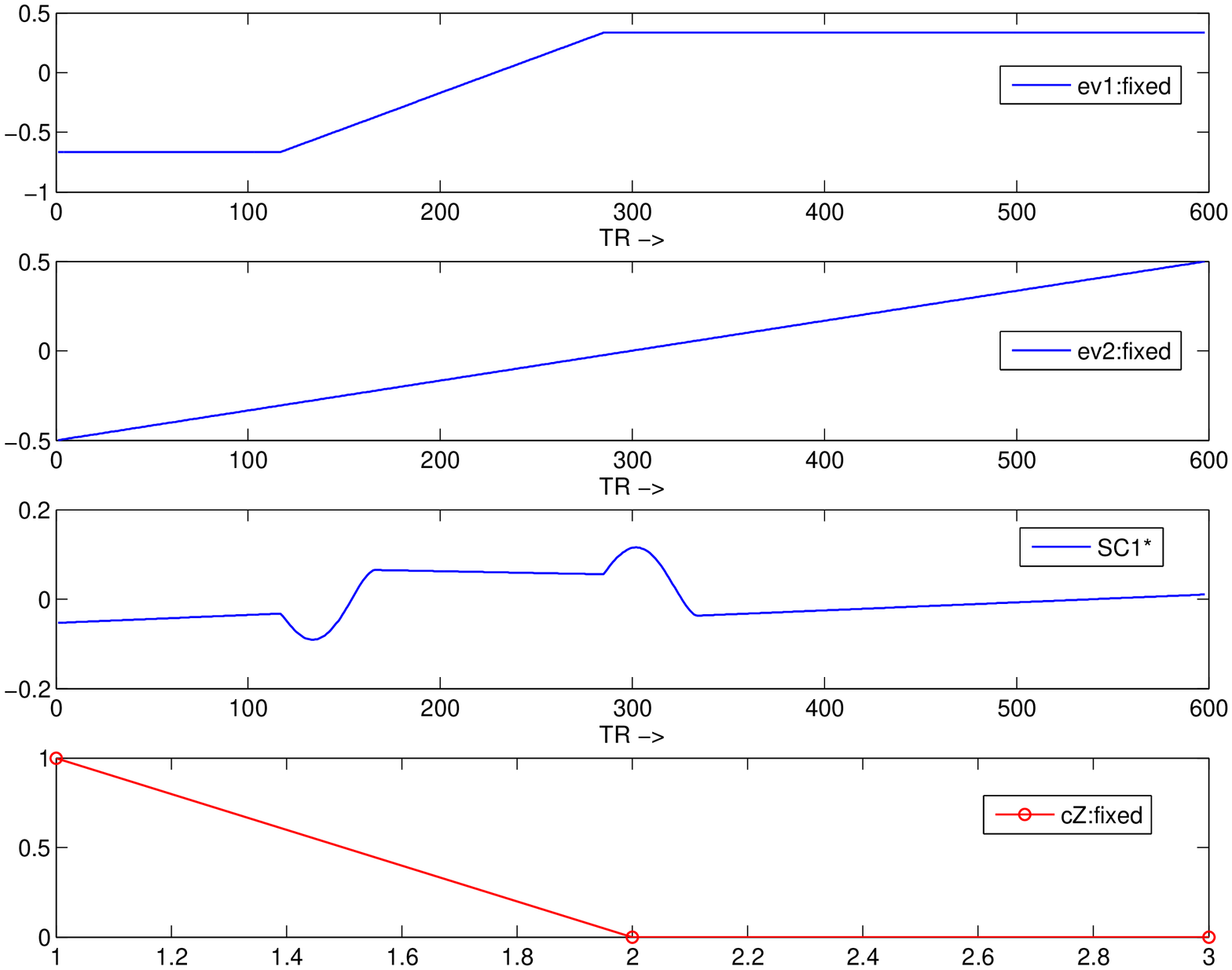}
}

\\

\subfigure[]
{
\hspace{-2cm}
\label{case1Ac}
\includegraphics[width = 75mm, angle = -90]{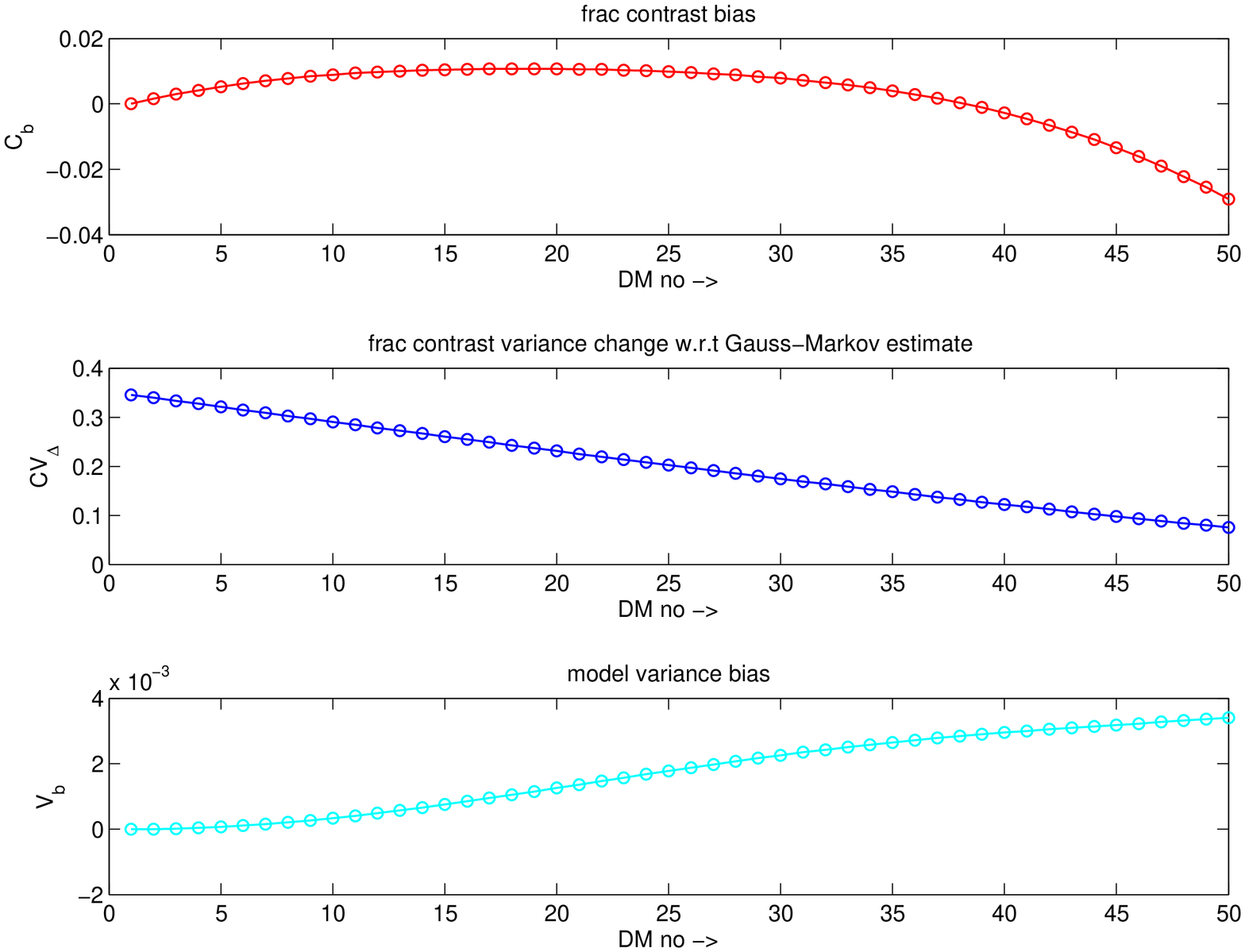}
}

&

\subfigure[]
{
\hspace{-1cm}
\label{case1Ad}
\includegraphics[width = 75mm, angle = -90]{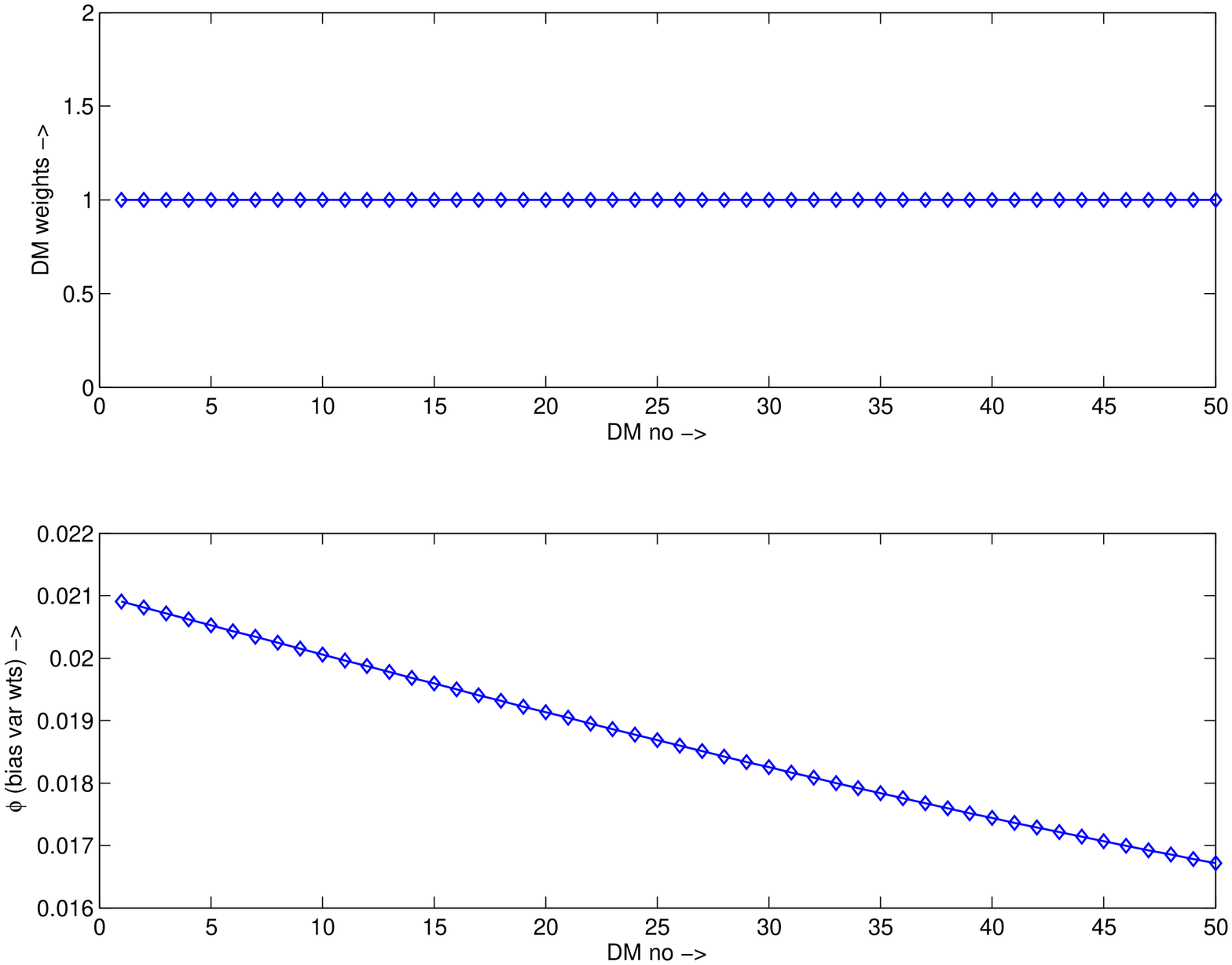}
}

\end{tabular}
\caption{ Example 2: (a) Initial design matrix (DM) initialized using the automatic strategy proposed in section \ref{initializationstrategy}. The first two columns were fixed at their initial values and the contrast was fixed at [1;0;0] (b) Estimated optimal DM. (c) Performance curves showing the fractional contrast bias $C_b$, contrast variance change w.r.t Gauss-Markov estimate $CV_{\Delta}$ and model variance bias $V_b$ (d) For all $i$, the DM weights $w_i = 1$ implying equal likelihood of observing any of the specified DM's.  The bias-variance scalings $\phi_i$ were chosen using the automatic strategy proposed in section \ref{intelligentphi} using $k = 10^3$. }
\label{case1AA}
\end{figure}

\begin{figure}[htbp]
\centering
\begin{tabular}{cc}

\subfigure[]
{
\hspace{-2cm}
\label{case1Ae}
\includegraphics[width = 75mm, angle = -90]{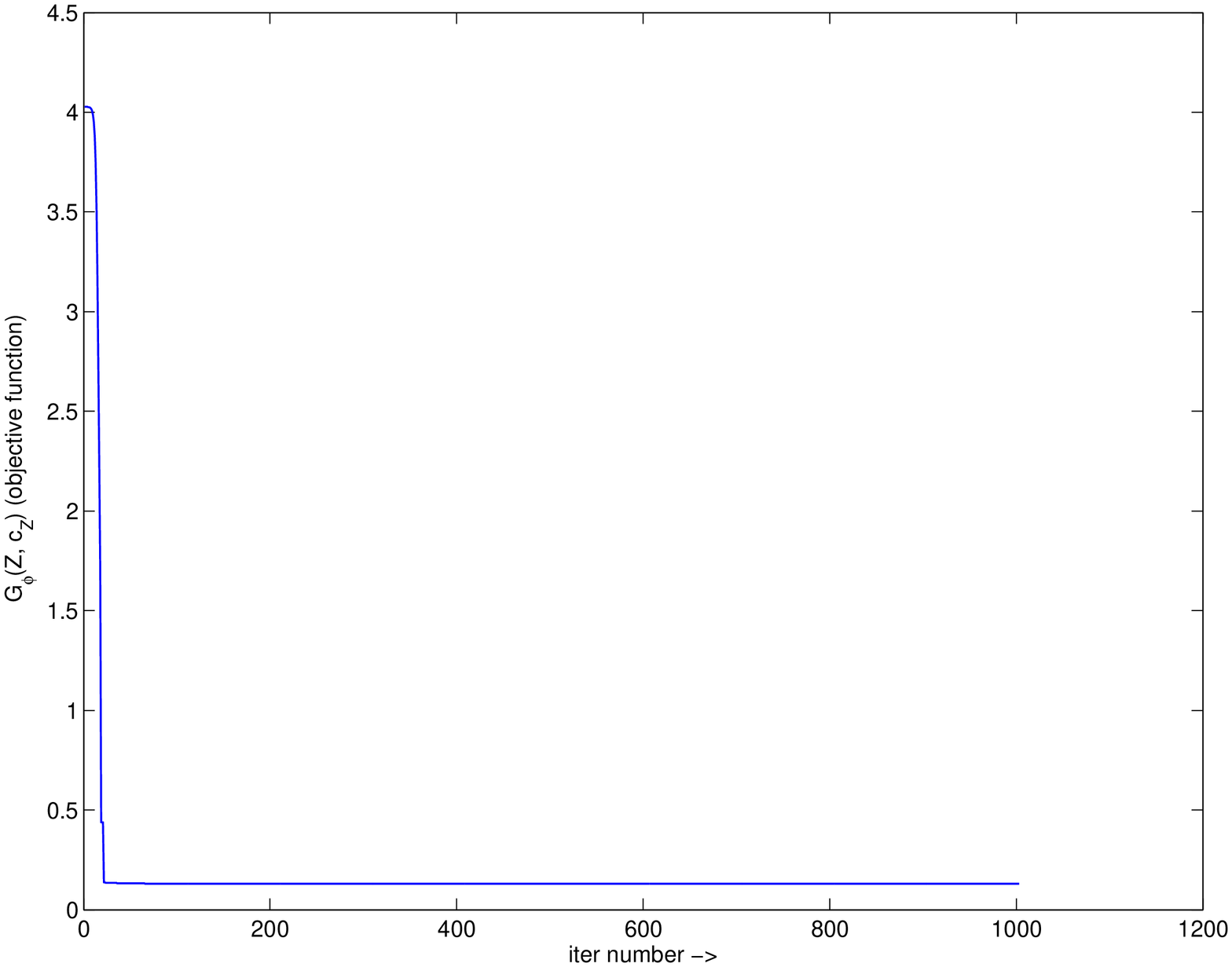}
}

&

\subfigure[]
{
\hspace{-1cm}
\label{case1Af}
\includegraphics[width = 75mm, angle = -90]{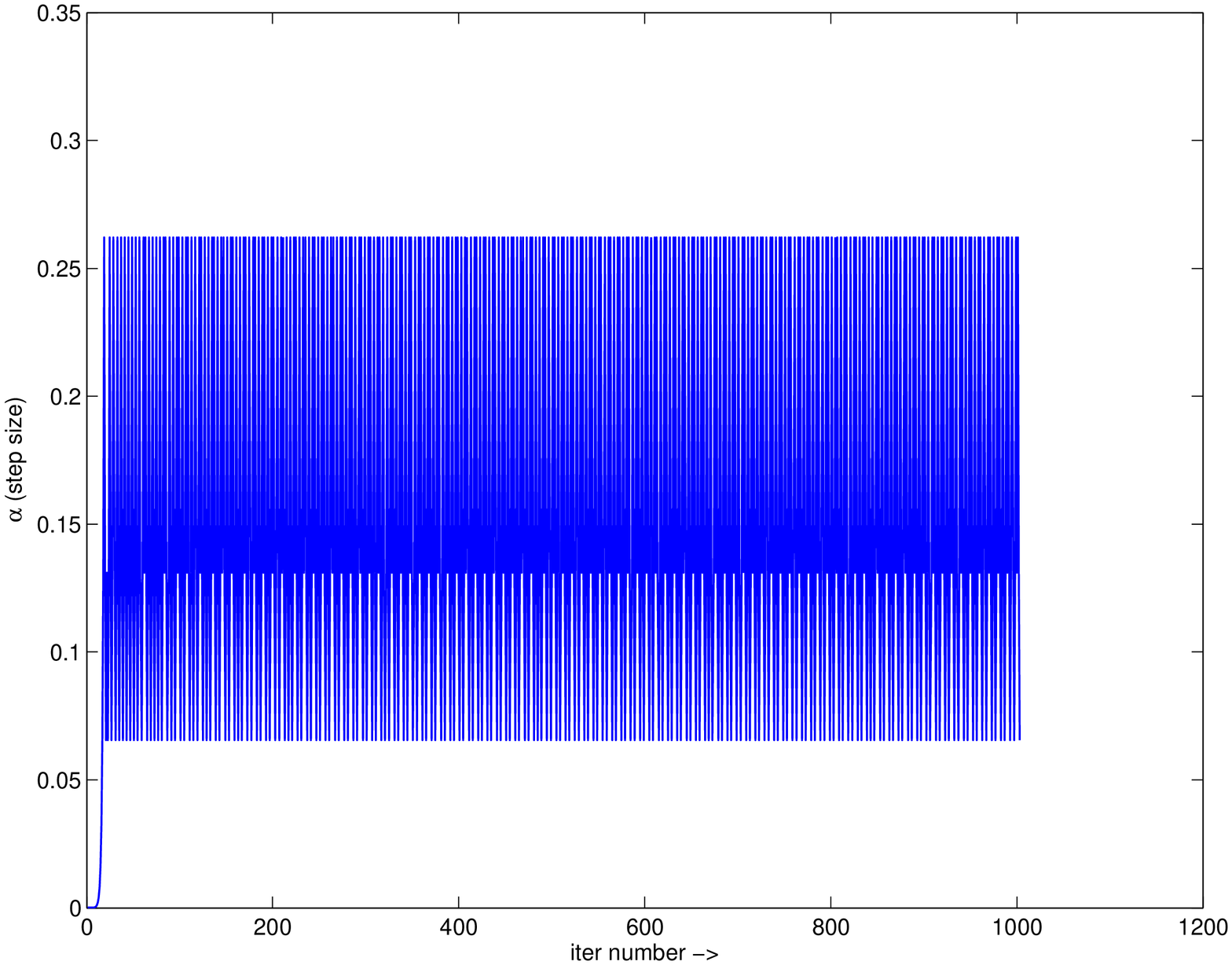}
}

\\

\subfigure[]
{
\hspace{-2cm}
\label{case1Ag}
\includegraphics[width = 75mm, angle = -90]{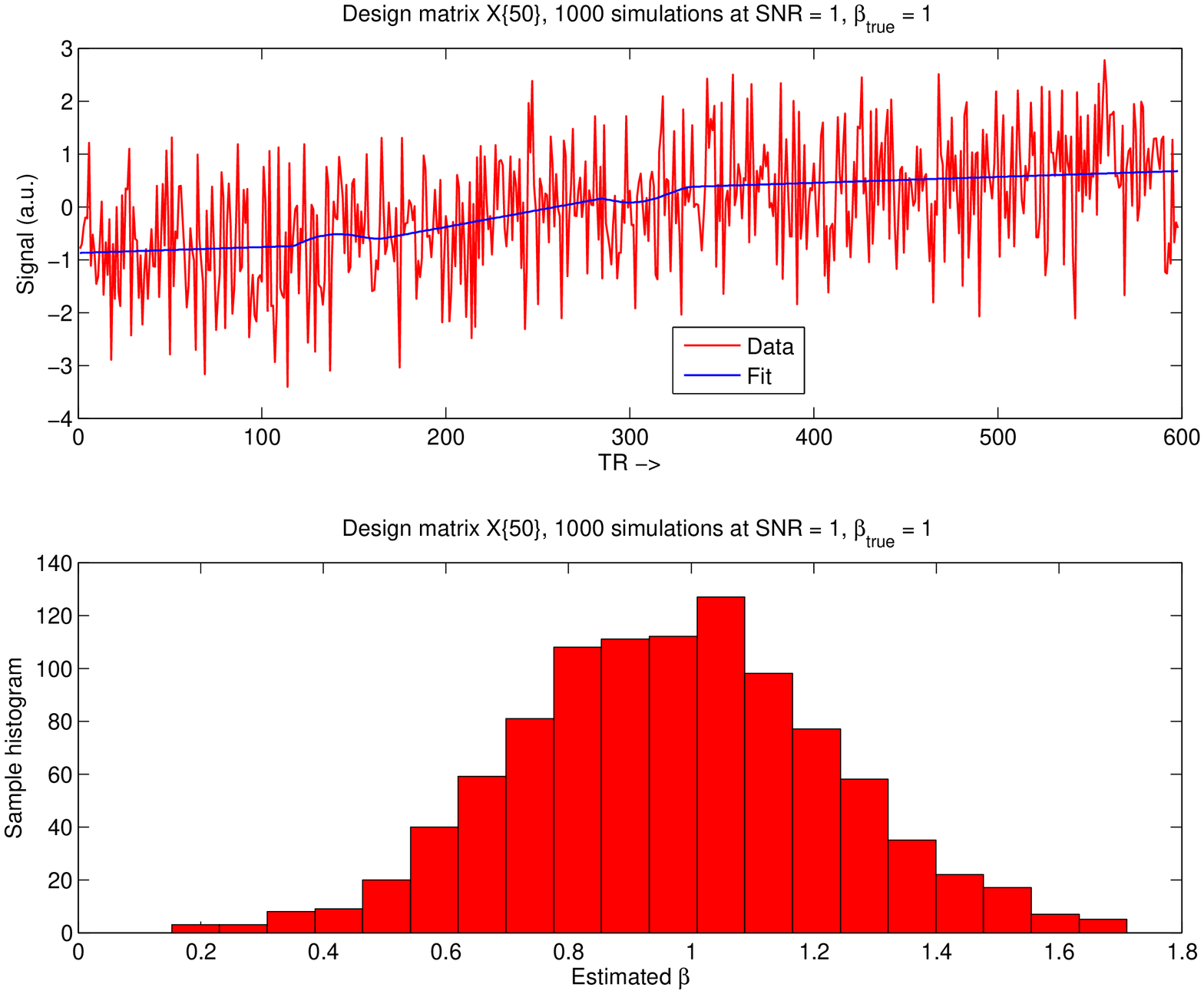}
}

&

\subfigure[]
{
\hspace{-1cm}
\label{case1Ah}
\includegraphics[width = 75mm, angle = -90]{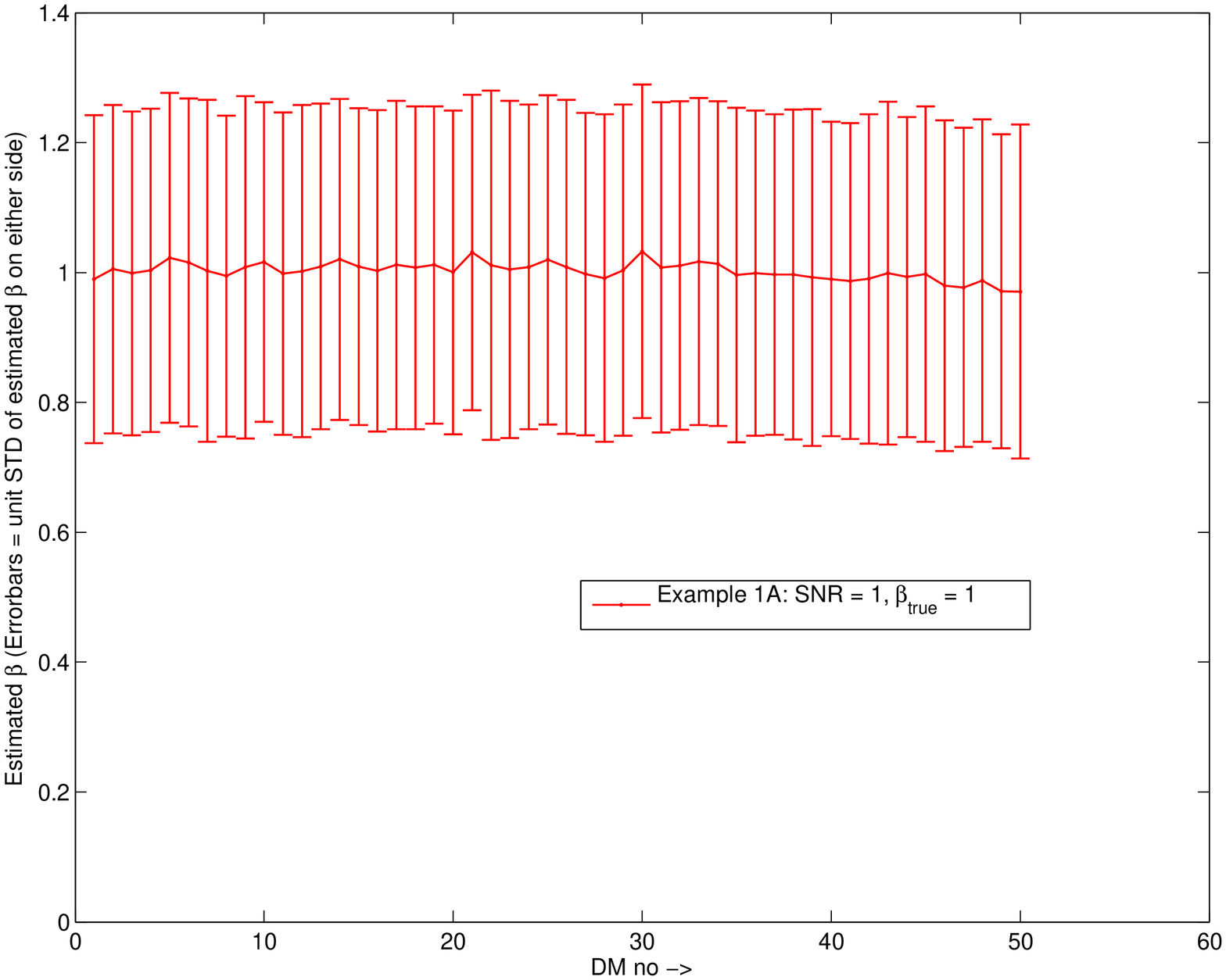}
}

\end{tabular}
\caption{  Example 2: (a) Figure showing the evolution of objective function values $G_\phi(Z,c_Z)$ over algorithm iterations. Notice how the function value stabilizes as convergence is reached (b) Figure showing the variation in the step size $\alpha$ over algorithm iterations. Step size controlling parameter $\theta$ in Algorithm 1 was set to $\theta = 2$. For each design matrix (DM) $X_i$ entered into optimization, 1000 simulated data-sets were generated at SNR $\frac{\beta_i}{\sigma_i}$. A GLM analysis was run on each of these data-sets using the optimized DM. Figure (c) shows an example of simulated data for DM $X_{50}$ at SNR $\frac{\beta_{50}}{\sigma_{50}}$ and the GLM fit using the optimal DM. It also shows the distribution of $c_Z^T \hat{\gamma}$ over 1000 simulations. Figure (d) is a summary errorbar plot showing $\hat{E}(c_Z^T \hat{\gamma})$ over 1000 simulations for data generated from each DM. The error bars represent unit standard deviation of $c_Z^T \hat{\gamma}$ (\textbf{not} standard deviation of $\hat{E}(c_Z^T \hat{\gamma})$ )  to quantify the variance in estimation via simulation.}
\label{case1AB}
\end{figure}

This example is the same as Example 1 except for two differences: 
\begin{enumerate}
\item First we choose $\phi_i$ automatically using the strategy proposed in Section \ref{intelligentphi}. We used Option A to initialize $\phi$ using $k = 10^3$. 
\item Second, we initialize the columns of $Z$ automatically using the strategy proposed in Section \ref{initializationstrategy}.
\end{enumerate}
The results are shown in Figures \ref{case1AA} and \ref{case1AB}.

\subsection{Example 3}
\begin{figure}[htbp]
\centering
\begin{tabular}{cc}

\subfigure[]
{
\hspace{-2cm}
\label{case2a}
\includegraphics[width = 75mm, angle = -90]{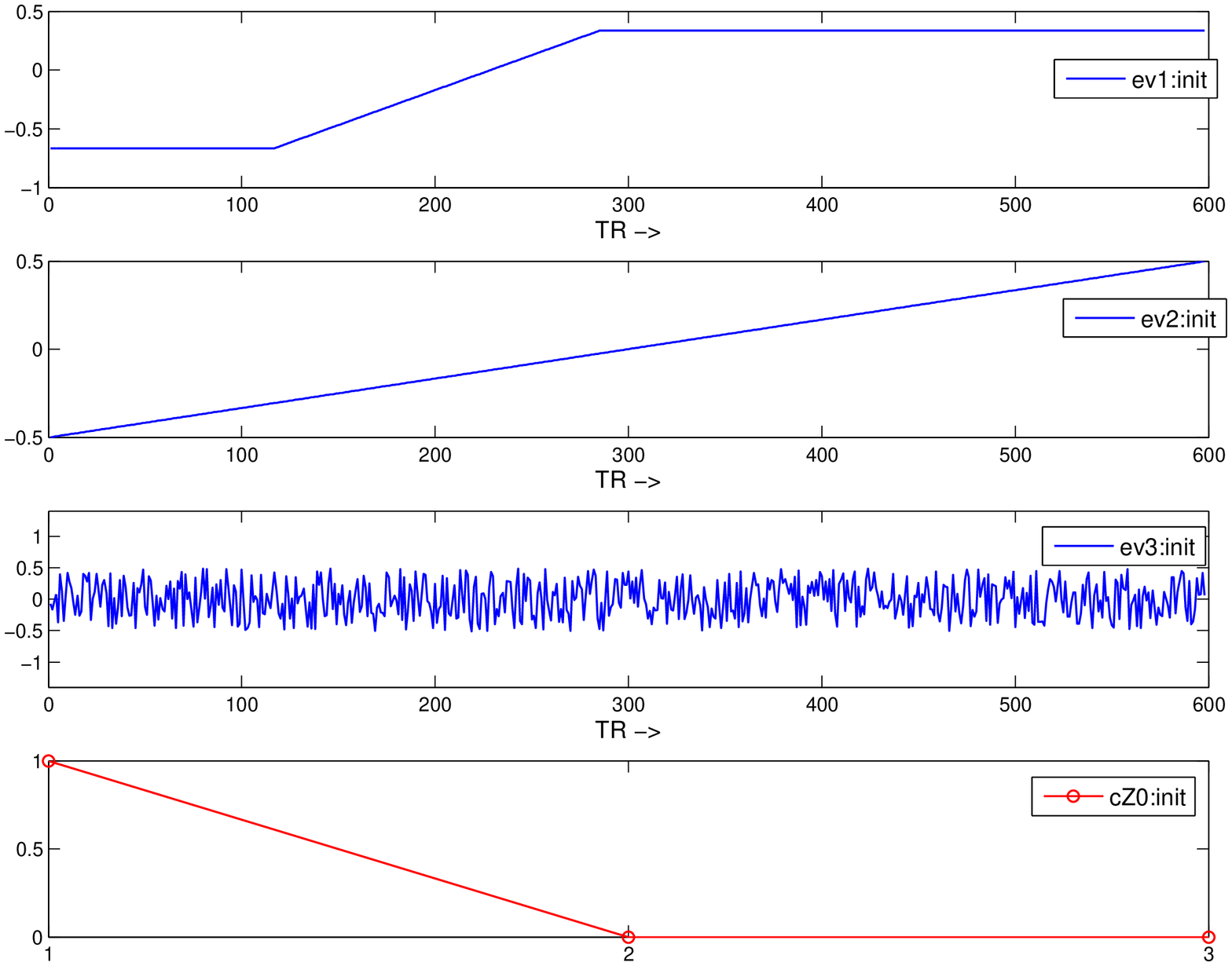}
}

&

\subfigure[]
{
\hspace{-1cm}
\label{case2b}
\includegraphics[width = 75mm, angle = -90]{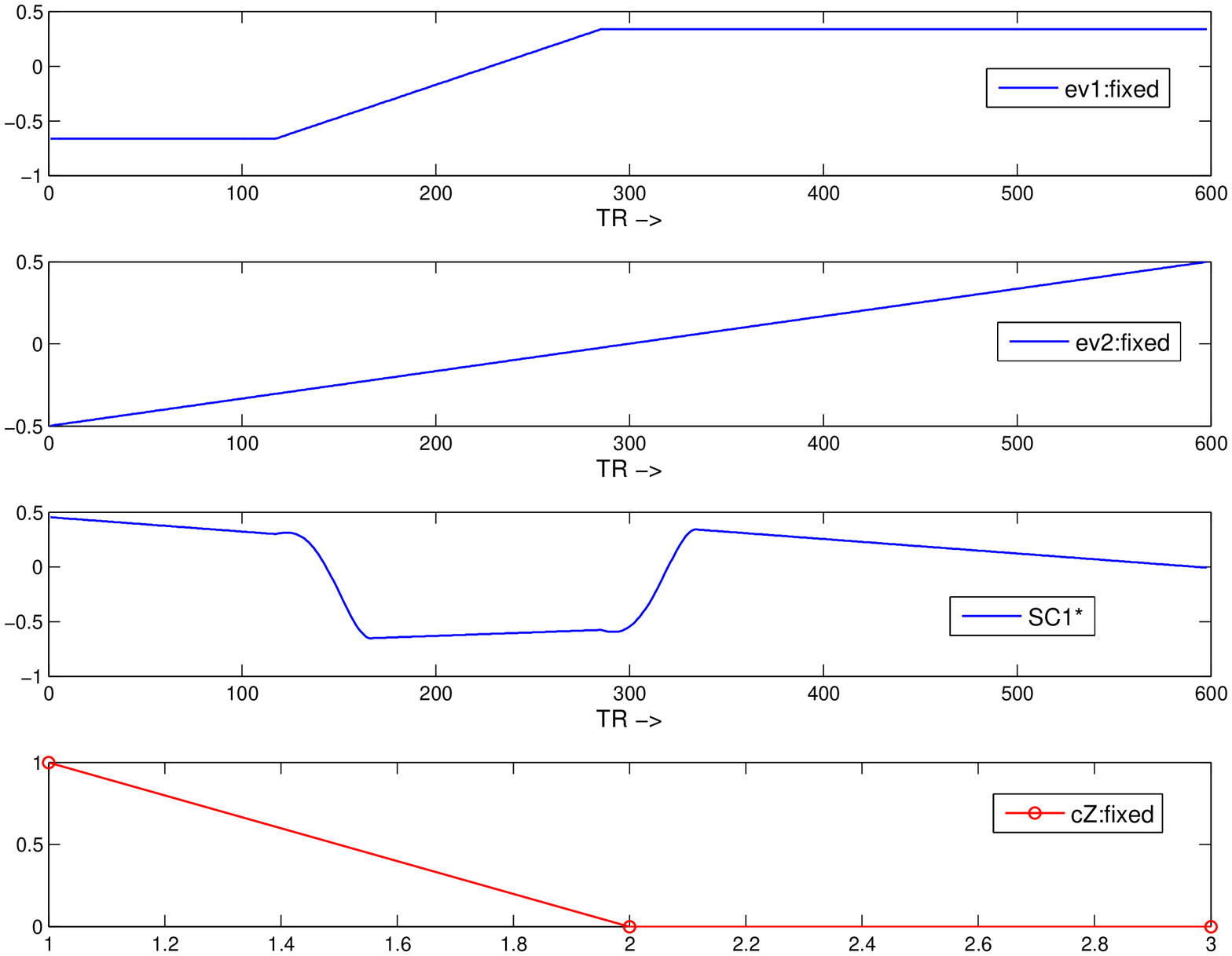}
}

\\

\subfigure[]
{
\hspace{-2cm}
\label{case2c}
\includegraphics[width = 75mm, angle = -90]{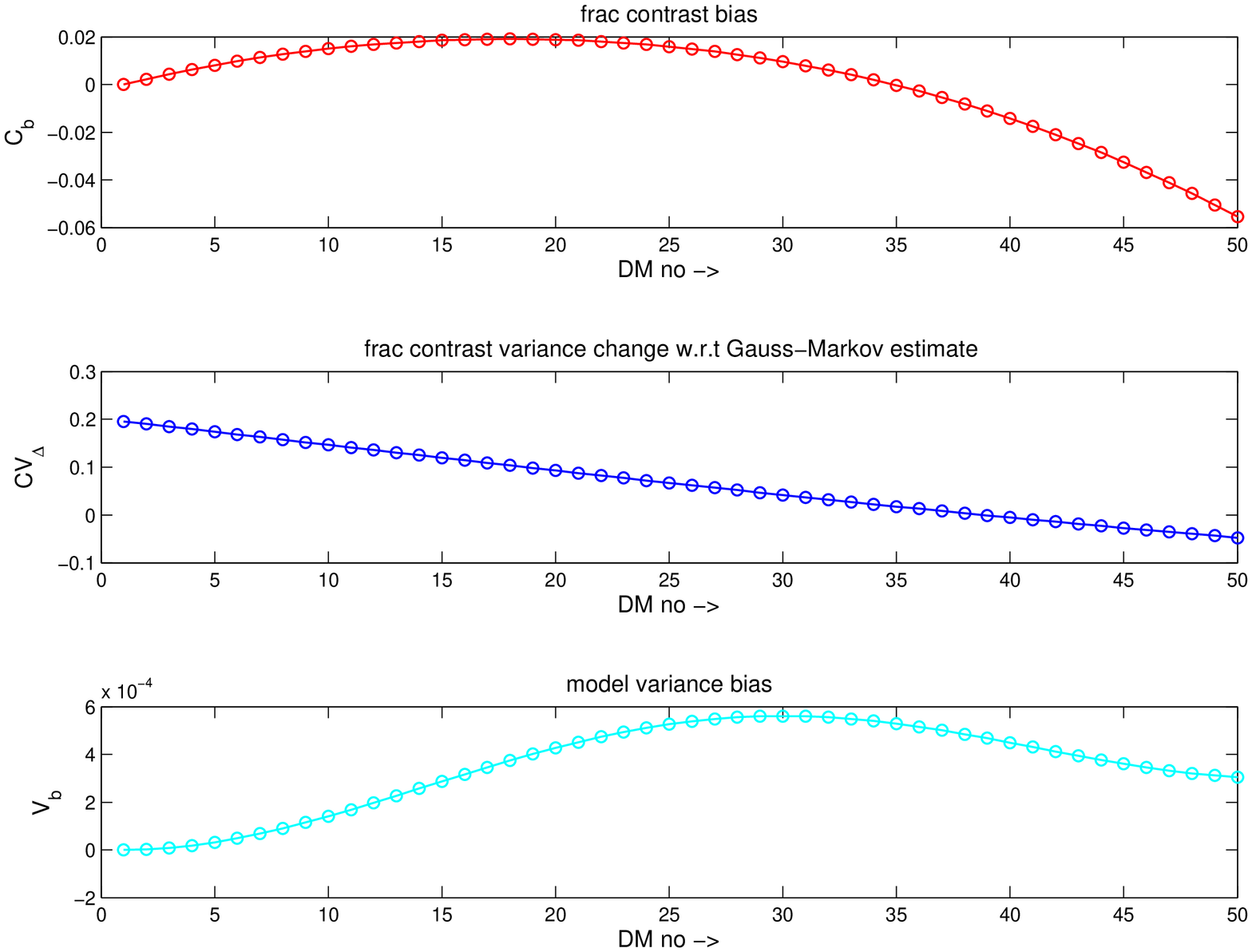}
}

&

\subfigure[]
{
\hspace{-1cm}
\label{case2d}
\includegraphics[width = 75mm, angle = -90]{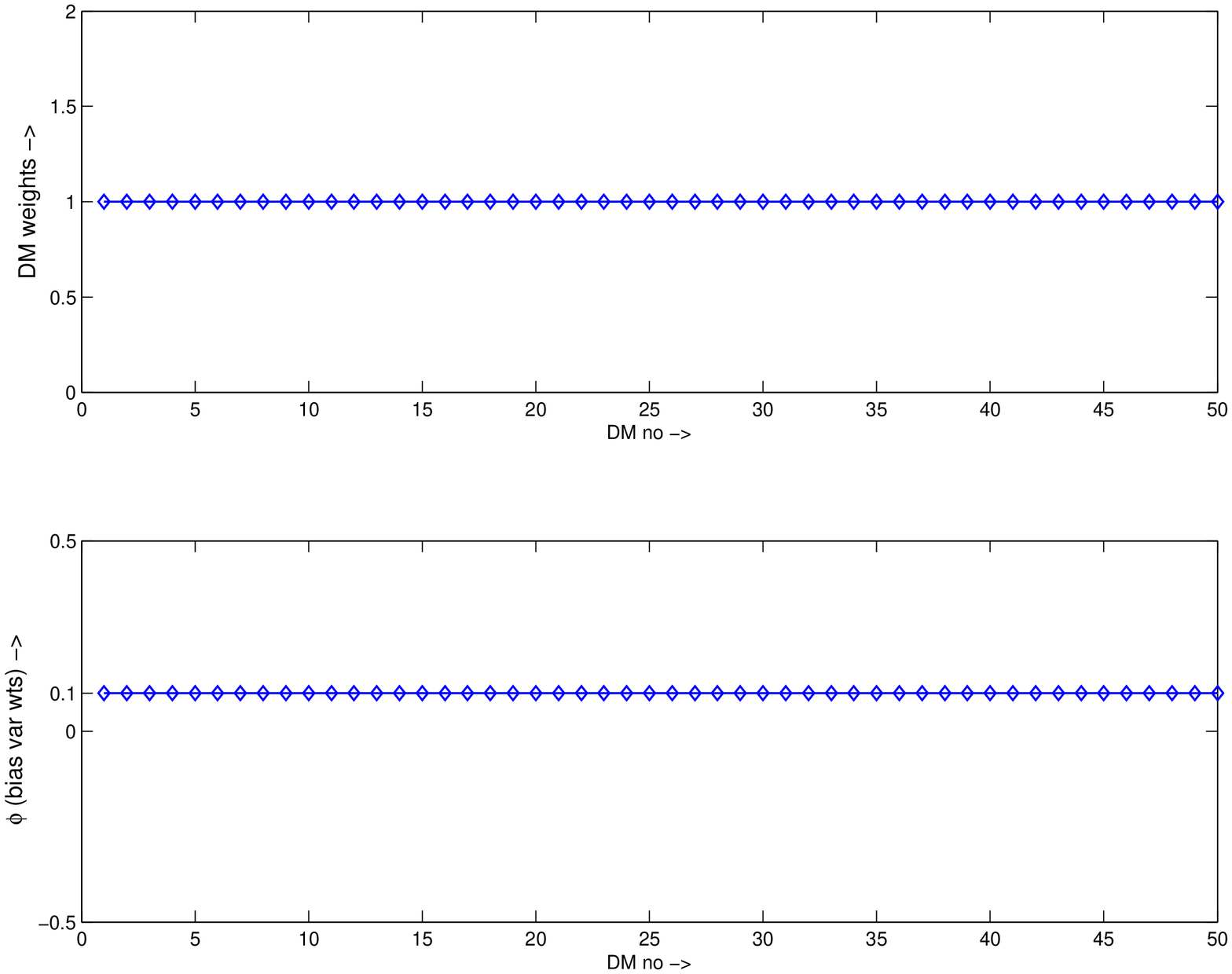}
}

\end{tabular}
\caption{  Example 3: (a) Initial design matrix (DM) along with random initialization of the 3rd column. The first two columns were fixed at their initial values and the contrast was fixed at [1;0;0] (b) Estimated optimal DM. Notice how the 3rd column converges to a non-random profile (c) Performance curves showing the fractional contrast bias $C_b$, contrast variance change w.r.t Gauss-Markov estimate $CV_{\Delta}$ and model variance bias $V_b$ (d) In this example $w_i = 1$ and $\phi_i$ = 0.1 indicating a higher weighting to bias term during optimization.}
\label{case2A}
\end{figure}

\begin{figure}[htbp]
\centering
\begin{tabular}{cc}

\subfigure[]
{
\hspace{-2cm}
\label{case2e}
\includegraphics[width = 75mm, angle = -90]{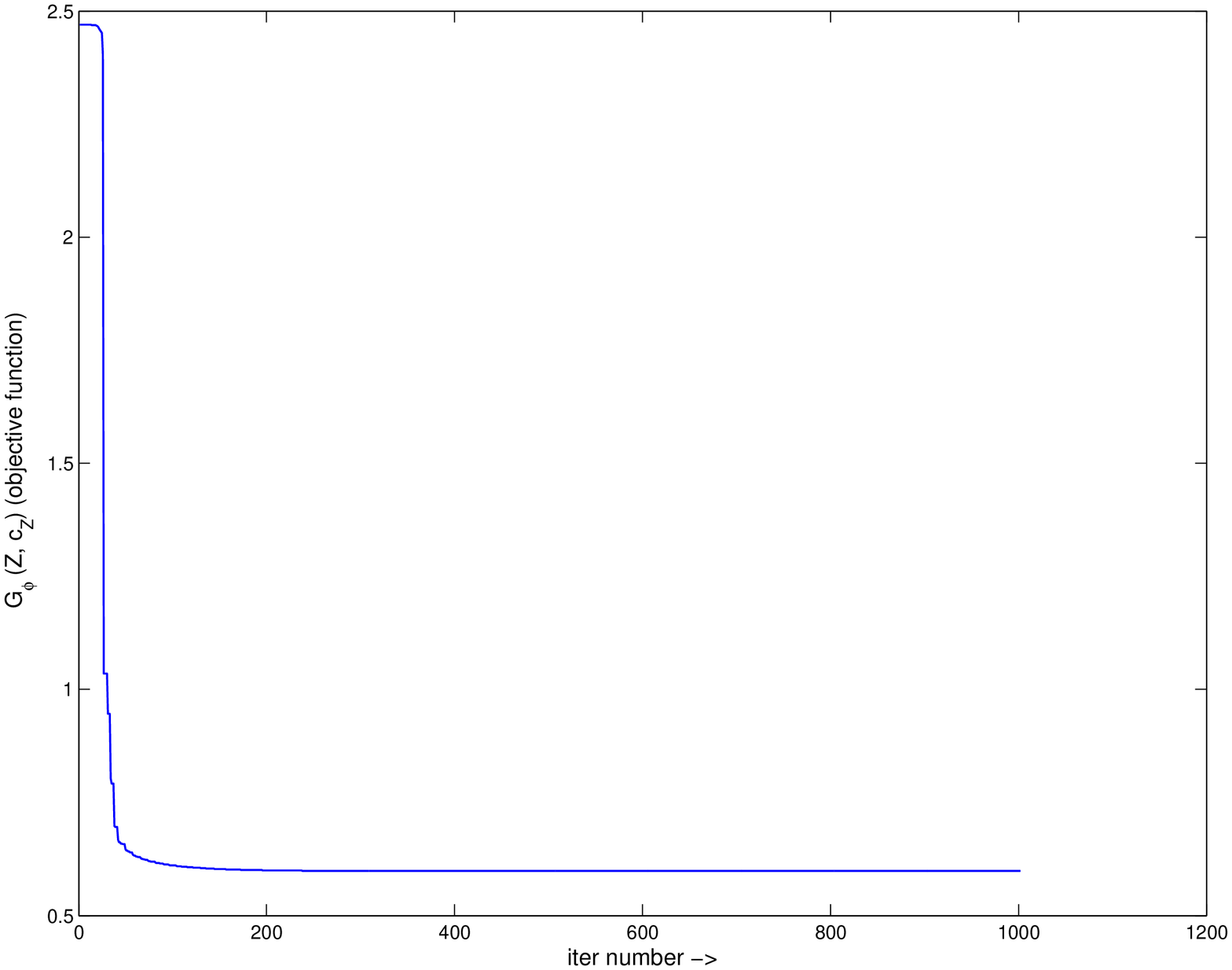}
}

&

\subfigure[]
{
\hspace{-1cm}
\label{case2f}
\includegraphics[width = 75mm, angle = -90]{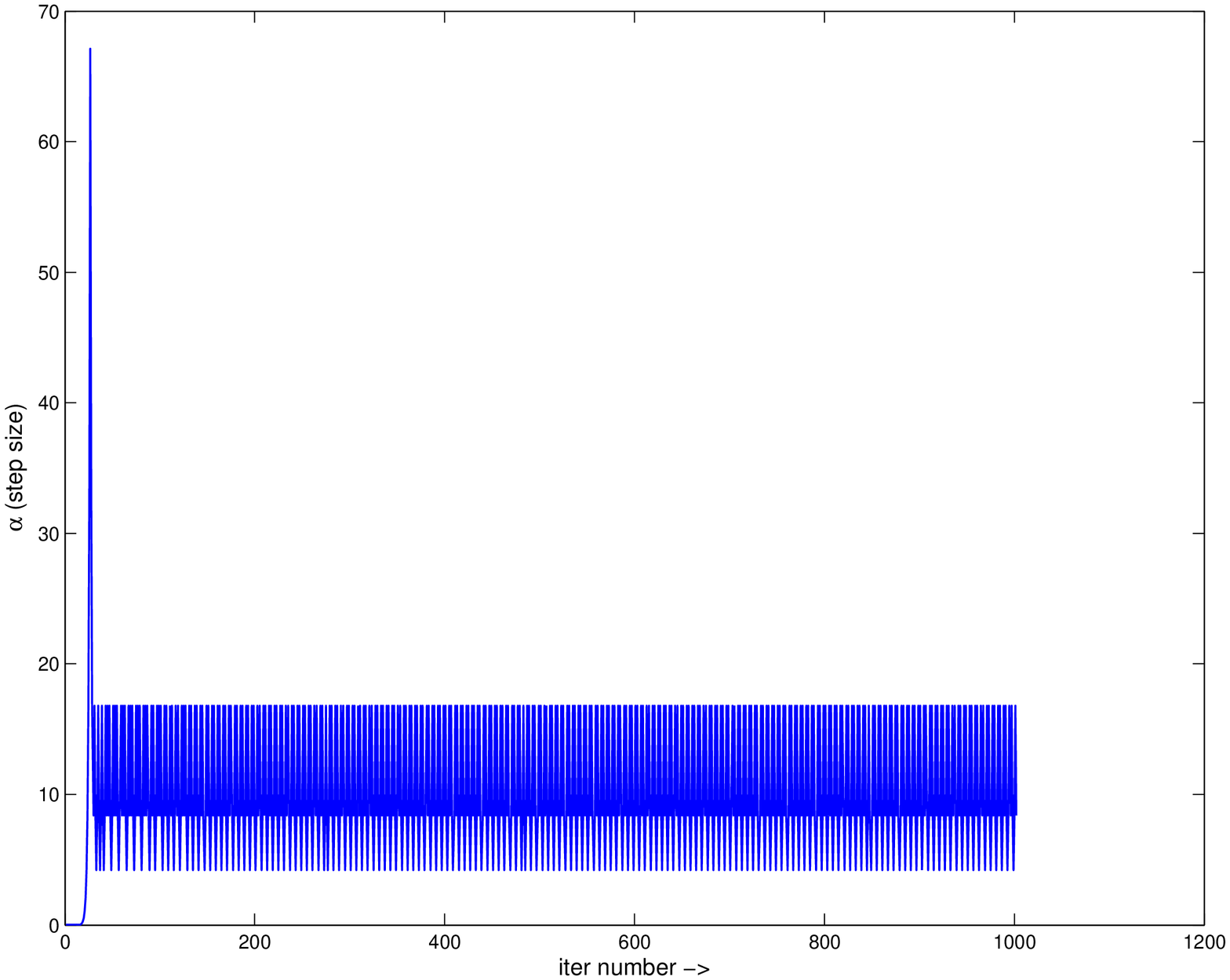}
}

\\

\subfigure[]
{
\hspace{-2cm}
\label{case2g}
\includegraphics[width = 75mm, angle = -90]{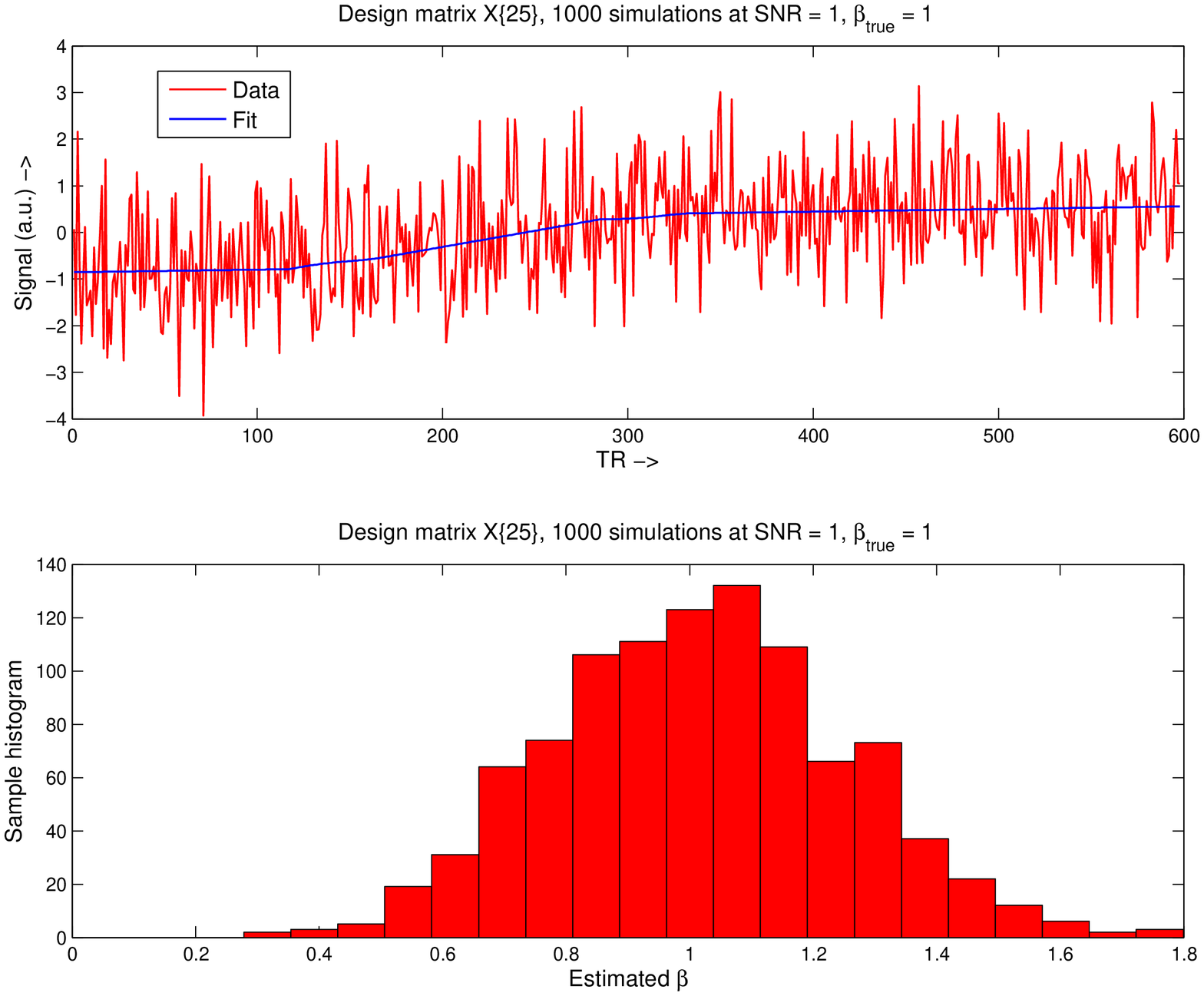}
}

&

\subfigure[]
{
\hspace{-1cm}
\label{case2h}
\includegraphics[width = 75mm, angle = -90]{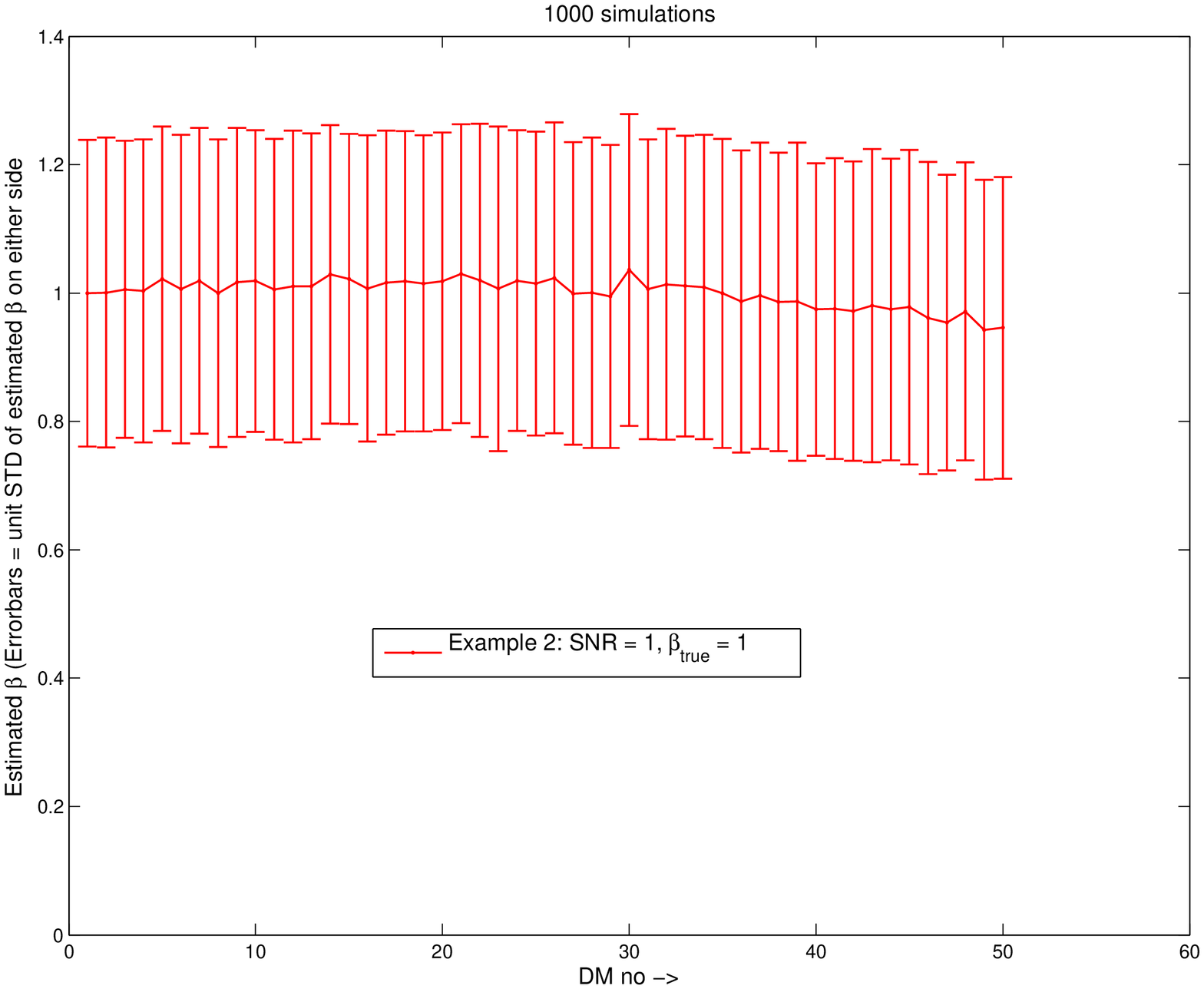}
}

\end{tabular}
\caption{  Example 3: (a) Figure showing the evolution of objective function values $G_\phi(Z,c_Z)$ over algorithm iterations. Notice how the function value stabilizes as convergence is reached (b) Figure showing the variation in the step size $\alpha$ over algorithm iterations. Step size controlling parameter $\theta$ in Algorithm 1 was set to $\theta = 2$. For each design matrix (DM) $X_i$ entered into optimization, 1000 simulated data-sets were generated at SNR $\frac{\beta_i}{\sigma_i}$. A GLM analysis was run on each of these data-sets using the optimized DM. Figure (c) shows an example of simulated data for DM $X_{25}$ at SNR $\frac{\beta_{25}}{\sigma_{25}}$ and the GLM fit using the optimal DM. It also shows the distribution of $c_Z^T \hat{\gamma}$ over 1000 simulations. Figure (d) is a summary errorbar plot showing $\hat{E}(c_Z^T \hat{\gamma})$ over 1000 simulations for data generated from each DM. The error bars represent unit standard deviation of $c_Z^T \hat{\gamma}$ (\textbf{not} standard deviation of $\hat{E}(c_Z^T \hat{\gamma})$ )  to quantify the variance in estimation via simulation.}
\label{case2B}
\end{figure}

In this example, we attempt to investigate the effect of bias-variance weightings on the performance curves. We put a higher emphasis on reducing bias by choosing $\phi = 0.1$. Please note that the relationship between the value of $\phi_i$ and the importance of bias or variance terms is non-linear (see section \ref{intelligentphi}). The first two columns of $Z$ were fixed as before and the contrast $c_Z$ was fixed at [1;0;0]. We also chose $w_i = 1$ to give equal weights to all design matrices. The unconstrained column in $Z$ was initialized randomly with elements drawn from a uniform distribution $U(0,1)$. The results are shown in figure \ref{case2A} and \ref{case2B}.


\subsection{Example 4}
\begin{figure}[htbp]
\centering
\begin{tabular}{cc}

\subfigure[]
{
\hspace{-2cm}
\label{case3a}
\includegraphics[width = 75mm, angle = -90]{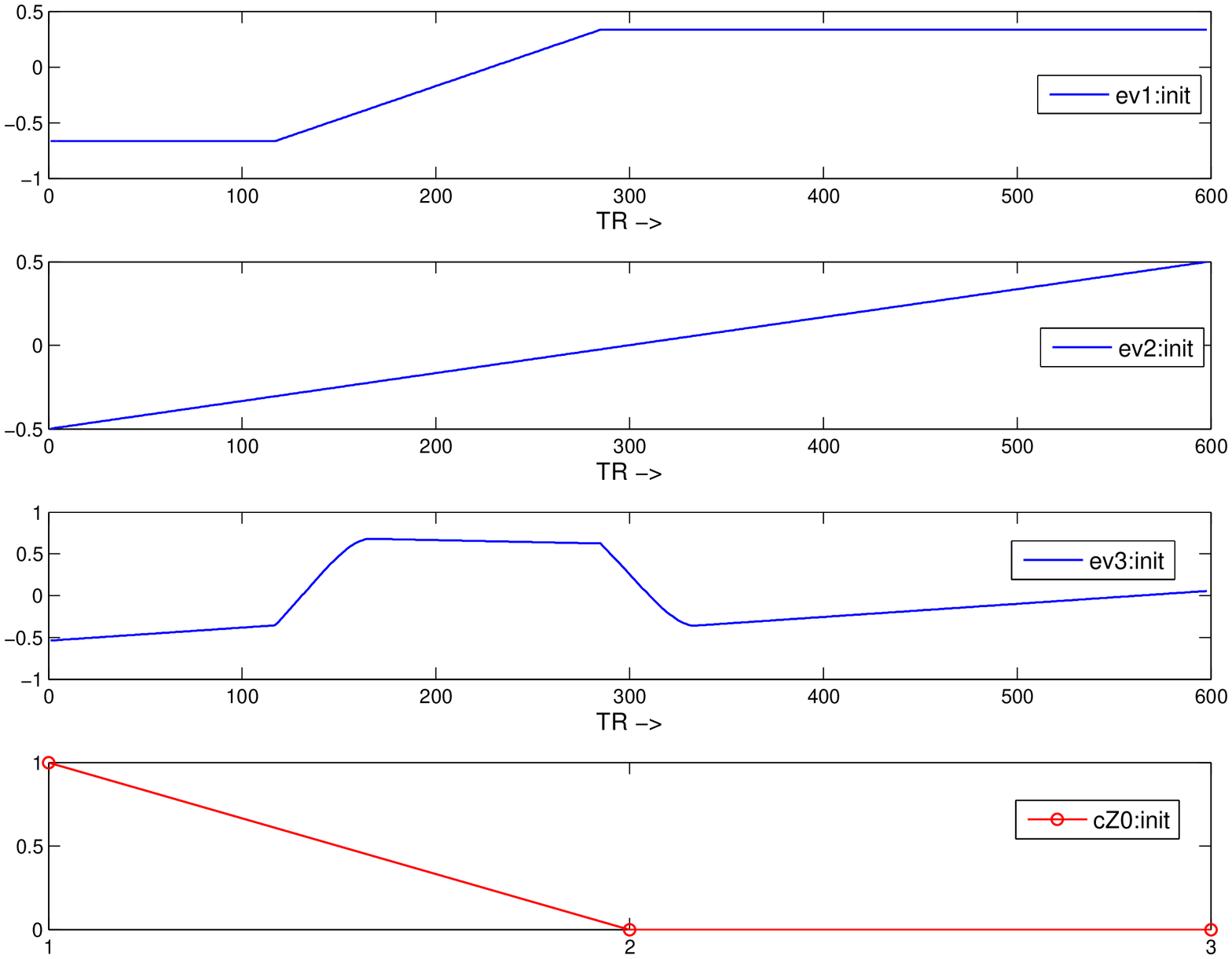}
}

&

\subfigure[]
{
\hspace{-1cm}
\label{case3b}
\includegraphics[width = 75mm, angle = -90]{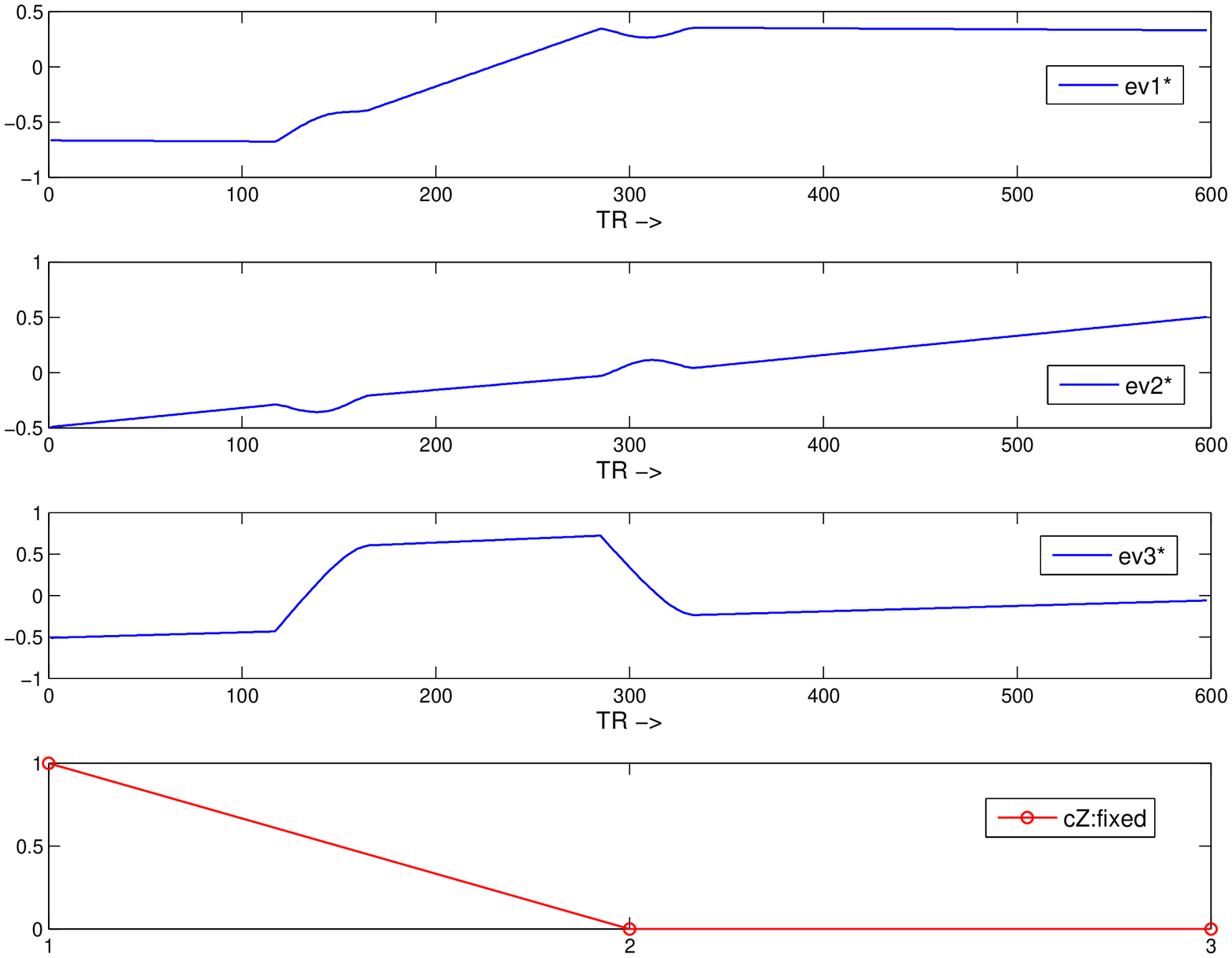}
}

\\

\subfigure[]
{
\hspace{-2cm}
\label{case3c}
\includegraphics[width = 75mm, angle = -90]{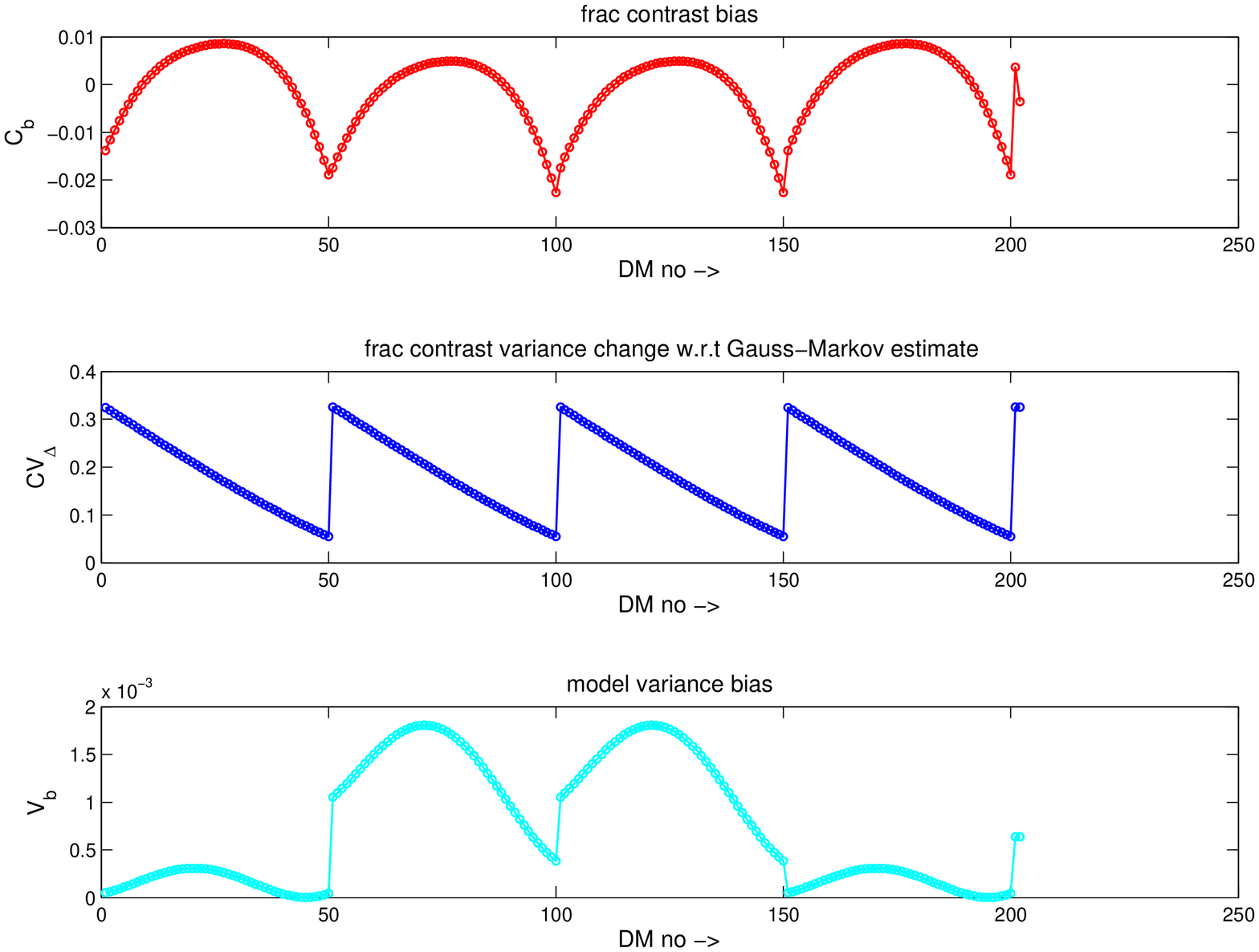}
}

&

\subfigure[]
{
\hspace{-1cm}
\label{case3d}
\includegraphics[width = 75mm, angle = -90]{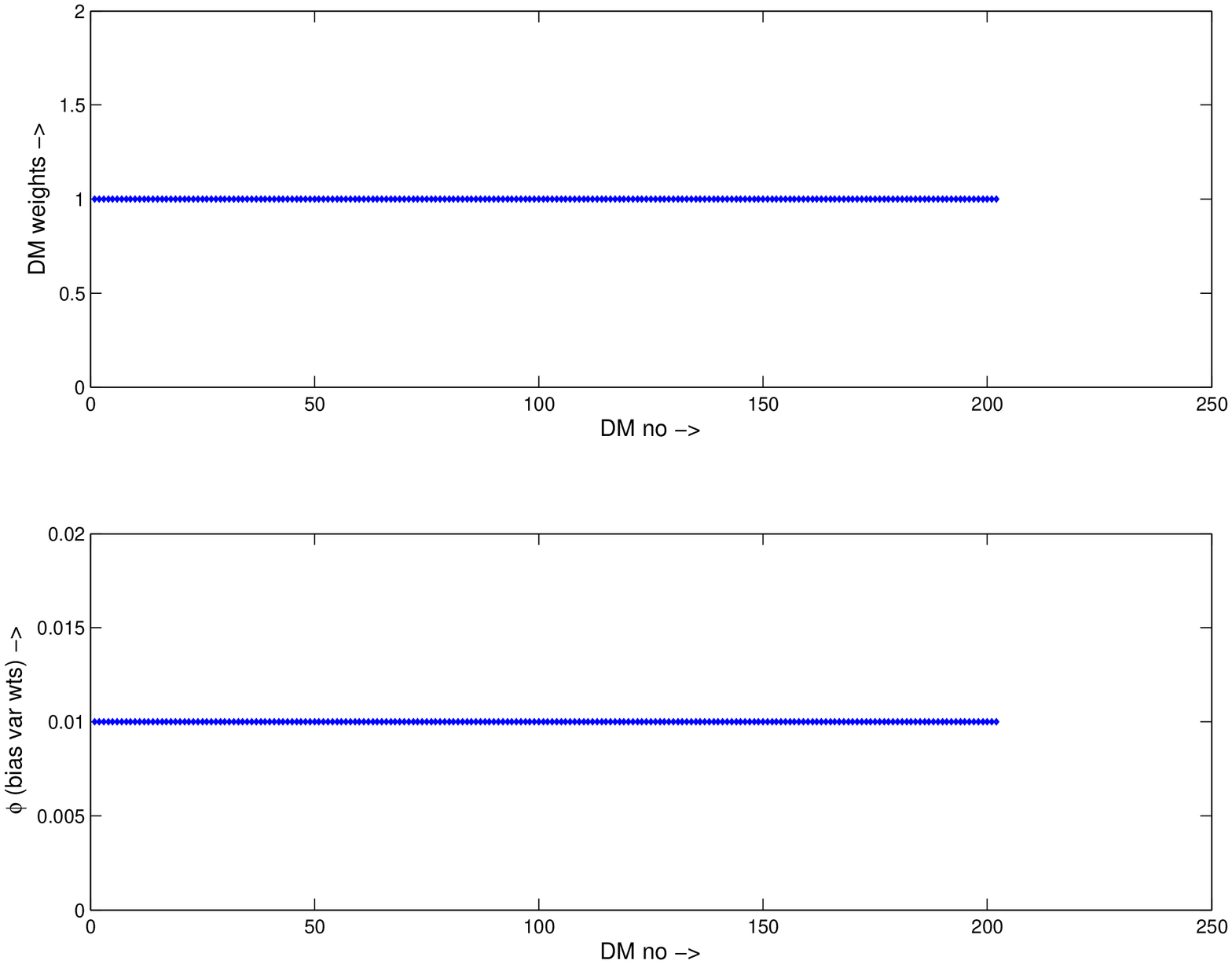}
}

\end{tabular}
\caption{  Example 4: (a) Initial design matrix (DM). The 3rd column was initialized using the 3rd col of $\hat{Z}$ found in Example 1. The first two columns were left unconstrained in this case but the contrast was fixed at [1;0;0]. The set of potential DM's from Example 1 and 2 was augmented by adding more DM's at SNR's of $[1;-0.5], [-1;0.5],[-1;-0.5]$. A "null" DM was also added at SNR $[0;1]$ and $[0;-1]$ indicating that "pure drift" should be matched to size "0". (b) Estimated optimal DM. Notice how the unconstrained columns 1 and 2 converge to non-intuitive shapes (c) Performance curves showing the fractional contrast bias $C_b$, contrast variance change w.r.t Gauss-Markov estimate $CV_{\Delta}$ and model variance bias $V_b$ (d) In this example $w_i = 1$ and $\phi_i$ = 0.01 indicating a higher weight to the bias term during optimization. }
\label{case3A}
\end{figure}

\begin{figure}[htbp]
\centering
\begin{tabular}{cc}

\subfigure[]
{
\hspace{-2cm}
\label{case3e}
\includegraphics[width = 75mm, angle = -90]{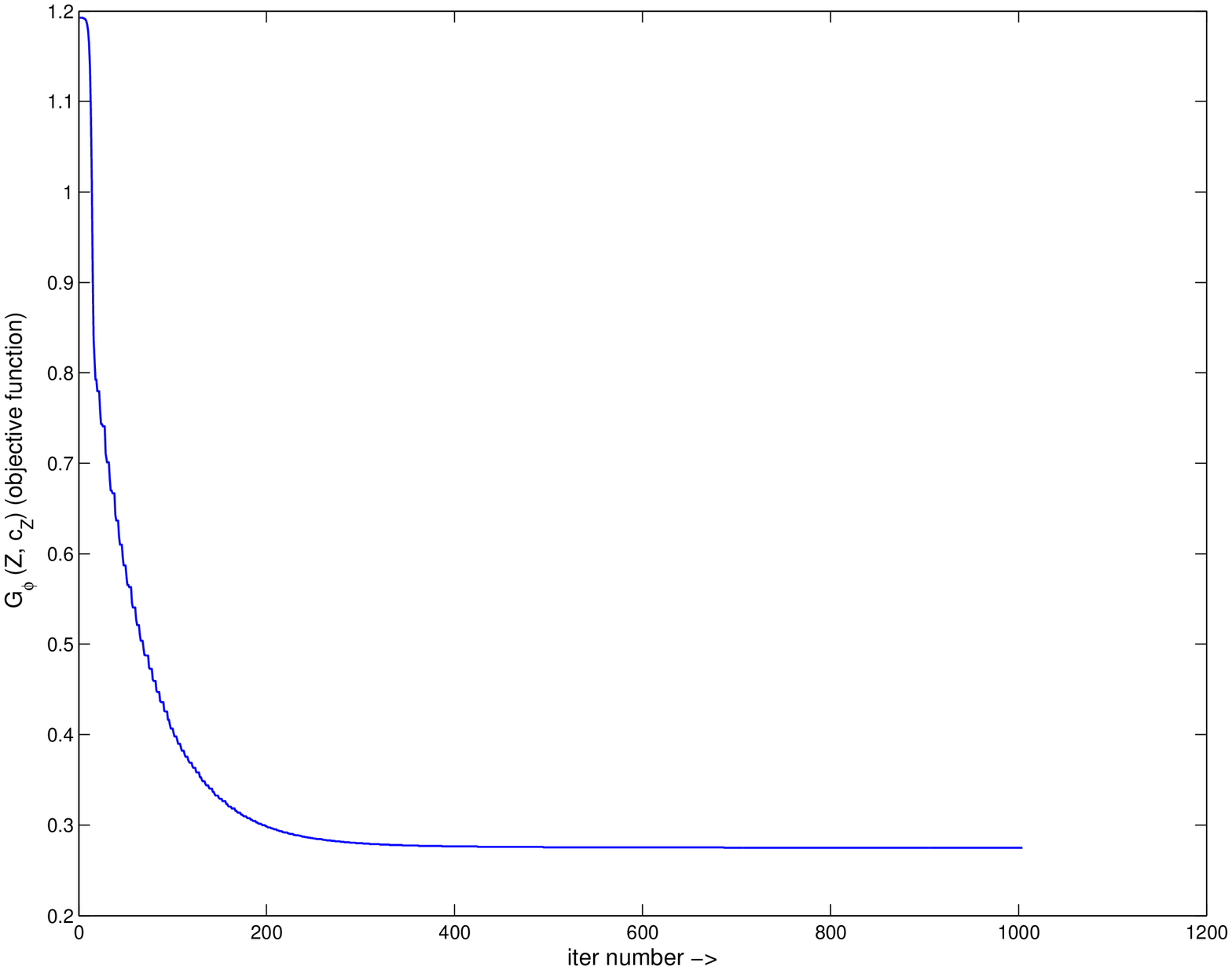}
}

&

\subfigure[]
{
\hspace{-1cm}
\label{case3f}
\includegraphics[width = 75mm, angle = -90]{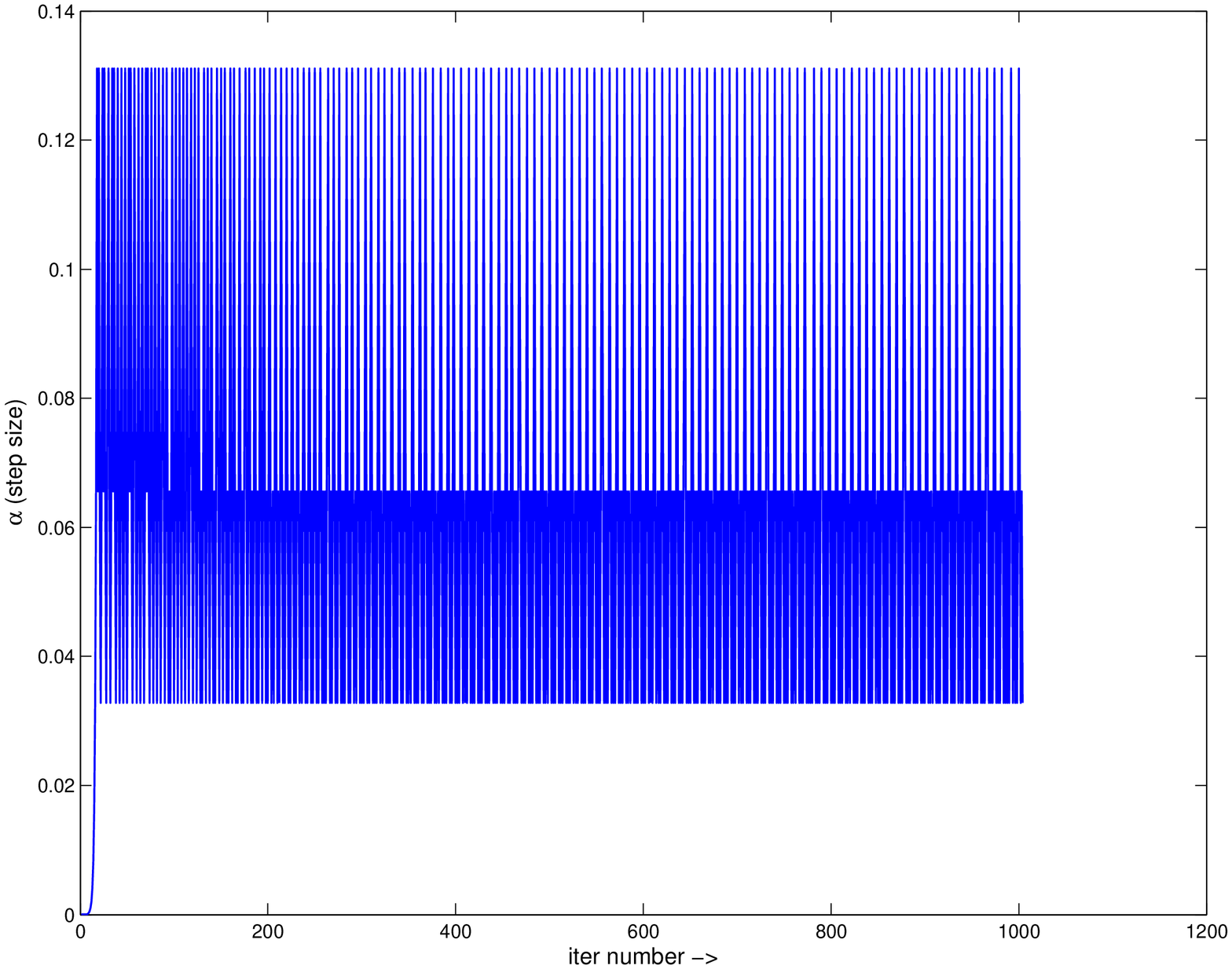}
}

\\

\subfigure[]
{
\hspace{-2cm}
\label{case3g}
\includegraphics[width = 75mm, angle = -90]{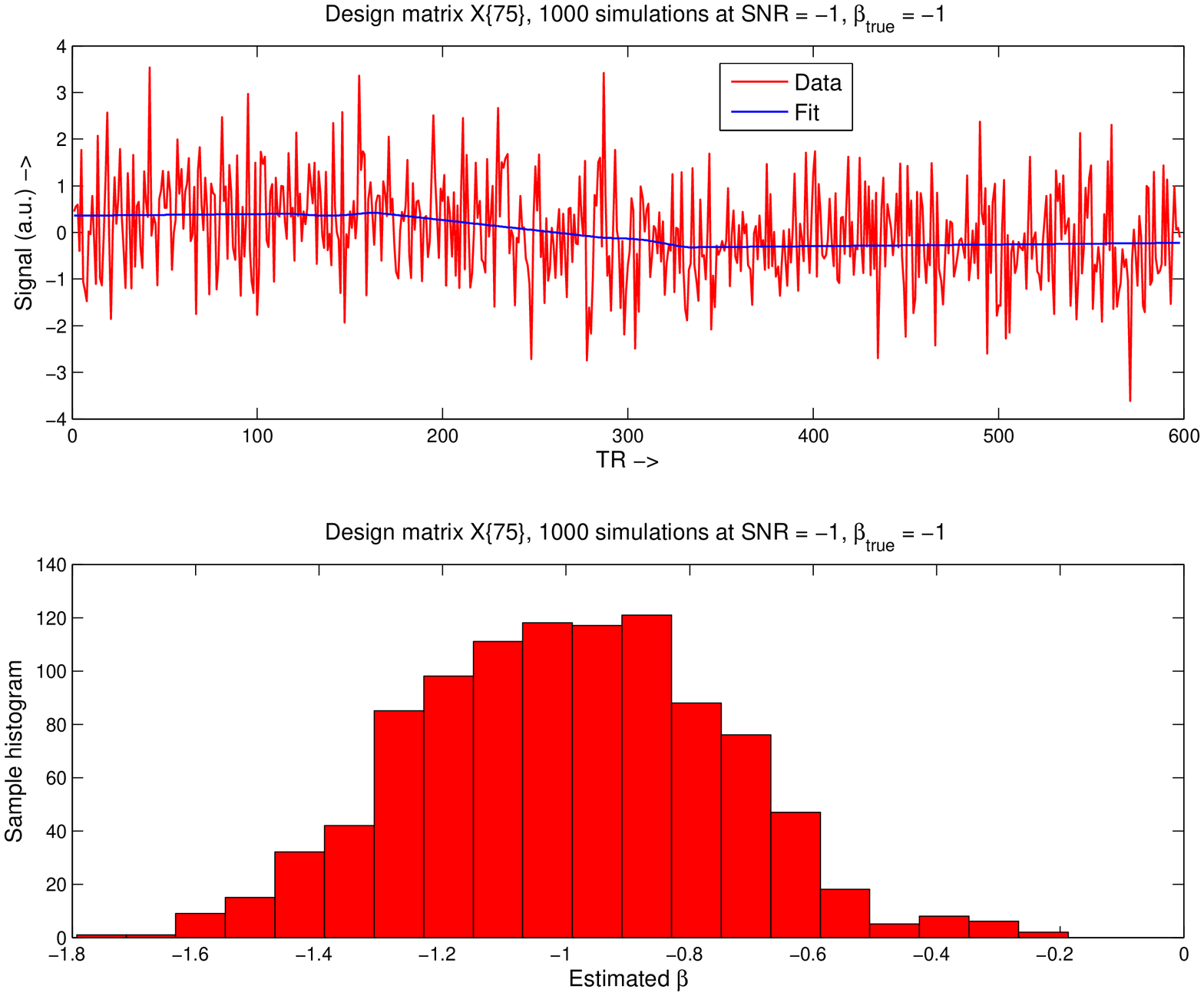}
}

&

\subfigure[]
{
\hspace{-1cm}
\label{case3h}
\includegraphics[width = 75mm, angle = -90]{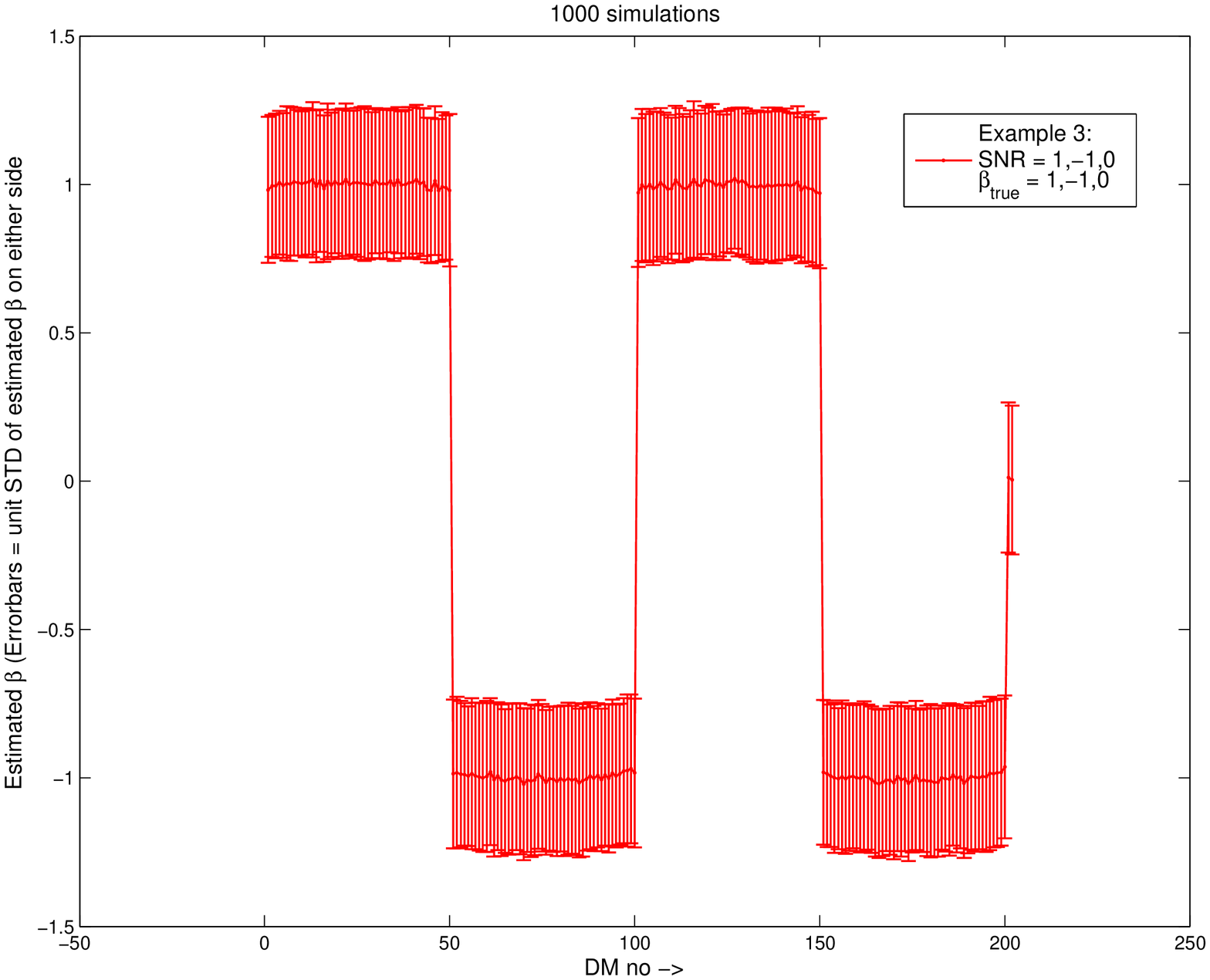}
}

\end{tabular}
\caption{  Example 4: (a) Figure showing the evolution of objective function values $G_\phi(Z,c_Z)$ over algorithm iterations. Notice how the function value stabilizes as convergence is reached (b) Figure showing the variation in the step size $\alpha$ over algorithm iterations. Step size controlling parameter $\theta$ in Algorithm 1 was set to $\theta = 2$ (c) , (d) For each design matrix (DM) entered into optimization, 1000 simulated data-sets were generated at SNR $\frac{\beta_i}{\sigma_i}$. A GLM analysis was run on each of these data-sets using the optimized DM. Figure (c) shows an example of simulated data for DM $X_{75}$ at SNR $\frac{\beta_{75}}{\sigma_{75}}$ and the GLM fit using the optimal DM. It also shows the distribution of $c_Z^T \hat{\gamma}$ over 1000 simulations. Figure (d) is a summary errorbar plot showing $\hat{E}(c_Z^T \hat{\gamma})$ over 1000 simulations for data generated from each DM. The error bars represent unit standard deviation of $c_Z^T \hat{\gamma}$ (\textbf{not} standard deviation of $\hat{E}(c_Z^T \hat{\gamma})$ )  to quantify the variance in estimation via simulation.}
\label{case3B}
\end{figure}

In this example, we attempt to investigate the effect of bias-variance weightings on the performance curves as well as the effect of optimizing the entire design matrix $Z$. We put a much higher emphasis on reducing bias by choosing $\phi = 0.01$. As before, note that the relationship between the value of $\phi_i$ and the importance of bias or variance terms is non-linear (see section \ref{intelligentphi}). The contrast $c_Z$ was fixed at [1;0;0] but the design matrix $Z$ was left unconstrained. We also chose $w_i = 1$ to give equal weights to all design matrices. The potential DM set from before was augmented with the following additional DM's
\begin{enumerate}
\item $X_i, c_{X_i}$ same as before but $\frac{\beta_i}{\sigma_i} = [-1;0.5]$
\item $X_i, c_{X_i}$ same as before but $\frac{\beta_i}{\sigma_i} = [1;-0.5]$
\item $X_i, c_{X_i}$ same as before but $\frac{\beta_i}{\sigma_i} = [-1;-0.5]$
\item $X_0, c_{X_0}$ same as before but $\frac{\beta_0}{\sigma_0} = [0;1]$
\end{enumerate}
This was done to constrain the sample space to make $Z$ unbiased for sign changes relative to drift (1,2, and 3) as well as unbiased for pure drift (4). The 3rd column of $Z$ was initialized to the optimal solution found in Example 1.
The results are shown in figure \ref{case3A} and \ref{case3B}.

\subsection{Example 5}
\begin{figure}[htbp]
\centering
\begin{tabular}{cc}

\subfigure[]
{
\hspace{-2cm}
\label{case4a}
\includegraphics[width = 75mm, angle = -90]{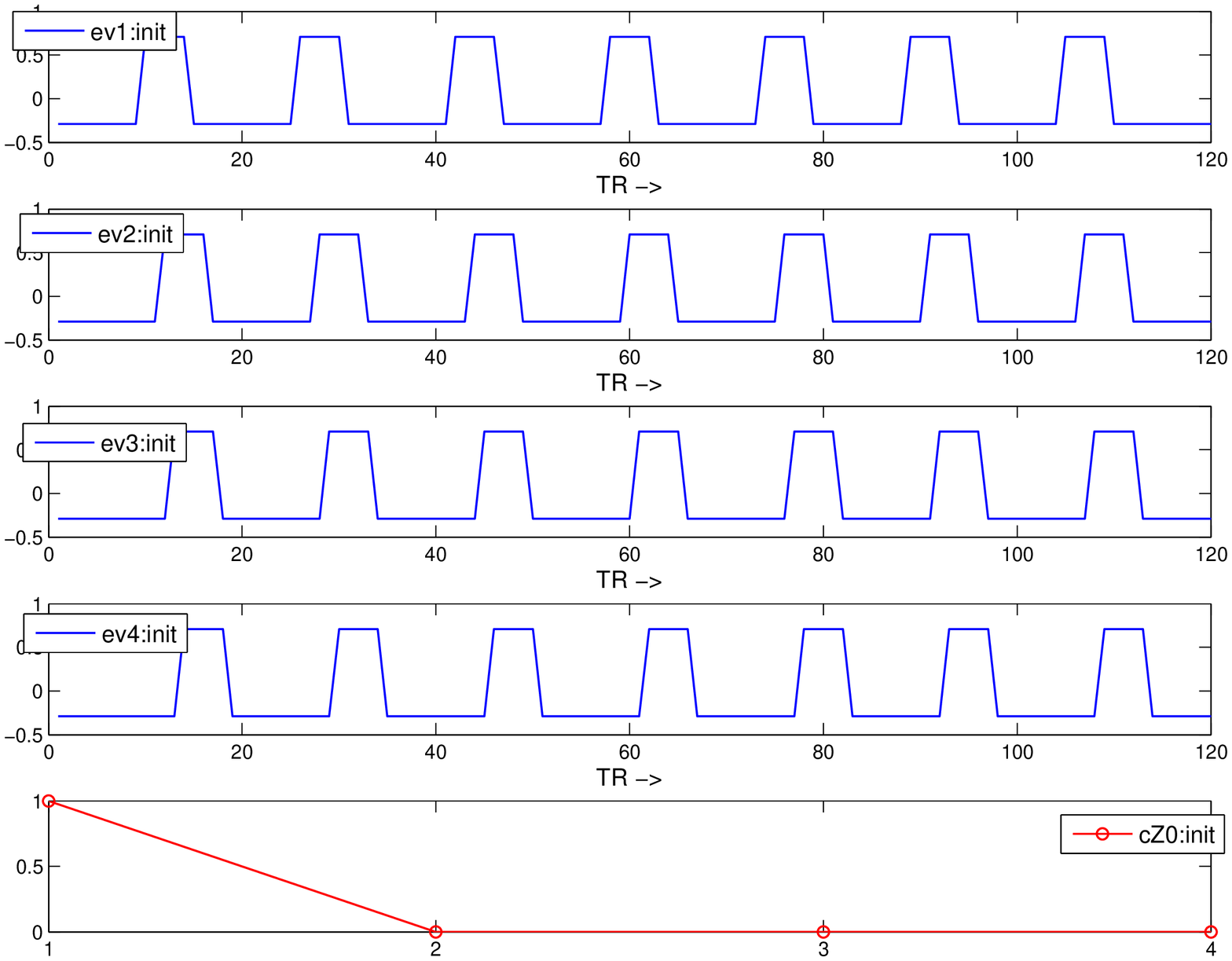}
}

&

\subfigure[]
{
\hspace{-1cm}
\label{case4b}
\includegraphics[width = 75mm, angle = -90]{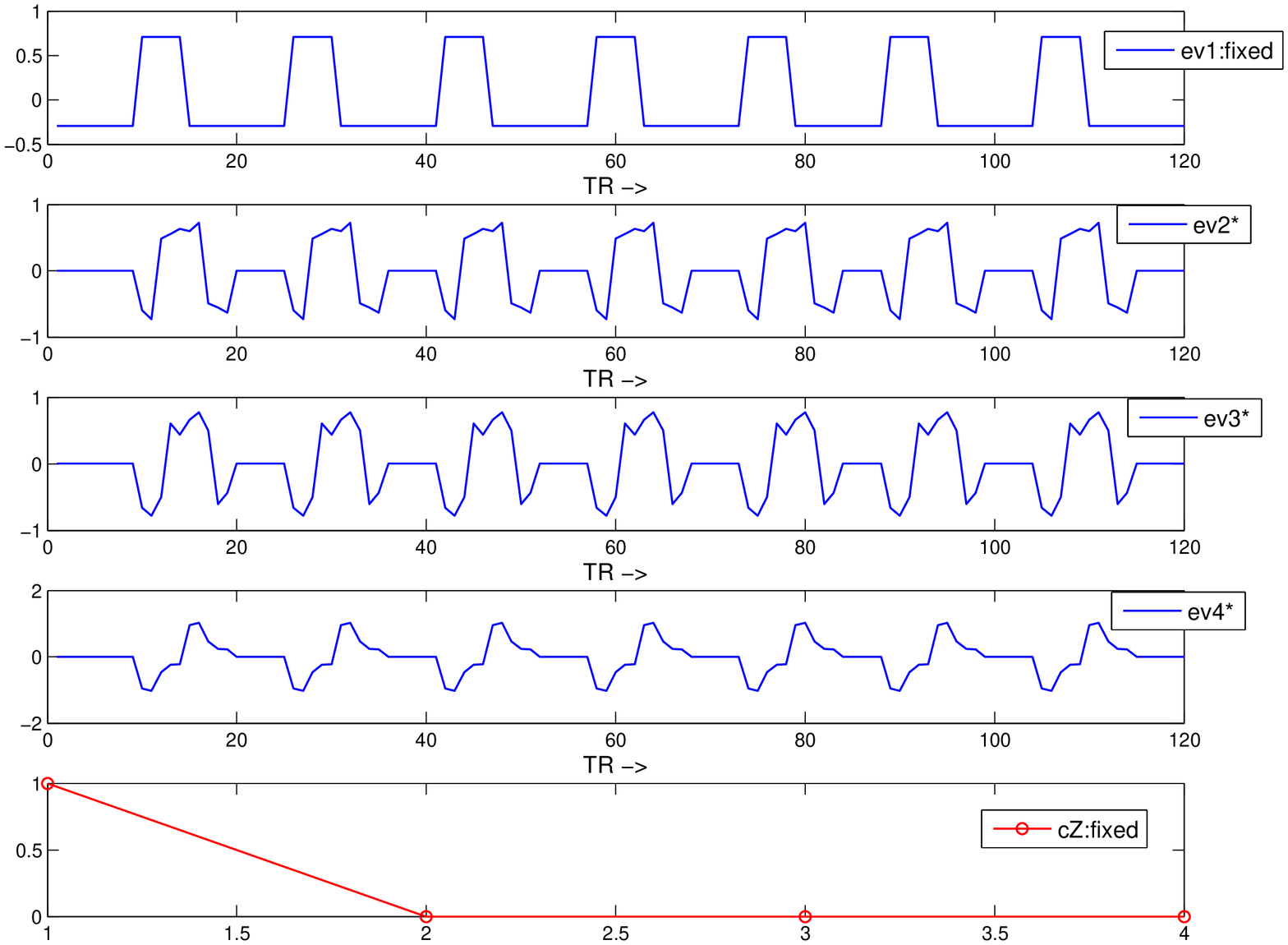}
}

\\

\subfigure[]
{
\hspace{-2cm}
\label{case4c}
\includegraphics[width = 75mm, angle = -90]{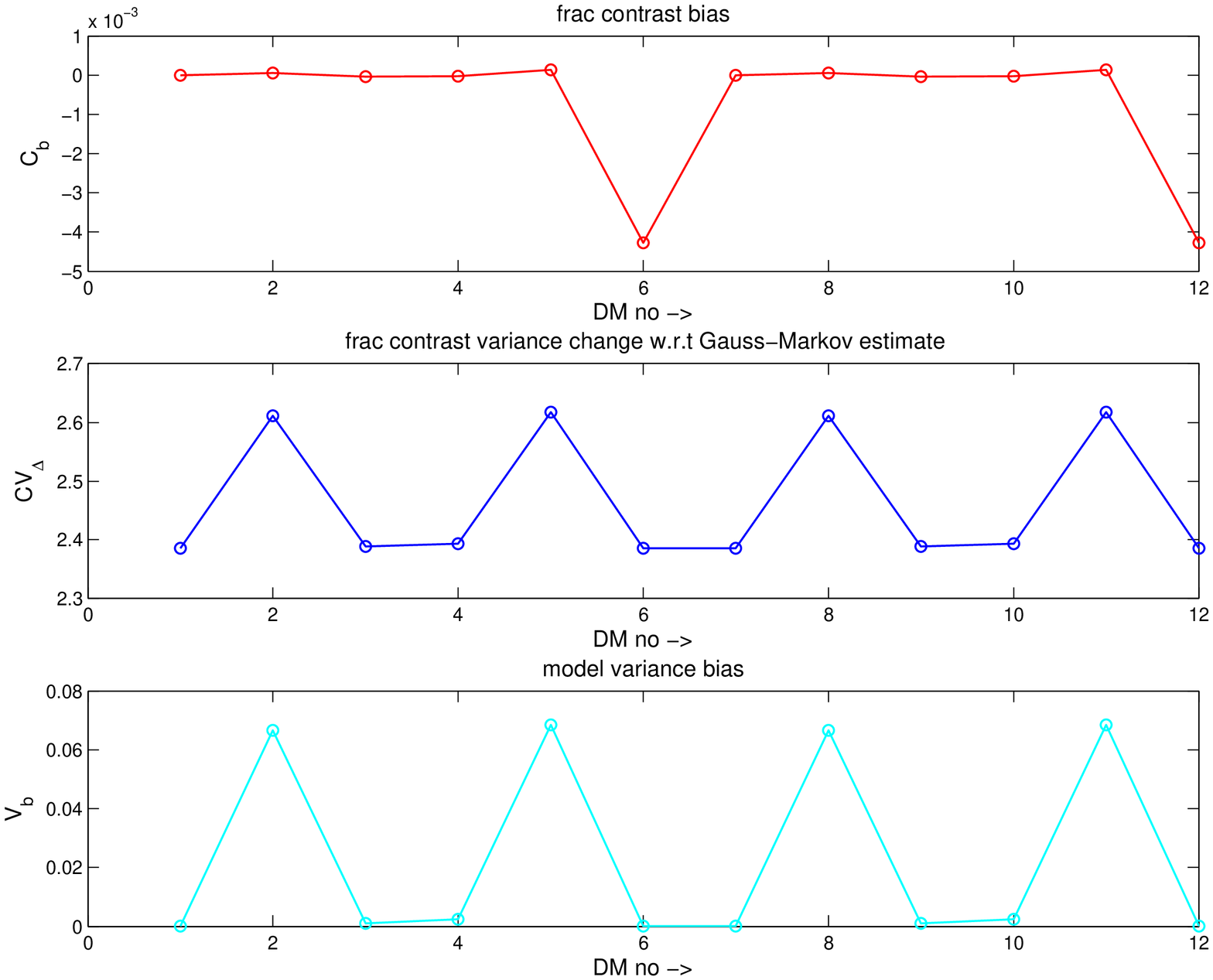}
}

&

\subfigure[]
{
\hspace{-1cm}
\label{case4d}
\includegraphics[width = 75mm, angle = -90]{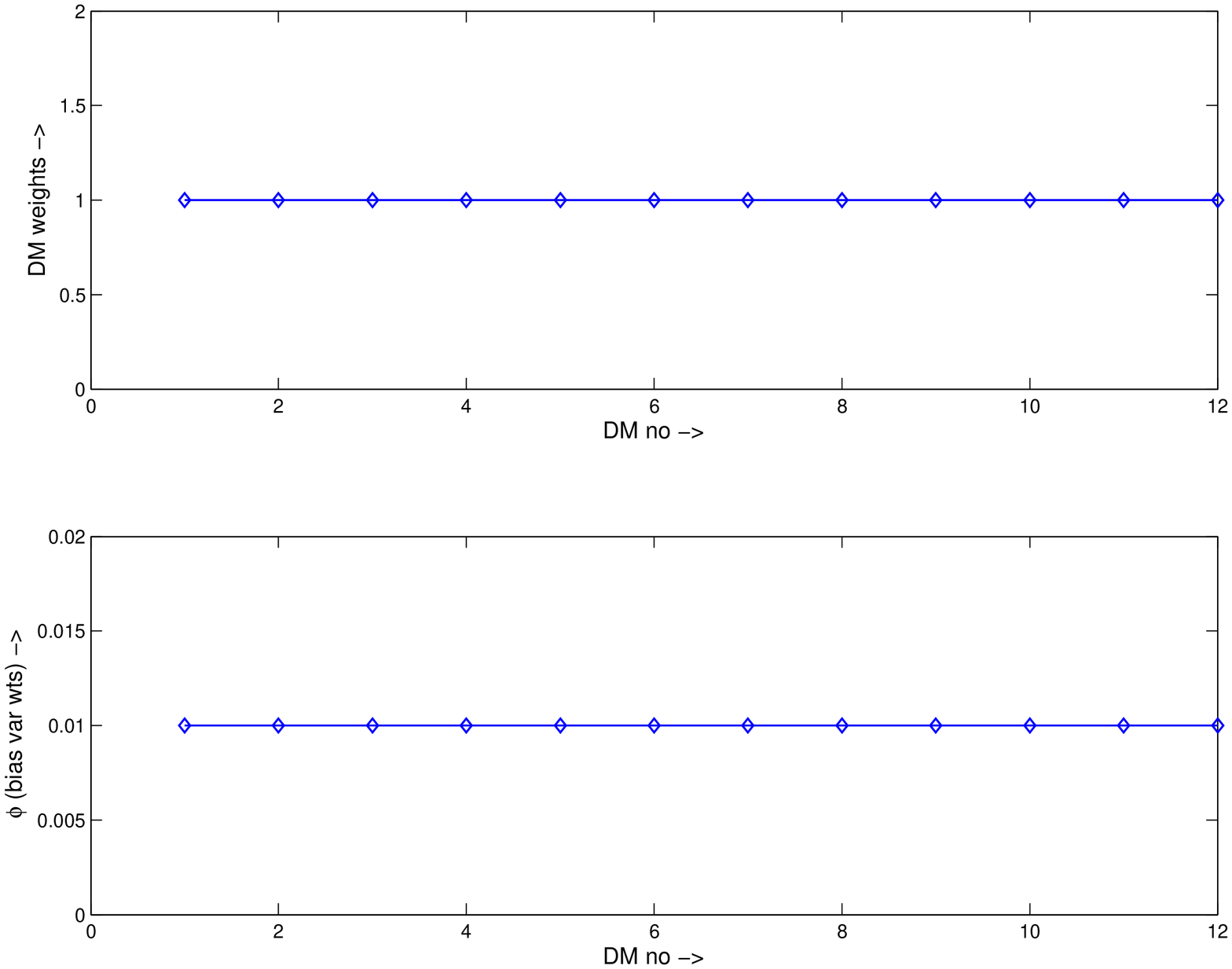}
}

\end{tabular}
\caption{  Example 5: (a) Initial design matrix (DM) along with initialization of columns 2 to 4 of $Z_0$ using $X_2$ to $X_4$ respectively. The first column in $Z$ was fixed to the basic block design EV,  columns 2 to 4 in $Z$ were left unconstrained and the contrast was fixed at [1;0;0;0] (b) Estimated optimal DM showing optimal columns 2 to 4 (c) Performance curves showing the fractional contrast bias $C_b$, contrast variance change w.r.t Gauss-Markov estimate $CV_{\Delta}$ and model variance bias $V_b$ (d) In this example $w_i = 1$ and $\phi_i$ = 0.01 indicating a predominant weight to bias term during optimization.}
\label{case4A}
\end{figure}

We use a standard block design EV to illustrate application of the proposed technique. The EV consists of blocks of 0's and 1's (see \ref{case4a})The data might contain shifted or unshifted responses. The user estimates a maximum shift of 6 timepoints and would like to 
capture the variable data using an optimized design matrix. The user predominantly wants to control bias when the entire matrix $Z$ is optimized for a fixed contrast $c_Z=[1;0;0;0]$. It is also desired to keep the main EV fixed as per the experimental paradigm and that the optimized contrast yield an unbiased estimate both for "positive" and "negative" activation. For this example, we chose $p = 4$ for the size of the optimal DM, the weights $w_i = 1$ and bias variance weights $\phi_i = 0.01$. The set of potential DM's in this case capturing both "positive" and "negative" responses is thus:

\begin{enumerate}
\item For $i = 1,\ldots,6$: $X_i$ = basic block design EV shifted to the right by $i$ timepoints, $c_{X_i}$ = [1] and $\frac{\beta_i}{\sigma_i}$ = [1]
\end{enumerate}
\begin{enumerate}
\item For $i = 6,\ldots,12$: $X_{i} = X_{i - 6}$, $c_{X_i}$ = [1] and $\frac{\beta_i}{\sigma_i}$ = [-1]
\end{enumerate}

\begin{figure}[htbp]
\centering
\begin{tabular}{cc}

\subfigure[]
{
\hspace{-2cm}
\label{case4e}
\includegraphics[width = 75mm, angle = -90]{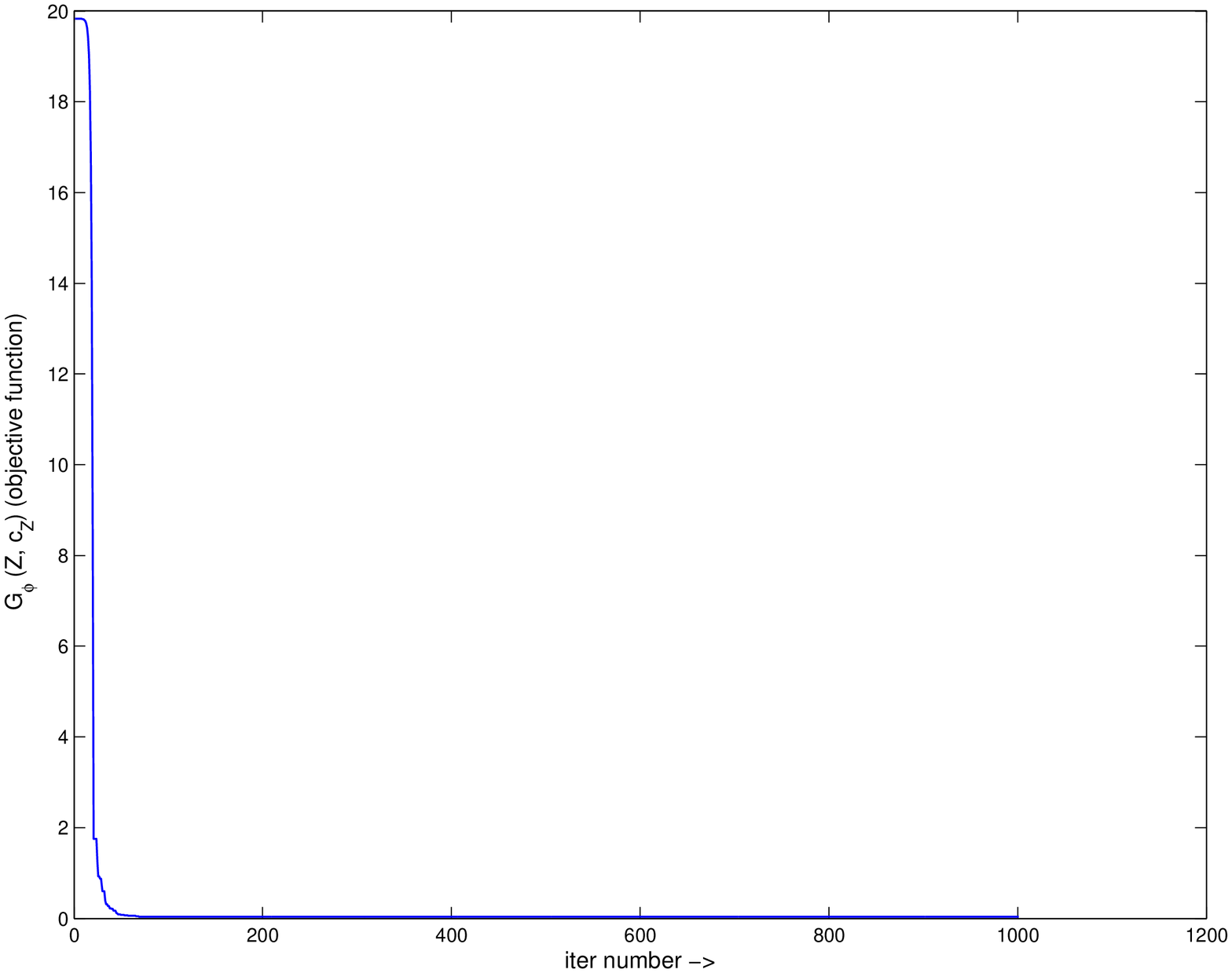}
}

&

\subfigure[]
{
\hspace{-1cm}
\label{case4f}
\includegraphics[width = 75mm, angle = -90]{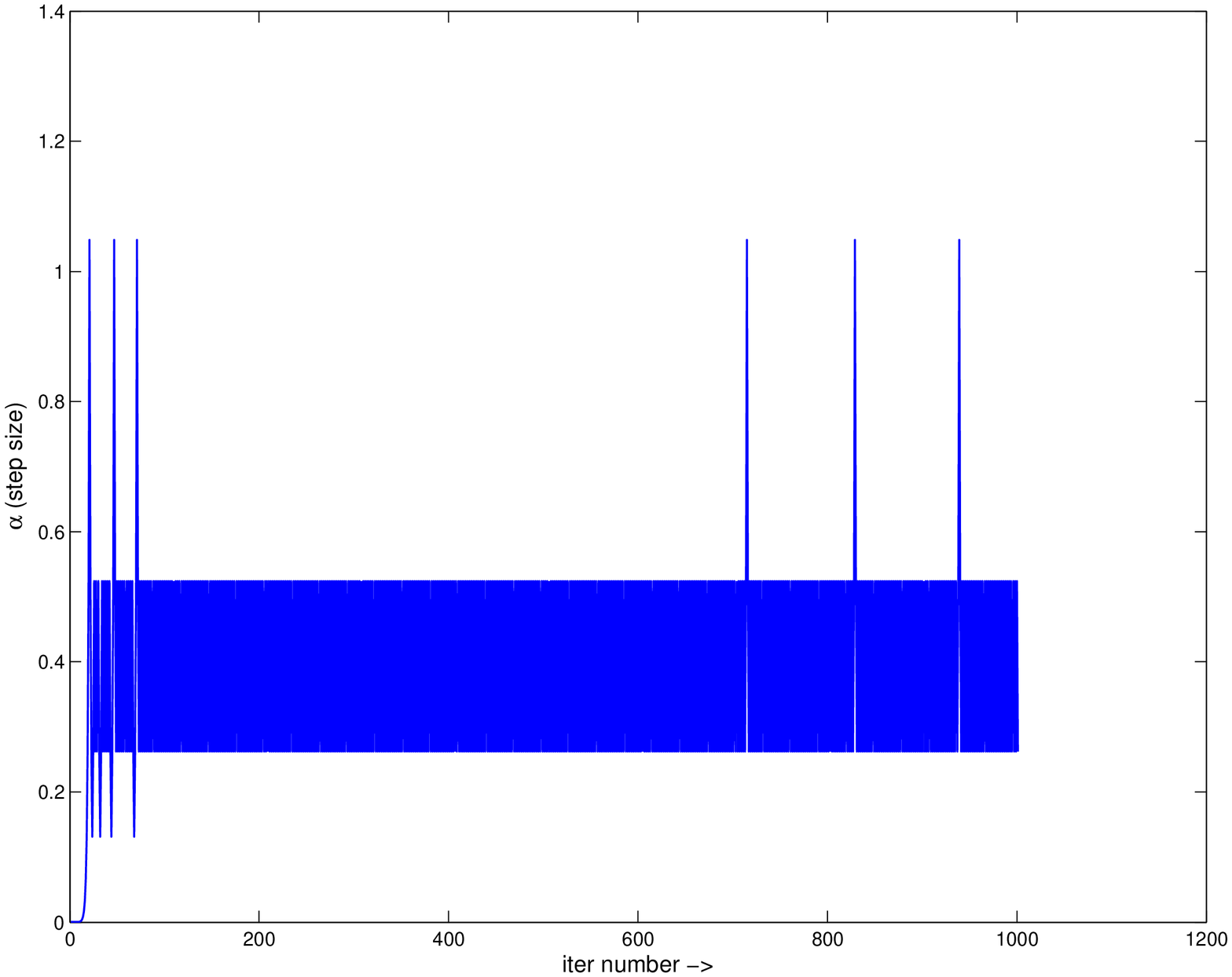}
}

\\

\subfigure[]
{
\hspace{-2cm}
\label{case4g}
\includegraphics[width = 75mm, angle = -90]{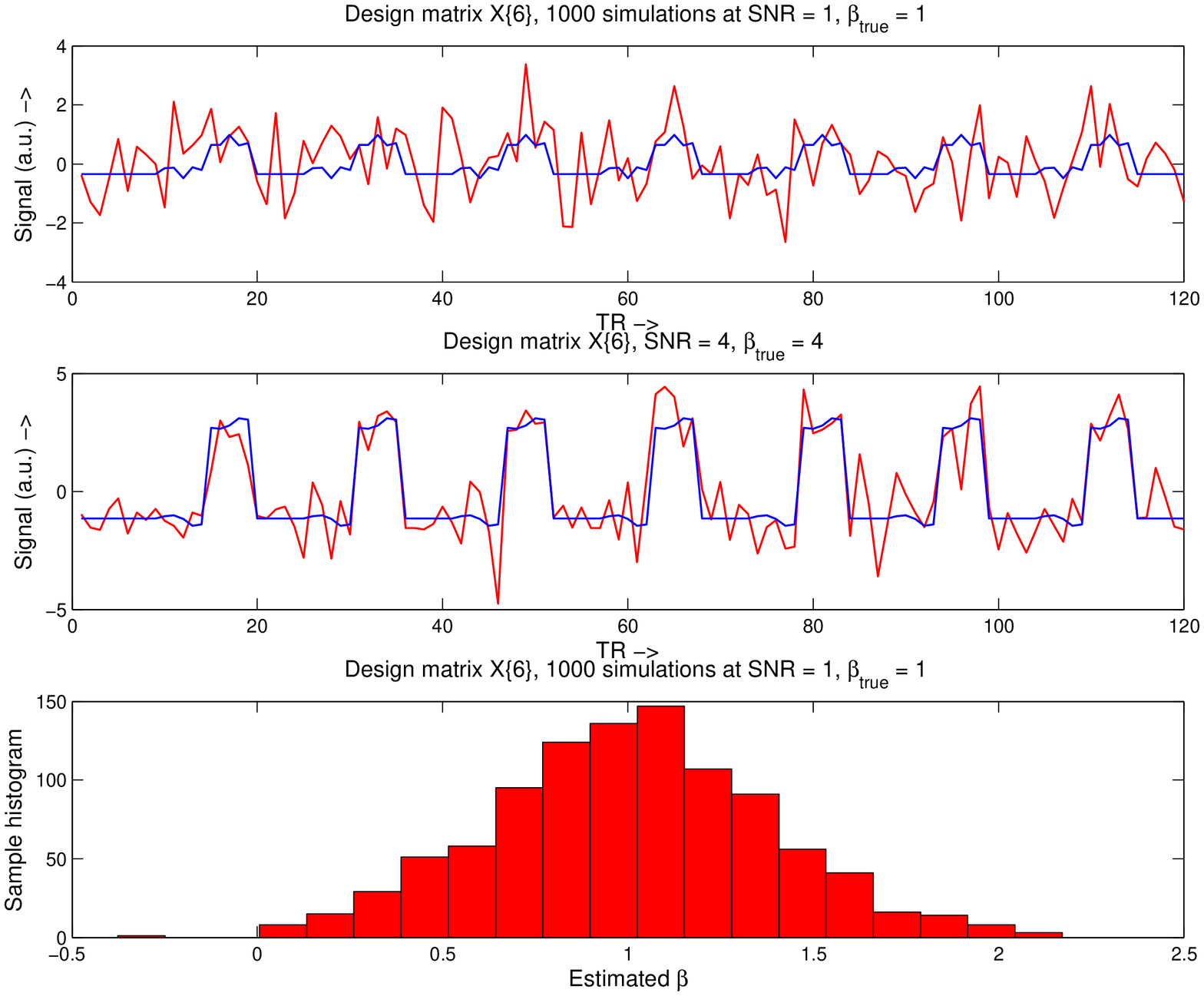}
}

&

\subfigure[]
{
\hspace{-1cm}
\label{case4h}
\includegraphics[width = 75mm, angle = -90]{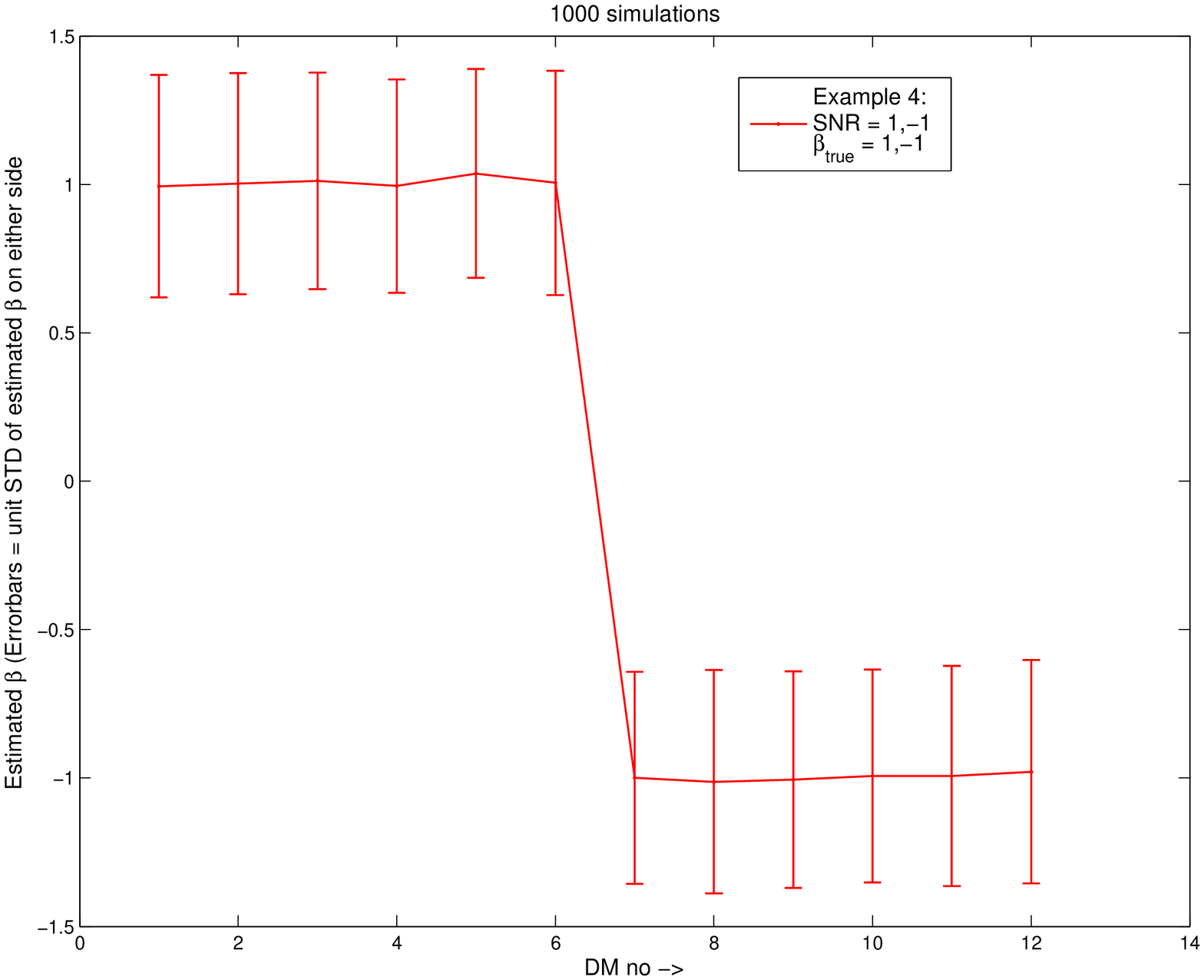}
}

\end{tabular}
\caption{  Example 5: (a) Figure showing the evolution of objective function values $G_\phi(Z,c_Z)$ over algorithm iterations. Notice how the function value stabilizes as convergence is reached (b) Figure showing the variation in the step size $\alpha$ over algorithm iterations. Step size controlling parameter $\theta$ in Algorithm 1 was set to $\theta = 2$ (c) , (d) For each design matrix (DM) entered into optimization, 1000 simulated data-sets were generated at SNR $\frac{\beta_i}{\sigma_i}$. A GLM analysis was run on each of these data-sets using the optimized DM. Figure (c) shows an example of simulated data for DM $X_{6}$ at SNR $\frac{\beta_{6}}{\sigma_{6}}$ and the GLM fit using the optimal DM. We also show a GLM fit at a higher SNR of 4 for illustration purposes. It also shows the distribution of $c_Z^T \hat{\gamma}$ over 1000 simulations. Figure (d) is a summary errorbar plot showing $\hat{E}(c_Z^T \hat{\gamma})$ over 1000 simulations for data generated from each DM. The error bars represent unit standard deviation of $c_Z^T \hat{\gamma}$ (\textbf{not} standard deviation of $\hat{E}(c_Z^T \hat{\gamma})$ )  to quantify the variance in estimation via simulation.}
\label{case4B}
\end{figure}

The results are shown in figure \ref{case4A} and \ref{case4B}.

\subsection{Example 6}
It is well known that there is no single haemodynamic response function (HRF) that captures the impulse response properties of all voxels in the brain. Here we illustrate the application of the technique developed in this paper to enable the simultaneous capture of a set of plausible HRF shapes.
Plausible HRF shapes can be generated using any reasonable parameterization of HRF. Here we generate HRF shapes using the 5 parameter half-cosine parameterization as used in \cite{FLOBS:2004}. The details of this parameterization of HRF are given in the Appendix \ref{hrf_parameterization}. 200 plausible HRF shapes
were generated by sampling the 5 parameters from a uniform distribution as follows:

\begin{eqnarray}
    h_1 = U(1s , 3s) \nonumber \\
    h_2 = U(3s , 7s) \nonumber \\
    h_3 = U(3s , 7s) \nonumber \\
    h_4 = U(3s , 9s) \nonumber \\
    f = U(0,0.5) 
\end{eqnarray}
The units of $h_1,\ldots,h_4$ are in sec (s) and $f$ is dimensionless. $U(a,b)$ denotes the uniform distribution on $[a,b]$. Data was generated by sampling at every 0.1 s.

\begin{figure}[htbp]
\begin{center}
\includegraphics[width = 75mm, angle = -90]{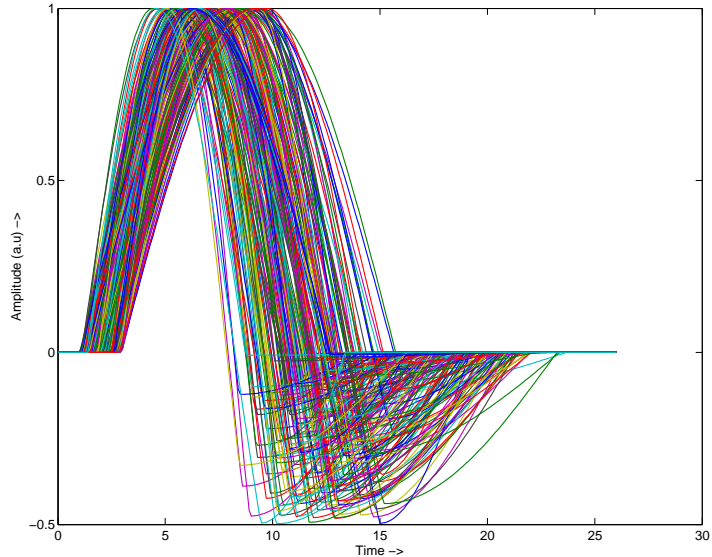}
\caption{200 samples of HRF drawn from a 5 parameter half-cosine parameterization of Haemodynamic Response Function (HRF) that were used as input to the optimization process in Example 6.}
\label{hrfsamples}
\end{center}
\end{figure}

\begin{figure}[htbp]
\centering
\begin{tabular}{cc}

\subfigure[]
{
\hspace{-2cm}
\label{case5a}
\includegraphics[width = 75mm, angle = -90]{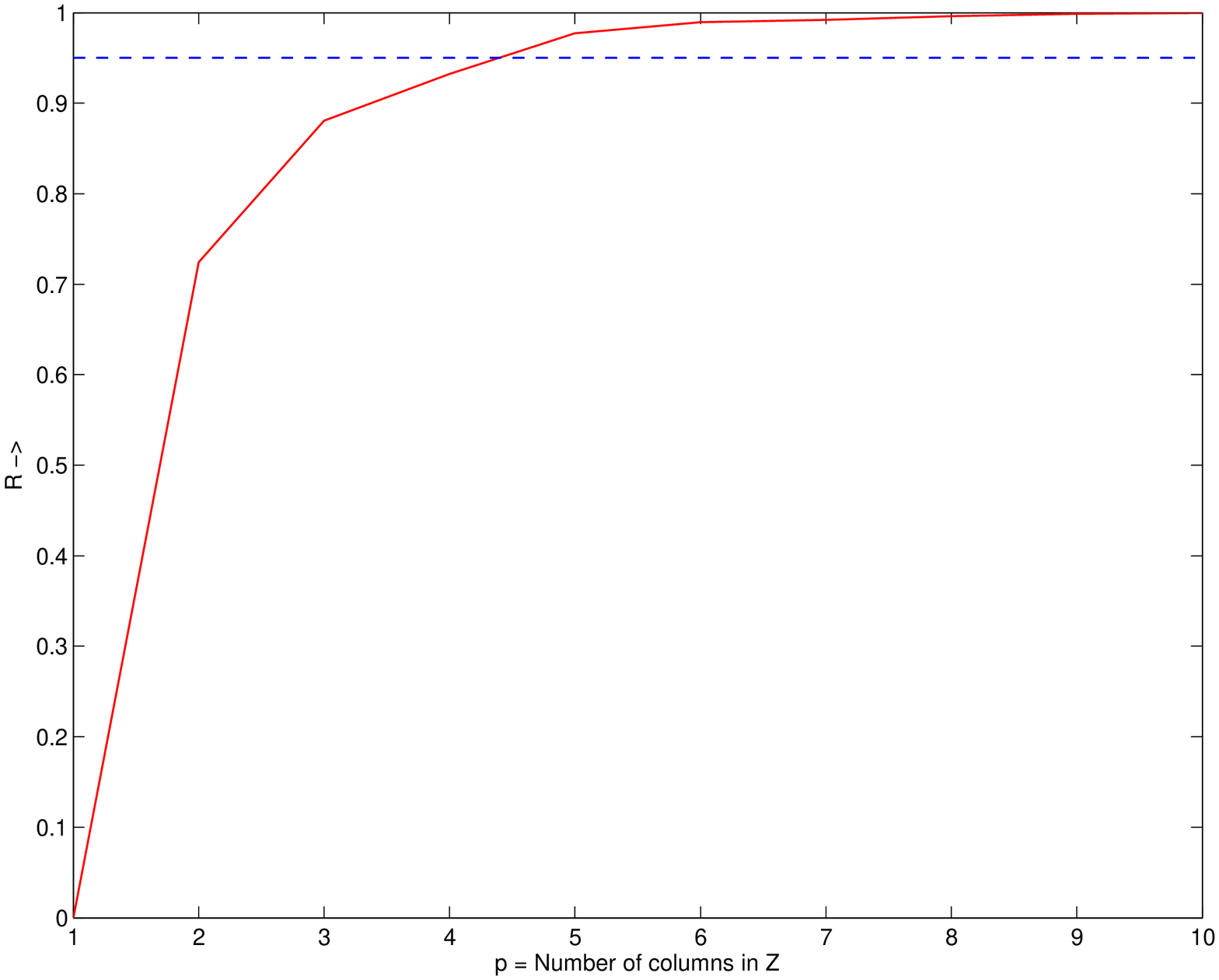}
}

&

\subfigure[]
{
\hspace{-1cm}
\label{case5b}
\includegraphics[width = 75mm, angle = -90]{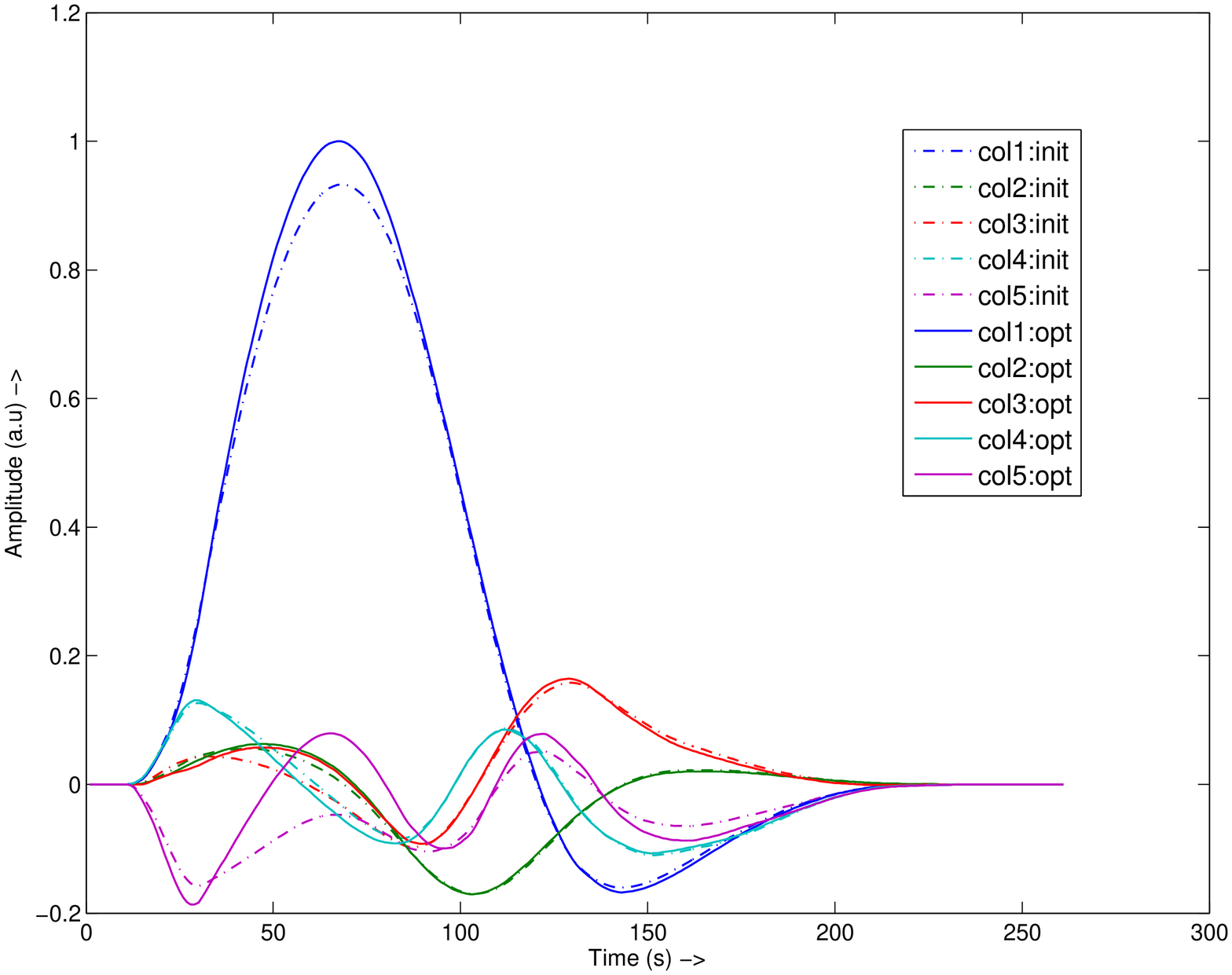}
}

\\

\subfigure[]
{
\hspace{-2cm}
\label{case5c}
\includegraphics[width = 75mm, angle = -90]{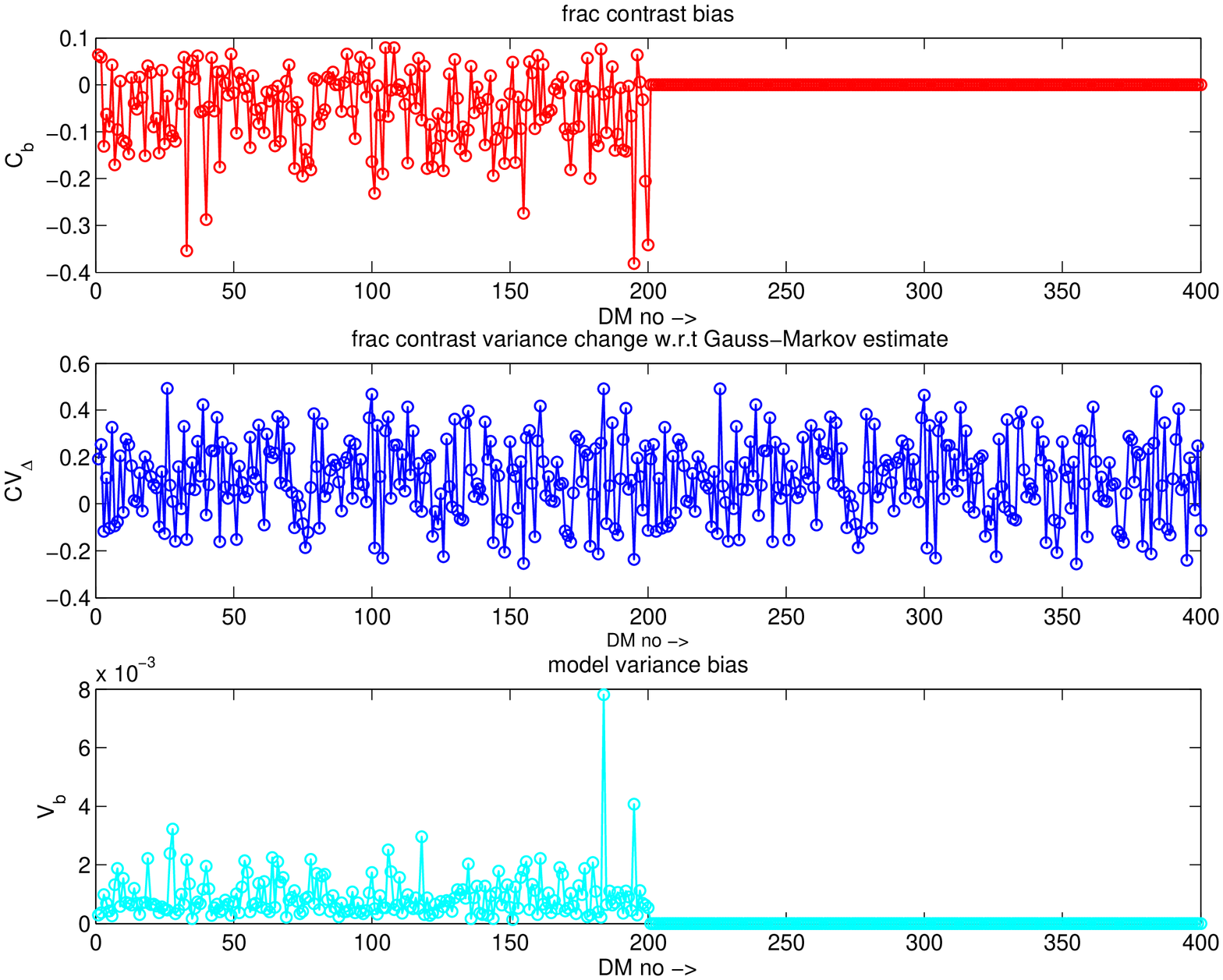}
}

&

\subfigure[]
{
\hspace{-1cm}
\label{case5d}
\includegraphics[width = 75mm, angle = -90]{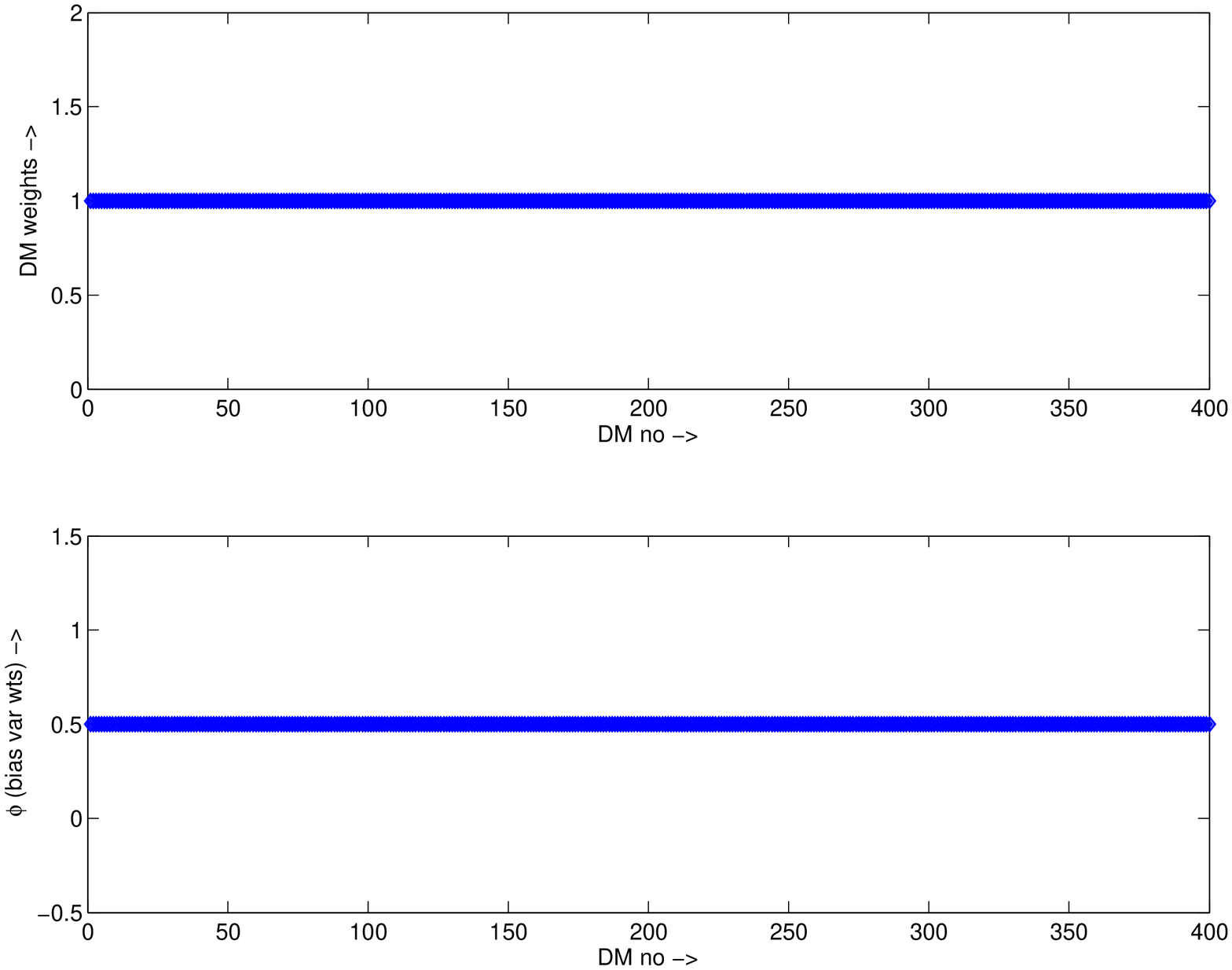}
}

\end{tabular}
\caption{  Example 6: (a) $R$ versus $p$ curve for determining the optimal number of columns in $Z$. Using a cutoff of $R_c = 0.95$ the optimal number of columns in $Z$ was determined to be $p_{opt} = 5$. (b) The 5 columns of $Z$ were left unconstrained and the contrast was fixed at $[1;0;0;0]$. The automatic initialization strategy described before was used to initialize the 5 columns of $Z_0$ (dotted lines). The optimized columns are shown in the same figure using solid lines. (c) Performance curves showing the fractional contrast bias $C_b$, contrast variance change w.r.t Gauss-Markov estimate $CV_{\Delta}$ and model variance bias $V_b$ (d) In this example $w_i = 1$ and $\phi_i$ = 0.5 indicating an equal weight to bias and variance terms during optimization.}
\label{case5A}
\end{figure}

\begin{figure}[htbp]
\centering
\begin{tabular}{cc}

\subfigure[]
{
\hspace{-2cm}
\label{case5e}
\includegraphics[width = 75mm, angle = -90]{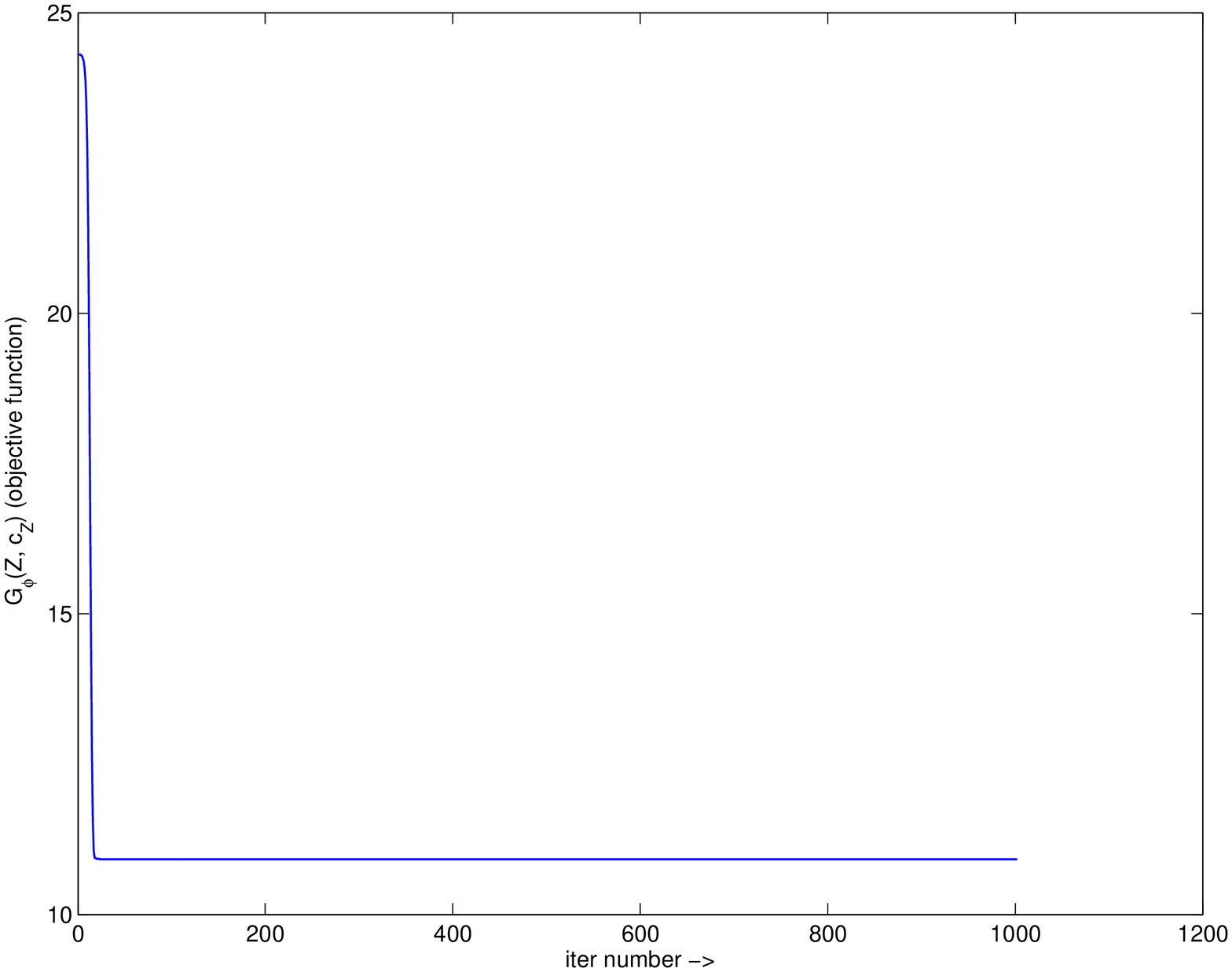}
}

&

\subfigure[]
{
\hspace{-1cm}
\label{case5f}
\includegraphics[width = 75mm, angle = -90]{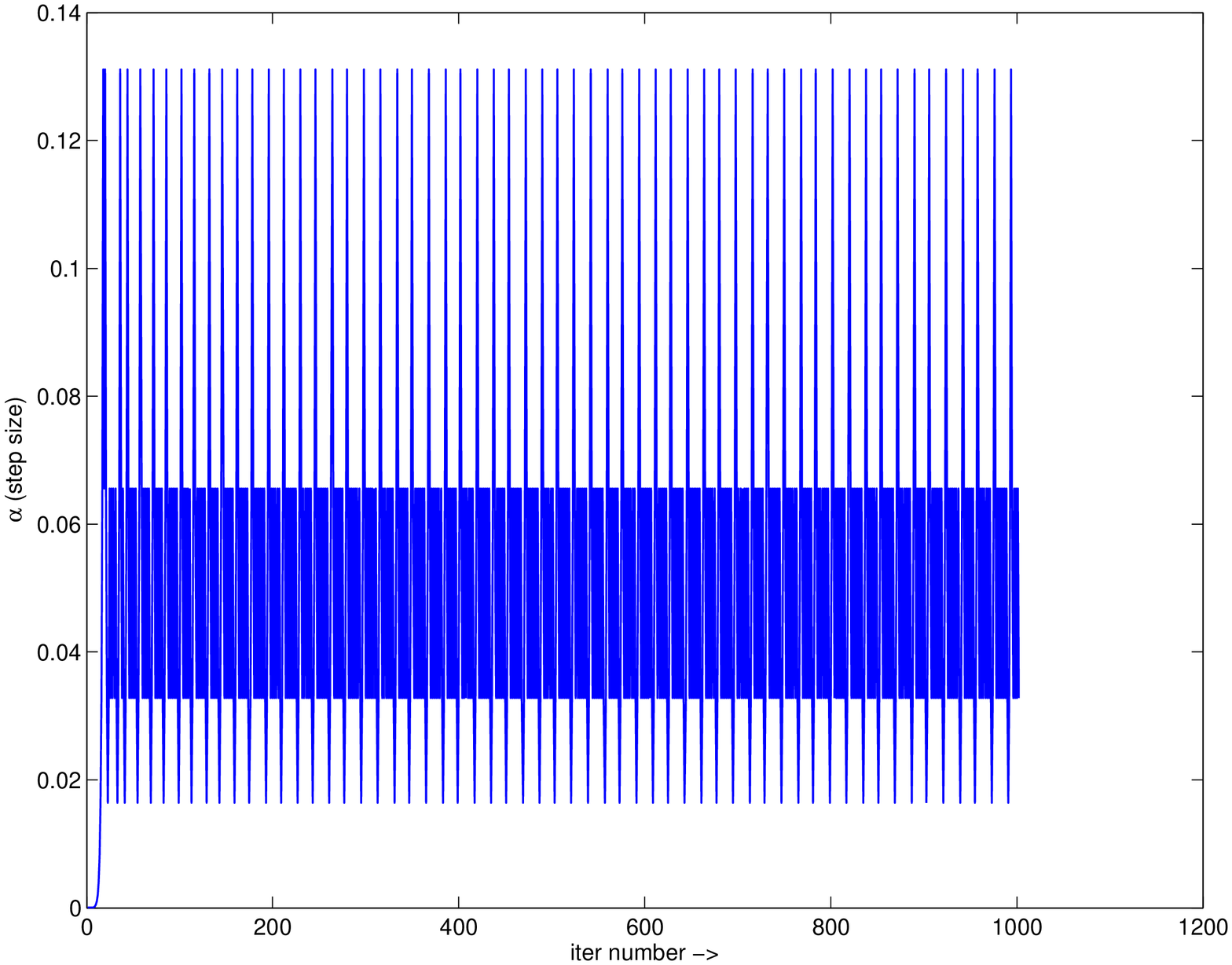}
}

\\

\subfigure[]
{
\hspace{-2cm}
\label{case5g}
\includegraphics[width = 75mm, angle = -90]{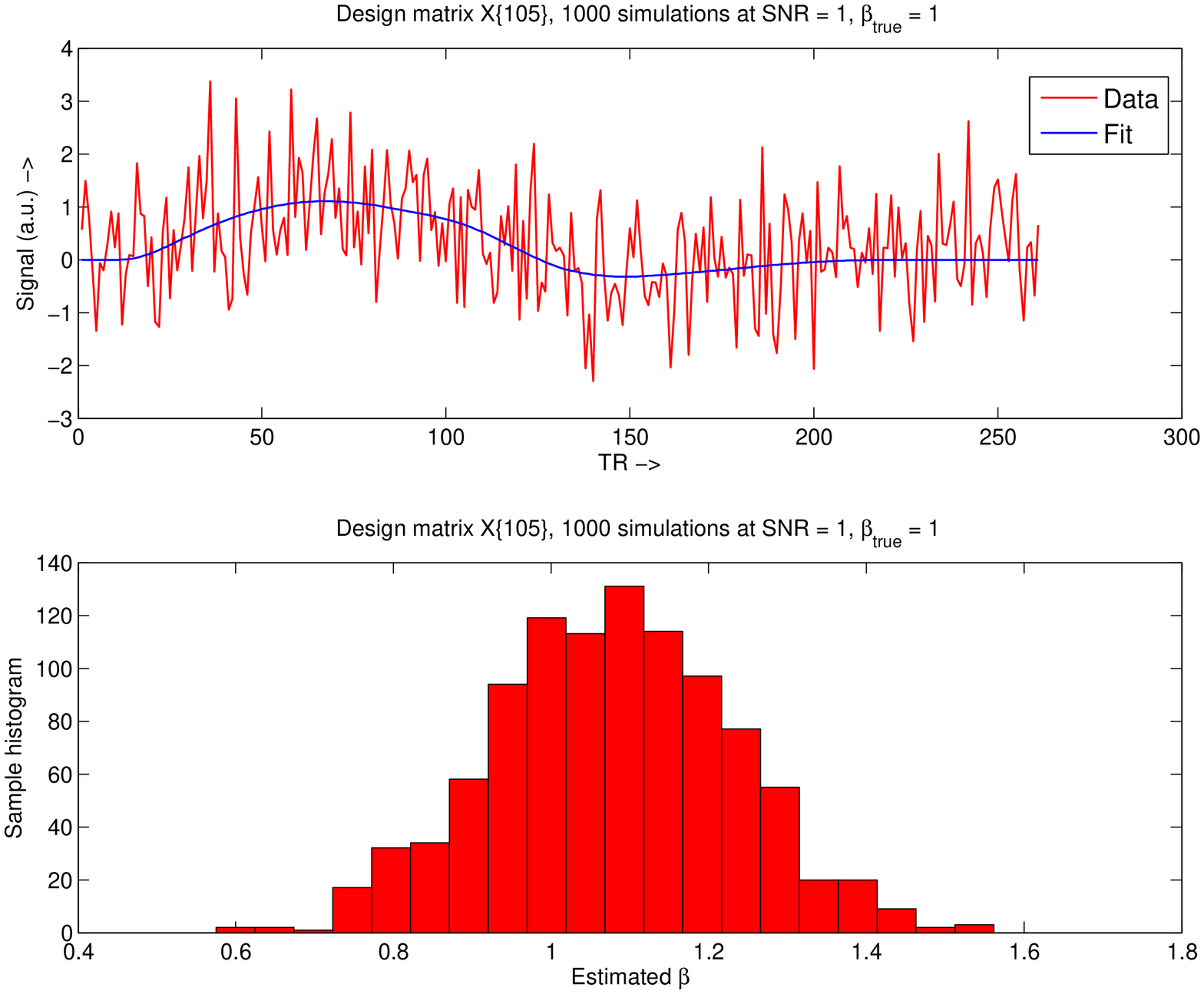}
}

&

\subfigure[]
{
\hspace{-1cm}
\label{case5h}
\includegraphics[width = 75mm, angle = -90]{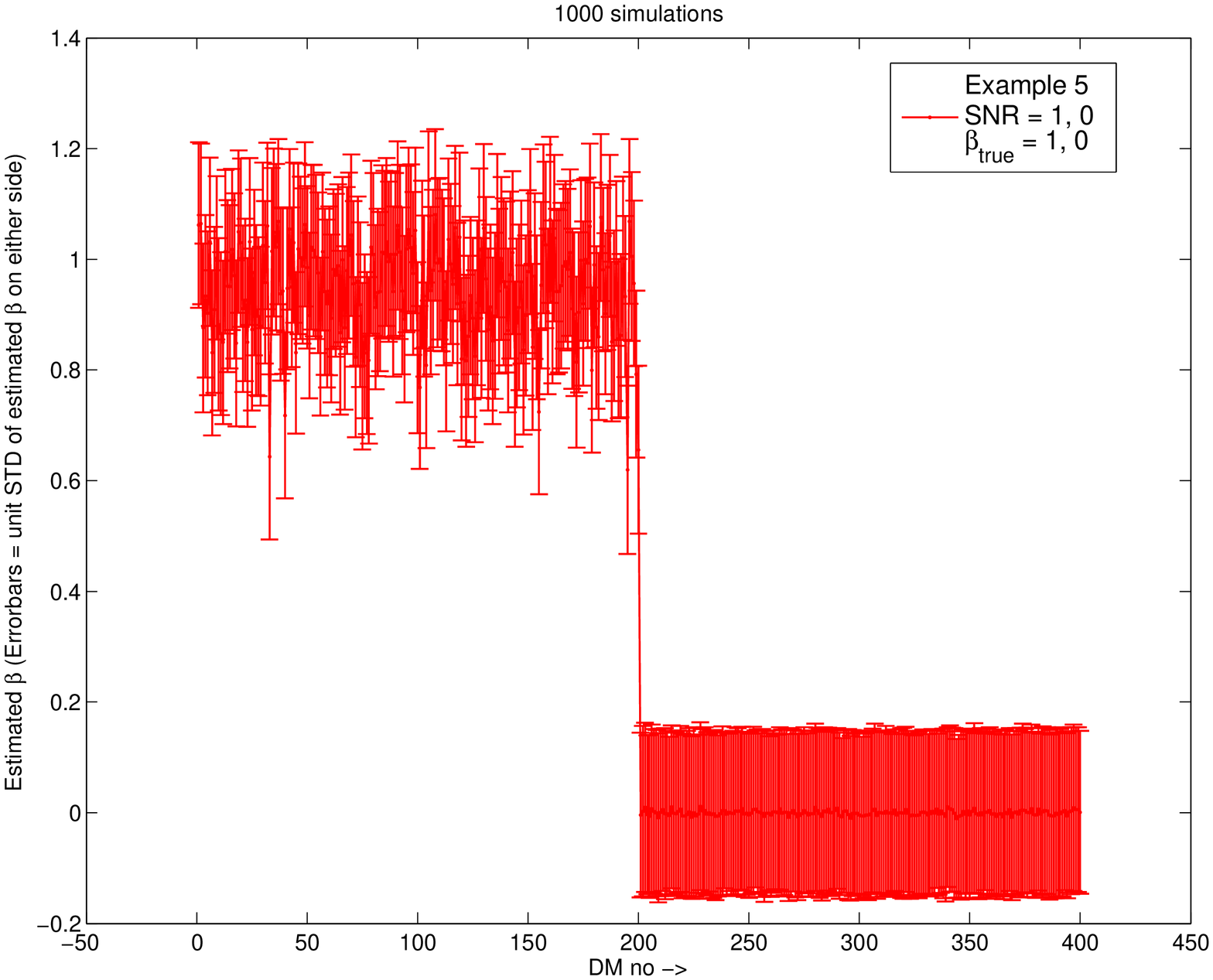}
}

\end{tabular}
\caption{  Example 6: (a) Figure showing the evolution of objective function values $G_\phi(Z,c_Z)$ over algorithm iterations. Notice how the function value stabilizes as convergence is reached (b) Figure showing the variation in the step size $\alpha$ over algorithm iterations. Step size controlling parameter $\theta$ in Algorithm 1 was set to $\theta = 2$ (c) , (d) For each design matrix (DM) entered into optimization, 1000 simulated data-sets were generated at SNR $\frac{\beta_i}{\sigma_i}$. A GLM analysis was run on each of these data-sets using the optimized DM. Figure (c) shows an example of simulated data for DM $X_{105}$ at SNR $\frac{\beta_{105}}{\sigma_{105}}$ and the GLM fit using the optimal DM. It also shows the distribution of $c_Z^T \hat{\gamma}$ over 1000 simulations. Figure (d) is a summary errorbar plot showing $\hat{E}(c_Z^T \hat{\gamma})$ over 1000 simulations for data generated from each DM. The error bars represent unit standard deviation of $c_Z^T \hat{\gamma}$ (\textbf{not} standard deviation of $\hat{E}(c_Z^T \hat{\gamma})$ )  to quantify the variance in estimation via simulation.}
\label{case5B}
\end{figure}

The automatic initialization strategy described before was used for initialization. The weights $w_i$ were set to 1 and $\phi_i$ were set to their default values of 0.5.
Optimization results are shown in figure \ref{case5A} and \ref{case5B}. Once the set of "optimal" HRF capturing functions are found, they can be entered into any GLM analysis as follows:
\begin{enumerate}
\item Convolve the experimental EV with each of the "optimal" HRF capturing functions.
\item Next enter the resulting DM into a GLM analysis and use the contrast $c_Z$ from the optimization above to capture the "size" of underlying signal optimally.
\end{enumerate}

\section{Application to real fMRI data}
In this section, we describe a case study in detail and show how the approach can be applied in practice.

\subsection{Data acquisition}
This case study describes a buprenorphine infusion scan at 0.2 mg/70kg dosage.
Elimination half-lives for intravenous administration of buprenorphine (0.3 mg) range between 1.2 to 7.2 hours (mean = 2.2 hours), while the terminal half-life is ~3 hours (Bullingham, et al. 1980).  Onset of buprenorphine (intravenous, 0.3 mg to 0.6 mg dose) is within five minutes and analgesic effects lasts for ~6 to 8 hours (Downing, et al. 1977).

In the 25 minute infusion scan, a 5 minute baseline was collected prior to the first infusion of buprenorphine or placebo.  Four infusions, totaling 8 ml, were performed at minutes 5, 7, 9 and 11.  Each 2 ml infusion was performed at a rate of 0.1 ml/sec and controlled by an automatic microinjector (Medrad Spectris, Colombus, OH).  

All data were collected on a 3 Tesla Siemens Trio scanner with an 8-channel phased array head coil (Erlangen, Germany).  Infusion data were collected using a gradient echo-echo planar pulse sequence (GE-EPI) at a 3.5 x 3.5 x 3.5 mm3 resolution.  GE-EPI Parameters: Time of Repetition (TR) = 2500 msecs, Time of Echo (TE) = 30 msecs, Field of View (FOV) = 224x224, Flip Angle (FA) = 90¡, num of Slices = 41 axial slices and num of Volumes = 600.  The acquisition time for the infusion scan was 25 mins and 5 secs.  T1-weighted structural images were acquired using a 3-D magnetization-prepared rapid gradient echo (MPRAGE) sequence at a resolution of 1.33 x 1.0 x 1.0 mm3.  MPRAGE Parameters: TR = 2100 msecs, TE = 2.74 msecs, Time of Inversion (TI) = 1100 msecs,  FA = 12¡, num of Slices = 128 sagittal slices (Mugler and Brookeman 1990).

\subsection{Data Analysis}
\subsubsection{Preprocessing}
Single subject data analysis was performed using FMRIB Software Library (FSL) (\url{http://www.fmrib.ax.ac.uk/fsl}).  The first two volumes were removed to account for MR signal instability in the initially acquired volumes. Raw fMRI data was preprocessed using the following steps 1) Skull stripping using a Brain Extraction Tool, 2) Motion correction using FMRIBÕs Linear Motion Correction tool (MCFLIRT), 3) Spatial smoothing with a 5 mm FWHM spatial filter. 

\subsubsection{Model Free analysis}
Following this preprocessing, a projection pursuit analysis using the algorithm ADIS (Automated Decomposition Into Sources) was performed to extract spatial maps of minimum entropy to identify patterns of activity in the brain. ADIS (\url{http://arxiv.org/abs/0902.4879}, arXiv:0902.4879v1 [stat.CO]) is a probabilistic and constrained projection pursuit software that outperforms conventional ICA algorithms in several benchmark tests. Data was analyzed as follows:
\begin{enumerate}
\item	Dimension reduction via probabilistic PCA. A lower bound for the latent dimension was determined via a bootstrap approach. A cross-validation analysis was used to estimate the true latent dimension.
\item The PCA-reduced data was decomposed into a set of minimum entropy spatial maps and their associated timecourses using "negentropy" based projection pursuit. The ADIS optimization core was used to perform constrained optimization.The resulting $z$-maps were thresholded at $z > 3$ and $z < -3$.
\end{enumerate}

The purpose of running ADIS was to demonstrate the existence of multiple infusion response profiles even for single subject data.
\begin{figure}[htbp]
\begin{center}
\includegraphics[width = 75mm, angle = -90]{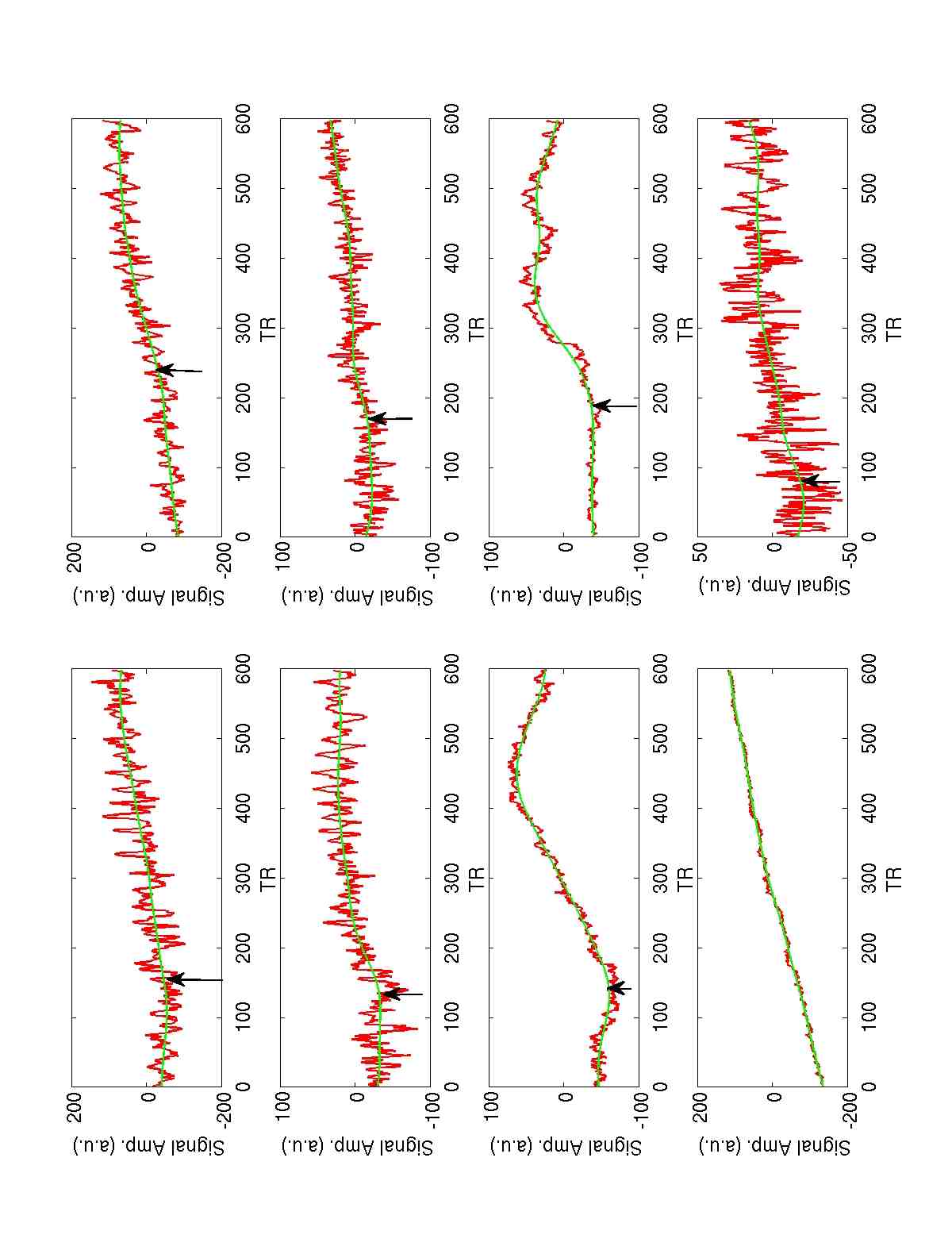}
\caption{Some infusion response profiles extracted by ADIS illustrating the variability of the signal of interest. The black arrow indicates the approximate timepoint at which the signal starts going up from baseline. Note that the individual infusion responses are potentially corrupted by the linear drift which was identified as a global component in the brain.}
\label{figbup}
\end{center}
\end{figure}

ADIS produced a total of 16 components. Figure \ref{figbup} shows associated timecourses for some minimum entropy spatial maps. The linear drift was found to be present globally throughout the brain superimposed on the various response profiles.
Thus an example of a potential design matrix would be the one such as shown in Figure \ref{fig1} with the first column a ramped step change modeling the infusion response and the second column a covariate of no interest as the linear drift. The contrast of interest to the investigator will be $[1;0]$. Clearly the data contains multiple response profiles differing in mainly their rise from baseline. There could be many more variations possible in general datasets. In the study of pharmacological fMRI responses, the exact response of the brain is not easily modeled because of pharmacokinetics and pharmacodynamics differences between the circulating system and the brain. Henc, it is feasible to expect different responses across brain structures. It seems appropriate to use a "ramp" model with potential delays to accomodate responses in different brain structures.

\subsubsection{Optimal DM computation}

Subsequent to this data "inspection" stage we carried out an analysis to compute the optimal design matrix. Two main EVs were proposed for this analysis as shown in \ref{fig1}. The EV1 captures an "infusion response" while EV2 represents pure "linear drift". For the purposes of optimization we proposed 723 potential DMs with the intention of capturing delays upto 180 timepoints. 
Let $X_0$ be the DM shown in \ref{fig1}. Define
\begin{eqnarray}
X_i(:,1) = X_0(:,1) \mbox{ shifted to the right by } i \mbox{ timepoints } \\
X_i(:,2) = X_0(:,2) \mbox{ for all } i
\end{eqnarray}
These 723 DMs are as follows:
\begin{enumerate}
\item The first 180 DMs were $X_{1}$, \ldots, $X_{180}$ at $\frac{\beta_i}{\sigma_i} = [1;0.5]$ and $c_{X_i} = [1;0]$
\item The next 180 DMs were $X_{1}$, \ldots, $X_{180}$ at $\frac{\beta_i}{\sigma_i} = [-1;0.5]$ and $c_{X_i} = [1;0]$
\item The next 180 DMs were $X_{1}$, \ldots, $X_{180}$ at $\frac{\beta_i}{\sigma_i} = [1;-0.5]$ and $c_{X_i} = [1;0]$
\item The next 180 DMs were $X_{1}$, \ldots, $X_{180}$ at $\frac{\beta_i}{\sigma_i} = [-1;-0.5]$ and $c_{X_i} = [1;0]$
\item DM 721 was $X_{0}$ at $\frac{\beta_i}{\sigma_i} = [0;1]$ and $c_{X_i} = [1;0]$
\item DM 722 was $X_{0}$ at $\frac{\beta_i}{\sigma_i} = [0;-1]$ and $c_{X_i} = [1;0]$
\item DM 723 was $X_{0}$ at $\frac{\beta_i}{\sigma_i} = [0;0]$ and $c_{X_i} = [1;0]$
\end{enumerate}

\begin{figure}[htbp]
\centering
\begin{tabular}{cc}

\subfigure[]
{
\hspace{-2cm}
\label{casebupa}
\includegraphics[width = 75mm, angle = -90]{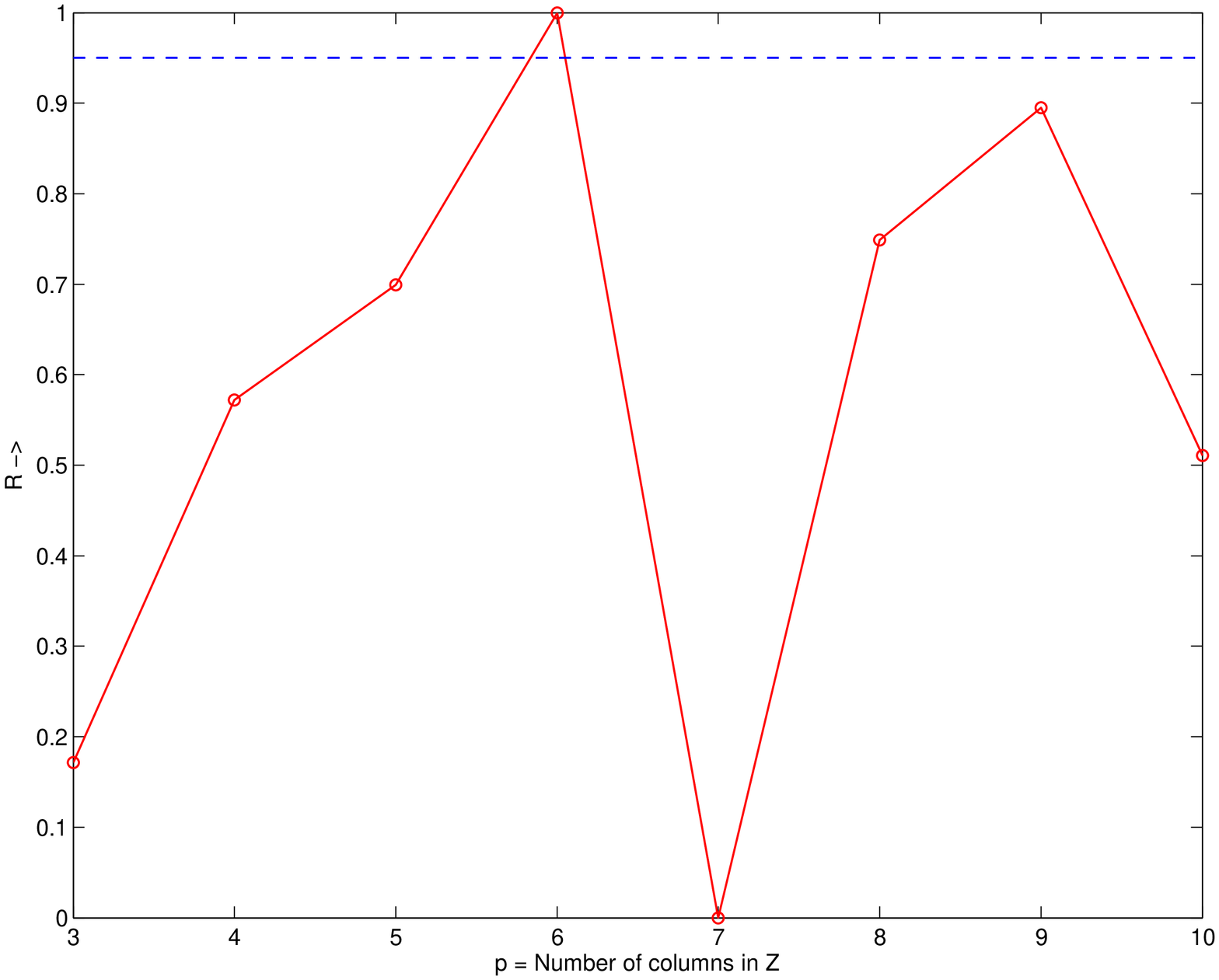}
}

&

\subfigure[]
{
\hspace{-1cm}
\label{casebupb}
\includegraphics[width = 75mm, angle = -90]{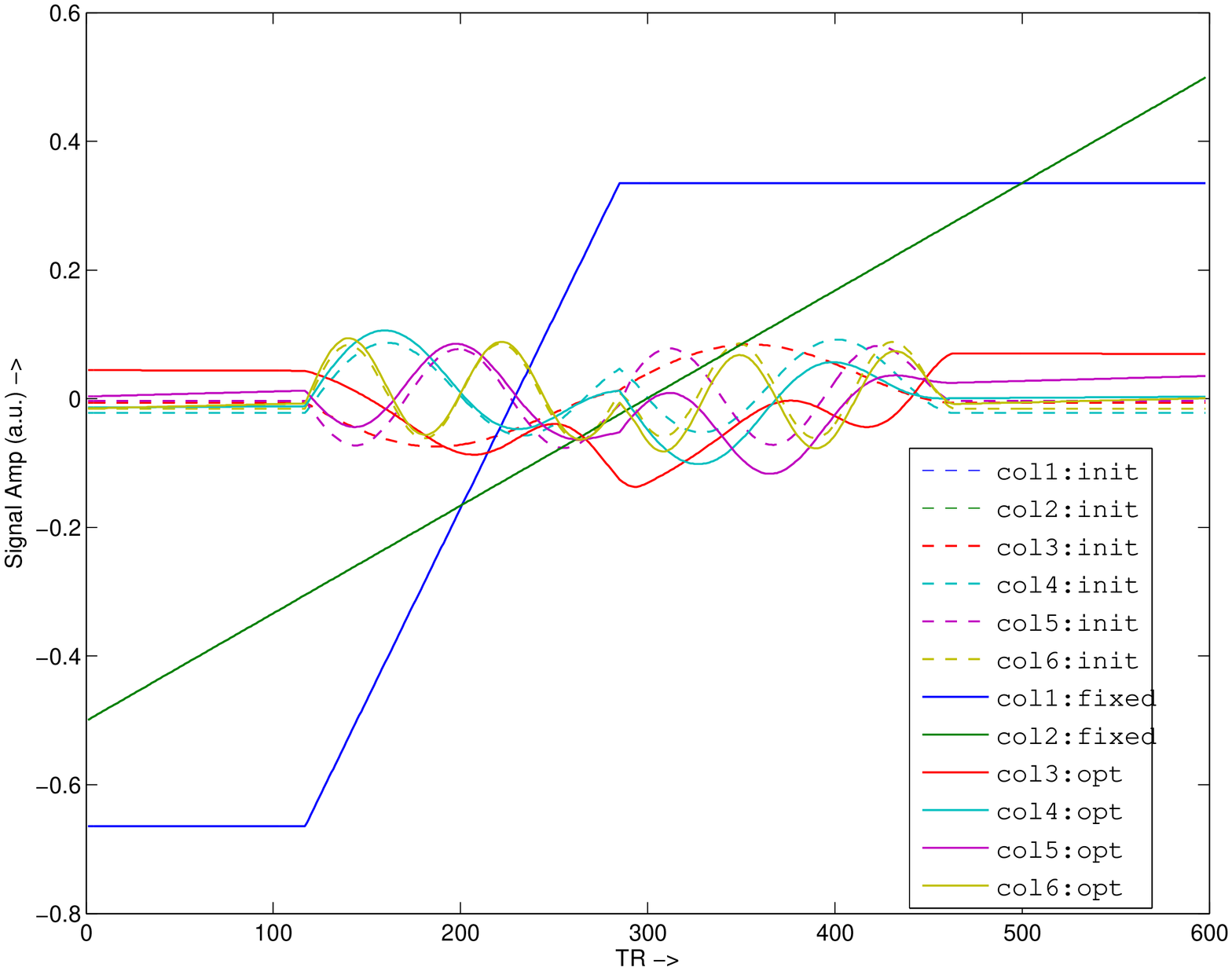}
}

\\

\subfigure[]
{
\hspace{-2cm}
\label{casebupc}
\includegraphics[width = 75mm, angle = -90]{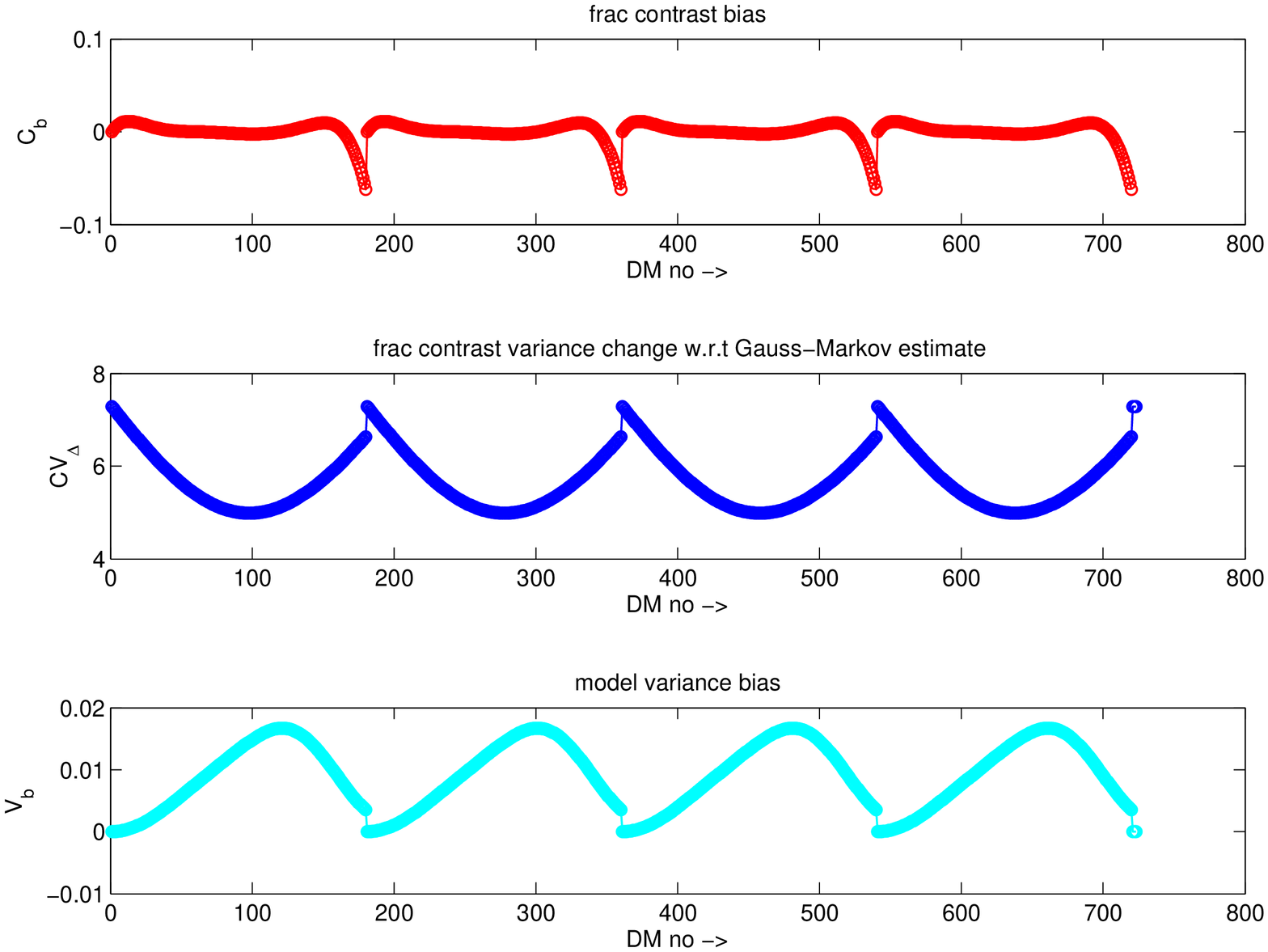}
}

&

\subfigure[]
{
\hspace{-1cm}
\label{casebupd}
\includegraphics[width = 75mm, angle = -90]{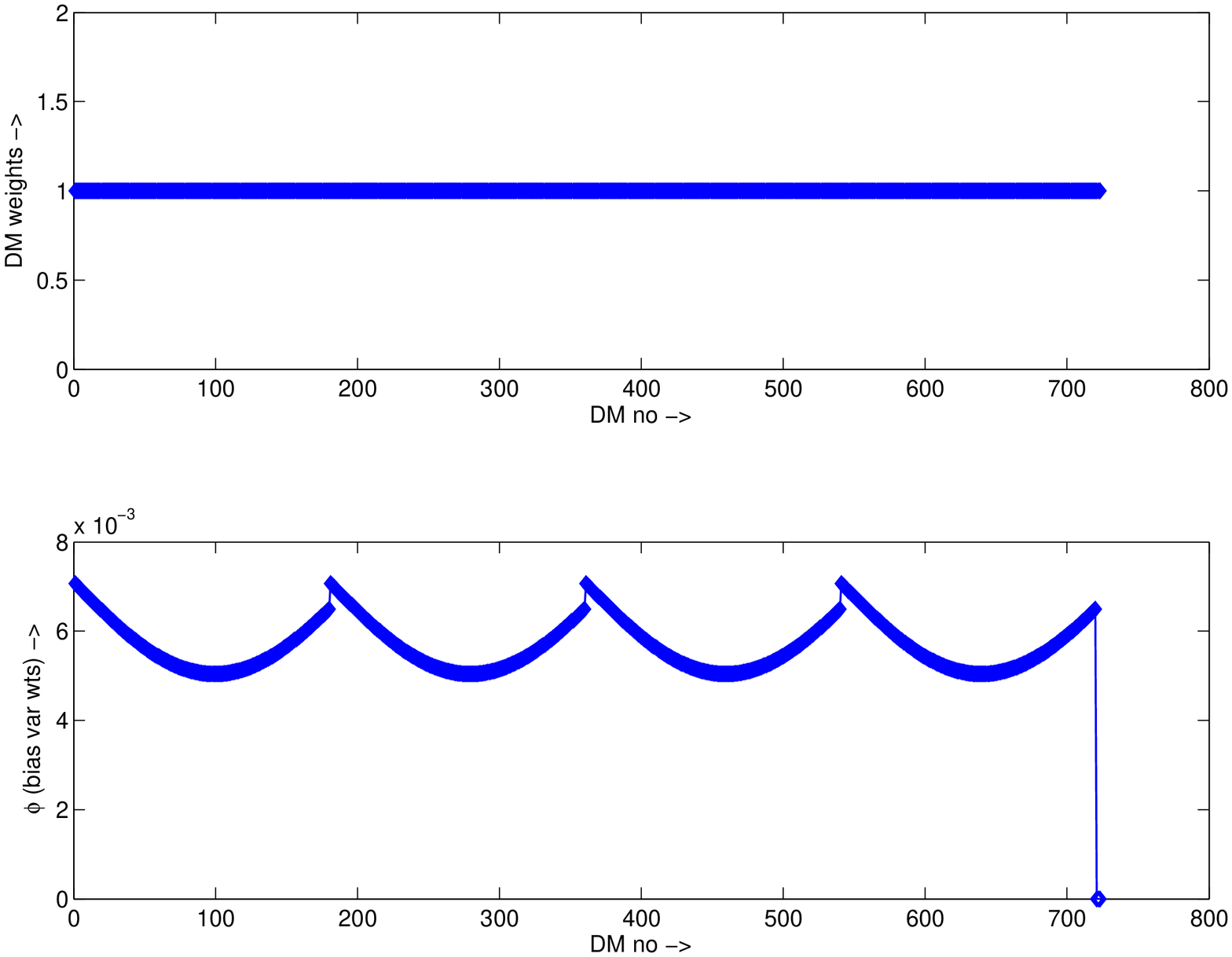}
}

\end{tabular}
\caption{  fMRI case study: (a) $R$ versus $p$ curve for determining the optimal number of columns in $Z$. Using a cutoff of $R_c = 0.95$ the optimal number of columns in $Z$ was determined to be $p_{opt} = 6$. (b) The first 2 columns of $Z$ were constrained and the others were left unconstrained during optimization. The contrast was fixed at $[1;0;0;0;0;0]$. The automatic initialization strategy described before was used to initialize the 6 columns of $Z_0$ (dotted lines). The optimized columns are shown in the same figure using solid lines. (c) Performance curves showing the fractional contrast bias $C_b$, contrast variance change w.r.t Gauss-Markov estimate $CV_{\Delta}$ and model variance bias $V_b$ (d) In this example $w_i = 1$ and $\phi_i$ was chosen using the automatic initialization strategy described before.}
\label{casebupA}
\end{figure}

\begin{figure}[htbp]
\centering
\begin{tabular}{cc}

\subfigure[]
{
\hspace{-2cm}
\label{casebupe}
\includegraphics[width = 75mm, angle = -90]{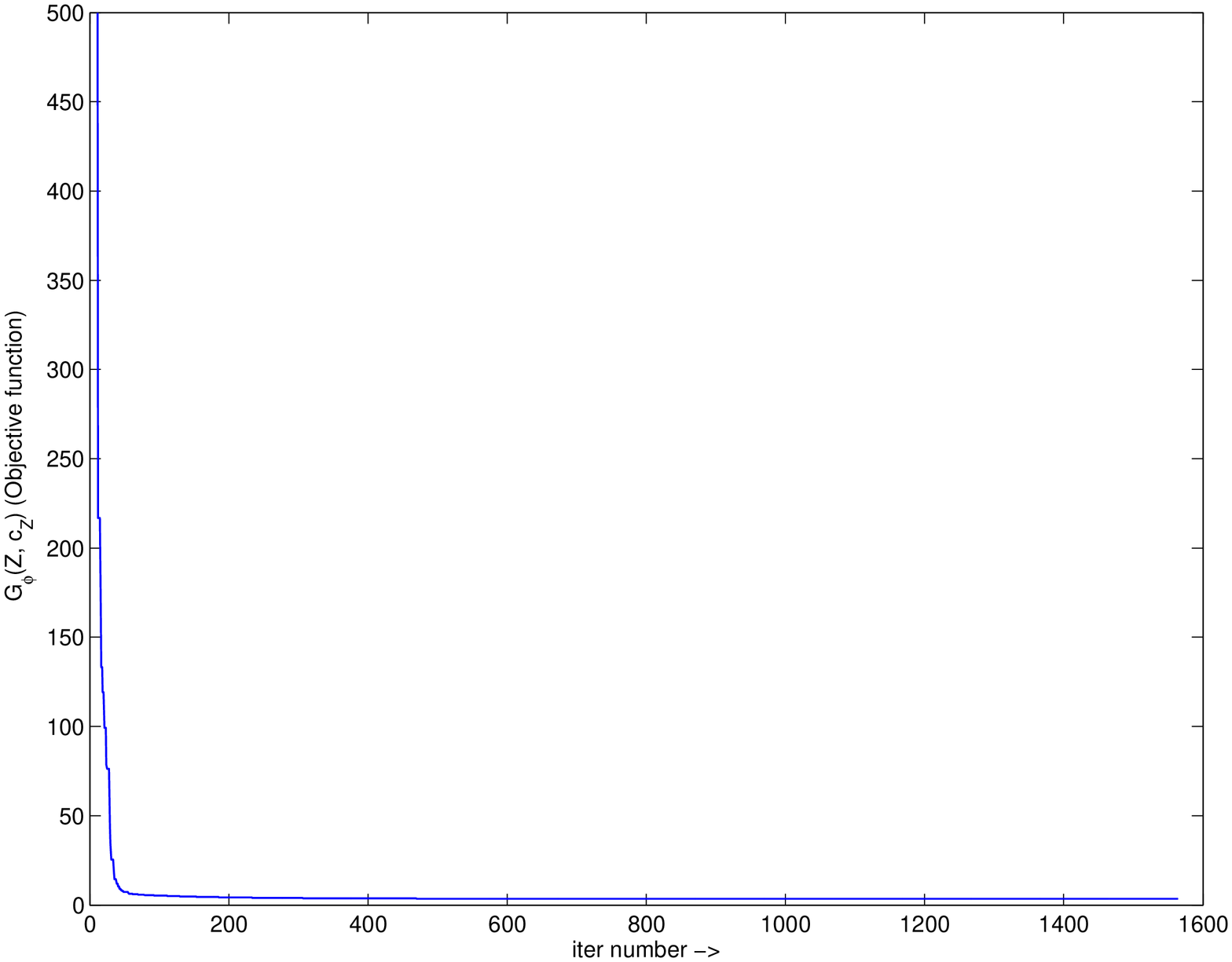}
}

&

\subfigure[]
{
\hspace{-1cm}
\label{casebupf}
\includegraphics[width = 75mm, angle = -90]{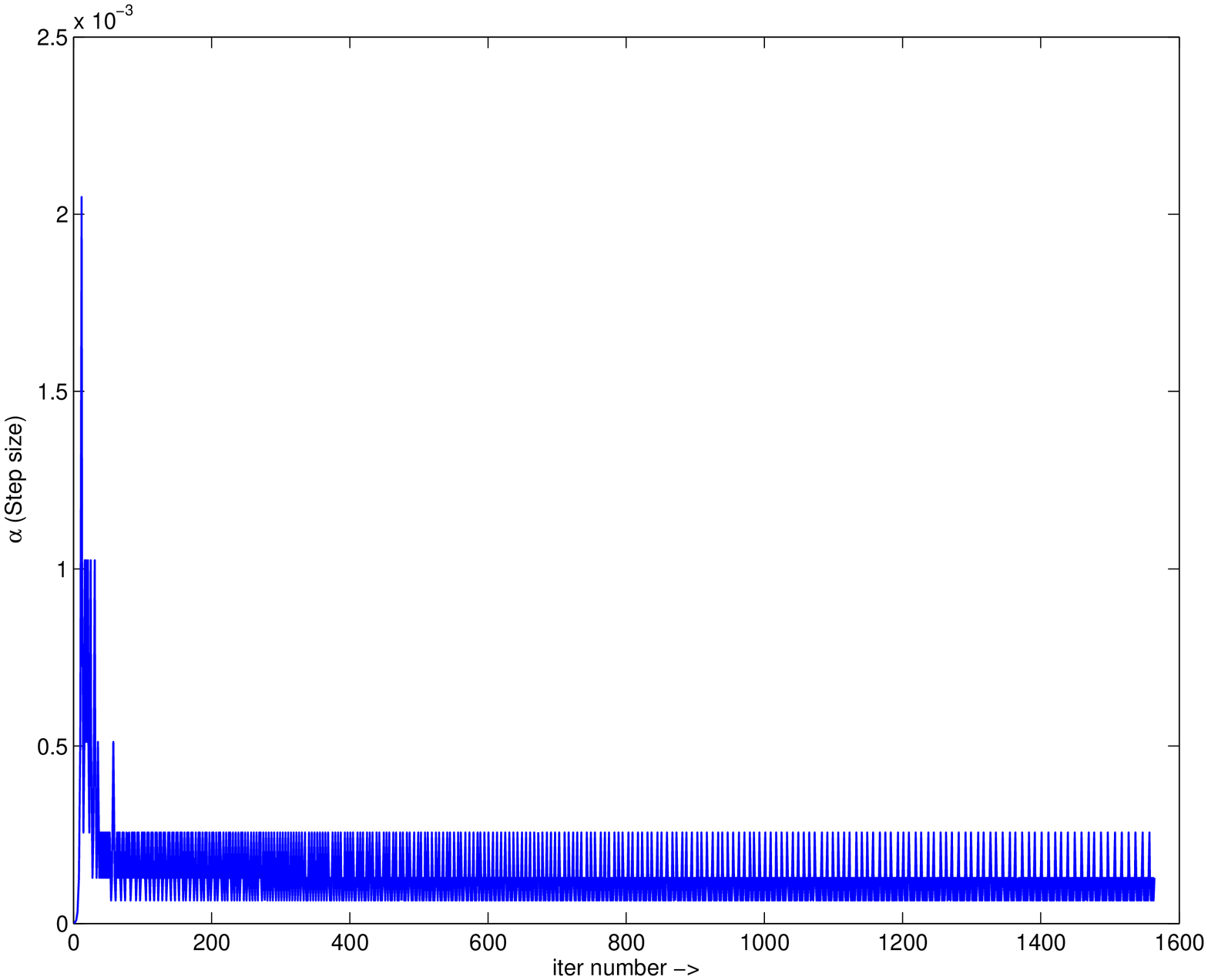}
}

\\

\subfigure[]
{
\hspace{-2cm}
\label{casebupg}
\includegraphics[width = 75mm, angle = -90]{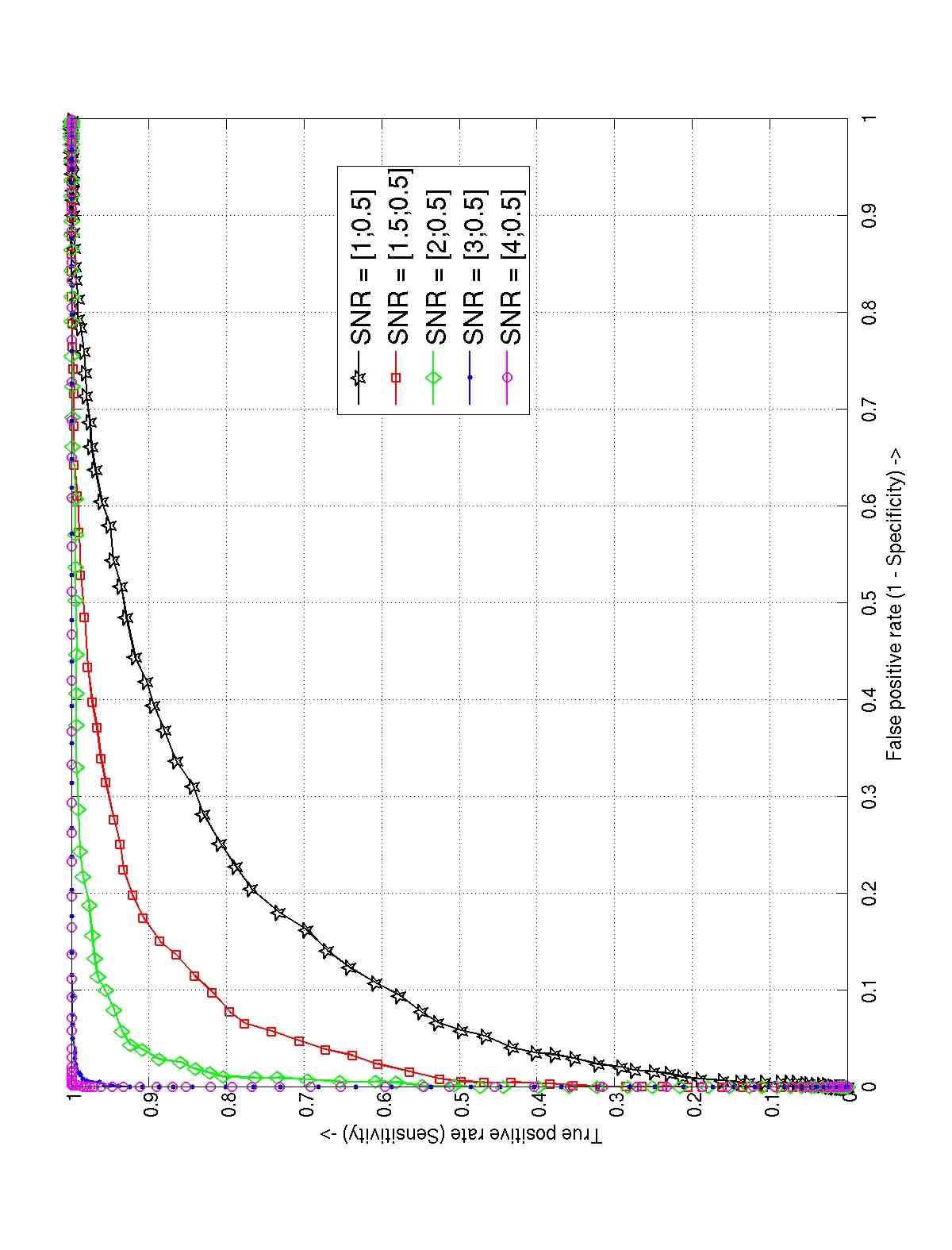}
}

&

\subfigure[]
{
\hspace{-1cm}
\label{casebuph}
\includegraphics[width = 75mm, angle = -90]{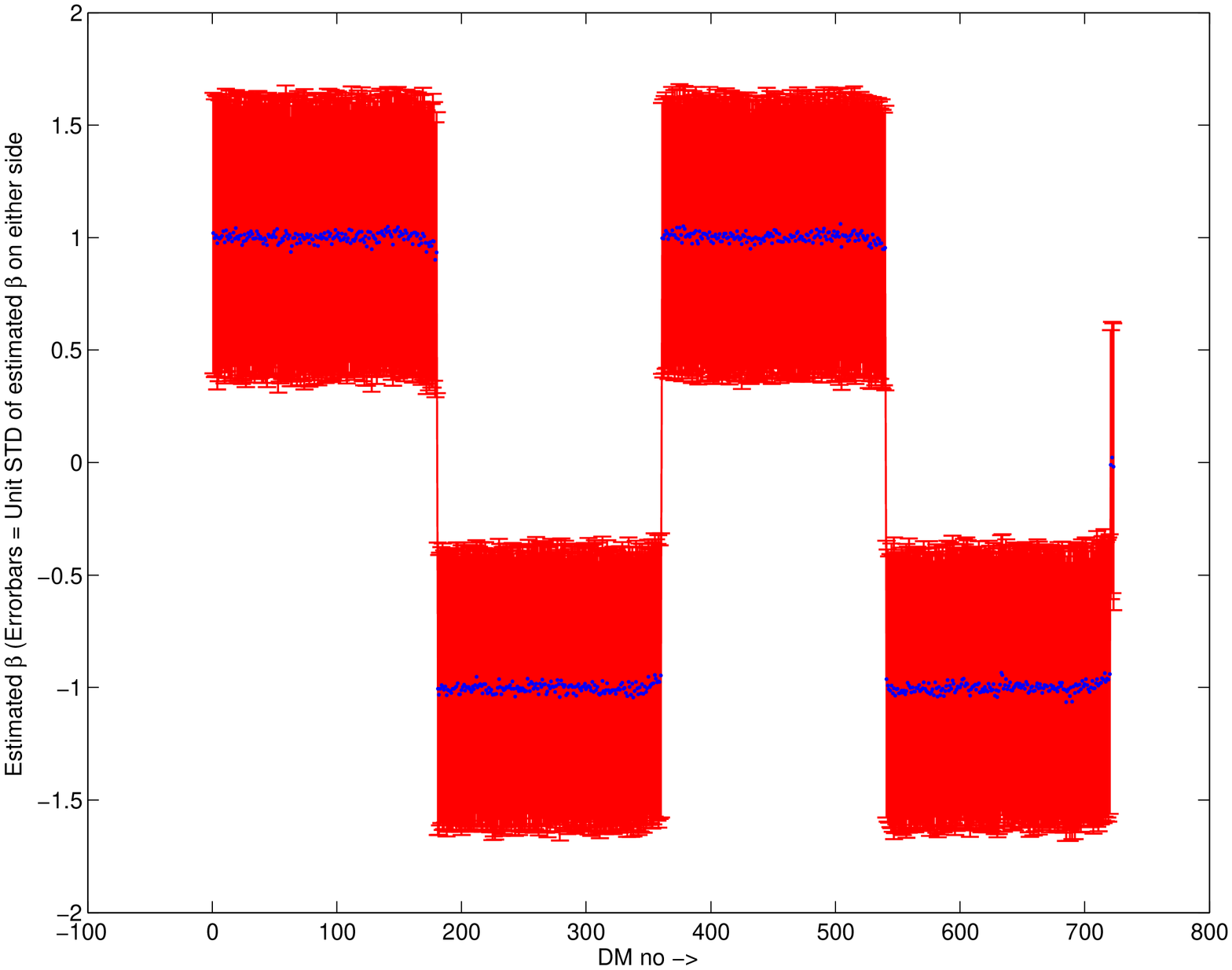}
}

\end{tabular}
\caption{  fMRI case study: (a) Figure showing the evolution of objective function values $G_\phi(Z,c_Z)$ over algorithm iterations. Notice how the function value stabilizes as convergence is reached (b) Figure showing the variation in the step size $\alpha$ over algorithm iterations. Step size controlling parameter $\theta$ in Algorithm 1 was set to $\theta = 2$. For each design matrix (DM) entered into optimization, 1000 simulated data-sets were generated at SNR $\frac{\beta_i}{\sigma_i}$. A GLM analysis was run on each of these data-sets using the optimized DM. Figure (c) shows the ROC curve for data generated from the design matrix $X_{180}$ at various SNR values. Figure (d) is a summary errorbar plot showing $\hat{E}(c_Z^T \hat{\gamma})$ over 1000 simulations for data generated from each DM. The error bars represent unit standard deviation of $c_Z^T \hat{\gamma}$ (\textbf{not} standard deviation of $\hat{E}(c_Z^T \hat{\gamma})$ )  to quantify the variance in estimation via simulation.}
\label{casebupB}
\end{figure}

(1) to (4) above instruct the optimization process to be sensitive to both "activation" and "deactivation" with both "positive" and "negative" linear drift. We also explicitly instruct the optimization process to match pure "positive" or "negative" linear drift to an "infusion response" of size 0 using (5) and (6) above. (7) instructs the optimization to match "infusion response" to size 0 in the absence of any signal (either infusion response or linear drift). We choose a conservative SNR of 1 for the signal of interest and a linear drift amplitude 50\% that of the main signal. The computed optimal DM and its performance are shown in Figures \ref{casebupA}-\ref{casebupB}

For the optimization, we fixed the first two columns of $Z$ to the EVs shown in \ref{fig1}. The contrast was also fixed to be $[1;\ldots;0]$ corresponding to the main infusion response.


\subsubsection{GLM analysis}

\begin{figure}[htbp]
\begin{center}
\includegraphics[width = 100mm, angle = -90]{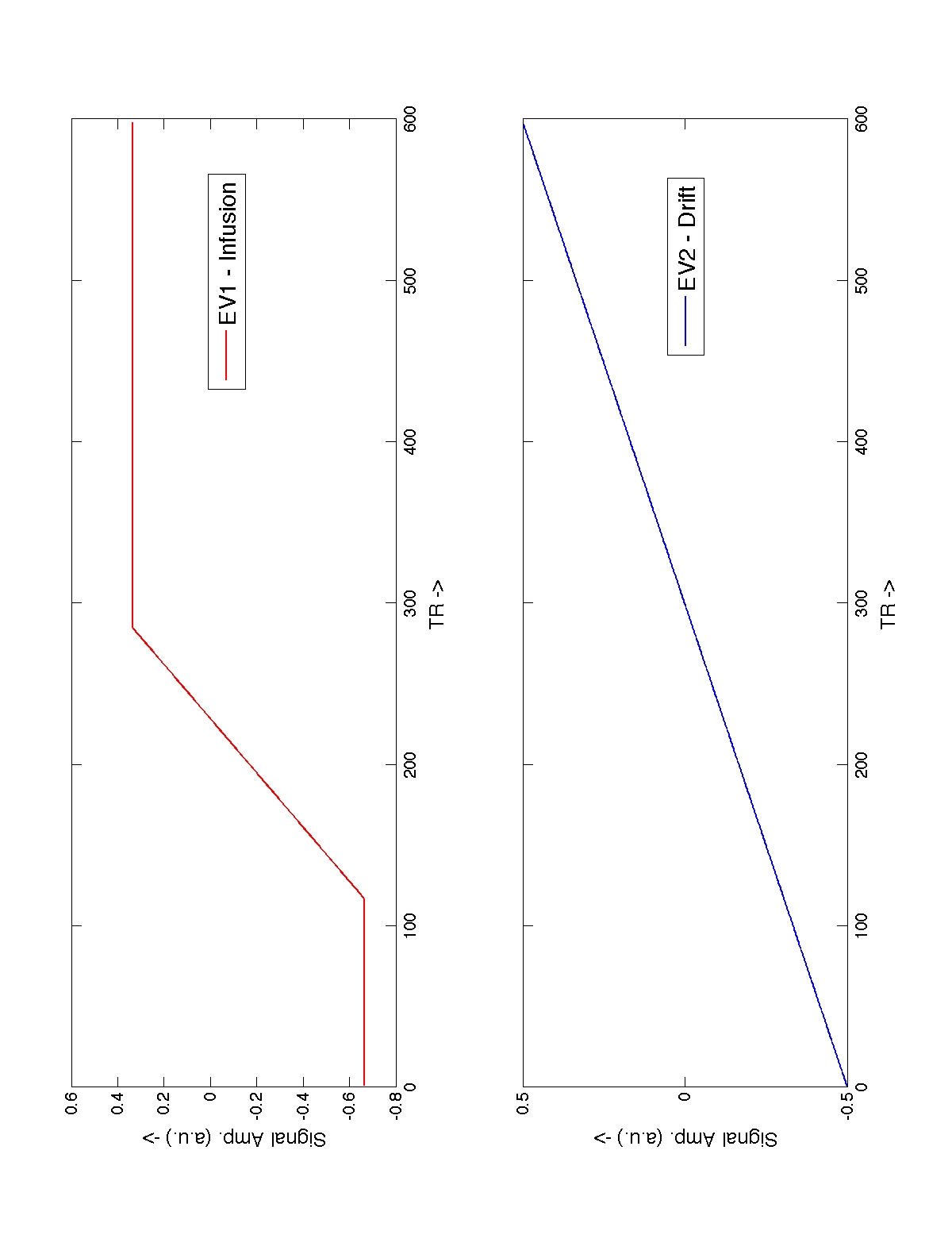}
\caption{Example infusion design matrix. The top figure shows a typical "infusion response" and the bottom figure shows a confounding "linear drift".}
\label{fig1}
\end{center}
\end{figure}

Subsequent to the computation of optimal DM, two GLM analyses were carried out using FSL's tool FEAT:
(1) using an optimized design matrix
(2) using a design matrix containing 3 EVs. The 2 EVs shown in \ref{fig1} and an additional EV representing the temporal derivative of the "infusion EV".

\section{Results}

Figures \ref{convdiag} - \ref{fig17} illustrate the results of validation tests for \hyperlink{alg1}{Algorithm 1}. It was found that the maximum difference in objective function using \hyperlink{alg1}{Algorithm 1} and the more sophisticated optimization solver was of the order of $10^{-5}$ (see Table \ref{table1}).

\subsection{Case studies 1-6}
In Examples 1-4, the size of the optimal DM was chosen to be $p = 3$. The weights $w_i$ were chosen to be $1$ for all $i$ to reflect equal likelihood of potential DMs in all Examples.

\subsubsection{Example 1}
Performance curves for Example 1 are shown in Figures \ref{case1A} - \ref{case1B}. In this example $\phi_i$ was set to its default value of $0.5$. Columns 1 and 2 of the $Z$ were fixed during optimization and the unconstrained column in $Z$ was initialized by drawing from a uniform distribution $U(0,1)$.

It is seen that the contrast bias $|C_{bi}|$ is maintained at $<0.05$ for $i = 1,\ldots,36$. For $i =  37,\ldots,50$, $|C_{bi}|$ increases to around 0.12 for $i=50$. At the same time the contrast variance change w.r.t. Gauss-Markov estimate $CV_{\Delta i}$ decreases monotonically from $0.096$ for $i = 1$ to $0.002$ for $i = 20$ and becomes negative from $i = 21,\ldots,50$ implying a lower variance than the Gauss-Markov estimator. The model variance bias $V_{bi}$ is maintained at $< 2 \times 10^{-4}$ for all $i$.
\subsubsection{Example 2}
Performance curves for Example 2 are shown in Figures \ref{case1AA} - \ref{case1AB}. In this example, $\phi_i$ was chosen automatically as described in section \ref{intelligentphi}. Columns 1 and 2 of the $Z$ were fixed during optimization and the unconstrained column in $Z$ was initialized at the optimal solution from Example 1.

It is seen from Figure \ref{case1Ad} that lower weights $\phi_i$ were assigned for increasing $i$ (approximately $20\%$ lower for $i = 50$ compared to $i = 1$). With this choice of $\phi_i$ it was seen that the contrast bias $|C_{bi}|$ was reduced to $< 0.01$ for all $i < 36$. For $i = 37\ldots,50$, $|C_{bi}|$ increases to a maximum of $0.03$ for $i = 50$. The value of $CV_{\Delta i}$ decreases monotonically from 0.345 for $i = 1$ to 0.07 for $i = 50$ implying an increased variance compared to Example 1. The model variance bias $V_{bi}$ was maintained at $< 4 \times 10^{-3}$ for all $i$.
\subsubsection{Example 3}
Performance curves for Example 3 are shown in Figures \ref{case2A} - \ref{case2B}. In this example, $\phi_i$ was chosen to be $0.1$ for all $i$. Columns 1 and 2 of the $Z$ were fixed during optimization and the unconstrained column in $Z$ was initialized by drawing from a uniform distribution $U(0,1)$. 

It was found that $|C_{bi}|$ was maintained at $< 0.02$ for $i = 1,\ldots,41$. For $i = 42,\ldots,50$, $|C_{bi}|$ increases to $0.055$ for $i = 50$. It was also found that $CV_{\Delta i}$ decreases monotonically from $0.195$ for $i = 1$ to $0.003$ for $i = 38$ and becomes negative from $i = 39,\ldots,50$ indicating a lower variance than the Gauss-Markov estimator. The model variance bias $V_{bi}$ was maintained $<1 \times 10^{-3}$ for all $i$.

\subsubsection{Example 4}
Performance curves for Example 4 are shown in Figures \ref{case3A} - \ref{case3B}.
In this example, $\phi_i$ was chosen to be $0.01$. Optimization was initialized using solution found in Example 1. The sample space of $Z$ was left unconstrained but the set of potential DMs was augmented to explicitly instruct the optimization to enable detection of signals in the presence of confounds, as well as treat "drift" signals as "null" data.

It was found that $|C_{bi}|$ was reduced to $<0.02$ for all $i$ and $CV_{\Delta i}$ decreased monotonically from around $0.32$ to around $0.05$ for each of the four segments $i = 1,\ldots,50$, $i = 51,\ldots,100$, $i = 101,\ldots,150$ and $i = 151,\ldots,200$. The model variance bias $V_{bi}$ was maintained $< 2 \times 10^{-3}$ for all $i$.

\subsubsection{Example 5}
Performance curves for Example 5 are shown in Figures \ref{case4A} - \ref{case4B}.
This example illustrated a block design experiment where the user wants to primarly control bias. $\phi_i$ was chosen to be 0.01 in this example indicating a preferential reduction of bias. The first column of the optimal DM was fixed to the primary block EV. The optimization process was initialized using shifted versions of the primary EV. 

It was found that the optimal DM reduced $|C_{bi}|$ to $< 0.0042$ for $i$ while $CV_{\Delta i}$ was around  2.38 for $i = 1,3,4,6,7,9,10$ and 2.61 for $i = 2,5,8,11$ indicating an increased variance relative to the Gauss-Markov estimator. The model variance bias $V_{bi}$ was maintained $<2.4 \times 10^{-3}$ for $i = 1,3,4,6,7,8,10$ and $<6.7 \times 10^{-2}$ for $i = 2,5,8,11$.

\subsubsection{Example 6}
Peformance curves for Example 6 are shown in Figures \ref{case5A} - \ref{case5B}.
This example illustrated the construction of a set of optimal HRF capturing functions that enable capture of locally variable HRF functions in the brain. The optimization process was explicitly indicated to match "null" data with signal size of "0" optimally using additional DMs from $i = 201,\ldots,400$ at $\frac{\beta_i}{\sigma_i} = 0$.
$\phi_i$ was set to its default value of $0.5$. The size of optimal DM was found to be $p = 5$ using \hyperlink{alg2}{Algorithm 2}. The 5 columns of optimal DM were left unconstrained during the optimization process and initialized using the procedure described in section \ref{initializationstrategy}.

It was found that the mean absolute value of contrast bias was maintained at $|C_{bi}| <0.075$ and the mean value of $CV_{\Delta i}$ was maintained at $<0.107$ for all HRF shapes $i = 1,\ldots,200$. For the "null" data $|C_{bi}| = 0$ and the mean $CV_{\Delta i}$ was maintained at $<0.106$ for $i = 201,\ldots,400$. The model variance bias $V_{bi}$ was reduced to $< 8 \times 10^{-3}$ for all $i$.

\subsection{fMRI case study}
This case study deals with the optimal capture of signals for an fMRI infusion study. An initial model free exploration of data revealed the presence of multiple infusion profiles differing in their time to take off from baseline as illustrated in Figure \ref{figbup}. For the DM optimization, $\phi_i$ was initialized automatically using the strategy describe in section \ref{intelligentphi}. The first two columns of $Z$ were fixed to the EVs described in Figure \ref{fig1}. The optimal number of columns in $Z$ were estimated using \hyperlink{alg2}{Algorithm 2} to be $p = 6$. DM optimization was initialized automatically using the procedure described in section \ref{initializationstrategy}. The set of potential DMs were augmented with additional DMs (total 723 potential DMs) to guarantee detection of "positive" and "negative" activation in the presence of confounding drift as well as to match both the "drift" and "null" data to a signal size "0".

Performance curves for the optimal DM are shown in Figures \ref{casebupA} - \ref{casebupB}. 
It was found that the mean absolute value of $|C_{bi}|$ was reduced to $0.0057$ and the mean value of $CV_{\Delta i}$ was $5.70$ for $i = 1,\ldots,720$ representing the "non-null data". For the "null" data from $i = 721,\ldots,723$, we found $|C_{bi}| \equiv 0$ and the mean $CV_{\Delta i}$ was $7.28$. The model variance bias $V_{bi}$ was reduced to $<1.7 \times 10^{-2}$ for all $i$. Figure \ref{casebupg} shows the ROC curve for the detection of signal generated from $X_{180}$ versus the "null" data generated from a "pure" drift signal at SNR $[0;0.5]$ over 1000 simulations from each. We chose $X_{180}$ because it is an extreme DM that produces the highest absolute contrast bias $|C_{bi}| \sim 0.063$. A simple decision rule based on the $T$ statistic calculated from equation \ref{tstatdef} was used for positive activation detection to generate the ROC curve. For a cutoff $T$-statistic $t_c$:
\begin{eqnarray} 
\mbox{Decide activation if: } T(\hat{\gamma}, \hat{\sigma_1}; \hat{Z}; \hat{c_Z})  \ge t_c \\
\mbox{Decide null if: }  T(\hat{\gamma}, \hat{\sigma_1}; \hat{Z}; \hat{c_Z}) < t_c 
\end{eqnarray}
where $\hat{\gamma}$ and $\hat{\sigma_1}$ are the parameter vector and residual standard deviation respectively estimated using the optimal DM $\hat{Z}$ and contrast $\hat{c_Z}$. It is found that for SNR of $\frac{\beta}{\sigma} = [2;0.5]$, a sensitivity of 93.5\% was obtained at a specificity of 94.2\% and for SNR of $\frac{\beta}{\sigma} = [3;0.5]$, a sensitivity of 98.8\% was obtained at a specificity of 98.7\%. The real fMRI data was found to have approximately an $SNR > 4$ for the primary infusion response indicating a high sensitivity and specificity of signal detection using the optimal DM $\hat{Z}$.

Figure \ref{figfmri} shows the $t$-stat image corresponding to the "infusion response" overlaid onto the anatomical image (transformed into the "fmri" space) of the subject. For illustration purposes, we extracted raw timecourses from 6 sample points in the brain the details of which are shown in Table \ref{tblsampleprofiles}. Figure \ref{casefmriA} - \ref{casefmriC} show the raw timecourses from real fMRI data with full and partial model fits obtained using the optimal DM $\hat{Z}$. For illustration purposes we also show the full and partial model fits obtained using a "naive" DM such as that using the temporal derivative of the "infusion response" as a covariate. Sample point 1 represents a canonical 0-delay infusion response, while sample point 2 represents a 200 timepoint delayed infusion response. Sample points 3 and 4 represent infusion responses with delays of around 100 timepoints while sample points 5 and 6 represent infusion responses with delays of around 150 timepoints. It was found that using the optimal DM $\hat{Z}$ it was possible to detect all infusion responses (delayed or not)  in a robust and unbiased fashion.

\begin{figure}[htbp]
\begin{center}
\includegraphics[width = 100mm, angle = -90]{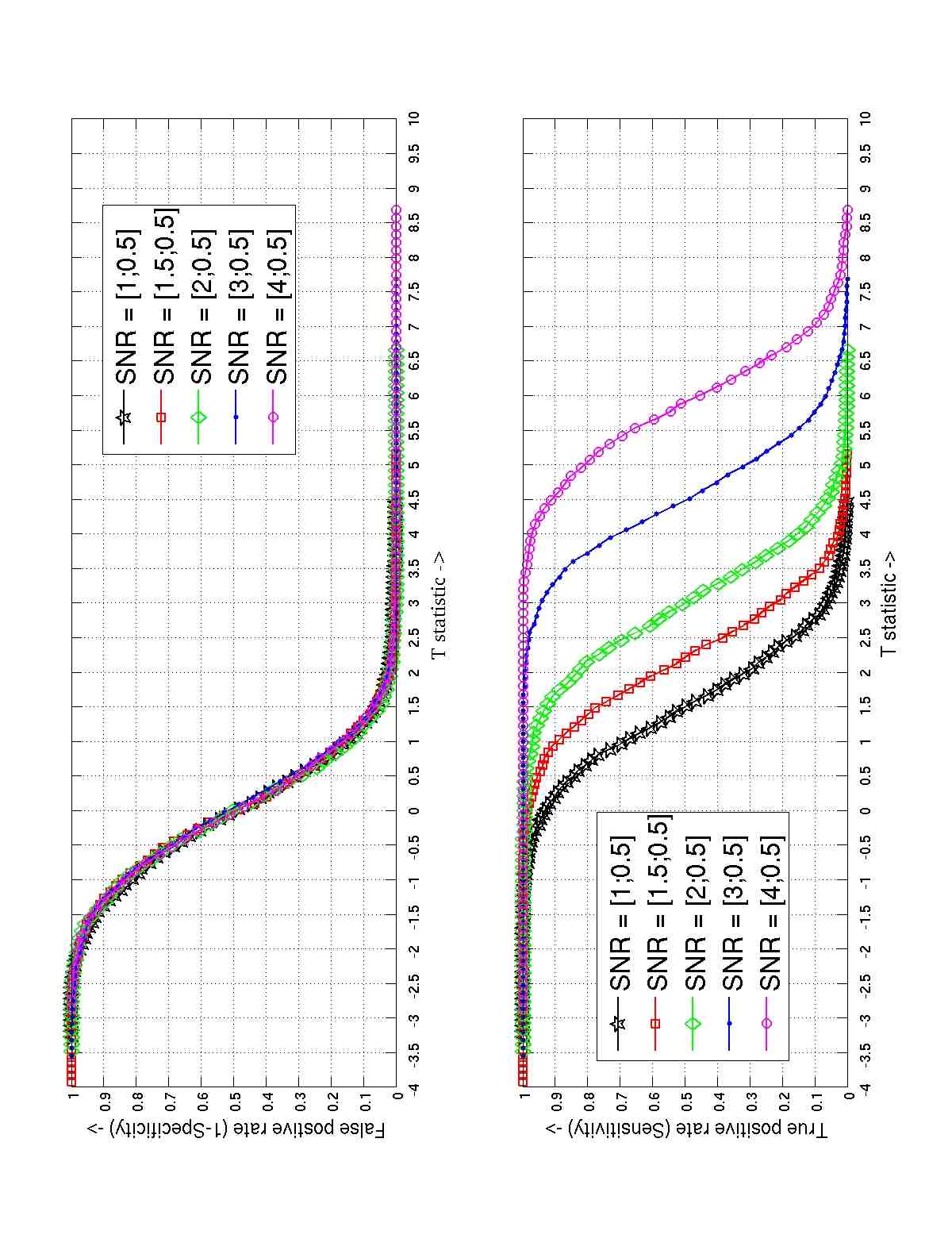}
\caption{Figure shows the variation of False positive rate (1 - Specificity) and True positive rate (Sensitivity) with the $T$ statistic threshold for data generated as follows: (1) "Activation" generated at various SNRs from an extreme DM - $X_{180}$, that produces the highest absolute contrast bias $|C_{b}| \sim 0.063$ of all 723 DMs used in optimizaton (2) "Null" data generated from a "pure drift" signal at SNR $[0;0.5]$ which is a covariate of no interest and hence should be matched to a signal size of "0".}
\label{figroc2}
\end{center}
\end{figure}

\begin{figure}[htbp]
\begin{center}
\includegraphics[width = 180mm]{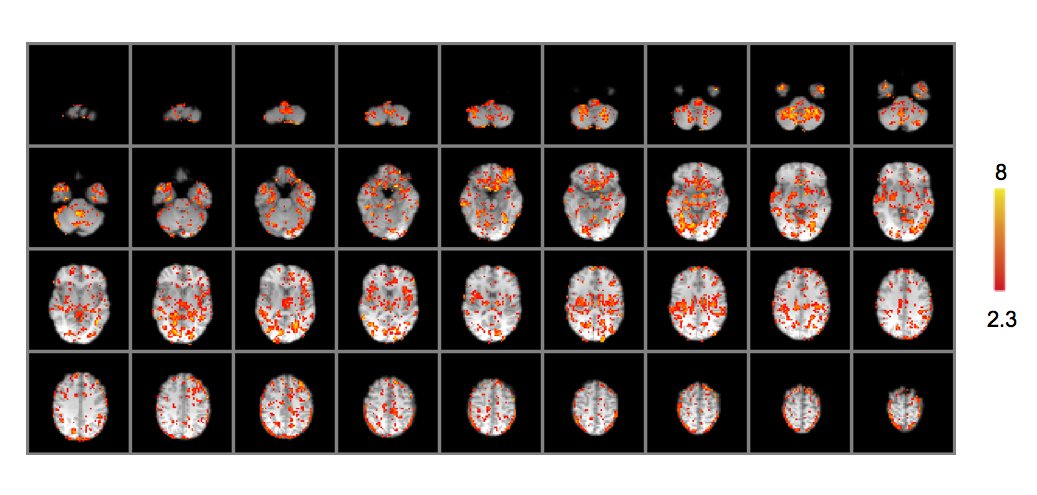}
\caption{fMRI data was analyzed using GLM with the optimal DM. Figure shows $t$-stat map for the contrast of interest $[1;0;0;0;0;0]$ representing the "infusion response". The anatomical image for the sample subject was transformed to native space for $t$-map display purposes. The $t$-map was thresholded at an uncorrected threshold of $2.3$. From the ROC curve shown in Figure \ref{figroc2}, a specificity of $\sim 99\%$ and a sensitivity of $>98\%$ is obtained at SNRs $\ge 3$. Real fMRI data had an SNR $> 4$ and hence a cutoff of 2.3 on the $t$-statistic map performs well.}
\label{figfmri}
\end{center}
\end{figure}

\begin{table}[htdp]
\begin{center}
\begin{tabular}{|c|c|c|c|}
\hline
Sample & Coords & $t$-stat Opt DM & $t$-stat Der DM \\
\hline
1 & [42,41,29] & 4.84 & 8.56 \\
\hline
2 & [40,20,25] & 4.06 & -4.39 \\
\hline
3 & [42,28,24] & 5.89 & -0.86 \\
\hline
4 & [36,16,24] & 4.45 & -2.38 \\
\hline
5 & [18,26,24] & 3.27 & -1.67 \\
\hline
6 & [36,11,22] & 4.35 & -1.17 \\
\hline
\end{tabular}
\end{center}
\label{tblsampleprofiles}
\caption{Table shows 6 sample voxels, their co-ordinates and the GLM $t$-stat values for the contrast representing the "size" of infusion EV using the optimal design matrix (Opt DM) and the design matrix with the temporal derivative (Der DM).}
\end{table}%

\begin{figure}[htbp]
\centering
\begin{tabular}{cc}

\subfigure[]
{
\hspace{-2cm}
\label{fmri1}
\includegraphics[width = 75mm, angle = -90]{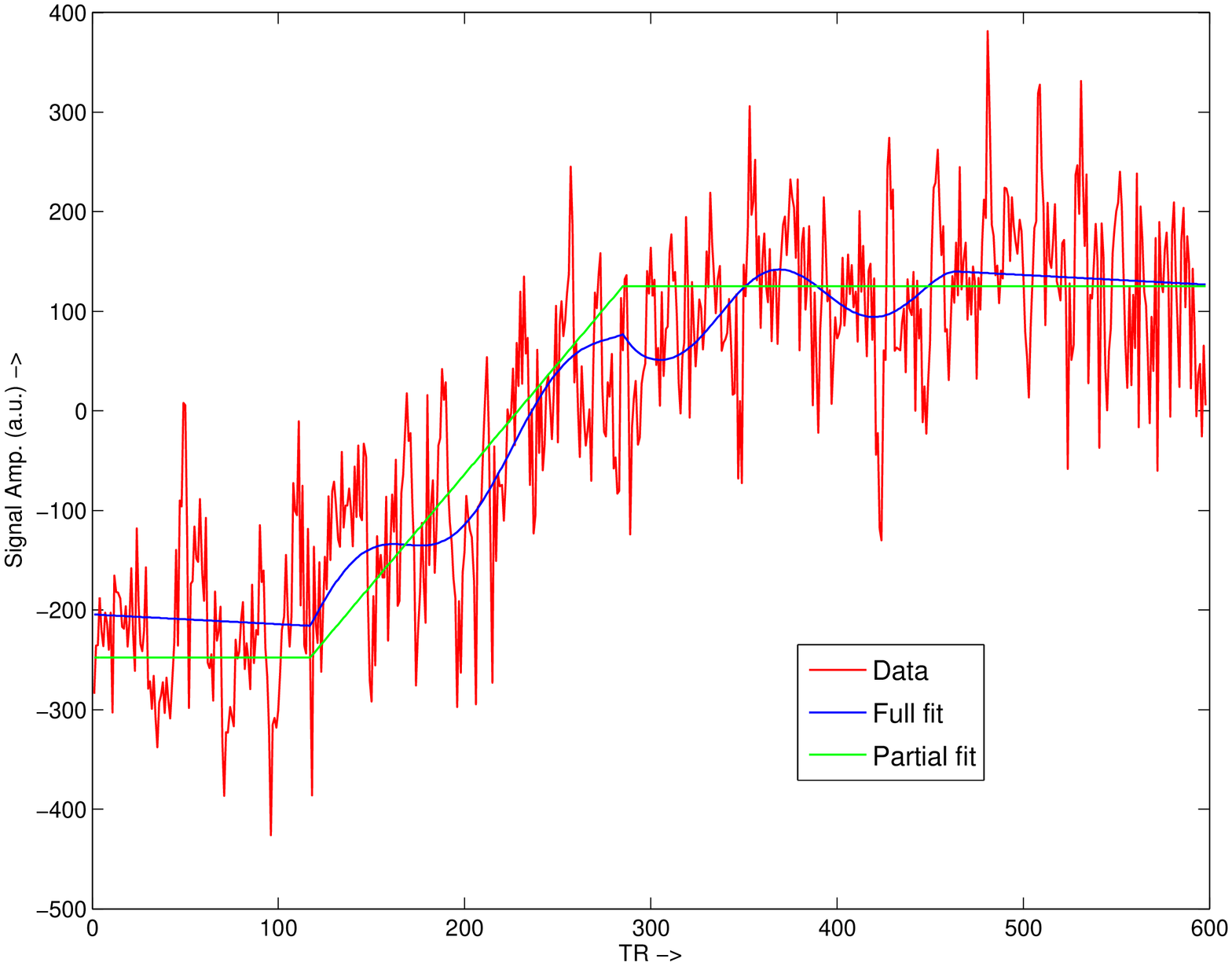}
}

&

\subfigure[]
{
\hspace{-1cm}
\label{fmri2}
\includegraphics[width = 75mm, angle = -90]{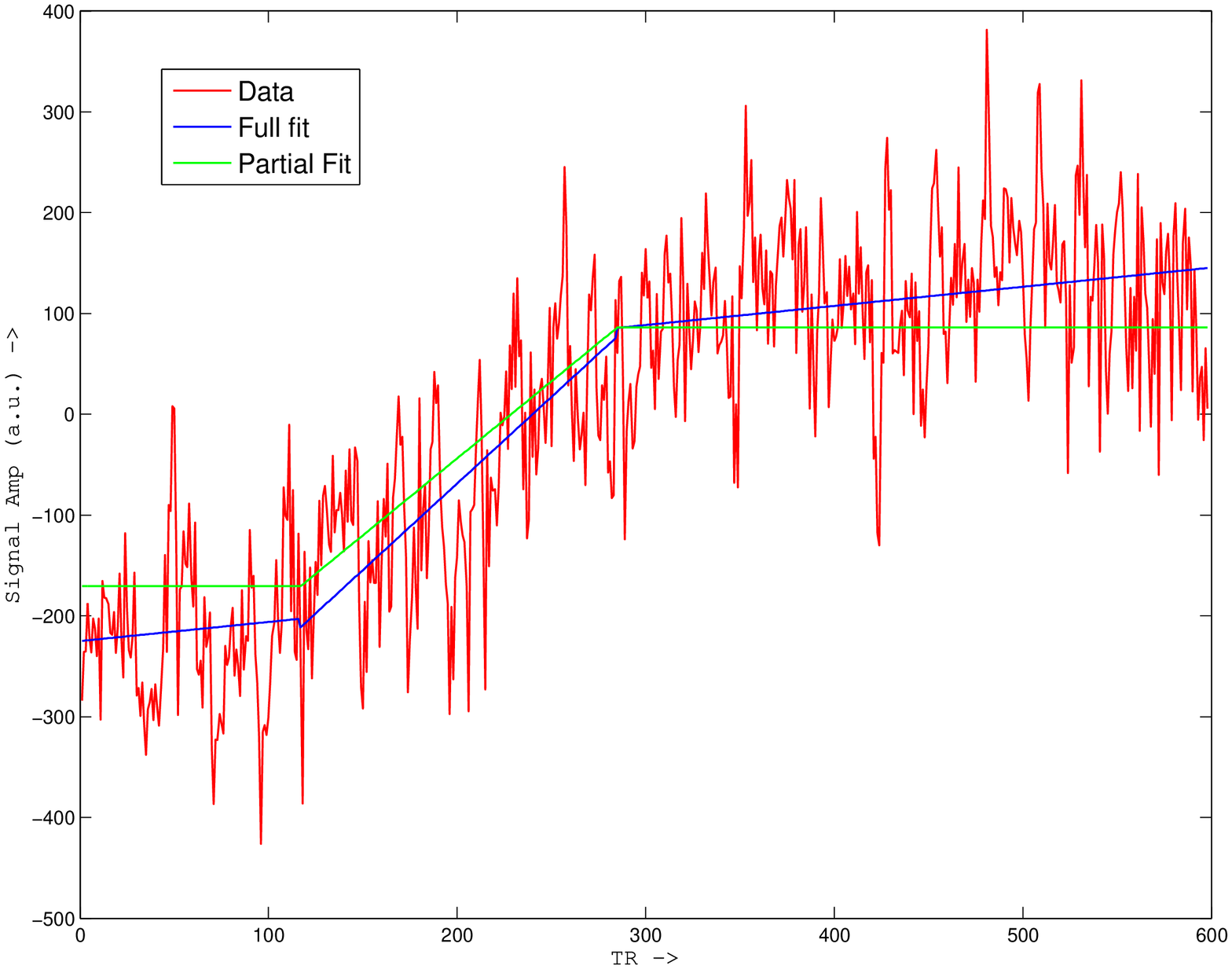}
}

\\

\subfigure[]
{
\hspace{-2cm}
\label{fmri3}
\includegraphics[width = 75mm, angle = -90]{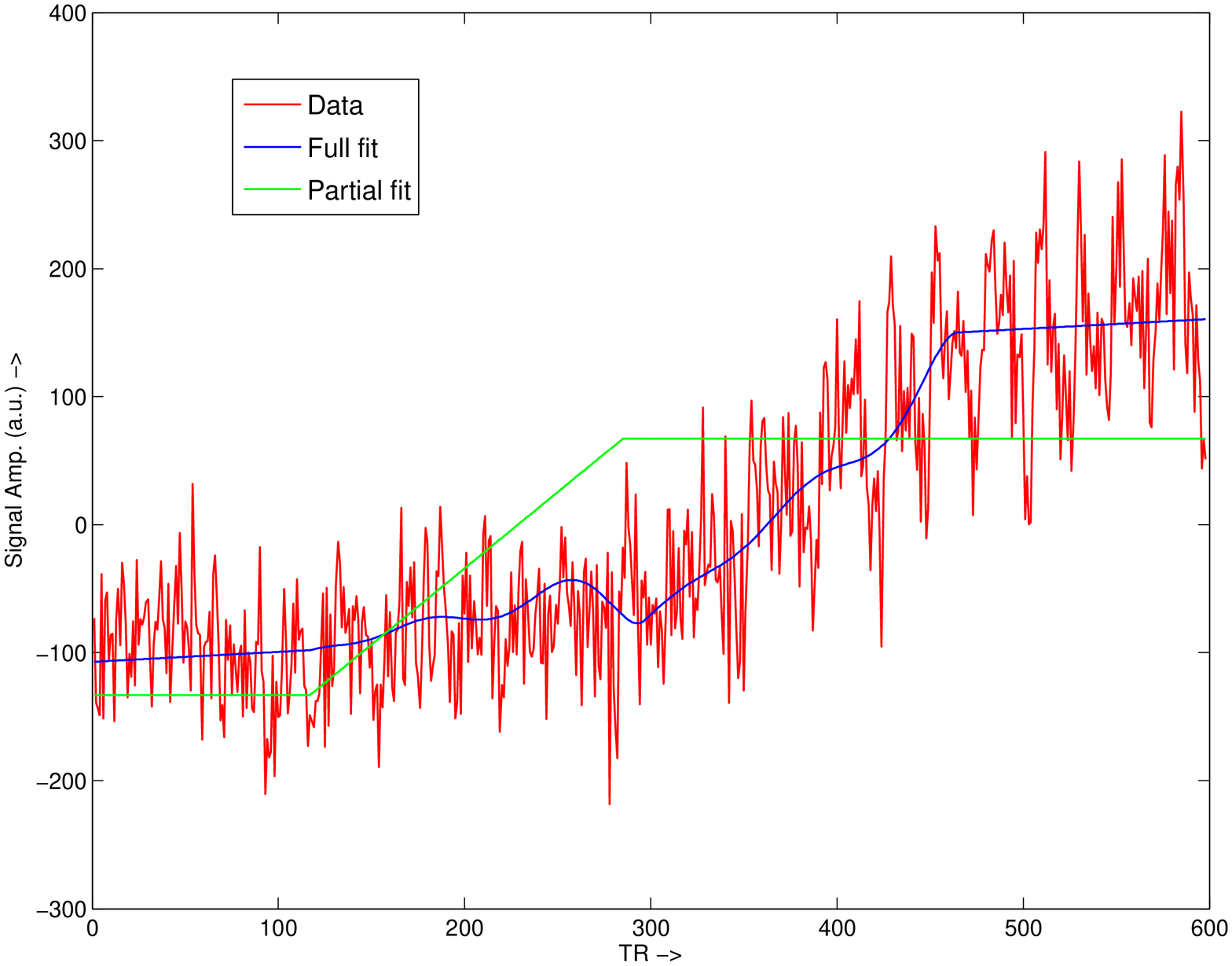}
}

&

\subfigure[]
{
\hspace{-1cm}
\label{fmri4}
\includegraphics[width = 75mm, angle = -90]{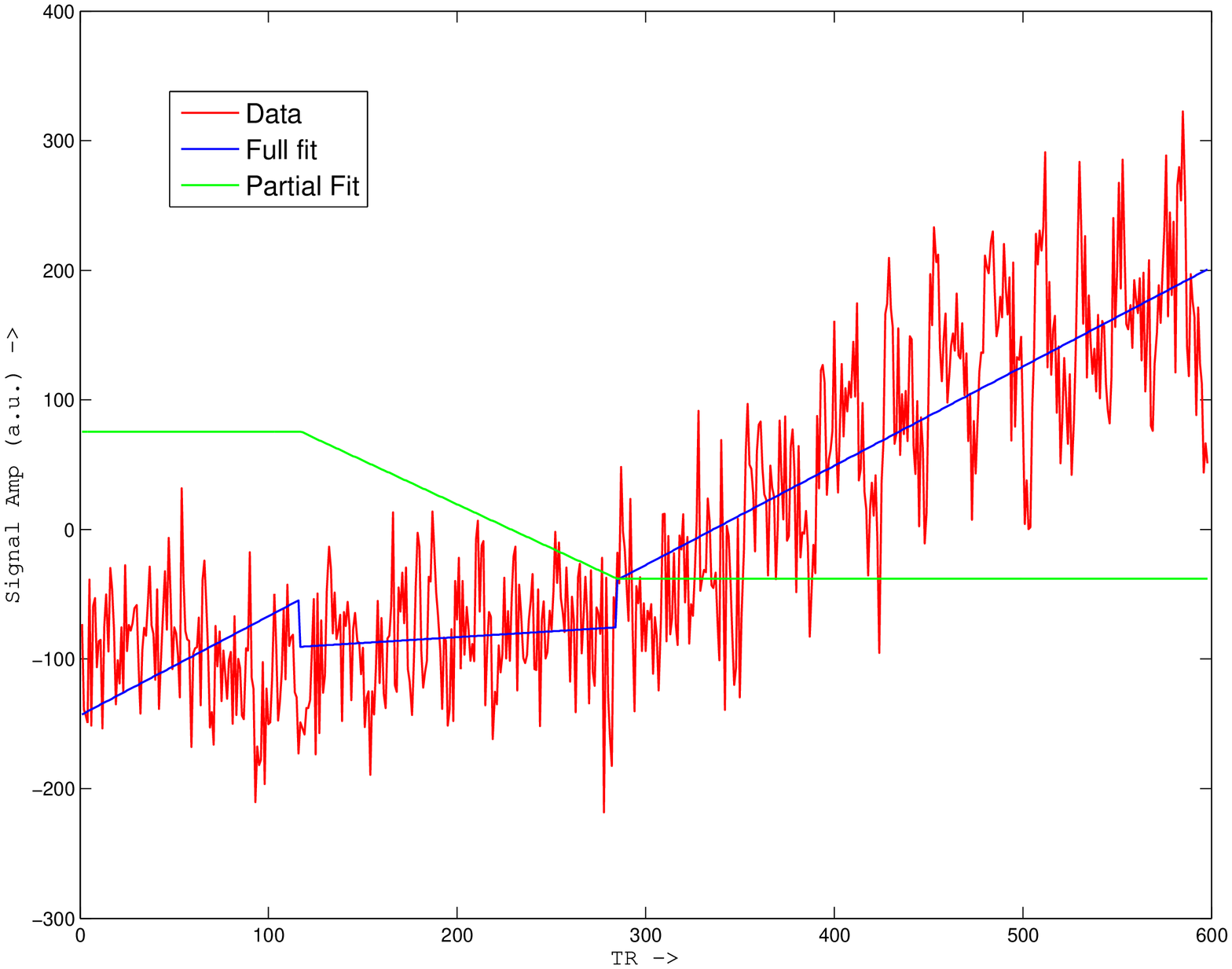}
}

\end{tabular}
\caption{fMRI data extracted from sample voxels 1 and 2. A GLM analysis was run using the optimal DM and a DM containing the 2 EVs shown in \ref{fig1} and the temporal derivative of the 1st EV (i.e., the infusion ev). Figures show the full model fit and partial fit corresponding to the contrast of interest. Figures (a) and (c) are for the optimal DM while (b) and (d) are for the "temporal derivative" based DM.}
\label{casefmriA}
\end{figure}

\begin{figure}[htbp]
\centering
\begin{tabular}{cc}

\subfigure[]
{
\hspace{-2cm}
\label{fmri5}
\includegraphics[width = 75mm, angle = -90]{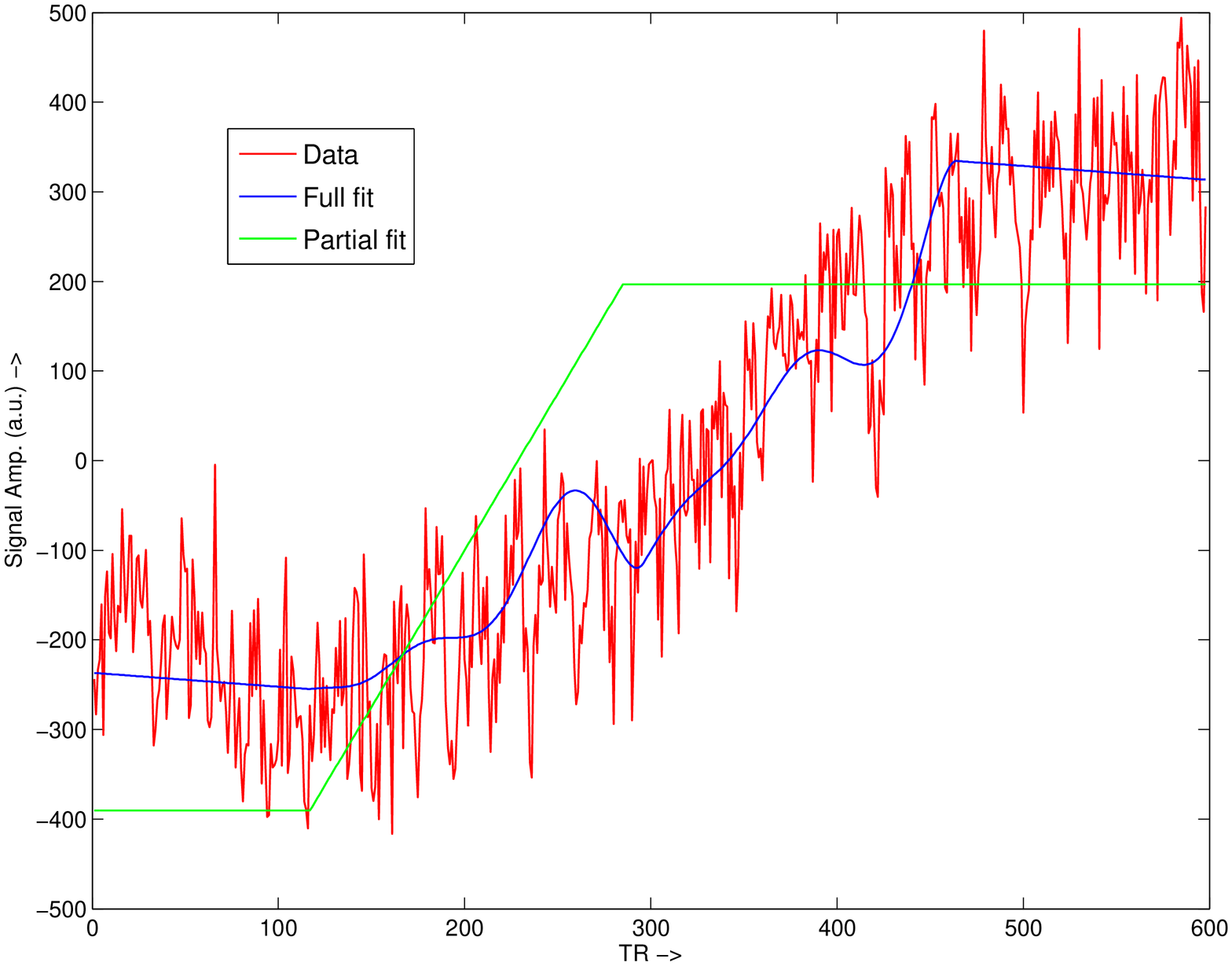}
}

&

\subfigure[]
{
\hspace{-1cm}
\label{fmri6}
\includegraphics[width = 75mm, angle = -90]{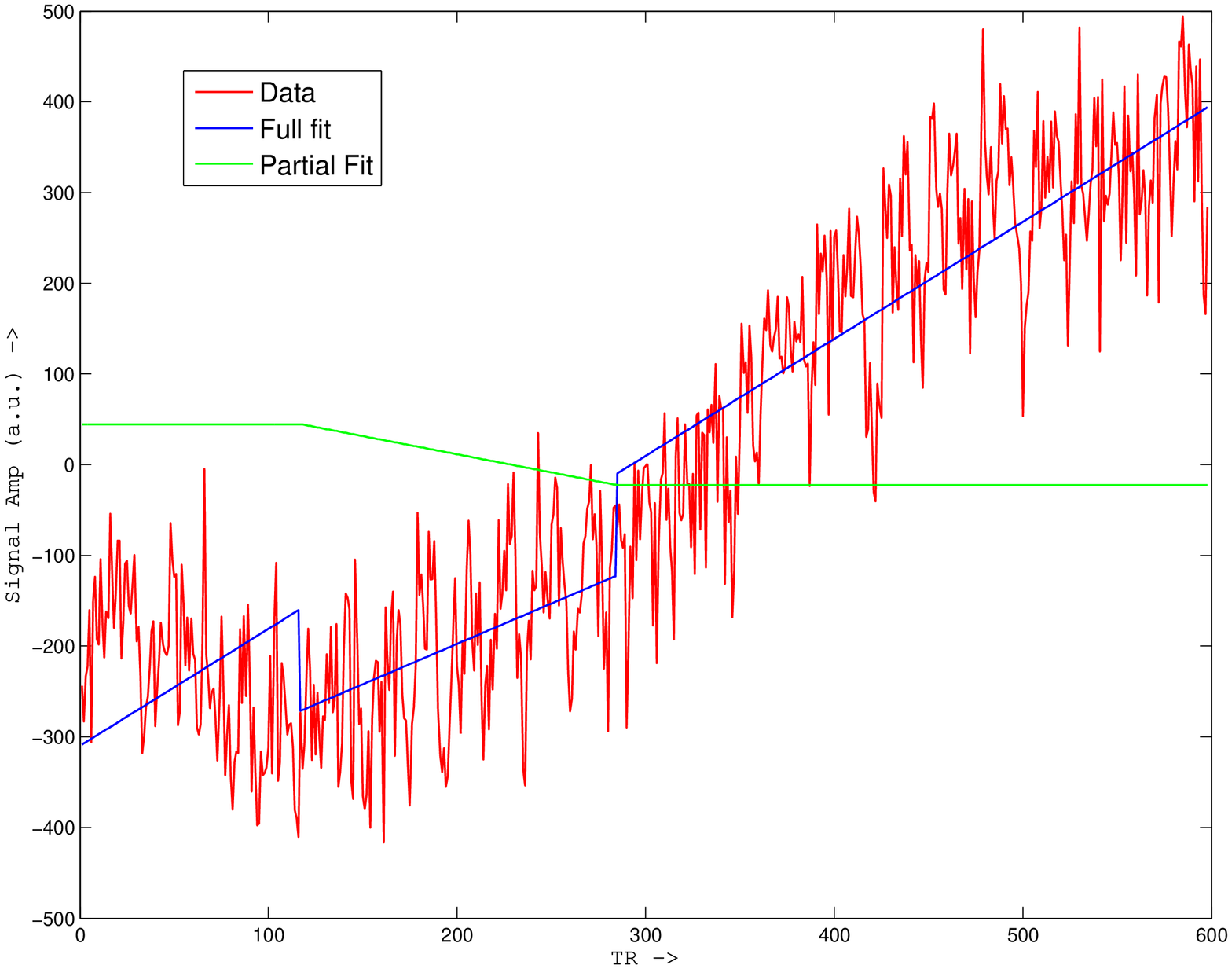}
}

\\

\subfigure[]
{
\hspace{-2cm}
\label{fmri7}
\includegraphics[width = 75mm, angle = -90]{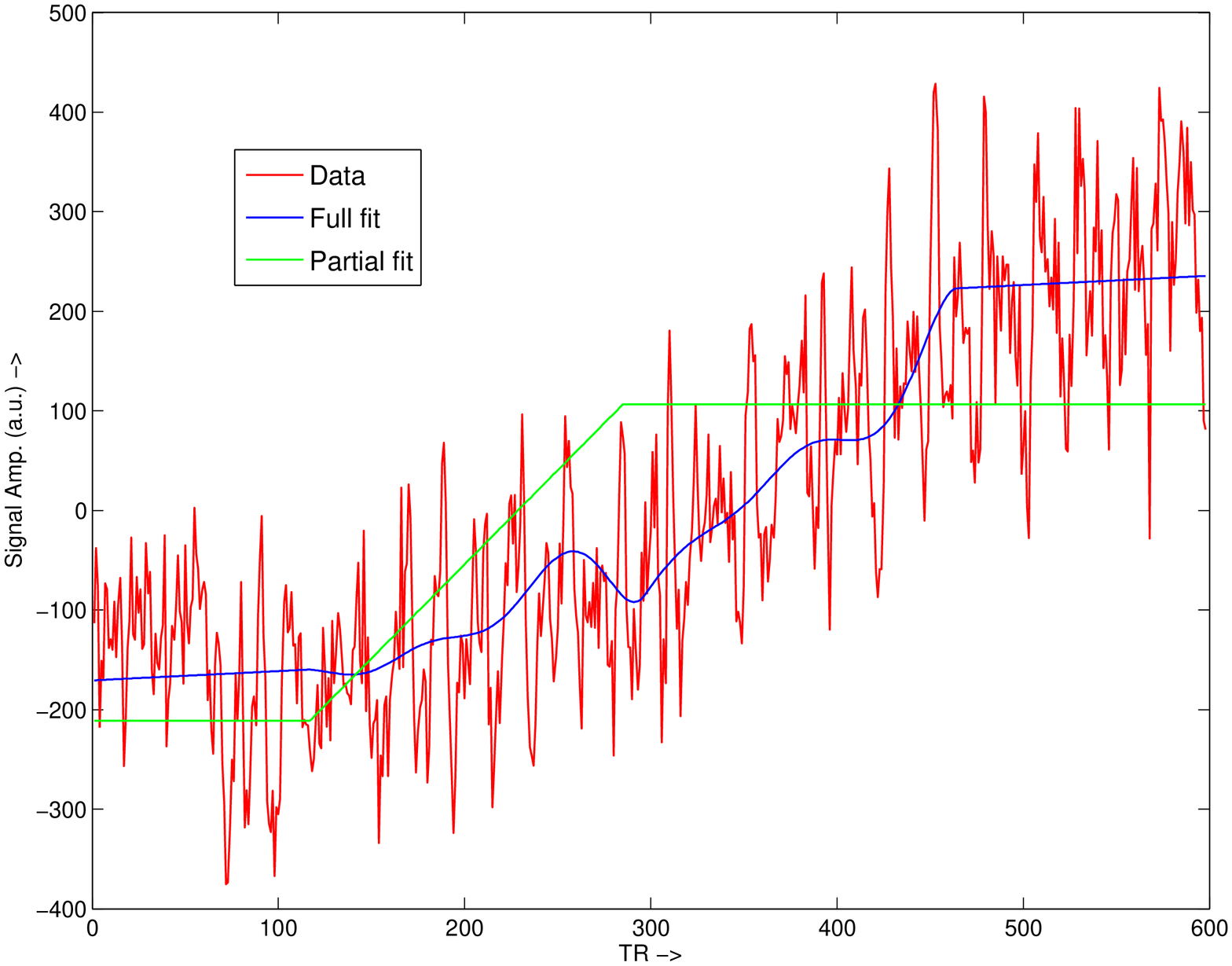}
}

&

\subfigure[]
{
\hspace{-1cm}
\label{fmri8}
\includegraphics[width = 75mm, angle = -90]{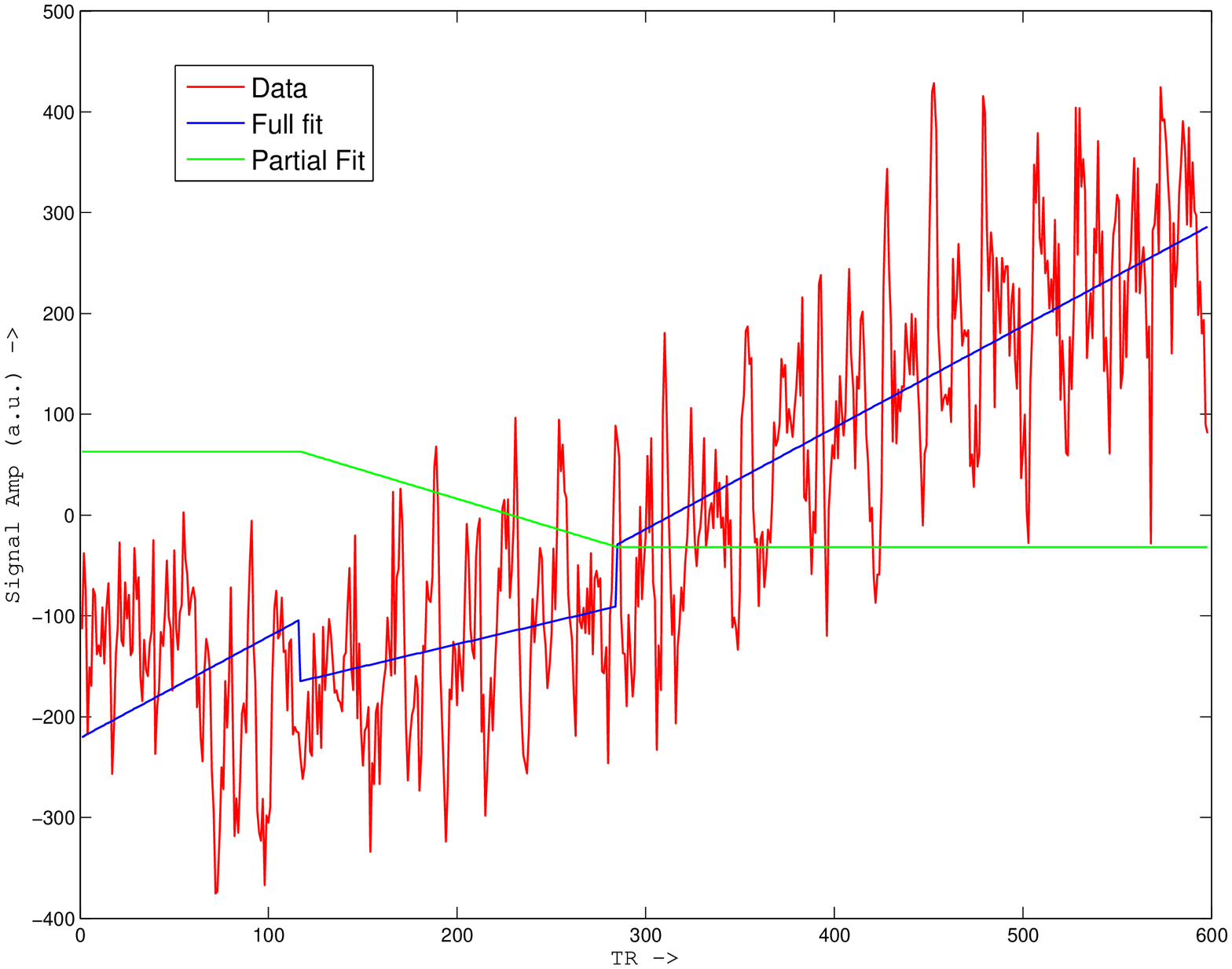}
}

\end{tabular}
\caption{fMRI data extracted from sample voxels 3 and 4. A GLM analysis was run using the optimal DM and a DM containing the 2 EVs shown in \ref{fig1} and the temporal derivative of the 1st EV (i.e., the infusion ev). Figures show the full model fit and partial fit corresponding to the contrast of interest. Figures (a) and (c) are for the optimal DM while (b) and (d) are for the "temporal derivative" based DM.}
\label{casefmriB}
\end{figure}

\begin{figure}[htbp]
\centering
\begin{tabular}{cc}

\subfigure[]
{
\hspace{-2cm}
\label{fmri9}
\includegraphics[width = 75mm, angle = -90]{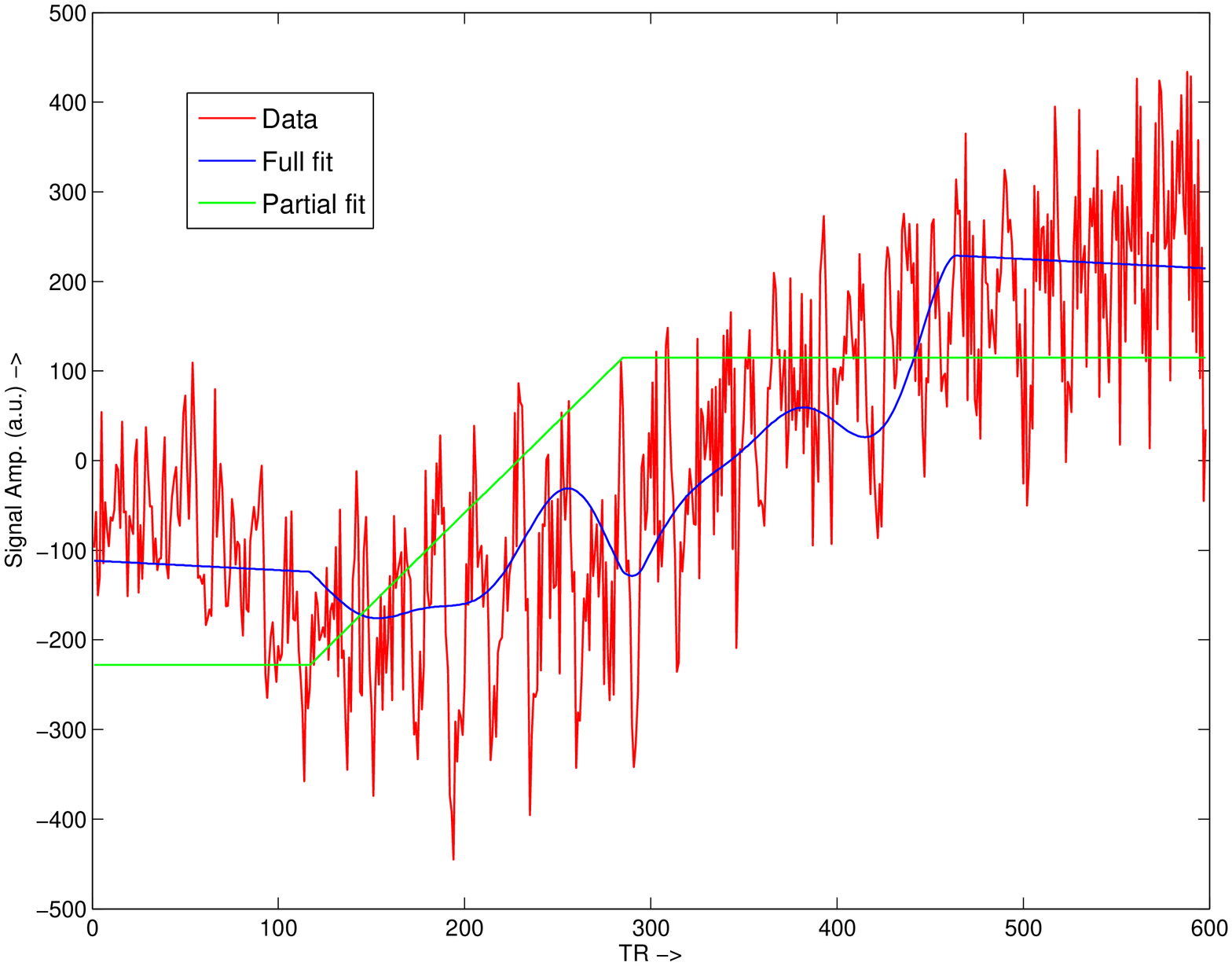}
}

&

\subfigure[]
{
\hspace{-1cm}
\label{fmri10}
\includegraphics[width = 75mm, angle = -90]{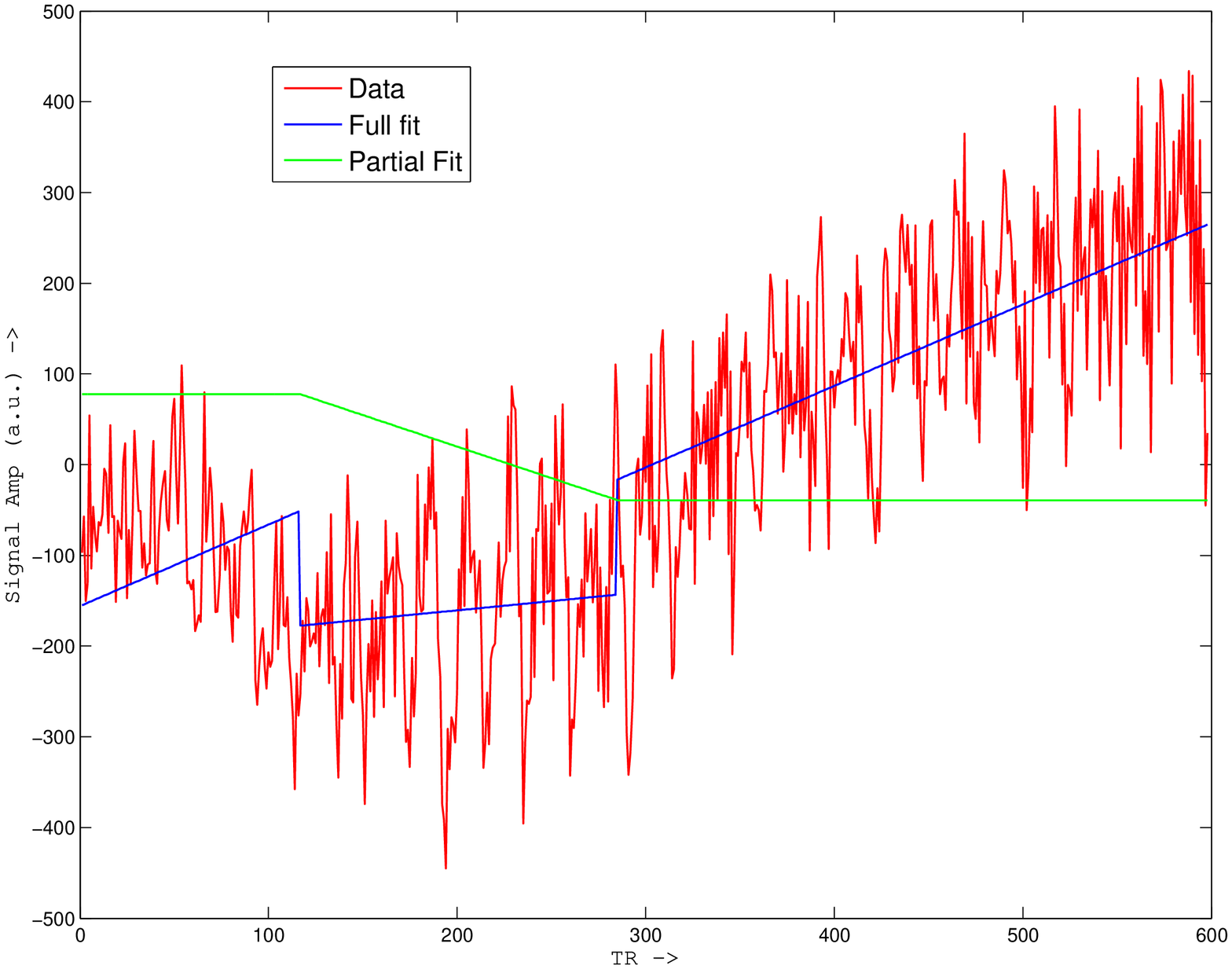}
}

\\

\subfigure[]
{
\hspace{-2cm}
\label{fmri11}
\includegraphics[width = 75mm, angle = -90]{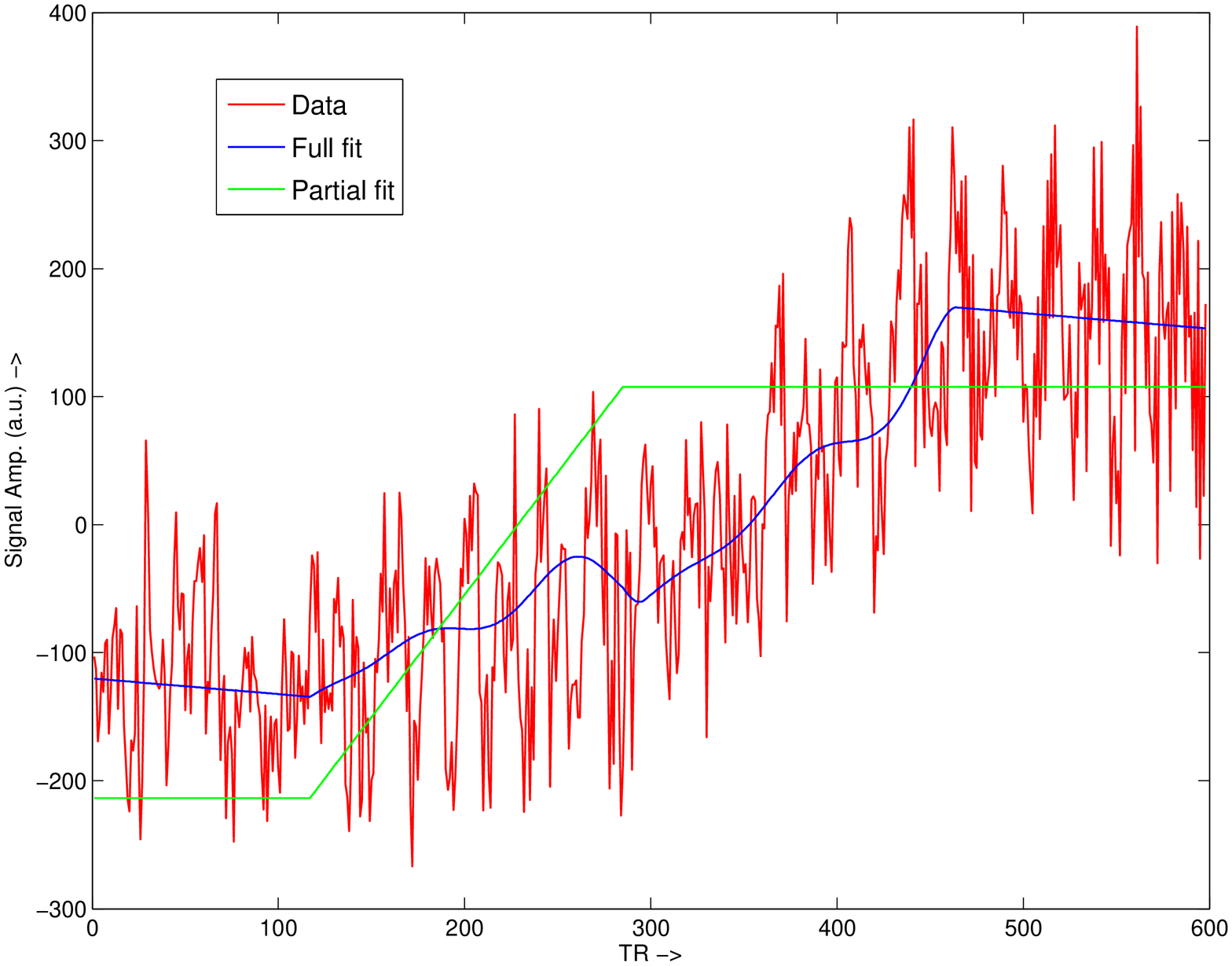}
}

&

\subfigure[]
{
\hspace{-1cm}
\label{fmri12}
\includegraphics[width = 75mm, angle = -90]{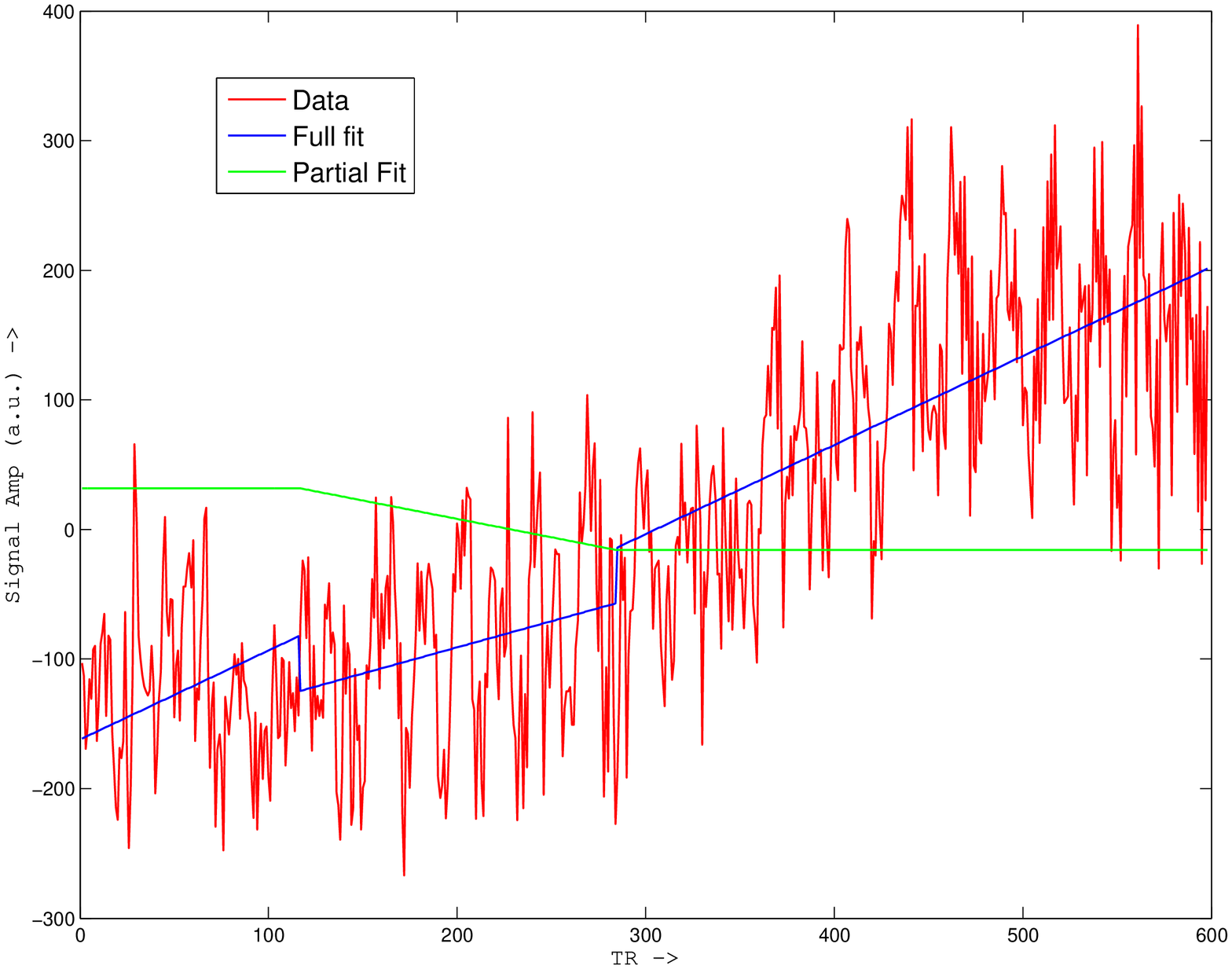}
}

\end{tabular}
\caption{fMRI data extracted from sample voxels 5 and 6. A GLM analysis was run using the optimal DM and a DM containing the 2 EVs shown in \ref{fig1} and the temporal derivative of the 1st EV (i.e., the infusion ev). Figures show the full model fit and partial fit corresponding to the contrast of interest. Figures (a) and (c) are for the optimal DM while (b) and (d) are for the "temporal derivative" based DM.}
\label{casefmriC}
\end{figure}


\section{Discussion}

The objectives of this paper were the development of a theoretical framework and a numerical algorithm to enable optimization of design matrices used in fMRI analyses. This optimization framework (SMART) allows a user to optimize an objective function capturing the bias-variance decomposition over a set of potential design matrices. Within this very general framework it is possible to specify weights measuring the expected frequency of occurrence of various design matrices as well as preferentially control for bias or variance, if desired. The sample space for optimization is controlled via constraints on the columns of the optimal design matrix and the associated contrast. 

We validated our numerical algorithm by comparing it with a more sophisticated optimization solver in two validation tests. We proposed a strategy for choosing the number of columns in the optimal design matrix as well as strategies for automatic selection of initial point for the optimization algorithm and for choosing local bias-variance
weightings $\phi_i$ based on user-specified objectives.

We then illustrated the application of the technique by considering 6 case studies. Our aim in these examples was to illustrate the degree of control that a user has in terms of controlling the optimal solution based on user specifications. The first 4 examples illustrated the variation of various control parameters in the algorithmic framework as applied to a phMRI study. Example 5 illustrated the application of the proposed technique to a block design case study. Example 6 addressed the important issue of locally variable HRF functions in fMRI data. The goal in example 6 was to come up with a set of HRF-modeling functions to capture a range of HRF shapes while maintaining optimality with respect to bias and variance of the primary contrast capturing the response amplitude of the underlying EV.

\begin{figure}[htbp]
\begin{center}
\includegraphics[width = 100mm, angle = -90]{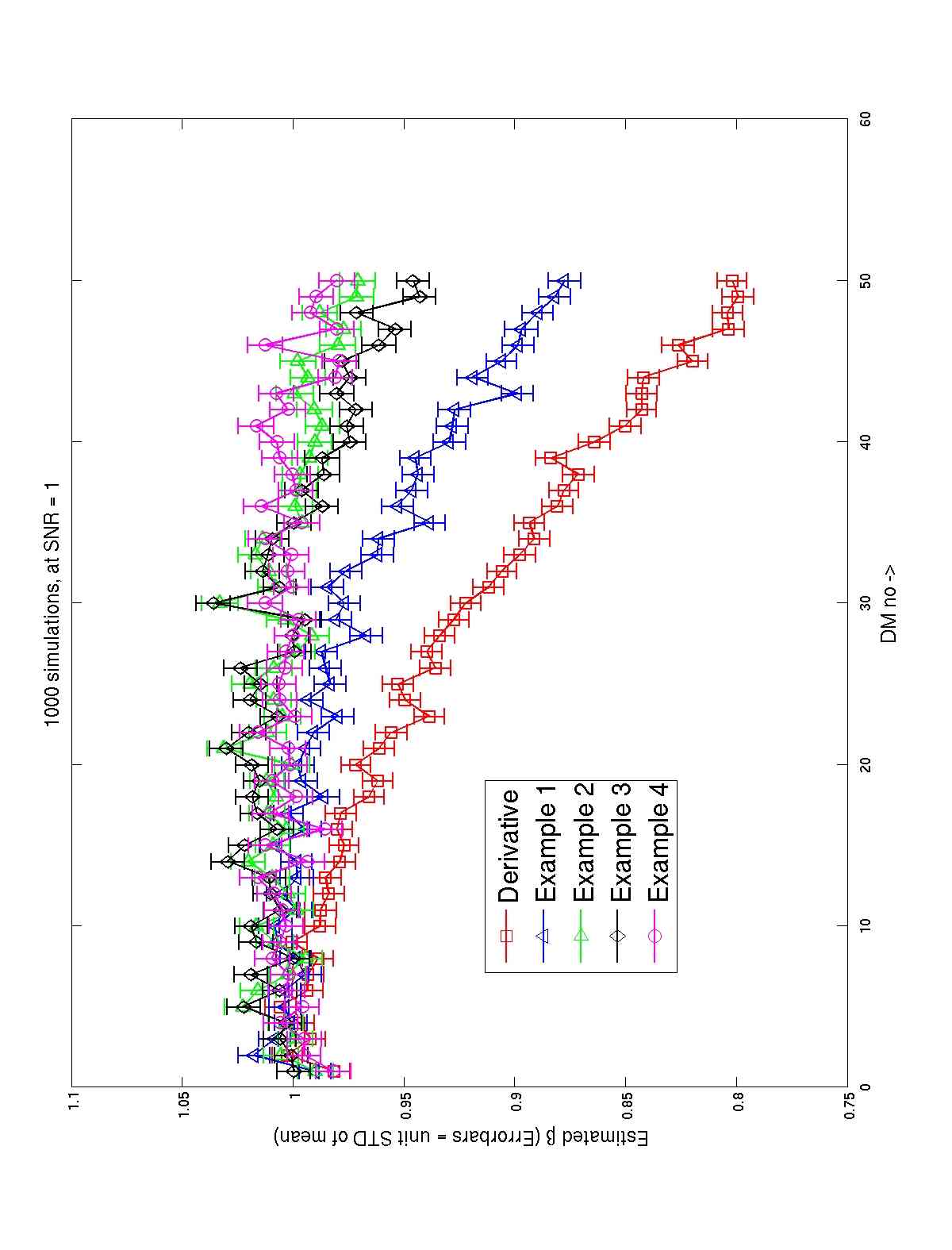}
\caption{For each design matrix (DM) $X_i$ from Examples 1, 2, 3 and 4 entered into optimization, 1000 simulated data-sets were generated at SNR $\frac{\beta_i}{\sigma_i}$. A GLM analysis was run on each of these data-sets using the optimized DMs for Example 1, Example 2, Example 3, Example 4 as well as a DM $X_D$ such that the first two columns of $X_D$ are the same as the first two columns in $\hat{Z}$ from Example 1 but the 3rd column is the temporal derivative of $X_D(:,1)$. Figure above shows a summary errorbar plot showing $\hat{E}(c_Z^T \hat{\gamma})$ over 1000 simulations for data generated from each DM and analyzed via GLM using optimized DM's from Example 1, 2, 3, 4 as well as using $X_D$. The errorbars represent the standard deviation of $\hat{E}(c_Z^T \hat{\gamma})$ so that bias can be statistically compared across the four cases.}
\label{figderivbias}
\end{center}
\end{figure}

We examined how the optimized design matrix derived using SMART compared to an alternative in which a temporal derivative is added as an additional EV to capture variation in signal onset. Figure \ref{figderivbias} compares the bias in parameter estimate at SNR = 1 over 1000 simulations for the temporal derivative approach as well as for the four SMART phMRI design matrices. A significant reduction in bias is observed when using the optimized design matrices in comparison to the temporal derivative approach, with Examples 2-4 performing best.

Finally, we applied the technique to a real infusion phMRI dataset. First, an "optimal"  design matrix was derived using SMART. Next, we examined the bias-variance properties and generated ROC curves for the estimated design matrix. Finally,  a GLM analysis  was performed on the phMRI dataset using this design matrix. It was found that this design matrix achieves very high ($>98\%$) sensitivity and specificity of signal detection over a range of signal variations in the data.


The use of a set of basis functions has been recommended before for capturing variability in fMRI signal shapes. For example, as relates to the capture of locally varying HRF shapes the work by FMRIB analysis group on FLOBS \cite{FLOBS:2004} is notable in that it constrains the basis set to have sensible HRF shapes. These approaches allow one to test the hypothesis about the presence or absence of signal by using F-statistics. However, limitations of such approaches include (1) the inability to measure the amplitude of the HRF signals and (2) The inability to combine amplitude measures from single-subject analyses into a group level analysis. Because the signal amplitude is never measured in these basis function approaches it is also not controlled for bias and variance. In contrast, the HRF-capturing functions developed in Example 6 illustrate how variable signal shapes can be captured using a single contrast while optimizing the bias and variance of the resulting signal amplitude estimates as per user defined objectives.

In this paper we considered an inverse problem. Instead of proposing a statistical estimator and studying its bias/variance properties, we define an objective function that captures the bias-variance decomposition over the set of potential signal shapes in the data and then explicitly optimize this objective function for both a design matrix and a contrast. This results in an optimized design matrix and contrast that automatically captures the amplitude of signals of interest. The resulting PE values can be easily carried over for a group level analysis.

\section{Conclusion}
We developed a theoretical framework (SMART) to enable calculation of optimal design matrices for fMRI analyses that simultaneously enables detection of multiple signal responses as well as controlling for bias and variance in the GLM estimation. This is achieved by optimizing for the contrasts and basis functions of the design matrix to minimize the bias-variance decomposition over a set of design matrices capturing anticipated variability in the data. Although the technique was developed with motivation from fMRI, its development is quite general and appears to be applicable to a variety of problems from many other disciplines.


\section{Appendix}
\subsection{Derivation of gradient equations}\label{gradient_derivation}

Consider the matrix $A \in \mathbf{R}^{n \times m}$, square matrices $B, C \in \mathbf{R}^{n \times n}$ and vectors $x \in \mathbf{R}^n$ and $y, z \in \mathbf{R}^m$. Recall that the matrix derivative with respect to matrix $A$ of a scalar quantity $f$ is denoted by $\frac{\partial f}{\partial A}$. The $ij$th element of $\frac{\partial f}{\partial A}$ 
is defined as:
\begin{equation}
\left[ \frac{\partial f}{\partial A} \right]_{ij} = \frac{\partial f}{\partial a_{ij} }
\end{equation}
where $a_{ij}$ is the $ij$th element of $A$.

Below we note some basic identities:

\begin{equation}\label{a1}
\left[ \frac{\partial}{\partial A} (x^T A y) \right]_{ij}= \left[ \frac{\partial}{\partial A} (y^T A^T x) \right]_{ij} = \left[ x y^T \right]_{ij}
\end{equation}

\begin{equation}\label{a2}
\left[ \frac{\partial}{\partial B} (x^T B^{-1} x) \right]_{ij} = \left[ -B^{-T} x x^T B^{-T} \right]_{ij}
\end{equation}

\begin{equation}\label{a21}
\left[ \frac{\partial}{\partial A} (y^T (A^T A)^{-1} z) \right]_{ij} = \left[ -A (A^T A)^{-1} (z y^T + y z^T) (A^T A)^{-1} \right]_{ij}
\end{equation}

\begin{equation}\label{a3}
\left[\frac{\partial}{\partial B} \mbox{trace}(B C) \right]_{ij} = \left[ C^T \right]_{ij}
\end{equation}

\begin{equation}\label{a3}
\left[\frac{\partial}{\partial B} \mbox{trace}(B^T C) \right]_{ij} = \left[ C \right]_{ij}
\end{equation}

\begin{equation}\label{a4}
\mbox{ trace }(B C) = \mbox{ trace } (C B)
\end{equation}

Recall, equation \ref{objfn}, the objective function under consideration:
\begin{eqnarray}
G_\phi(Z, c_Z) = c_Z^T (Z^T Z)^{-1} c_Z \left[  \sum_{i = 1}^m 2 \phi_i w_i + tr \left( P_Z H \Phi_V \Sigma H^T \right) \right] \nonumber \\  + \, c_Z^T (Z^T Z)^{-1} Z^T H \Phi_B H^T Z (Z^T Z)^{-1} c_Z - 2 c_Z^T (Z^T Z)^{-1} Z^T H \Phi_B \ell + \sum_{i = 1}^m w_i (2 - 2 \phi_i) (c_{X_i}^T \beta_i/\sigma_i)^2
\end{eqnarray}

The matrix gradient with respect to $Z$ can be written as:
\begin{eqnarray}\label{a5}
\frac{\partial}{\partial Z} G_\phi(Z, c_Z) = \frac{\partial}{\partial Z} \left\{ c_Z^T (Z^T Z)^{-1} c_Z  \right\}  \left[  \sum_{i = 1}^m 2 \phi_i w_i + tr \left( P_Z H \Phi_V \Sigma H^T \right) \right]  \nonumber \\ 
+  c_Z^T (Z^T Z)^{-1} c_Z \frac{\partial}{\partial Z} \left\{  \left[  \sum_{i = 1}^m 2 \phi_i w_i + tr \left( P_Z H \Phi_V \Sigma H^T \right) \right] \right\} \nonumber \\
+  \frac{\partial}{\partial Z} \left\{ c_Z^T (Z^T Z)^{-1} Z^T H \Phi_B H^T Z (Z^T Z)^{-1} c_Z  \right\} \nonumber \\
+ \frac{\partial}{\partial Z} \left\{    - 2 c_Z^T (Z^T Z)^{-1} Z^T H \Phi_B \ell \right\} \nonumber \\
+  \frac{\partial}{\partial Z}  \left\{   \sum_{i = 1}^m w_i (2 - 2 \phi_i) (c_{X_i}^T \beta_i/\sigma_i)^2 \right\}
\end{eqnarray}

For the sake of presentation clarity, we define the following smaller terms in equation \ref{a5}:
\begin{eqnarray}
\mbox{ \textbf{Term 1} } = \frac{\partial}{\partial Z} \left\{ c_Z^T (Z^T Z)^{-1} c_Z  \right\} \nonumber \\
\mbox{ \textbf{Term 2}} = \frac{\partial}{\partial Z } \left\{  \left[  tr \left( P_Z H \Phi_V \Sigma H^T \right) \right] \right\} \nonumber \\
\mbox{ \textbf{Term 3}} = \frac{\partial}{\partial Z } \left\{ c_Z^T (Z^T Z)^{-1} Z^T H \Phi_B H^T Z (Z^T Z)^{-1} c_Z  \right\} \nonumber \\
\mbox{ \textbf{Term 4}} =  \frac{\partial}{\partial Z } \left\{    - 2 c_Z^T (Z^T Z)^{-1} Z^T H \Phi_B \ell \right\} \nonumber \\
\mbox{ \textbf{Term 5}} =  \frac{\partial}{\partial Z}  \left\{   \sum_{i = 1}^m w_i (2 - 2 \phi_i) (c_{X_i}^T \beta_i/\sigma_i)^2 \right\}
\end{eqnarray}

\subsection{Term 1}

From \ref{a21} we have:
\begin{equation} \label{a7}
\frac{\partial}{\partial Z} \left\{ c_Z^T (Z^T Z)^{-1} c_Z  \right\} = -Z (Z^T Z)^{-1} (2 c_Z c_Z^T) (Z^T Z)^{-1}
\end{equation}

\subsection{Term 2}
\begin{eqnarray} \label{a8}
\frac{\partial}{\partial Z_{ij} } \left\{  \left[  tr \left( P_Z H \Phi_V \Sigma H^T \right) \right] \right\} = tr \left(  - \frac{\partial Z}{\partial Z_{ij}} (Z^T Z)^{-1} Z^T H \Phi_V \Sigma H^T \right) \nonumber \\ 
+  tr \left(  - Z \frac{\partial}{\partial Z_{ij}} \left\{ (Z^T Z)^{-1} \right\}  Z^T H \Phi_V \Sigma H^T \right) +  tr \left(  - Z (Z^T Z)^{-1} \frac{\partial Z^T}{\partial Z_{ij}} H \Phi_V \Sigma H^T \right)
\end{eqnarray}

Again,
\begin{equation} \label{a9}
tr \left(  - \frac{\partial Z}{\partial Z_{ij}} (Z^T Z)^{-1} Z^T H \Phi_V \Sigma H^T \right) = - \left[ H \Sigma^T \Phi_V^T H^T Z (Z^T Z)^{-1} \right]_{ij}
\end{equation}
and
\begin{equation} \label{a10}
 tr \left(  - Z (Z^T Z)^{-1} \frac{\partial Z^T}{\partial Z_{ij}} H \Phi_V \Sigma H^T \right) =  tr \left(  -  \frac{\partial Z^T}{\partial Z_{ij}} H \Phi_V \Sigma H^T Z (Z^T Z)^{-1} \right) = - \left[ H \Phi_V \Sigma H^T Z (Z^T Z)^{-1} \right]_{ij}
\end{equation}

Also,
\begin{eqnarray} \label{a11}
tr \left(  - Z \frac{\partial}{\partial Z_{ij}} \left\{ (Z^T Z)^{-1} \right\}  Z^T H \Phi_V \Sigma H^T \right) \nonumber \\
= tr \left( Z (Z^T Z)^{-1} \frac{\partial Z^T}{\partial Z_{ij}} Z (Z^T Z)^{-1} Z^T H \Phi_V \Sigma H^T + Z (Z^T Z)^{-1} Z^T \frac{\partial Z}{\partial Z_{ij}} (Z^T Z)^{-1} Z^T H \Phi_V \Sigma H^T \right) \nonumber \\
= tr \left(  \frac{\partial Z^T}{\partial Z_{ij}} Z (Z^T Z)^{-1} Z^T H \Phi_V \Sigma H^T Z (Z^T Z)^{-1} +  \frac{\partial Z}{\partial Z_{ij}} (Z^T Z)^{-1} Z^T H \Phi_V \Sigma H^T Z (Z^T Z)^{-1} Z^T \right) \nonumber \\
=  \left[ Z (Z^T Z)^{-1} Z^T H \Phi_V \Sigma H^T Z (Z^T Z)^{-1} +  Z (Z^T Z)^{-1} Z^T H \Sigma^T \Phi_V^T H^T Z (Z^T Z)^{-1} \right]_{ij}
\end{eqnarray}

Combining \ref{a8}, \ref{a9}, \ref{a10} and \ref{a11} we get:
\begin{eqnarray}\label{a12}
\frac{\partial}{\partial Z} \left\{  \left[  tr \left( P_Z H \Phi_V \Sigma H^T \right) \right] \right\} = - H \Sigma^T \Phi_V^T H^T Z (Z^T Z)^{-1} + Z (Z^T Z)^{-1} Z^T H \Phi_V \Sigma H^T Z (Z^T Z)^{-1} \nonumber \\ 
+  Z (Z^T Z)^{-1} Z^T H \Sigma^T \Phi_V^T H^T Z (Z^T Z)^{-1}  - H \Phi_V \Sigma H^T Z (Z^T Z)^{-1} \nonumber \\
= -(H \Phi_V \Sigma H^T + H \Sigma^T \Phi_V^T H^T) Z (Z^T Z)^{-1} + Z (Z^T Z)^{-1} Z^T (H \Phi_V \Sigma H^T + H \Sigma^T \Phi_V^T H^T) Z (Z^T Z)^{-1} \nonumber \\
= - P_Z (H \Phi_V \Sigma H^T + H \Sigma^T \Phi_V^T H^T) Z (Z^T Z)^{-1}
\end{eqnarray}

\subsection{Term 3}
\begin{eqnarray}\label{a13}
\frac{\partial}{\partial Z_{ij} } \left\{ c_Z^T (Z^T Z)^{-1} Z^T H \Phi_B H^T Z (Z^T Z)^{-1} c_Z  \right\} \nonumber \\
= c_Z^T  \frac{\partial}{\partial Z_{ij} } \left\{ (Z^T Z)^{-1} \right\} Z^T H \Phi_B H^T Z (Z^T Z)^{-1} c_Z  \nonumber \\ 
+ c_Z^T (Z^T Z)^{-1} \frac{\partial}{\partial Z_{ij} } \left\{ Z^T \right\} H \Phi_B H^T Z (Z^T Z)^{-1} c_Z  \nonumber \\ 
+ c_Z^T (Z^T Z)^{-1} Z^T H \Phi_B H^T  \frac{\partial}{\partial Z_{ij} } \left\{ Z \right\} (Z^T Z)^{-1} c_Z   \nonumber \\ 
+ c_Z^T (Z^T Z)^{-1} Z^T H \Phi_B H^T Z  \frac{\partial}{\partial Z_{ij} } \left\{ (Z^T Z)^{-1} \right\} c_Z  
\end{eqnarray}

Applying \ref{a2} and \ref{a21} repeatedly to each of the above terms we get:
\begin{eqnarray}\label{a14}
\frac{\partial}{\partial Z} \left\{ c_Z^T (Z^T Z)^{-1} Z^T H \Phi_B H^T Z (Z^T Z)^{-1} c_Z  \right\} \nonumber \\
= -Z (Z^T Z)^{-1} (   Z^T H \Phi_B H^T Z (Z^T Z)^{-1} c_Z c_Z^T + c_Z c_Z^T (Z^T Z)^{-1} Z^T H \Phi_B^T H^T Z  ) (Z^T Z)^{-1} \nonumber \\
+  H \Phi_B H^T Z (Z^T Z)^{-1} c_Z c_Z^T (Z^T Z)^{-1} \nonumber \\
+ H \Phi_B^T H^T Z (Z^T Z)^{-1} c_Z c_Z^T (Z^T Z)^{-1} \nonumber \\
- Z (Z^T Z)^{-1} ( c_Z  c_Z^T (Z^T Z)^{-1} Z^T H \Phi_B H^T Z + Z^T H \Phi_B^T H^T Z (Z^T Z)^{-1} c_Z c_Z^T ) (Z^T Z)^{-1}
\end{eqnarray}
Since $\Phi_B$ is diagonal, this can be simplified to:
\begin{eqnarray}\label{a15}
\frac{\partial}{\partial Z } \left\{ c_Z^T (Z^T Z)^{-1} Z^T H \Phi_B H^T Z (Z^T Z)^{-1} c_Z  \right\} \nonumber \\
= -2 Z (Z^T Z)^{-1} (   Z^T H \Phi_B H^T Z (Z^T Z)^{-1} c_Z c_Z^T + c_Z c_Z^T (Z^T Z)^{-1} Z^T H \Phi_B^T H^T Z  ) (Z^T Z)^{-1} \nonumber \\
+  2 H \Phi_B H^T Z (Z^T Z)^{-1} c_Z c_Z^T (Z^T Z)^{-1} \nonumber \\
\end{eqnarray}

\subsection{Term 4}

\begin{eqnarray}\label{a16}
 \frac{\partial}{\partial Z_{ij} } \left\{    - 2 c_Z^T (Z^T Z)^{-1} Z^T H \Phi_B \ell \right\} \nonumber \\
 = \left\{    - 2 c_Z^T  \frac{\partial}{\partial Z_{ij} }  \left\{ (Z^T Z)^{-1} \right\} Z^T H \Phi_B \ell \right\} \nonumber \\
 +  \left\{    - 2 c_Z^T (Z^T Z)^{-1} \frac{\partial}{\partial Z_{ij} }  \left\{ Z^T \right\} H \Phi_B \ell \right\} \nonumber \\
\end{eqnarray}

Application of \ref{a2} and \ref{a21} gives:
\begin{eqnarray}\label{a17}
 \frac{\partial}{\partial Z } \left\{    - 2 c_Z^T (Z^T Z)^{-1} Z^T H \Phi_B \ell \right\} \nonumber \\
= (-2) (-Z) (Z^T Z)^{-1} ( Z^T H \Phi_B \ell c_Z^T + c_Z \ell^T \Phi_B^T H^T Z ) (Z^T Z)^{-1} \nonumber \\
+ (-2) H \Phi_B \ell c_Z^T (Z^T Z)^{-1}
\end{eqnarray}

\subsection{Term 5}
Term 5 is a constant w.r.t $Z$ and $c_Z$ and so the gradient w.r.t $Z$ and $c_Z$ is 0.

\subsection{Combining terms}

Combining \ref{a5}, \ref{a7}, \ref{a12}, \ref{a15} and \ref{a17} gives:
\begin{eqnarray}
\frac{\partial}{\partial Z} G_\phi(Z, c_Z) = \left[   -Z (Z^T Z)^{-1} (2 c_Z c_Z^T) (Z^T Z)^{-1} \right]  \left[  \sum_{i = 1}^m 2 \phi_i w_i + tr \left( P_Z H \Phi_V \Sigma H^T \right) \right]  \nonumber \\ 
+  c_Z^T (Z^T Z)^{-1} c_Z \left[  - P_Z (H \Phi_V \Sigma H^T + H \Sigma^T \Phi_V^T H^T) Z (Z^T Z)^{-1} \right]\nonumber \\
+  \left[  -2 Z (Z^T Z)^{-1} (   Z^T H \Phi_B H^T Z (Z^T Z)^{-1} c_Z c_Z^T + c_Z c_Z^T (Z^T Z)^{-1} Z^T H \Phi_B^T H^T Z  ) (Z^T Z)^{-1} \right] \nonumber \\ +  \left[ 2 H \Phi_B H^T Z (Z^T Z)^{-1} c_Z c_Z^T (Z^T Z)^{-1}  \right] \\
+ \left[ (-2) (-Z) (Z^T Z)^{-1} ( Z^T H \Phi_B \ell c_Z^T + c_Z \ell^T \Phi_B^T H^T Z ) (Z^T Z)^{-1} 
+ (-2) H \Phi_B \ell c_Z^T (Z^T Z)^{-1} \right]
\end{eqnarray}

Noting that $\Phi_B$, $\Phi_V$ and $\Sigma$ are diagonal and given the fact that diagonal matrices commute, rearrangement gives:
\begin{eqnarray}\label{gradobjfnder}
\frac{\partial}{\partial Z} G_\phi(Z, c_Z) = \left[   -Z (Z^T Z)^{-1} (2 c_Z c_Z^T) (Z^T Z)^{-1} \right]  \left[  \sum_{i = 1}^m 2 \phi_i w_i + tr \left( P_Z H \Phi_V \Sigma H^T \right) \right]  \nonumber \\ 
-  2 \left( c_Z^T (Z^T Z)^{-1} c_Z \right)  \left[  P_Z ( H \Phi_V \Sigma H^T ) Z (Z^T Z)^{-1} \right]\nonumber \\
-2 Z (Z^T Z)^{-1} \left[ c_Z c_Z^T (Z^T Z)^{-1} Z^T H \Phi_B H^T Z \right] (Z^T Z)^{-1} \nonumber \\
-2 Z (Z^T Z)^{-1} \left[ Z^T H \Phi_B H^T Z (Z^T Z)^{-1} c_Z c_Z^T \right] (Z^T Z)^{-1} \nonumber \\
+  \left[ 2 H \Phi_B H^T Z (Z^T Z)^{-1} c_Z c_Z^T (Z^T Z)^{-1}  \right] \nonumber \\
+ 2 Z (Z^T Z)^{-1} ( Z^T H \Phi_B \ell c_Z^T) (Z^T Z)^{-1} \nonumber \\
+ 2 Z (Z^T Z)^{-1} ( c_Z \ell^T \Phi_B^T H^T Z ) (Z^T Z)^{-1} \nonumber \\
-2 H \Phi_B \ell c_Z^T (Z^T Z)^{-1} 
\end{eqnarray}

Thus the gradient with respect to $c_Z$ is easily computed as:
\begin{eqnarray}\label{gradobjfn2der}
\frac{\partial}{\partial c_Z} G_\phi(Z, c_Z) = 2 (Z^T Z)^{-1} c_Z \left[  \sum_{i = 1}^m 2 \phi_i w_i + tr \left( P_Z H \Phi_V \Sigma H^T \right) \right] \nonumber \\ 
+ 2 (Z^T Z)^{-1} Z^T H \Phi_B H^T Z (Z^T Z)^{-1} c_Z - 2 (Z^T Z)^{-1} Z^T H \Phi_B \ell 
\end{eqnarray}

Equations \ref{gradobjfnder} and \ref{gradobjfn2der} are the same as \ref{gradobjfn} and \ref{gradobjfn2} respectively.

\subsection{Exact solver details}\label{optimization_solver}

Our optimization algorithm solves the general problem:
\begin{eqnarray}\label{a1}
\mbox{ min }_x f(x) & \\
\mbox{s.t. } c_i(x) = 0, & i = 1,2,\ldots, m \\
\mbox{s.t. } g_j(x) \ge 0, & j = 1,2, \ldots L
\end{eqnarray}

where $x \in R^n$.

We convert the inequality constraints into equality constraints via slack variables as follows:
\begin{eqnarray} \label{a2}
g_j(x) - s_j = 0 & \\ 
s_j \ge 0, & j = 1,2, \ldots L
\end{eqnarray}

Thus the optimization problem becomes:
\begin{eqnarray}\label{a3}
\mbox{ min } f(x) & \\
\mbox{s.t. } c_i(x) = 0, & i = 1,2,\ldots, m \\
\mbox{s.t. } g_j(x) - s_j = 0, &  j = 1,2, \ldots L \\
s_j \ge 0 & 
\end{eqnarray}

This problem is now an equality constrained problem where the inequalities have been replaced by the bound constraints on the slack variables. Thus it suffices to consider equality constrained problems with bounds on independent variables as follows:
\begin{eqnarray}\label{a4}
\mbox{ min } f(x) & \\
\mbox{s.t. } c_i(x) = 0, & i = 1,2,\ldots, m \\
\mbox{s.t. } l_i \le x_i \le u_i, & i = 1,2, \ldots n
\end{eqnarray}
where $x \in R^n$.

Our code uses a trust region based augmented lagrangian approach to solve these bound constrained problems following closely the LANCELOT software package \cite{LANCELOT:1992}, \cite{Conn:1991}. The augmented lagrangian function for the above problem is defined as:
\begin{equation}\label{a5}
\mathcal{L}(x, \lambda, \mu) = f(x) - \sum_{i = 1}^m \lambda_i c_i(x) + \frac{\mu}{2} \sum_{i = 1}^m c_i(x)^2
\end{equation}

At each outer iteration $k$, given current values of $\lambda^k$ and $\mu_k$ we solve the subproblem:

\begin{eqnarray}\label{a6}
\mbox{ min } \mathcal{L}(x, \lambda^k, \mu_k) \\
\mbox{ s.t. } l_i \le x_i \le u_i
\end{eqnarray}

If $P$ is the projection operator defined as
\begin{equation}\label{a7}
[P(z, l , u)]_i = \left \{
\begin{array}{ccc}
l_i & \mbox{ if } & z_i \le l_i \\
z_i & \mbox{ if }  & l_i \le z_i \le u_i \\
u_i & \mbox{ if } & z_i \ge u_i 
\end{array}
\right.
\end{equation}
then the Karush-Kuhn-Tucker (KKT) optimality condition for \ref{a6} is given as \cite{Conn:1991}:
\begin{equation}\label{a8}
x - P(x - \nabla_x \mathcal{L}(x, \lambda^k, \mu_k),  l , u) = 0
\end{equation}

The outer iteration code is given in Framework 1. Note that the penalty parameter $\mu_k$  is updated based on a feasibility monitoring strategy that allows for a decrease in $\mu_k$ if sufficient accuracy is not achieved in solving the subproblem \ref{a6}.

\begin{algorithm}
\begin{algorithmic}[1]
\REQUIRE Initial point $x_{init}$,  $\lambda^0$, $\mu_0$, $\theta^h \in (1, \infty)$, $\theta_l \in (0, 1)$
\STATE Choose tolerances $\eta^*_{con}$ and $\eta^*_{grad}$. The default is $\eta^*_{con} = \eta^*_{grad} = 1e{-6}$. .
\STATE $\mu = \mu_0$, $\eta_{con} = 1/\mu_0^{0.1}$, $\eta_{grad} = 1/\mu_0$
\FOR{$k = 0,1,2,\ldots$}

\STATE $found = 0$
\WHILE{ $found \neq 1$}
\STATE Try to find $x_k$ such that \\ $||x_k - P(x_k - \nabla_x \mathcal{L}(x_k, \lambda^k, \mu_k),  l , u)||_\infty \le \eta_{grad}$ via F2 using starting point as $x_{k - 1}$.
	\IF{ above step is completed successfully }
	 	\STATE Set $found = 1$
	\ELSE 
		\STATE $\lambda^{k + 1} = \lambda_k$
		\STATE $\mu_{k + 1} = \theta_l \mu_k$
		\STATE $\eta_{con} = 1/\mu_k^{0.1}$
		\STATE $\eta_{grad} = 1/\mu_k$	
	\ENDIF
		
\ENDWHILE

		\IF{ $ || c(x_k) ||_\infty \le \eta_{con}$ }
			\IF{ $ || c(x_k) ||_\infty \le \eta^*_{con}$ and \\ $||x_k - P(x_k - \nabla_x \mathcal{L}(x_k, \lambda^k, 0),  l , u)||_\infty \le \eta^*_{grad}$ }
				\STATE Stop and return current solution $x_k$.
			\ENDIF
			\STATE $\lambda^{k + 1} = \lambda_k - \mu_k c(x_k)$
			\STATE $\mu_{k + 1} = \mu_k$
			\STATE $\eta_{con} = \eta_{con} / \mu_{k+1}^{0.9} $
			\STATE $\eta_{grad} = \eta_{grad} / \mu_{k + 1} $
		\ELSE
			\STATE $\lambda^{k + 1} = \lambda_k$
			\STATE $\mu_{k + 1} = \theta_h \mu_k$
			\STATE $\eta_{con} = 1/\mu_k^{0.1}$
			\STATE $\eta_{grad} = 1/\mu_k$		
		\ENDIF

\ENDFOR
\end{algorithmic}
\caption{F1: Outer Iteration}
\end{algorithm}

At each inner iteration we form a quadratic approximation to the augmented lagrangian and approximately solve the inequality constrained quadratic sub-problem:

\begin{eqnarray}\label{a9}
\mbox{ min }_p \,\,\, \frac{1}{2} p^T \nabla^2_{xx} \mathcal{L}(x, \lambda, \mu) p + \nabla_x \mathcal{L}(x, \lambda, \mu)^T p \\
\mbox{ s.t. } l_i \le x_i \le u_i \\
\mbox{ s.t. } || p ||_\infty \le \Delta 
\end{eqnarray}

The inner iteration code uses non-linear gradient projection \cite{byrd95limited} followed by Newton-CG-Steihaug conjugate gradient iterations \cite{Steihaug:1983}. Quasi-Newton updates are performed using either SR1 \cite{Conn:SR1} (recommended for non-convex functions) or BFGS \cite{Broyden:1970} (recommended for convex functions). For very large problems, we switch to the limited memory variants  \cite{Nocedal:LM} of these quasi-Newton approximations. The algorithm details are given in Framework 2.  The trust region update code is based on a standard progress monitoring strategy \cite{Nocedal:book} and is given in Framework 3.

\begin{algorithm}
\begin{algorithmic}[1]
\REQUIRE  $j_{max}$, $\eta_{grad}$, $\Delta$, $l$, $u$, $\lambda^k$, $\mu_k$, $\eta \in (0,1)$, $flag$
\STATE $found = 0$
\STATE $x = x_{k - 1}$, $j = 1$
\STATE Compute, $g = \nabla_x \mathcal{L}(x, \lambda^k, \mu_k)$
\STATE Estimate $B =  \nabla^2_{xx} \mathcal{L}(x, \lambda^k, \mu_k)$ using BFGS, SR1 or limited memory BFGS, limited memory SR1 quasi Newton Updates.
\WHILE{$found \neq 1$ and $j \le j_{max}$}
\STATE Calculate the Cauchy point $p_c$ for problem:
\begin{eqnarray}
\mbox{ min }_p \,\,\, \frac{1}{2} p^T B p + g^T p \\
\mbox{ s.t. } l - x\le p \le u - x \\
 \mbox{ s.t. } || p ||_\infty \le \Delta 
\end{eqnarray}
using non-linear gradient projection and calculate the current active set $\mathcal{A}$. Let $e_i$ be the unit vector with $1$ at position $i$ and zeros elsewhere. If $i_1, i_2, \ldots i_q \notin \mathcal{A}$ then
let $\tilde{Q} = [e_{i_1}, e_{i_2},\ldots, e_{i_q}]$.
\STATE $\tilde{g} = \tilde{Q}^T ( g + B \, p_c)$ and $\tilde{B} = \tilde{Q}^T B \tilde{Q}$
\STATE Compute the approximate solution $\hat{v}$ to the problem
\begin{eqnarray}
\mbox{ min }_v \,\,\, \frac{1}{2} v^T \tilde{B} v + \tilde{g}^T v \\
\mbox{ s.t. } l - x \le p_c + \tilde{Q} v \le u - x \\
 \mbox{ s.t. } || p_c + \tilde{Q} v ||_\infty \le \Delta 
\end{eqnarray}
using truncated conjugate gradient iteration (Newton-CG, Steihaug). If $flag = 1$ use preconditioned Newton-CG using the inexact-modified Cholesky factorization.
\STATE Compute $\hat{p} = p_c + \tilde{Q} \hat{v}$
\STATE Calculate $\delta_{\mathcal{L}} =  \mathcal{L}(x) - \mathcal{L}(x + \hat{p}) $, $\delta_m = 0.5 \hat{p}^T B \hat{p} + g^T \hat{p}$ and $\rho = \delta_{\mathcal{L}} / \delta_m$

\IF{ $\rho > \eta$ }
\STATE	$x = x + \hat{p}$
\ENDIF

\STATE Compute new trust region radius $\Delta$ using Framework F3.

\STATE Compute, $g = \nabla_x \mathcal{L}(x, \lambda^k, \mu_k)$ if $\rho > \eta$ holds otherwise use the previous value.
\STATE Estimate $B =  \nabla^2_{xx} \mathcal{L}(x, \lambda^k, \mu_k)$ using BFGS, SR1 or limited memory BFGS, limited memory SR1 quasi Newton Updates. Do the update even if $\rho < \eta$.

\IF{$||x - P(x - \nabla_x \mathcal{L}(x, \lambda^k, \mu_k),  l , u)||_\infty \le \eta_{grad}$}
\STATE $found = 1$
\ENDIF

\STATE $j = j + 1$

\ENDWHILE
\end{algorithmic}
\caption{F2: Inner Iteration}
\end{algorithm}

\begin{algorithm}
\begin{algorithmic}[1]
\REQUIRE $\rho$, $\hat{p}$, $\Delta$
\IF{$\rho > 0.75$}
	\IF{$ ||\hat{p}||_\infty \le 0.8 \Delta $}
		\STATE $\Delta = \Delta$
	\ELSE
		\STATE $\Delta = 2 \Delta$
	\ENDIF
\ENDIF
\IF{$0.1 \le \rho \le 0.75$}
	\STATE $\Delta = \Delta$
\ELSE
	\STATE $\Delta = 0.5 \Delta$
\ENDIF
\RETURN $\Delta$
\end{algorithmic}
\caption{F3:Trust Region Update}
\end{algorithm}

\subsection{Half-cosine parameterization of HRF}\label{hrf_parameterization}
Here we describe the equations used to generate plausible HRF shapes via a 5-parameter half-cosine parameterization. 

\begin{figure}[htbp]
\begin{center}
\includegraphics[width = 100mm, angle = -90]{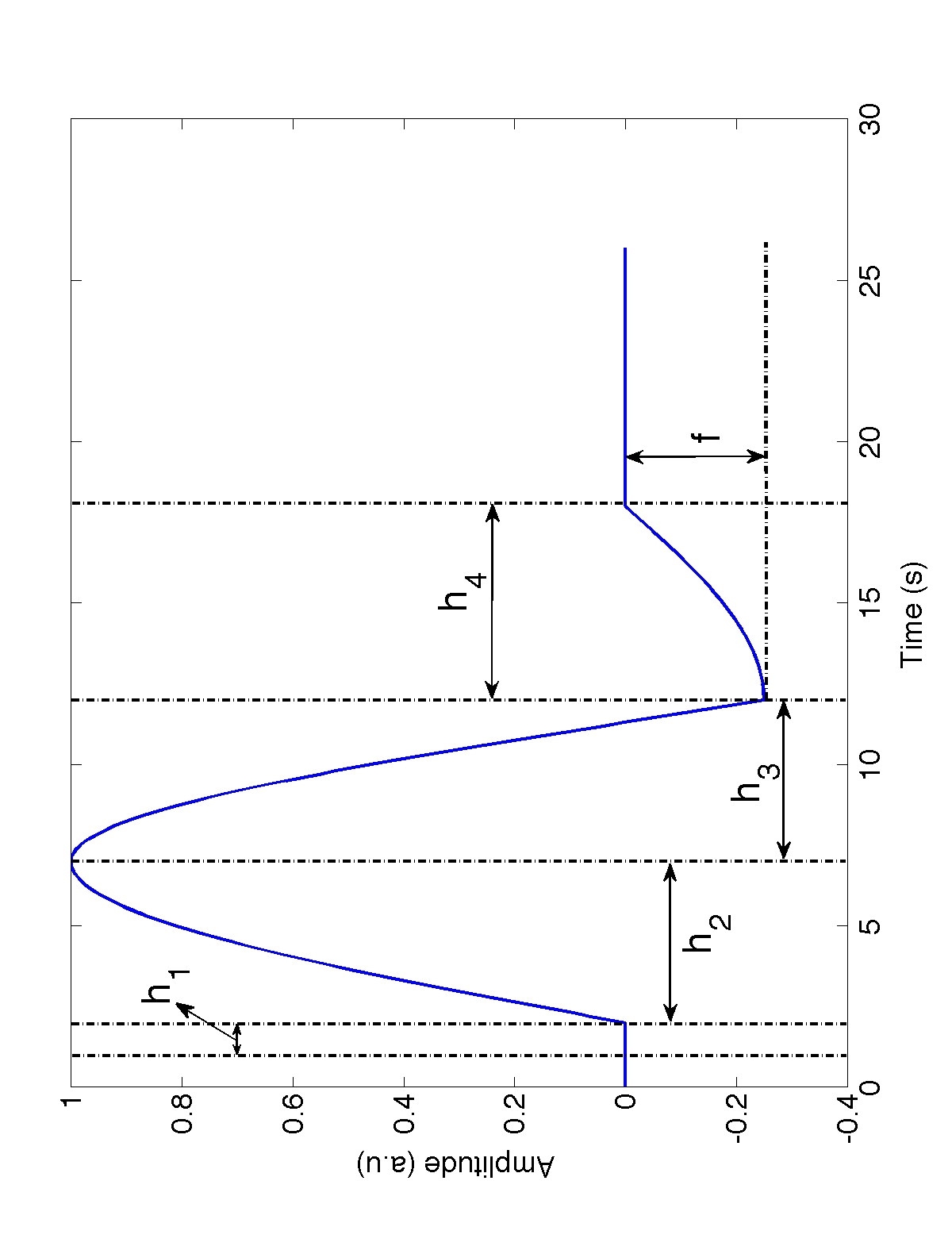}
\caption{5 parameter half-cosine parameterization of Haemodynamic Response Function (HRF). $h_1$ controls the time to first rise, $h_2$ controls the time to peak, $h_3$ controls the time to undershoot maximum and $h_4$ controls the time to return to baseline. The amplitude of undershoot is controlled
by the parameter $f$ while the height of rise from baseline is fixed at $1$. See \ref{HRFeqns} for the exact equation. }
\label{samplehrf}
\end{center}
\end{figure}

This parameterization is defined as follows:
\begin{eqnarray}\label{HRFeqns}
HRF(t) = \left\{ 
\begin{array}{ccc}
0 & \mbox{ if } & 0 \le t \le h_1\\
\cos \left[ \frac{\pi}{2} - \frac{\pi}{2 h_2} (t - h_1) \right] & \mbox{ if } & h_1 < t \le h_1 + h_2 \\
\cos \left[ \left( \frac{\pi}{2 h_3} + \frac{ \sin^{-1}(f_2)}{h_3} \right) (h_1 + h_2 - t) \right] & \mbox{ if } & h_1 + h_2 < t \le h_1 + h_2 + h_3 \\
f \cos \left[ \pi - \frac{\pi}{2 h_4} (t - h_1 - h_2 - h_3) \right] & \mbox{ if } & h_1 + h_2 + h_3 < t \le h_1 + h_2 + h_3 + h_4 \\
0 & \mbox{ if } & t > h_1 + h_2 + h_3 + h_4
\end{array}
\right.
\end{eqnarray}

\newpage
\bibliographystyle{plain}
\bibliography{bibliography_optglm_arxiv.bib}

\end{document}